\documentclass[12pt]{iopart}
\expandafter\let\csname equation*\endcsname\relax
\expandafter\let\csname endequation*\endcsname\relax
\usepackage{amsmath}
\usepackage[british]{babel}
\usepackage{newtxtext}
\usepackage[T1]{fontenc}
\usepackage{ae,aecompl}
\usepackage[utf8]{inputenc}
\usepackage{lmodern}
\usepackage{graphicx}
\usepackage{color}
\usepackage{changepage}
\usepackage[perpage]{footmisc}
\usepackage{iopams}
\usepackage[normalem]{ulem}
\usepackage{xtab,booktabs}

\begin{document}

\title[Nucleosynthesis in CCSNe of 11.2 and 17.0~M$_{\odot}$ progenitors]{Nucleosynthesis in 2D Core-Collapse Supernovae of 11.2 and 17.0~M$_{\odot}$ Progenitors: Implications for Mo and Ru Production}

\author{
M Eichler$^{1}$,
K Nakamura$^{2,3}$,
T Takiwaki$^{4}$,
T Kuroda$^{1,5}$,
K Kotake$^{3}$,
M Hempel$^{5}$,
R Cabez\'{o}n$^{6}$,
M Liebend\"orfer$^{5}$
and F-K Thielemann$^{5,7}$}

\address{$^{1}$ Institut f\"ur Kernphysik, Technische Universit\"at Darmstadt, D-64289 Darmstadt, Germany}
\ead{marius.eichler@physik.tu-darmstadt.de}
\address{$^{2}$ Faculty of Science and Engineering, Waseda University, Tokyo 169-8555, Japan}
\address{$^{3}$ Department of Applied Physics, Fukuoka University, Fukuoka 814-0180, Japan}
\address{$^{4}$ National Astronomical Observatory of Japan, Tokyo 181-8588, Japan}
\address{$^{5}$ Department of Physics, University of Basel, Klingelbergstr. 82, CH-4056 Basel, Switzerland}
\address{$^{6}$ Scientific Computing Core, sciCORE, University of Basel, Klingelbergstr. 61, CH-4056 Basel, Switzerland}
\address{$^{7}$ GSI Helmholtzzentrum f\"ur Schwerionenforschung GmbH, D-64291 Darmstadt, Germany}

\begin{abstract}
Core-collapse supernovae are the first polluters of heavy elements in the galactic history. As such, it is important to study the nuclear compositions of their ejecta, and understand their dependence on the progenitor structure (e.g., mass, compactness, metallicity). Here, we present a detailed nucleosynthesis study based on two long-term, two-dimensional core-collapse supernova simulations of a 11.2~M$_{\odot}$ and a 17.0~M$_{\odot}$ star. We find that in both models nuclei well beyond the iron group (up to $Z\approx44$) can be produced, and discuss in detail also the nucleosynthesis of the p-nuclei $^{92,94}$Mo and $^{96,98}$Ru. While we observe the production of $^{92}$Mo and $^{94}$Mo in slightly neutron-rich conditions in both simulations, $^{96,98}$Ru can only be produced efficiently via the $\nu$p-process. Furthermore, the production of Ru in the $\nu$p-process heavily depends on the presence of very proton-rich material in the ejecta. This disentanglement of production mechanisms has interesting consequences when comparing to the abundance ratios between these isotopes in the solar system and in presolar grains.
\end{abstract}

\noindent{\it Keywords\/}: nucleosynthesis, supernovae: general, stars: abundances
\vspace{0.5cm}

\noindent Accepted for publication in: \jpg \\
\maketitle

\section{Introduction}
\label{sec:introduction}
Core-collapse supernovae (CCSNe) mark the end of the lives of massive stars (M~$>~8$~M$_{\odot}$), and the nuclear compositions of their ejecta can be constrained by direct and indirect observations.
For instance, CCSNe appear earlier in the galactic history than the other main contributors to explosive nucleosynthesis (i.e., Type Ia SNe and neutron star mergers), which means that atmospheres of extremely metal-poor (EMP) stars carry the signatures of only one or a few CCSNe and therefore give us valuable insights about the typical compositions of their ejecta. The most metal-poor star currently known is SMSS0313-6708 \cite{keller2014}, with an upper limit to its metallicity around~$\mathrm{[Fe/H]}=-7$, and it is thought that its composition is the result of the ejecta of a single supernova mixing with the interstellar medium \cite{keller2014,ishigaki2014,bessell2015,nordlander2017}. In addition, light curves of CCSNe are powered by the radioactive decay of copiously produced unstable isotopes, such as $^{56}$Ni, $^{57}$Ni, and $^{44}$Ti, which means that the yields for these isotopes can be well constrained. In particular, from SN~1987A reliable numbers are available for the ejected masses of $^{56,57}$Ni and $^{44}$Ti. Seitenzahl \etal \cite{seitenzahl2014} give values for SN~1987A of M$\left(^{56}\mathrm{Ni}\right)~=~\left(7.1~\pm~0.3\right)~\times~10^{-2}$~M$_{\odot}$, M$\left(^{57}\mathrm{Ni}\right)~=~\left(4.1~\pm~1.8\right)~\times~10^{-3}$~M$_{\odot}$, and M$\left(^{44}\mathrm{Ti}\right)~=~\left(0.55~\pm~0.17\right)~\times~10^{-4}$~M$_{\odot}$. Similar yields of $^{56}$Ni have also been determined in other CCSN light curves (see, e.g., \cite{smartt2009}), while the amount of ejected $^{44}$Ti in SN~1987A is still uncertain and can vary by a factor of a few if an asymmetric explosion is considered (e.g., \cite{boggs2015}).

The most abundant nuclei in compositions of CCSN ejecta are $\alpha$-elements, followed by iron group nuclei, which are produced as a result of explosive burning of the lighter $\alpha$-elements (He, C, O, Mg, Si), but other nucleosynthesis processes are possible. Behind the supernova shock wave a baryonic wind, driven by neutrino absorptions, ejects additional matter. This \textit{neutrino-driven wind} has long been considered a good candidate for the r-process, but detailed studies have shown that the required conditions cannot be met \cite{huedepohl2010,fischer2010,arconesthielemann2013}, although slightly neutron-rich conditions could be achieved \cite{martinez2012,roberts2012,wanajo2011}. In these conditions ($0.4 \lesssim Y_e \lesssim 0.5$, depending also on the entropy) a weak r-process could operate \cite{arconesbliss2014}, characterized by lower neutron densities than the main r-process, while electron fractions above~0.5 yield favourable conditions for the $\nu$p-process \cite{froehlich2006,pruet2006,wanajo2006}. In both cases, however, nuclear species beyond the iron group elements can be synthesized, adding a specific nucleosynthesis signature to the nuclear composition of the supernova ejecta. One of the remaining puzzles in this context is the cosmic origin of the solar $^{92,94}$Mo and $^{96,98}$Ru abundances, which have historically been attributed to the p-process \cite{arnould2003}. The sensitivities of the isotopic abundance ratios to parameters of the neutrino-driven wind and nuclear reaction rates have been explored in Arcones \& Bliss \cite{arconesbliss2014} and Bliss \etal \cite{bliss2016}, while Travaglio \etal \cite{travaglio2015} showed that they can also be produced by the $\gamma$-process in type Ia SNe. However, all attempts to reproduce the solar isotopic abundance ratios between $^{92}$Mo and $^{94}$Mo have failed so far \cite{fisker2009}. Before the $\nu$p-process was known, Hoffman \etal \cite{hoffman1996} reported on the synthesis of $^{92}$Mo in slightly neutron-rich conditions ($Y_e~\approx~0.485$) in the neutrino-driven wind.
The detection of these Mo and Ru isotopes in presolar grains of type SiC X, which are thought to carry the isotopic signature of CCSNe, suggests that they at least co-originate from CCSNe.

Large studies of CCSNe comprising many different progenitor masses do not only serve the purpose of investigating the explosion characteristics of the different progenitors, but also help understand the explosive nucleosynthesis of $\alpha$-elements beyond Si and their distribution in the galaxy through space and time.
Ideally, these studies would be performed with three-dimensional CCSN simulations including full neutrino transport and general relativity, since the true explosion properties of CCSNe depend on three-dimensional, large-scale effects that play a role in and before the explosion, such as rotation, convection, or the standing accretion shock instability (e.g., \cite{janka2016,hix2016,mueller2016,takiwaki2016} and references therein). In one- or two-dimensional simulations, these effects can only be treated approximately. However, due to the extremely high computational cost of realistic 3D simulations, explodability and/or nucleosynthesis studies of many progenitor masses in three-dimensional CCSN simulations are not affordable at this point.

Therefore, past CCSN nucleosynthesis research had to rely on spherically symmetric models, using artificial energy deposition to trigger CCSN explosions, such as the piston or thermal bomb methods (e.g., \cite{woosleyweaver1995,thielemann1996,Umeda2002,hegerwoosley2010,thielemannproc2011,limongichieffi2012}).
Recently, it has become feasible to perform CCSN simulations for the full progenitor mass range in spherical symmetry and using models that go beyond these traditional methods, albeit still relying on artificial energy deposition by neutrinos. Perego \etal \cite{perego2015} have introduced a method called \textsc{PUSH} where the energy reservoir of the heavy-flavour neutrinos ($\nu_{\mu}$, $\overline{\nu}_{\mu}$, $\nu_{\tau}$, and $\overline{\nu}_{\tau}$) is tapped in regions heated by the electron neutrinos. The advantage of this approach is that it contains a detailed neutrino transport and that it does not directly affect the electron fraction $Y_e$, thus making it a suitable tool for nucleosynthesis studies. \textsc{PUSH} relies on a small number of free parameters which have been calibrated on the observables of SN~1987A. Another approach was presented in \cite{ugliano2012,sukhbold2016}, where the PNS in the centre of the SN was excised and the explosion was prescribed by a parametrized neutrino luminosity at the inner boundary of the computational domain. The parameters of this method (such as core-neutrino emission) were also calibrated on observables of SN~1987A as well as the Crab SN.

Larger samples of axisymmetric CCSN simulations are also becoming available \cite{takiwaki2014,mueller2015,dolence2015,oconnorcouch2015,bruenn2016,summa2016,pan2016}, with the most complete set so far reported by Nakamura \etal \cite{nakamura2015} (101 progenitors of solar metallicity, 247 ultra metal-poor and 30 zero-metal progenitors), using the Newtonian code \textsc{ZEUS-MP} \cite{hayes2006,iwakami2008} together with the \textit{isotropic diffusion source approximation} (IDSA \cite{liebendoerfer2009}) scheme for electron and anti-electron neutrino transport, and complemented by a leakage scheme for the heavy-flavour neutrinos.

In this work, we perform detailed full-network nucleosynthesis calculations for two long-term CCSN simulations presented in \cite{nakamura2015}, using a 11.2~M$_{\odot}$ and a 17.0~M$_{\odot}$ progenitor with solar metallicities from the Woosley \etal \cite{woosley2002} series. We find unusually low Ni masses for both models, which can be attributed to the axisymmetric nature of the simulations and to late-time accretion. With the low Ni yields, our CCSNe would classify as faint SNe (e.g., \cite{turatto1998,zampieri2003,pastorello2004,pastorello2006,nomoto2006,nomoto2013}). Furthermore we find large yields of elements beyond the iron group up to the relatively neutron-deficient stable isotopes of Mo and Ru. For other recent results of nucleosynthesis yields in two-dimensional CCSNe see also \cite{wanajo2017}. The uncertainties that arise with the commonly used post-processing approach have been discussed in detail by Harris \etal \cite{harris2017} on the example of four 2D CCSN models, while the impact of the dimensionality of an electron-capture SN simulation on the nuclear yields has been explored by Wanajo \etal \cite{wanajo2011}.

This paper is structured as follows: section~\ref{sec:method} describes the model used for the simulations and the nucleosynthesis procedure, the results are presented in section~\ref{sec:results} and discussed in section~\ref{sec:discussion}, and section~\ref{sec:conclusions} summarizes our findings.


\section{Method}
\label{sec:method}
We will describe the two models that we used for the nucleosynthesis calculations in section~\ref{sec:simulations}. Our considerations about a suitable ejection criterion are presented in section~\ref{sec:ejeccrits} and in appendix~\ref{app:ejeccrit}, and the setup of the nuclear network is the subject of section~\ref{sec:nucleomethod}.


\subsection{Simulations}
\label{sec:simulations}
Our CCSN models are based on long-term simulations of axisymmetric neutrino-driven explosions. 
The progenitor stars are non-rotating, solar metallicity models from \cite{woosley2002} with zero-age main sequence masses of $11.2$ and $17.0$~M$_{\odot}$.
These progenitor models retain their hydrogen envelope and are classified as red supergiant stars. They are therefore expected to explode as Type II supernovae.

The numerical code we employ for the core-collapse simulations is essentially the same as described in \cite{nakamura2015}, except for some minor revisions. 
The spatial domain covers 100,000~km from the center with a resolution of $n_r \times n_\theta = 1008 \times 128$ zones.
In order to evolve the electron fraction, we solve the spectral transport of electron and anti-electron neutrinos, using the isotropic diffusion source approximation (IDSA \cite{liebendoerfer2009}). All base-line weak interactions (such as charged current reactions), neutral current reactions including iso-energetic neutrino scattering on nuclei and nucleon, are included with the rates taken from Ref.~\cite{bruenn1985}.
Regarding heavy-lepton neutrinos, we employ a leakage scheme (see \cite{takiwaki2014}). 
In the high-density regime, we use the equation of state (EOS) of Lattimer \& Swesty \cite{lattimerswesty1991} with a nuclear incompressibility of $K~=~220$~MeV. 
At low densities, we employ an EOS accounting for photons, electrons, positrons, and the ideal gas contribution of silicon.
During our long-term CCSN simulation, we follow explosive nucleosynthesis by solving a simple nuclear network consisting of 13 alpha-nuclei to take into account the energy feedback into the hydrodynamic evolution. The details of the $\alpha$-network are discussed in Ref.~\cite{nakamura2014}.

Both of our CCSN models successfully revive their shock. The axis-symmetric nature of the simulations causes a preference of the outflow towards the polar directions (see figures~1~and~2). This behaviour is usually not observed in 3D simulations of regular CCSNe, since they do not have an imposed symmetry. Explosion times and energies are also systematically different, as is shown e.g., in Refs~\cite{janka2016,hix2016,takiwaki2014,lentz2015,melson2015}.
When the shock reaches the outer boundary of the computational domain, these models have diagnostic explosion energies of $1.9 \times 10^{50}$ and $1.2 \times 10^{51}$ ergs.

From the simulation we extract 129'024 tracer particles for our post-processing approach. Their initial distribution is determined by the $n_r \times n_\theta = 1008 \times 128$ zones, equidistant in the angular direction with an angular spacing of 1.41 degrees, and logarithmic in the radial direction with an outer radius of 100,000~km. Due to the density profile of the progenitor, this means that the tracer particles represent different masses depending on their initial distance to the center.
The computed domain includes $1.9 \,$M$_{\odot}$ ($4.1 \,$M$_{\odot}$) of the $11.2 ~$M$_{\odot}$ ($17.0 ~ $M$_{\odot}$) progenitor, corresponding to a region up to the bottom of the helium layer.


\subsection{Ejected Matter}
\label{sec:ejeccrits}
In a first step, it is important to know which particles are successfully ejected, i.e., become gravitationally unbound in the course of a SN. One key quantity for this discussion is the specific total energy (without rest masses) at the end of the simulation, which is defined as the sum of (specific) thermal energy, kinetic energy, and gravitational potential $\phi$:

\begin{equation}
e_{\rm tot} = e_{\rm th} + \frac{v^2}{2} + \phi \hspace{2px}.
\end{equation}

\begin{figure*}
\includegraphics[width=0.51\textwidth]{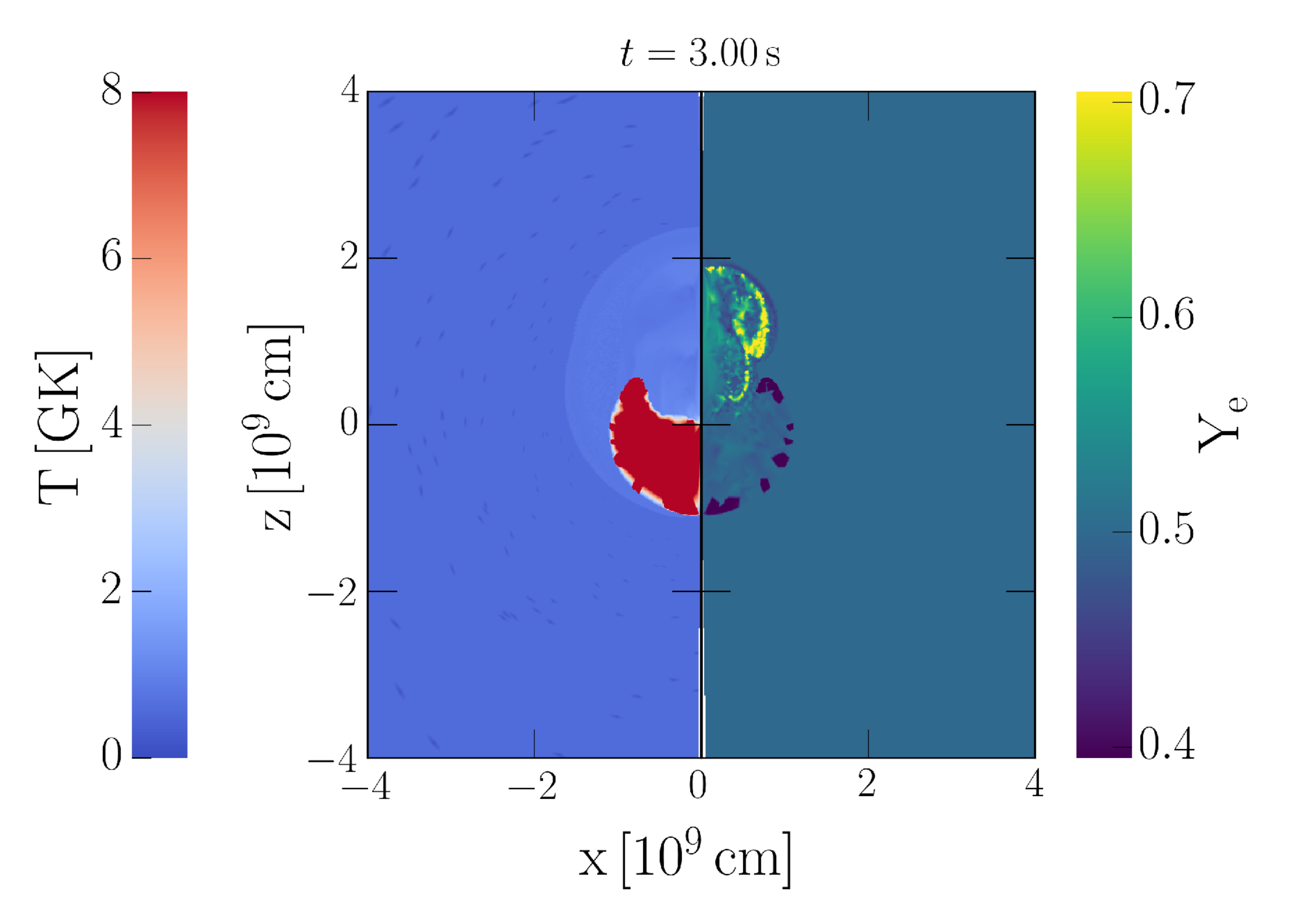}
\includegraphics[width=0.51\textwidth]{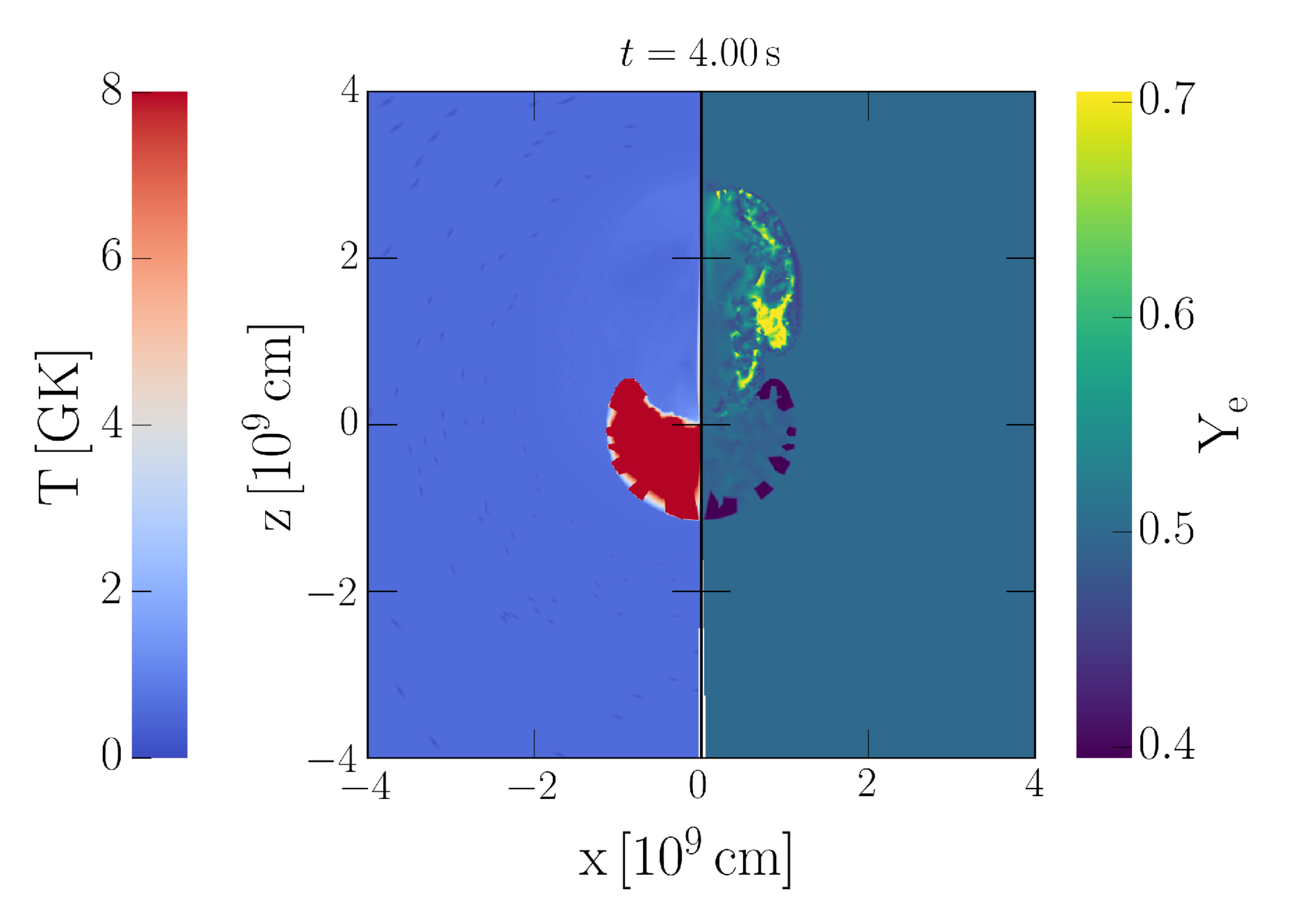}\\
\includegraphics[width=0.51\textwidth]{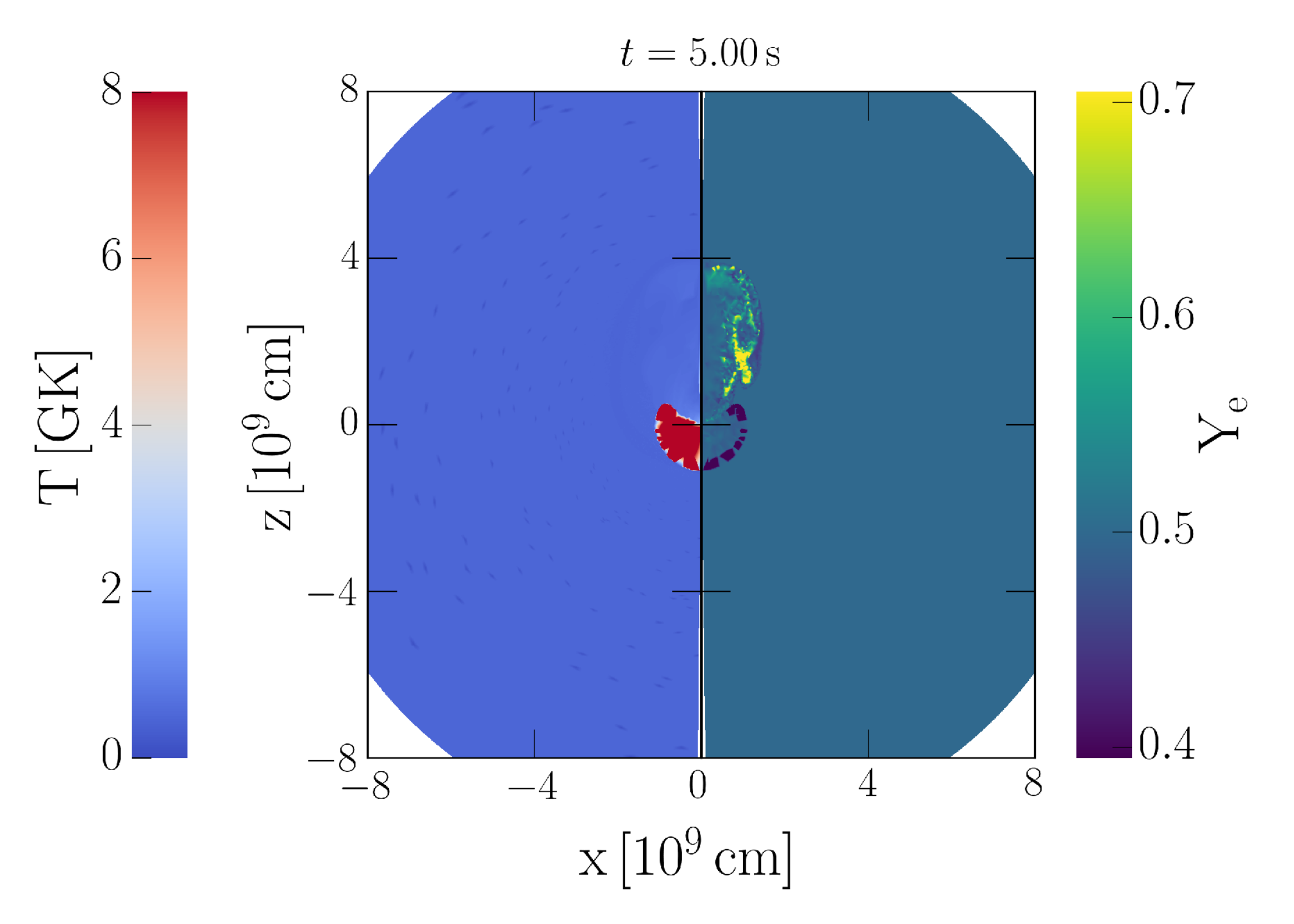}
\includegraphics[width=0.51\textwidth]{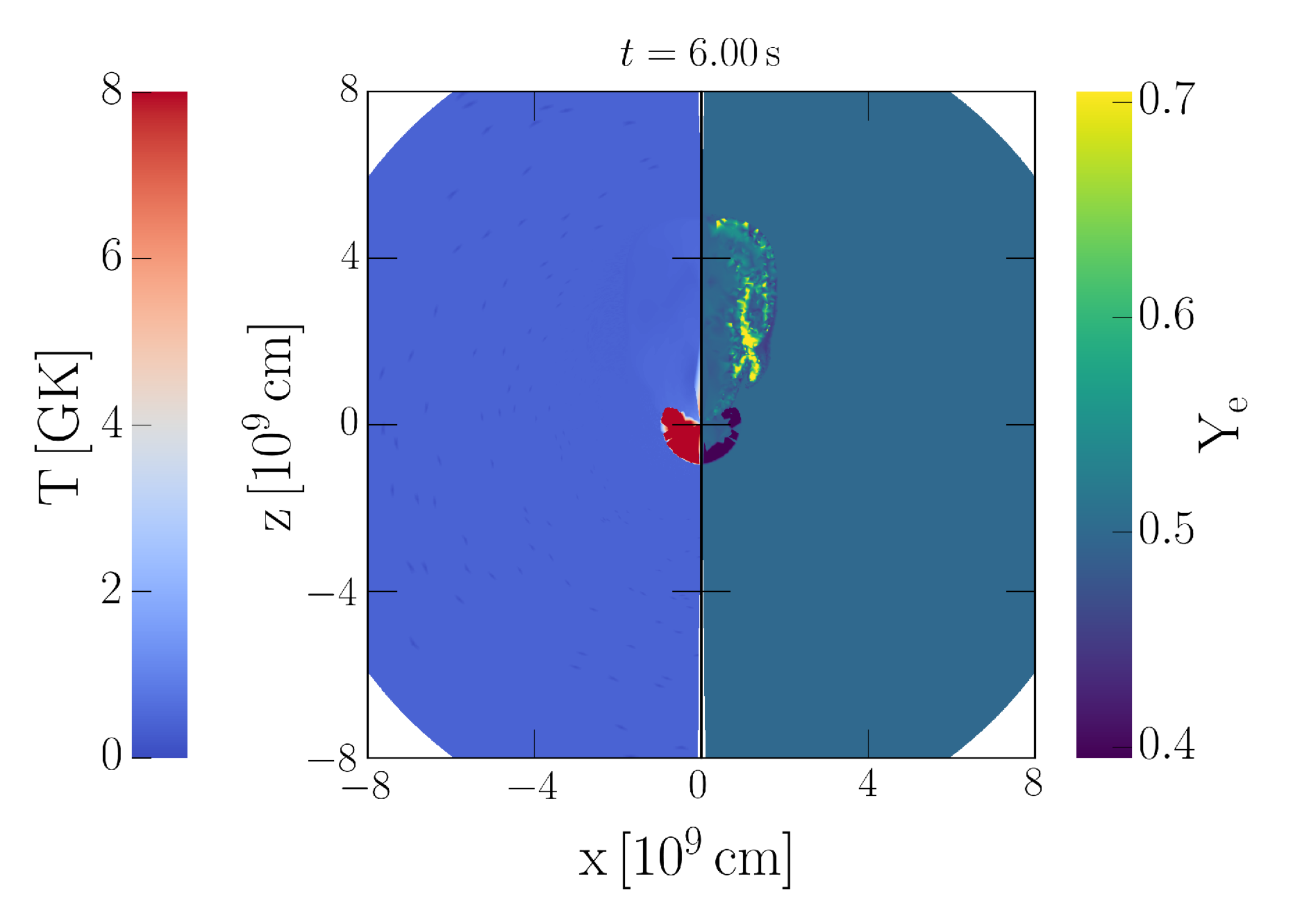}\\
\includegraphics[width=0.51\textwidth]{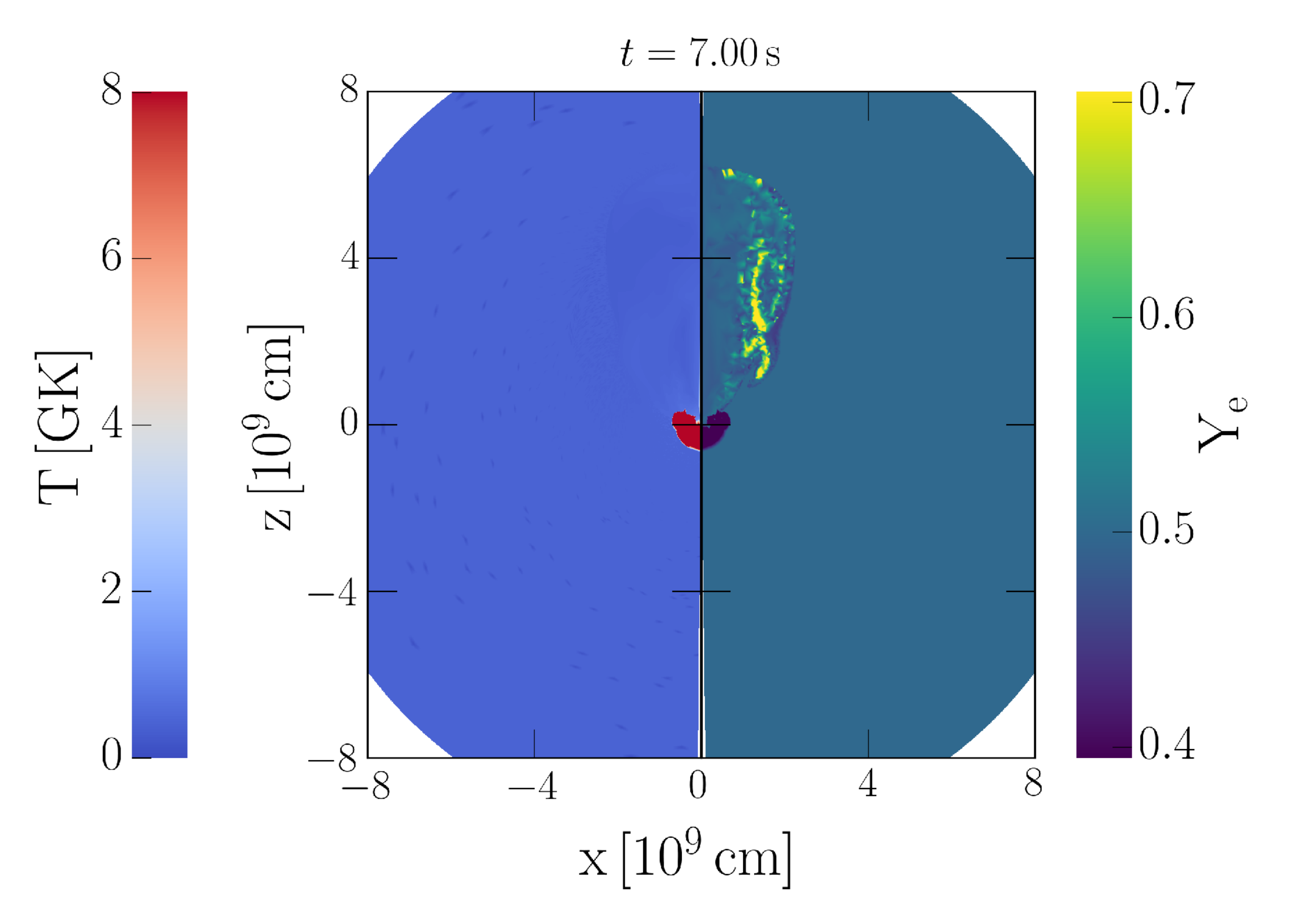}
\includegraphics[width=0.51\textwidth]{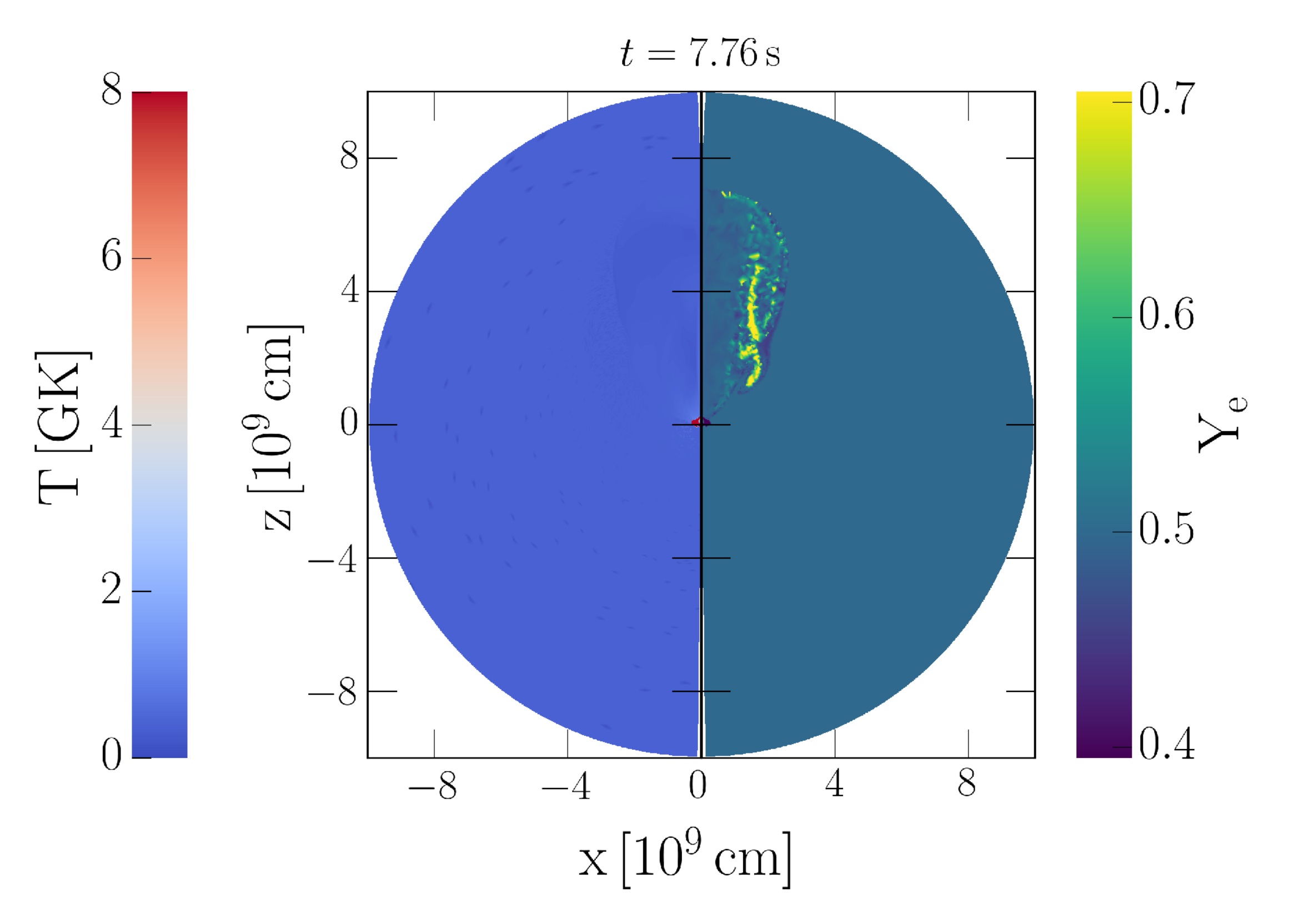}
\caption{Snapshots of the tracer particles in the CCSN simulation of the 11.2~M$_{\odot}$ progenitor showing the evolution of temperature and electron fraction. The unipolar nature of the explosion is clearly noticeable. Notice the varying scale in $x$- and $z$-axis.}
\label{fig:11.2_TempYe}
\end{figure*}

\begin{figure*}
\includegraphics[width=0.51\textwidth]{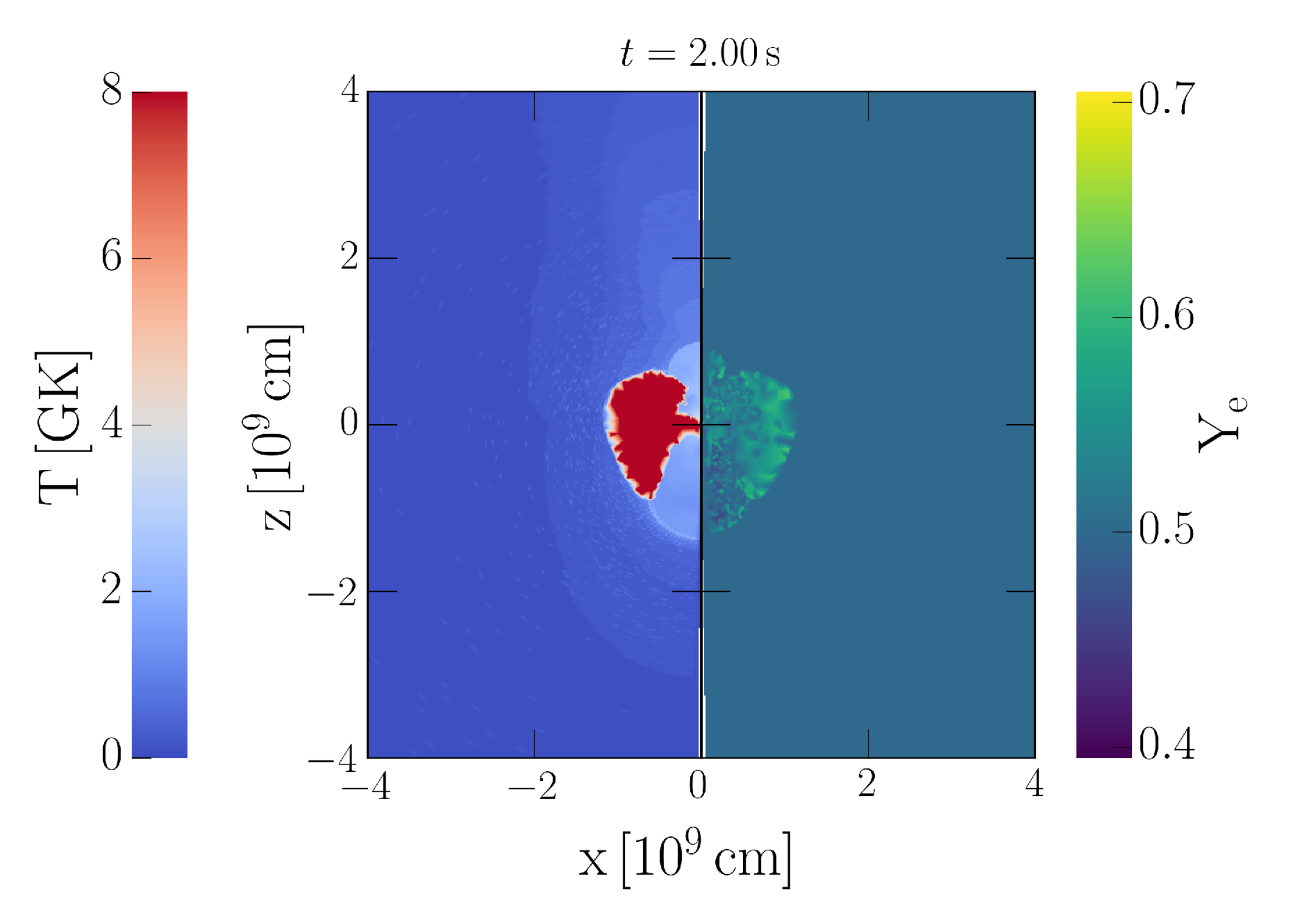}
\includegraphics[width=0.51\textwidth]{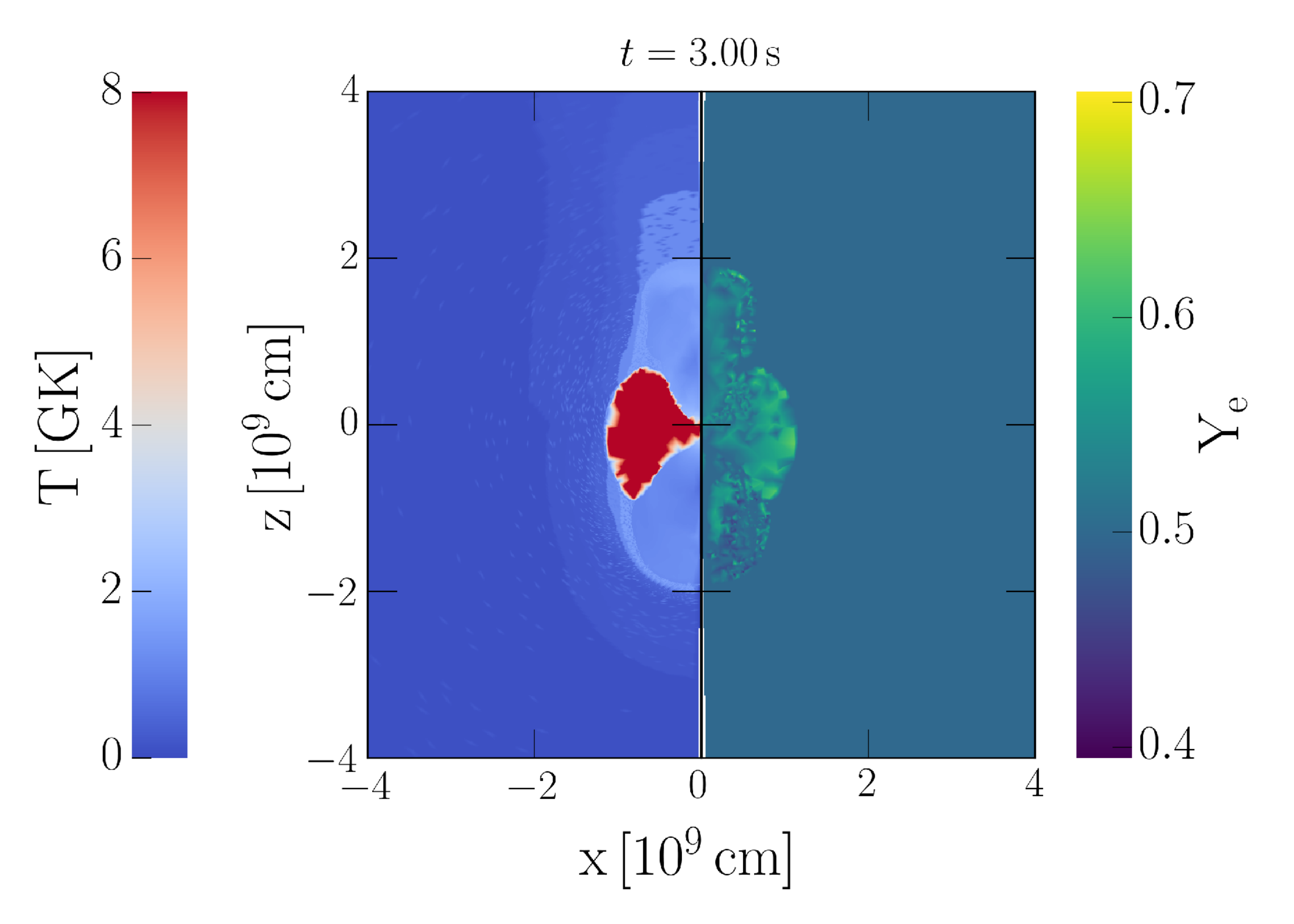}\\
\includegraphics[width=0.51\textwidth]{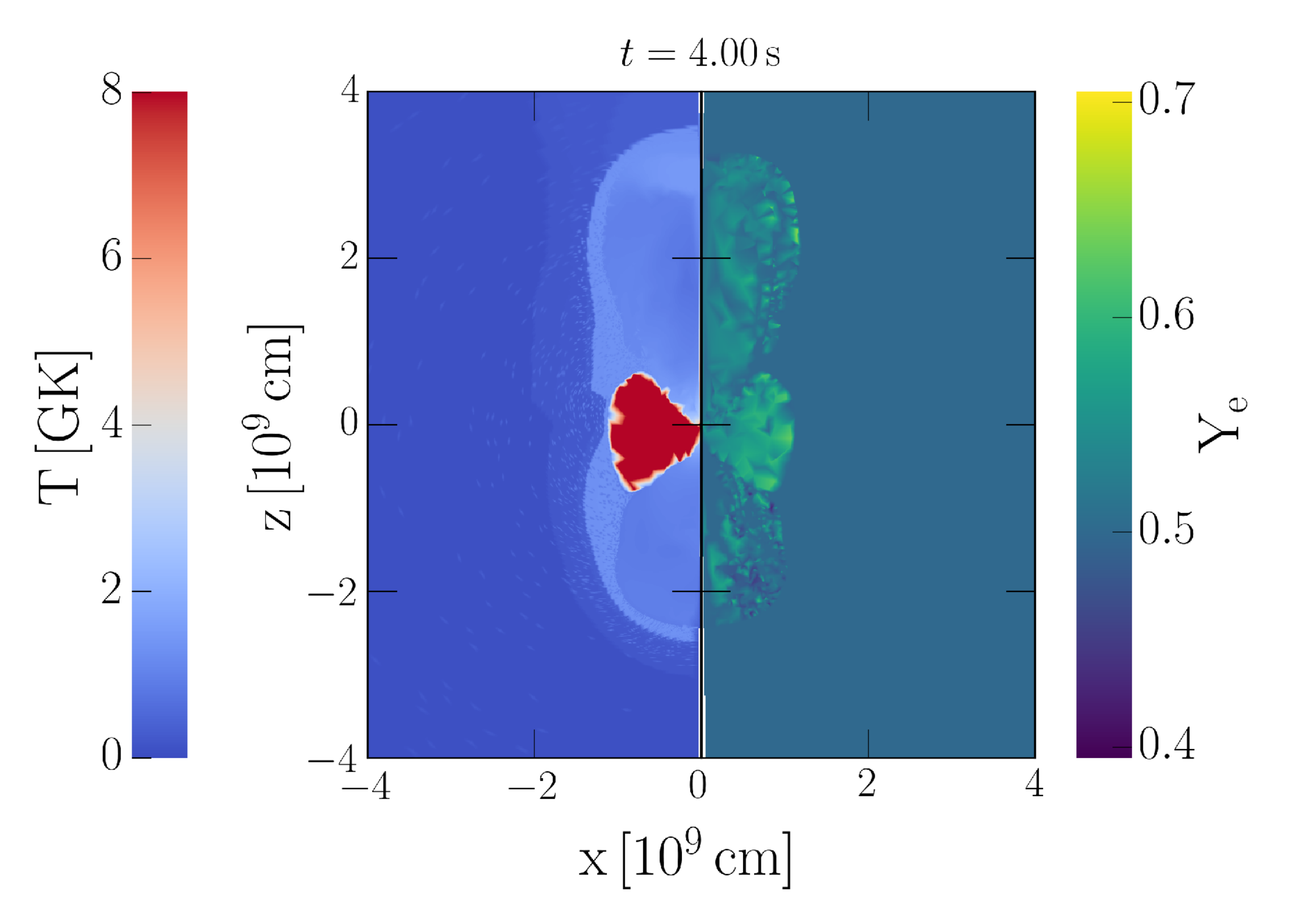}
\includegraphics[width=0.51\textwidth]{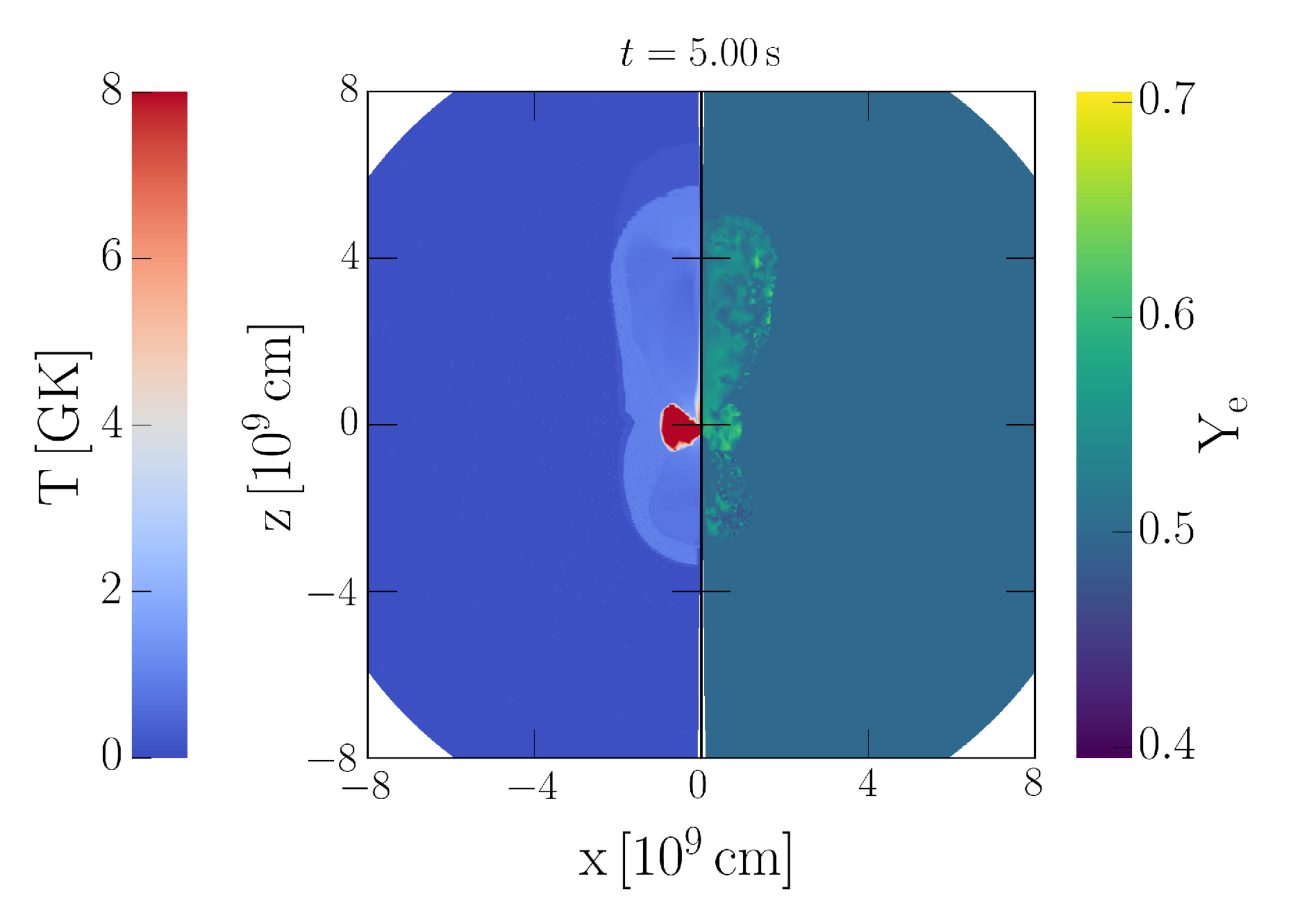}\\
\includegraphics[width=0.51\textwidth]{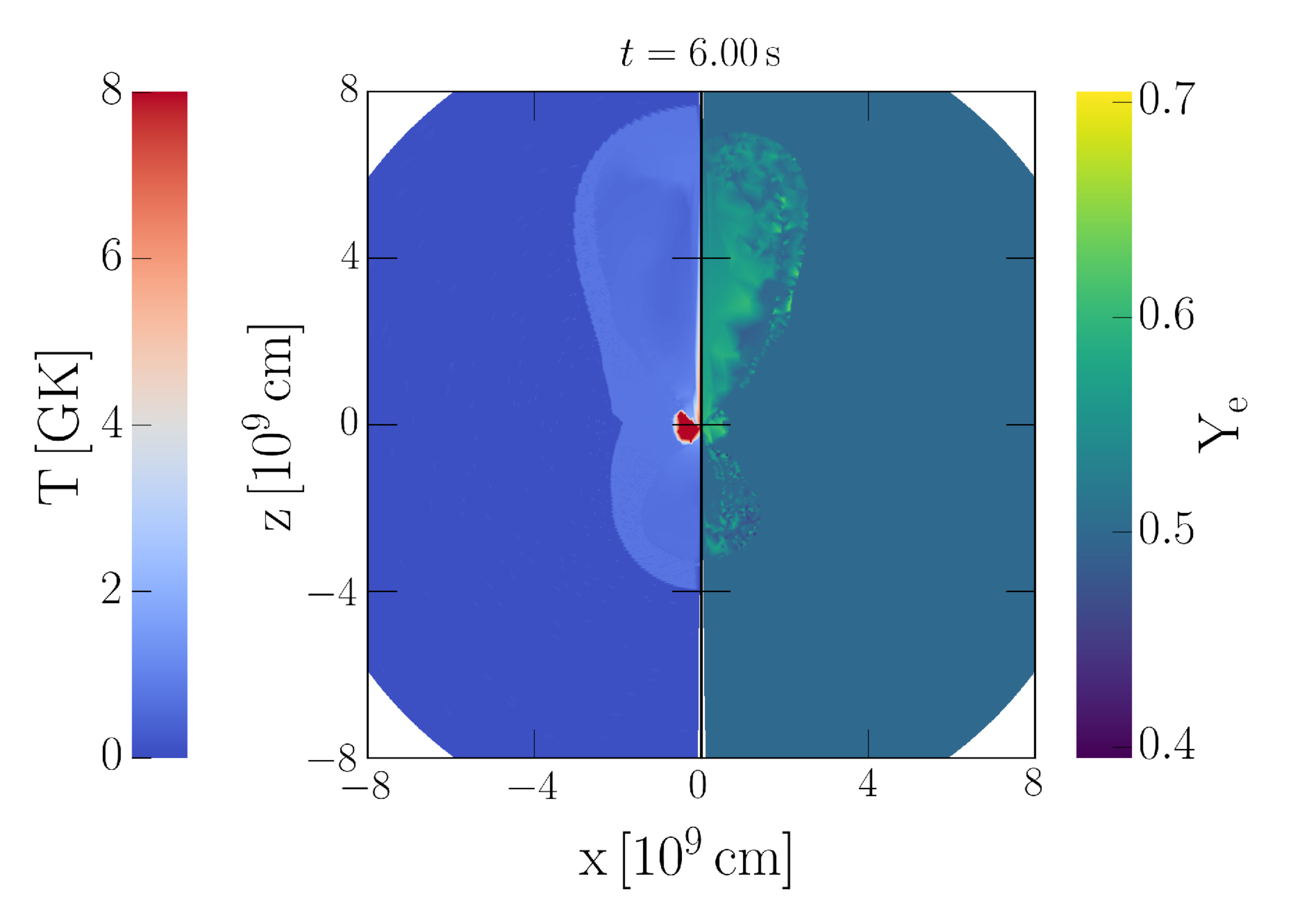}
\includegraphics[width=0.51\textwidth]{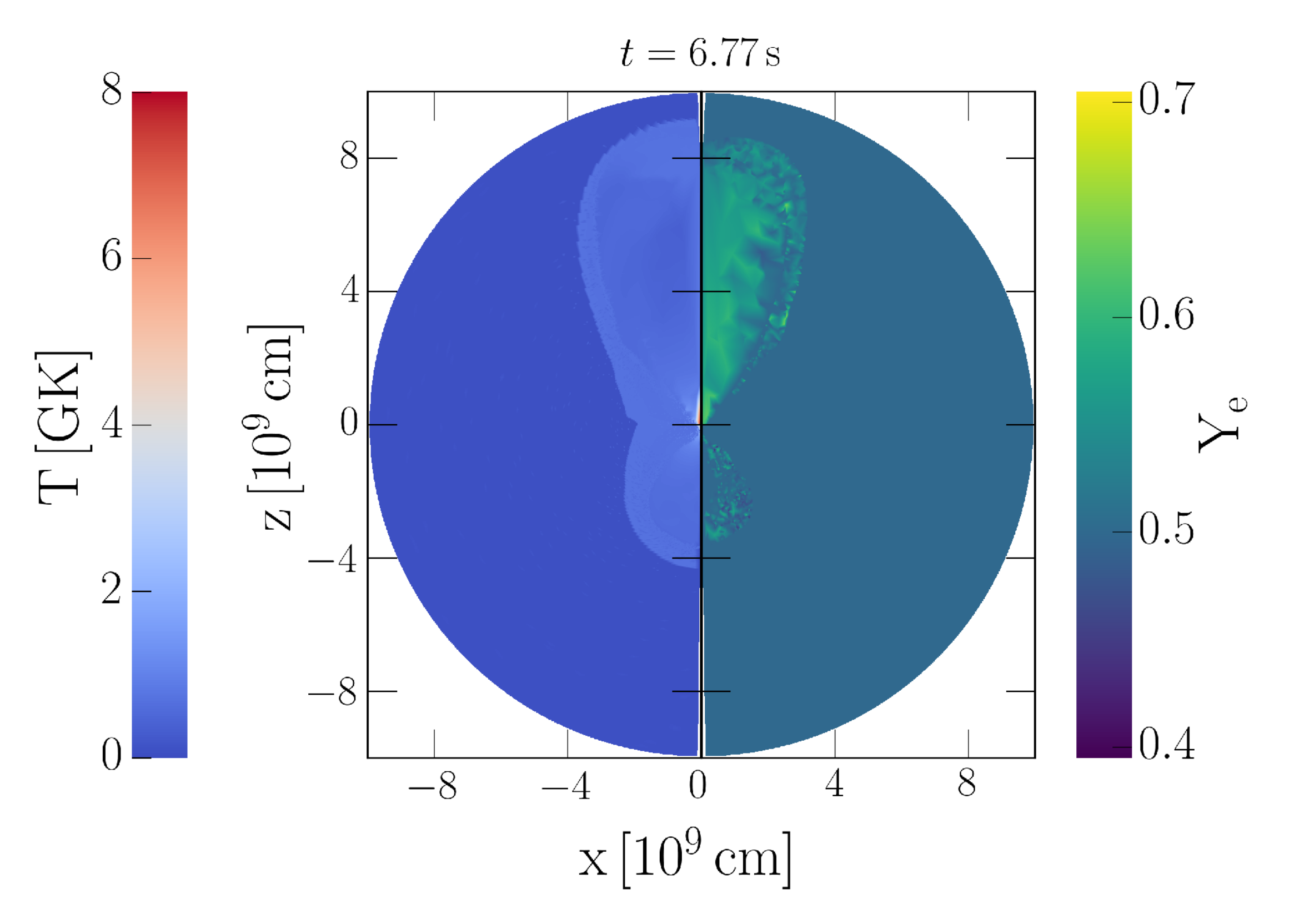}
\caption{Same as figure~\ref{fig:11.2_TempYe}, but for the 17.0~M$_{\odot}$ model.} 
\label{fig:17.0_TempYe}
\end{figure*}

Our simulations stop when the outgoing shock reaches the boundary of the computational domain. The simulations have been performed up to 7.76~s for the 11.2~M$_{\odot}$ progenitor and 6.77~s for the 17.0~M$_{\odot}$ progenitor. At this point, the shock proceeds outwards in the polar directions (only in the positive $z$-direction for the 11.2~M$_{\odot}$ model), while material around the equator is temporarily accelerated outwards by a weak shock front, but eventually stops and falls back towards the PNS. We consider the tracer particles within an angle of 30 degrees around the north pole (11.2) or both poles (17.0) to be ejected. Figures~\ref{fig:11.2_TempYe}~\&~\ref{fig:17.0_TempYe} show that at the end of the simulation, almost all the particles that reach a temperature above 1~GK are within that area and there are no tracers with temperatures above 2~GK anymore. Thus, the vast majority of the ejected particles with a high peak temperature are ejected along the polar directions. In order to account for the shock propagation after the end of the simulation, we add all tracers outside of the computational domain for the simulation, but within the same angle of 30~degrees around the north pole (11.2) or both poles (17.0), respectively. In the following, we refer to these criteria as $\theta^{30+}$ (11.2~M$_{\odot}$) and $\theta^{30}$ (17.0~M$_{\odot}$).


Different prescriptions for the ejected matter also alter the nuclear yields of the ejecta. However, we found that for the heavier nuclear species that are produced by explosive nucleosynthesis the changes in yields are negligible, even considering an upper and lower limit in ejected mass. A detailed discussion on the ejecta criterion can be found in \ref{app:ejeccrit}.

Unless stated otherwise, the nucleosynthesis yields are calculated using the $\theta^{30}$ criterion for the 17.0~M$_{\odot}$ progenitor, and the $\theta^{30+}$ criterion for the 11.2~M$_{\odot}$ model. We account for the fact that the shock will move through the outer layers of the star (which are not included in the computational domain), unbinding material from the helium and hydrogen shells in the process. However, as it has lost a lot of its energy, it does not heat up the material to the temperatures required for explosive nucleosynthesis with substantial He-destruction. Explosive He-burning had been assumed in the past to permit an r-process, based on highly active ($\alpha,n$) reactions, when the shock wave of explosive burning is passing these layers. However, the substantial amounts of $^{13}$C required for such an outcome, have never materialized in realistic models of massive stars \cite{cowan1991}. Therefore, we add a quarter of the progenitor material (a half of the progenitor in the case of the 17.0~M$_{\odot}$ simulation) starting from an enclosed mass coordinate of 1.92~M$_{\odot}$ (4.07~M$_{\odot}$) (corresponding to the outer boundary of the computational domain in the simulation) to the ejecta. The factor $1/4$ ($1/2$) is derived from the opening angle(s) of $120^{\circ}$, within which material is considered successfully ejected. Transferred into a 3D model, this would encompass one (two) solid angle(s) of $\pi$~sr, which corresponds to a quarter (a half) of the volume of the star. This procedure adds about 2.22~M$_{\odot}$ (4.89~M$_{\odot}$) of progenitor material which mainly consists of helium and hydrogen. Since this material is at a large distance from the PNS, it only reaches very low peak temperatures and is not directly affected by neutrinos, with neither the $3 \alpha$ process nor $\alpha$-captures on pre-existing seed nuclei efficient enough to significantly alter the composition. We therefore assume that these outer ejecta carry the unaltered progenitor composition.


\subsection{Post-processing procedure}
\label{sec:nucleomethod}
We perform our nucleosynthesis calculations with the full nuclear network {\sc Winnet} \cite{winteler2012} in a post-processing approach. We include 2713 isotopes up to proton number $Z = 60$ (Nd), covering the neutron-deficient as well as the neutron-rich side of the valley of $\beta$-stability. In precedent tests using the full elemental range (up to $Z = 110$) we have found no r-process in any of our trajectories. The reaction rates are based on experimentally known rates where available and predictions otherwise. The n-, p-, and alpha-captures are taken from Rauscher \& Thielemann \cite{rauscher2000}, who used known nuclear masses where available and the \textit{Finite Range Droplet Model} \cite{moeller1995} for unstable nuclei far from stability. The $\beta$-decay rates are from the nuclear database \textit{NuDat2}\footnote{http://www.nndc.bnl.gov/nudat2/}, while electron-neutrino and electron-antineutrino absorptions on nucleons are taken into account, using the rates of Fr\"ohlich \etal \cite{froehlich2006}. The neutrino luminosities and average energies are provided by the simulation. For a given tracer particle at radius $r$, the neutrino fluxes can be calculated via

\begin{equation}
F_{\nu_e/\overline{\nu}_e} \left(r,t \right) = \frac{L_{\nu_e/\overline{\nu}_e} \left(t \right)}{4 \pi r^2},
\end{equation}
where $L_{\nu_e/\overline{\nu}_e}$ are the neutrino number luminosities. Note that the quantities used here are averaged over all angular directions and that the local luminosities in the polar directions, where the outflow occurs, can differ from the angle-averaged values (see~\ref{app:lumintop}).

For ejected particles that reach a peak temperature $T_9^{\rm peak}~>~8$, we start the post-processing at the point where the temperature drops below 8~GK for the last time, (as it is possible that the tracer particles experience several heating periods). For these particles we assume an initial composition determined by nuclear statistical equilibrium (NSE) with $Y_e$ values from the simulation. For all other particles, we follow the full hydrodynamical trajectory and we correlate the initial abundances for each particle (i.e., at core bounce) with the nuclear composition in the progenitor data for the corresponding radius:

\begin{equation}
Y \left(A,Z \right)_{i} = Y \left(A,Z,r=r_{i} \right)_{prog} \hspace{2px},
\end{equation}

where $Y\left(A,Z\right)_i$ and $Y\left(A,Z\right)_{prog}$ denote the (initial) abundances of nuclear species ($A,Z$) in the tracer particle~$i$ and the progenitor, respectively, and $r_i$ denotes the initial radius of the tracer particle. The initial composition is particularly important for tracer particles that do not reach high temperatures.

The hydrodynamical simulation provides data up to a simulation time of about $t=7$~s. In order to make sure that we do not miss any nucleosynthesis processes taking place after that time, we extrapolate for each tracer to the point where the temperature drops below $0.01$~GK, assuming an adiabatic expansion with constant velocity. The temperature is calculated at each timestep using the equation of state of Timmes \& Swesty \cite{timmes2000}.

As the final composition of the tracer particles is heavily dependent on their peak temperatures during the simulation (see e.g., \cite{thielemann1996}), we divide them into 20 bins according to their peak temperature in the simulation. Using this method, we can make bin-by-bin nucleosynthesis comparisons across different CCSN simulations (see section~\ref{sec:bins}). Table \ref{table:bins} lists the temperature bins, the number of tracer particles, and the summed-up tracer masses in each temperature bin for both models (N$_{\rm X}$ gives the number of tracer particles in the case of the progenitor with a mass of X~M$_{\odot}$), where the ejection criteria discussed in section~\ref{sec:ejeccrits} have been applied.

\begin{table}
\caption{Temperature bins, number of ejected tracer particles N, and the summed-up tracer masses M$_{\rm ej}$ in each bin. \label{table:bins}}
\begin{indented}
\item[] \begin{tabular}{@{}crrrr}
\br
T$_9$ & N$_{17.0}$ & M$_{\rm ej, 17.0}$/M$_{\odot}$ & N$_{11.2}$ & M$_{\rm ej, 11.2}$/M$_{\odot}$ \vspace{0.1cm} \\
\mr
< 0.8    & 11631 & $4.41 \times 10^{-1}$ & 7652 & $4.54 \times 10^{-2}$\\
0.8--0.9 & 1346  & $5.78 \times 10^{-2}$ & 649 &  $7.06 \times 10^{-3}$\\
0.9--1.0 & 1239  & $5.27 \times 10^{-2}$ & 697 &  $7.24 \times 10^{-3}$\\
1.0--1.1 & 1288  & $4.82 \times 10^{-2}$ & 753 &  $6.95 \times 10^{-3}$\\
1.1--1.2 & 1178  & $4.27 \times 10^{-2}$ & 759 &  $5.73 \times 10^{-3}$\\
1.2--1.4 & 1848  & $6.83 \times 10^{-2}$ & 741 &  $6.49 \times 10^{-3}$\\
1.4--1.6 & 1802  & $5.84 \times 10^{-2}$ & 861 &  $6.99 \times 10^{-3}$\\
1.6--1.8 & 1205  & $2.88 \times 10^{-2}$ & 598 &  $3.43 \times 10^{-3}$\\
1.8--2.0 & 1034  & $1.67 \times 10^{-2}$ & 402 &  $2.38 \times 10^{-3}$ \\
2.0--2.4 & 1269  & $1.82 \times 10^{-2}$ & 615 &  $2.93 \times 10^{-3}$\\
2.4--2.8 & 727   & $6.02 \times 10^{-3}$ & 511 &  $1.81 \times 10^{-3}$ \\
2.8--3.2 & 356   & $2.27 \times 10^{-3}$ & 387 &  $1.37 \times 10^{-3}$ \\
3.2--3.6 & 213   & $1.19 \times 10^{-3}$ & 389 &  $1.05 \times 10^{-3}$\\
3.6--4.2 & 305   & $2.16 \times 10^{-3}$ & 282 &  $7.17 \times 10^{-4}$ \\
4.2--4.8 & 196   & $1.57 \times 10^{-3}$ & 185 &  $4.40 \times 10^{-4}$\\
4.8--5.4 & 389   & $2.30 \times 10^{-3}$ & 160 &  $3.36 \times 10^{-4}$\\
5.4--6.2 & 114   & $5.68 \times 10^{-4}$ & 179 &  $2.28 \times 10^{-4}$\\
6.2--7.0 & 38    & $4.54 \times 10^{-4}$ & 131 &  $9.20 \times 10^{-5}$ \\
7.0--8.0 & 50    & $4.81 \times 10^{-4}$ & 35  &  $1.33 \times 10^{-5}$ \\
> 8.0    & 1413  & $3.13 \times 10^{-2}$ & 1927 & $1.07 \times 10^{-2}$\\
\br
\end{tabular}
\end{indented}
\end{table}

Since the tracer particles in the low-temperature bins are very homogeneous and not many charged-particle reactions are expected at these temperatures, we do not post-process all the tracers in the bins with $T_9 < 2.8$ in order to save CPU time. Instead we perform nucleosynthesis calculations for 200~randomly selected particles and then calculate the isotopic yields by extrapolating to all tracers in each respective bin taking into account the (individual) mass each tracer particle represents. We do this for each bin up to $T_9^{\rm peak} = 2.8$ by first calculating an average mass fraction for each isotopic species, weighted by the tracer masses, and then multiplying with the total bin mass $M_{bin}$, obtained by summing up the masses of all the tracers in the bin:

\begin{equation}
\label{eq:bincomp}
M_{ej} \left(A,Z \right) = \frac{ \sum_{i} X_i \left(A,Z \right) M_i}{ \sum_{i} M_i} M_{bin} \hspace{2px}.
\end{equation}

Here, $X_i \left(A,Z \right)$ is the mass fraction of nucleus ($A,Z$) and $M_i$ is the mass corresponding to tracer particle $i$. The sums go over all post-processed tracer particles, while $M_{bin}$ is the sum over all particle masses in the bin. For the higher-temperature bins (T$_9 \geq 2.8$) all tracer particles are post-processed individually. Finally, the yields from the individual bins are added up to obtain the total nuclear yields.

The highest-temperature bin (T$_{\mathrm{9}}>8$) contains tracer particles that are ejected from the innermost regions above the PNS at different times in the simulation (see figures~\ref{fig:11.2_TempYe}~\&~\ref{fig:17.0_TempYe}). Since the neutrino and anti-neutrino luminosities change considerably over that time, the electron fractions of the tracer particles in the highest temperature bin span a broad range. We will therefore further divide these tracer particles into secondary bins according to their electron fraction in section~\ref{sec:yebins}.

\section{Results}
\label{sec:results}
In this section, we present our nucleosynthesis results for the two supernova models.

\subsection{Integrated nucleosynthesis yields}
\label{sec:yields}
\begin{figure}[t]
\includegraphics[width=0.54\textwidth]{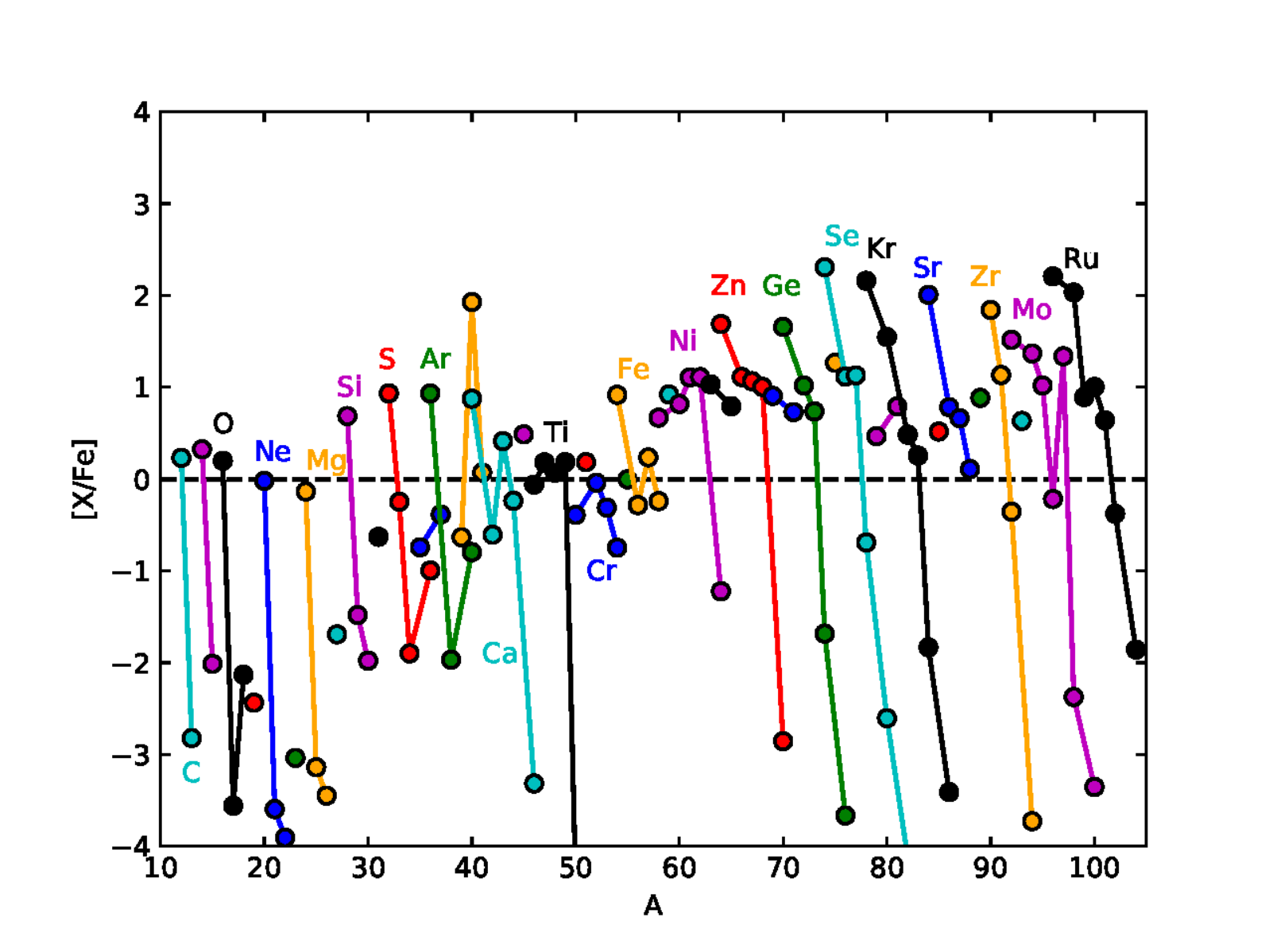}
\includegraphics[width=0.54\textwidth]{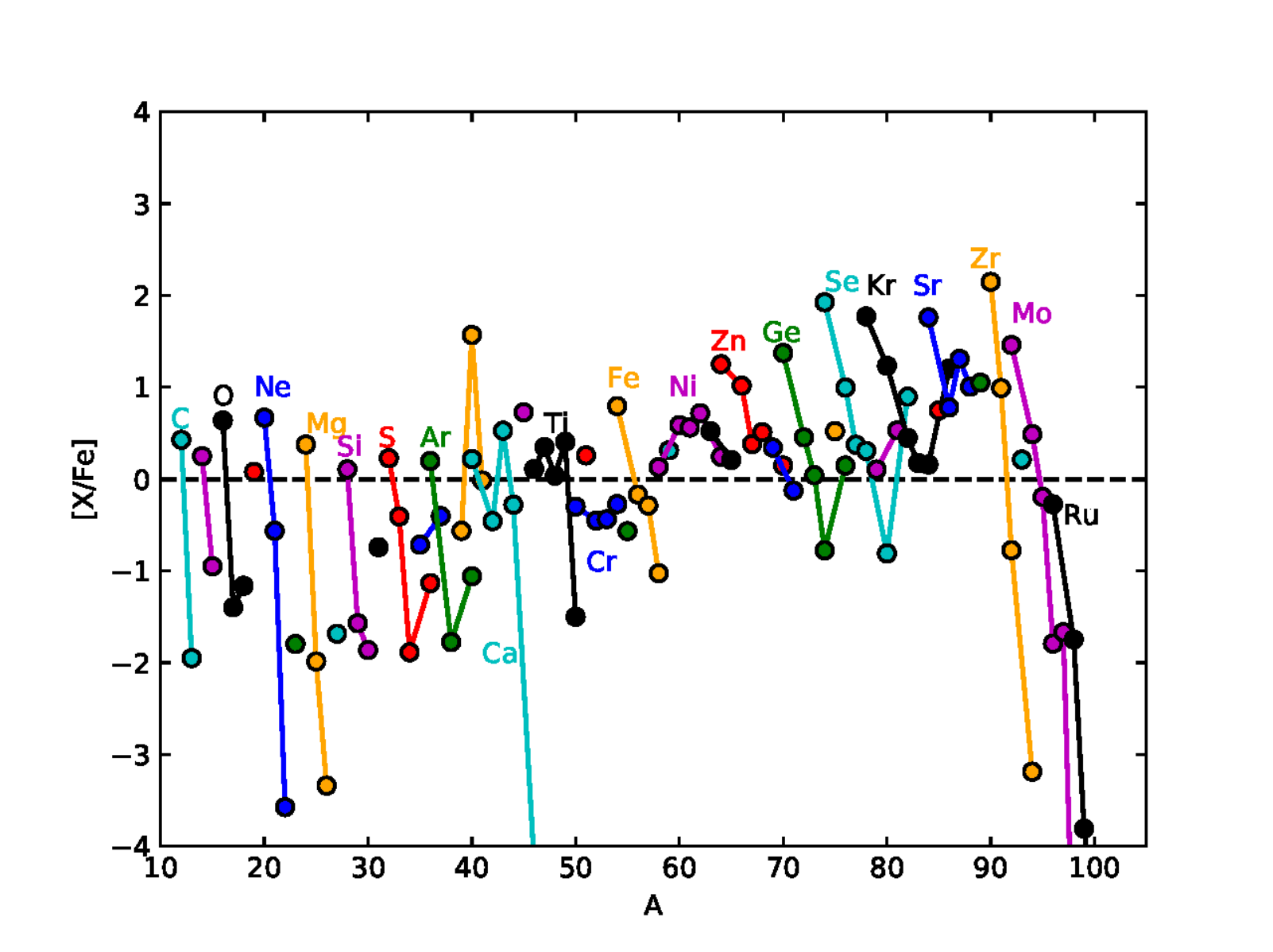}
\caption{Isotopic [X/Fe] values (from C to Ru) of the 11.2~M$_{\odot}$ model (left) and the 17.0~M$_{\odot}$ model (right) after the decay of all unstable nuclei with T$_{1/2} < 10^9$~yr. Isotopes of the same element are represented by connected data points of the same colour. The solar abundances are those of Lodders \etal \cite{lodders2009}.}
\label{fig:xoverfe}
\end{figure}

Figure~\ref{fig:xoverfe} shows the isotopic [X/Fe] distribution\footnote{[X/Fe] $= \log_{10} \left(Y_X/Y_{Fe} \right)_{\rm star} - \log_{10} \left(Y_X/Y_{Fe} \right)_{\odot}$} after decay to stability. Connected data points of the same colour represent different isotopes of the same element. All nuclei with T$_{1/2} < 10^9$~yr are considered to be completely decayed. The solar abundances are from Lodders \etal \cite{lodders2009}. The results are in agreement with the fact that the $\alpha$-elements O, Ne, Mg (which originate from hydrostatic burning and essentially are ejected in an unaltered way) are a dominant fraction of the ejecta and increase with stellar mass. On the other hand, $\alpha$-elements beyond Si are products of explosive burning and thus dependent on the strength of the explosion, i.e. the explosion energy. On average, integrated over initial stellar mass, all $\alpha$-elements (from O to Ti), are overproduced in CCSNe in comparison to Fe with [X/Fe] between 0.3 and 0.5. \cite{cayrel2004}. Only with the occurrence of supernovae type~Ia which very efficiently produce iron do the [X/Fe] values of these $\alpha$-elements approach 0 in younger stars (e.g., \cite{kobayashi2009}; our sun has [X/Fe]~$=0$ for all elements, by definition). The high abundances of neutron-deficient isotopes for Z~$>32$ (Ge and beyond) point towards the presence of a $\nu$p-process in our simulations. The $\nu$p-process is a mechanism of the rapid proton capture process that allows to bypass the $^{64}$Ge waiting point by means of antineutrino captures on free protons, converting them to neutrons which can be captured by $^{64}$Ge and other neutron-deficient isotopes in a ($n,p$)~reaction \cite{froehlich2006,pruet2006,wanajo2006}. In fact we do observe this behaviour in the most proton-rich particles in the 11.2~M$_{\odot}$ model, but we find that the majority of the heaviest isotopes is produced in neutron-rich conditions at high temperatures. A detailed discussion of these results is included in Sections~\ref{sec:yebins}~\&~\ref{sec:mo92}. The large positive [$^{92,94}$Mo/Fe] and [$^{96,98}$Ru/Fe] values (the latter only in the 11.2~M$_\odot$ case) are a very interesting aspect of our results, because the origin of these isotopes in our galaxy is not yet very well understood. 


The full compositions of the ejecta for both simulations, integrated over all ejected tracer particles, are also presented in tabular form in~\ref{app:tables}. Only nuclei with an ejected mass of at least $10^{-15}$~M$_{\odot}$ are included. In our approach we run each calculation until the temperature drops below $0.01$~GK, which means that the calculations for the individual tracer particles do not end exactly at the same time. Therefore, the yields in the tables do not correspond to one specific snapshot in time, which has a significance for unstable isotopes with half-lives of an order T$_{1/2} = 1000$~s and shorter. The $^{56}$Ni yields of $3.03 \times 10^{-3}$~M$_{\odot}$ ($1.30 \times 10^{-2}$~M$_{\odot}$) are very low for CCSNe in this mass range (see e.g., \cite{thielemann1996}). Figure~\ref{fig:criteria_yields} in \ref{app:ejeccrit} demonstrates that it is produced in considerably larger amounts (an upper limit criterion based purely on the density evaluated at 5~s indicates $6.57 \times 10^{-2}$~M$_{\odot}$ of ejected $^{56}$Ni for the 17.0~M$_{\odot}$ model), but the majority of it is accreted onto the PNS in the later stages of the simulation. Wanajo \etal \cite{wanajo2017} also find a very low $^{56}$Ni ejecta mass for all their 2D CCSN models, which they relate to the two-dimensionality of the models and the missing late-time ejecta that are not followed in their case.

\subsection{Correlation between peak temperature and nuclear composition}
\label{sec:bins}
\begin{figure}[t]
\includegraphics[width=0.34\textwidth]{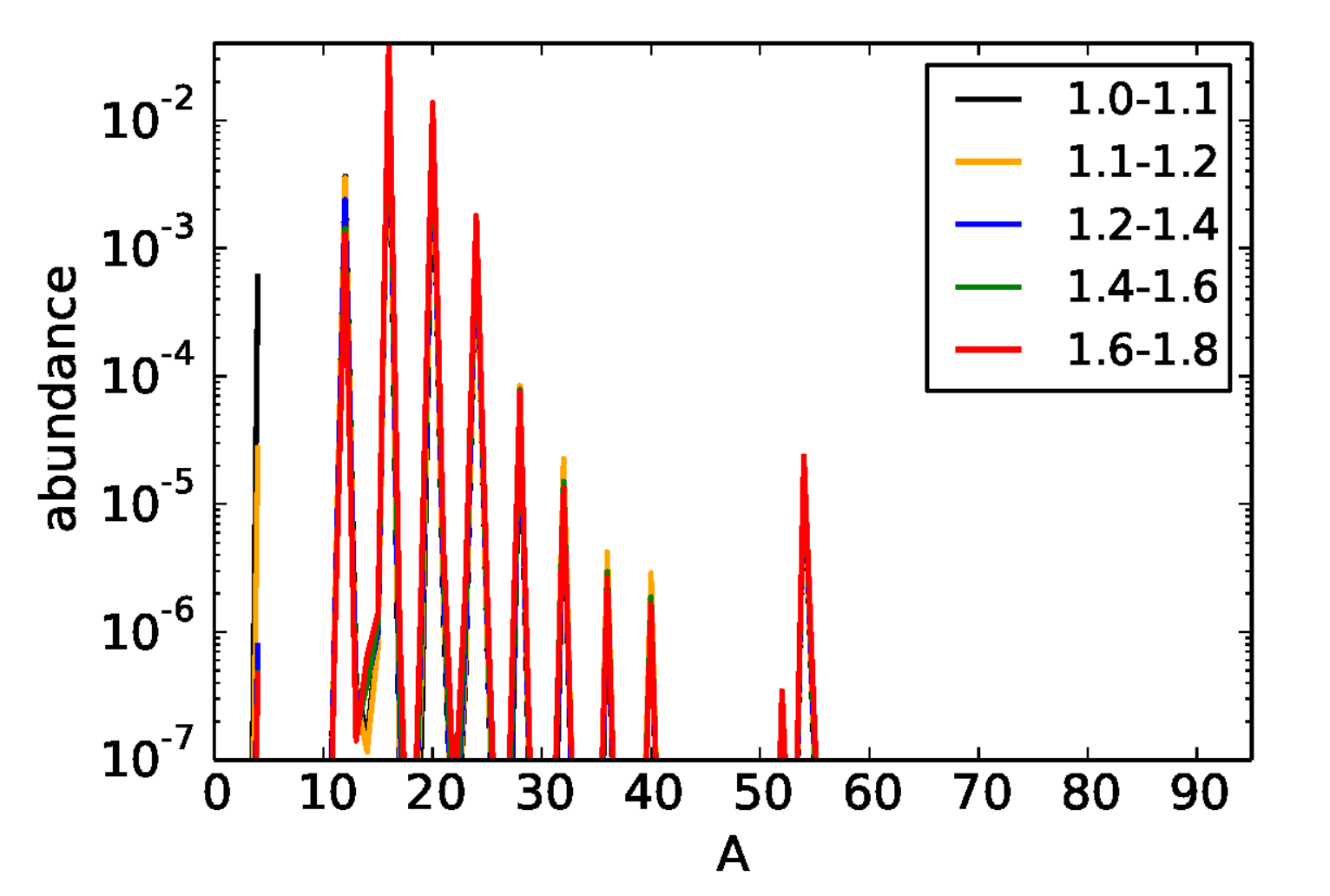}
\includegraphics[width=0.34\textwidth]{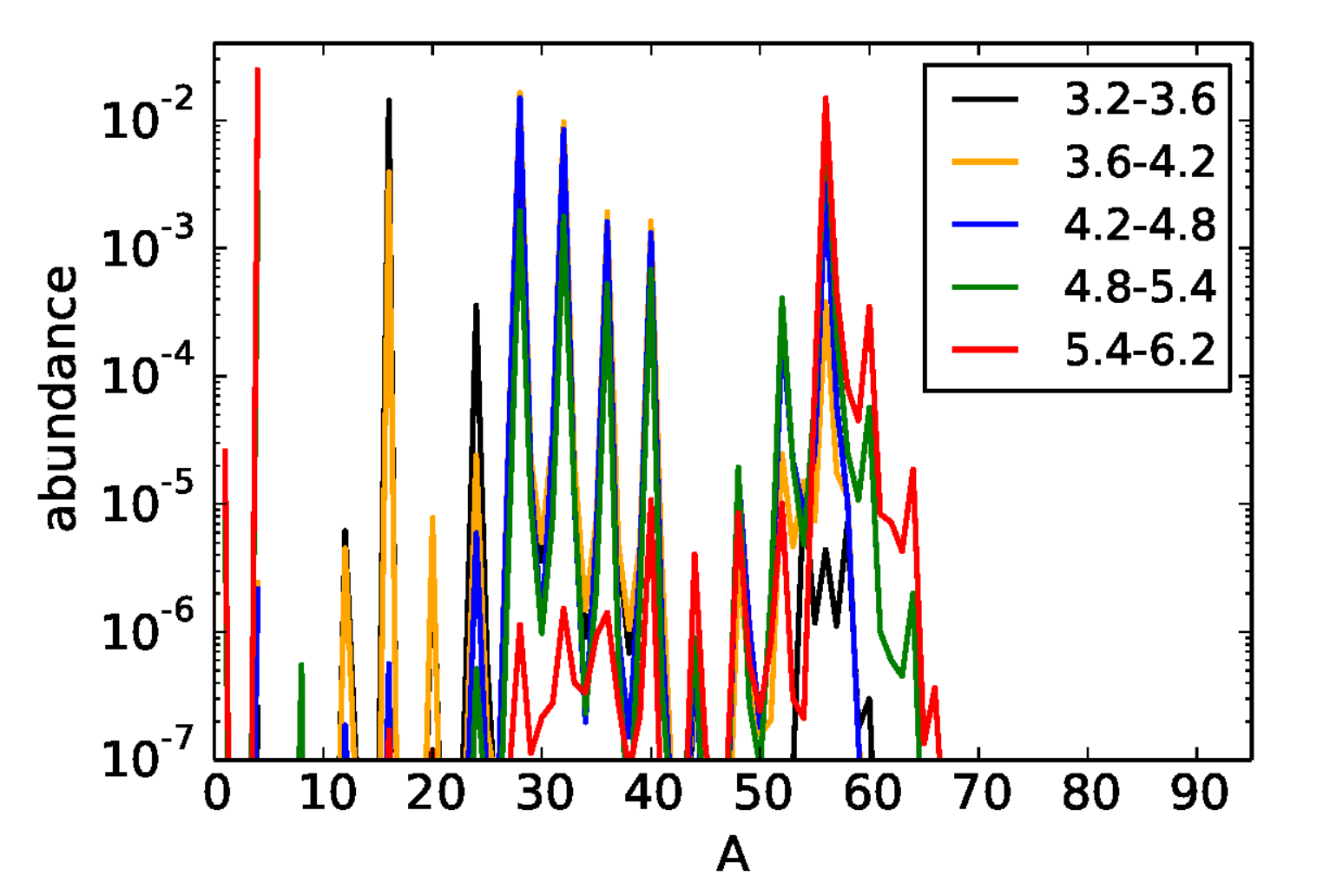}
\includegraphics[width=0.34\textwidth]{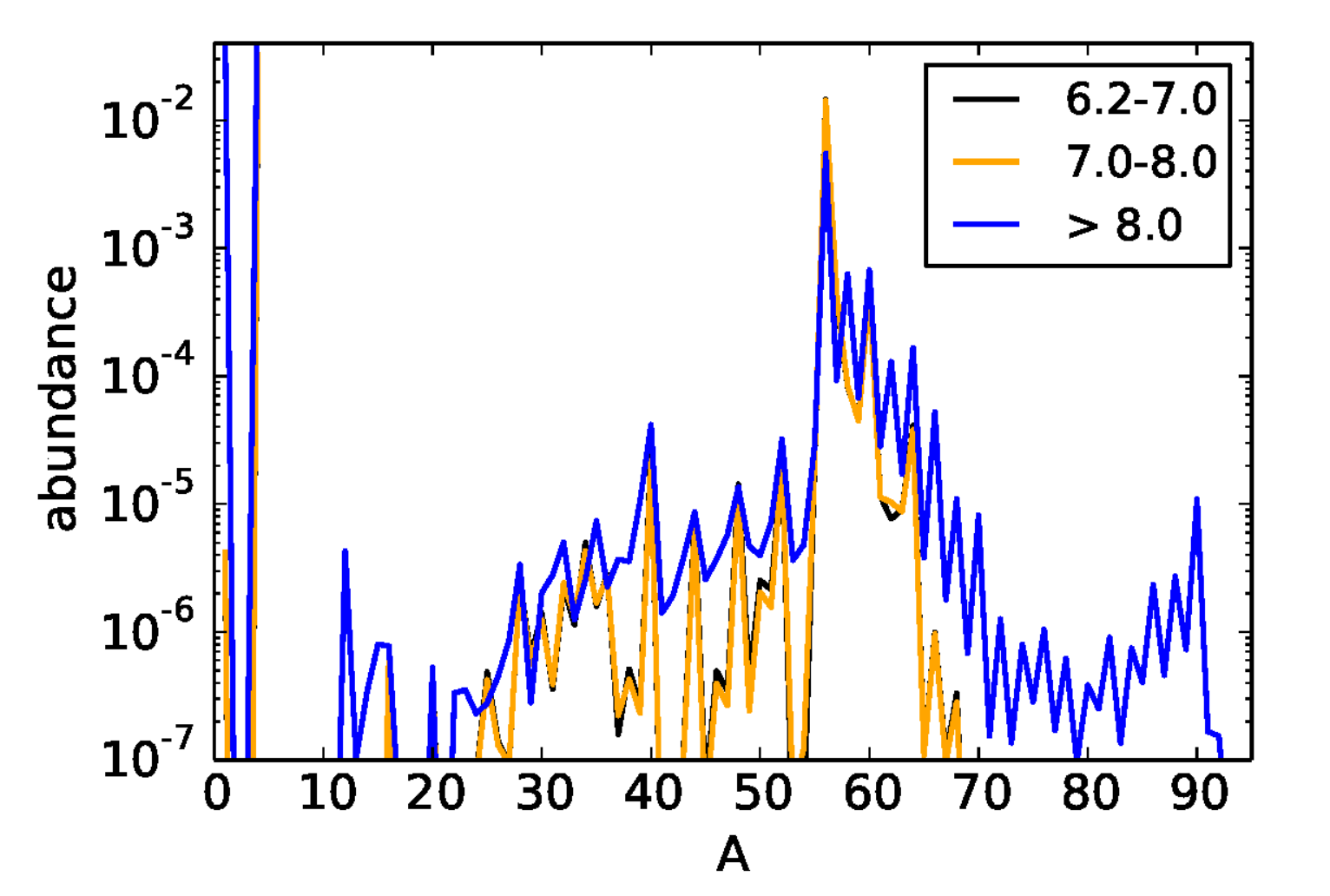}
\caption{Abundance patterns for individual peak temperature bins for the 17.0~M$_{\odot}$ model.}
\label{fig:17_diffbins} 
\end{figure}

Figure~\ref{fig:17_diffbins} compares the isotopic abundances in different temperature bins (see table~\ref{table:bins}) for the 17.0~M$_{\odot}$ model. For the low-temperature bins (left panel), the distributions are characterized by the progenitor abundances and the agreement among the different bins is very good, with the exception of $^4$He and the $\alpha$-nuclei $^{32}$S, $^{36}$Ar, and $^{40}$Ca. The temperature bins between $3.2~<~T_9~<~6.2$ (middle panel) mark a transition regime, in which the abundance distribution is shifted to heavier isotopes with increasing peak temperature. However, no nuclei beyond $A = 66$ are produced in large numbers. The right panel shows the three bins with the highest peak temperature. Compositions of tracer particles in these bins are determined by nuclear statistical equilibrium, as is evident from the identical compositions in the $6.2~<~T_9~<~7.0$ and the $7.0~<~T_9~<~8.0$ bins. The differing composition of the $T_9~\geq~8.0$ originates from the wide range of $Y_e$ values encountered in the hottest tracers (see section~\ref{sec:yebins}), whereas the other two bins only contain $Y_e \approx 0.5$ ejecta.

\begin{figure}[t]
\includegraphics[width=0.34\textwidth]{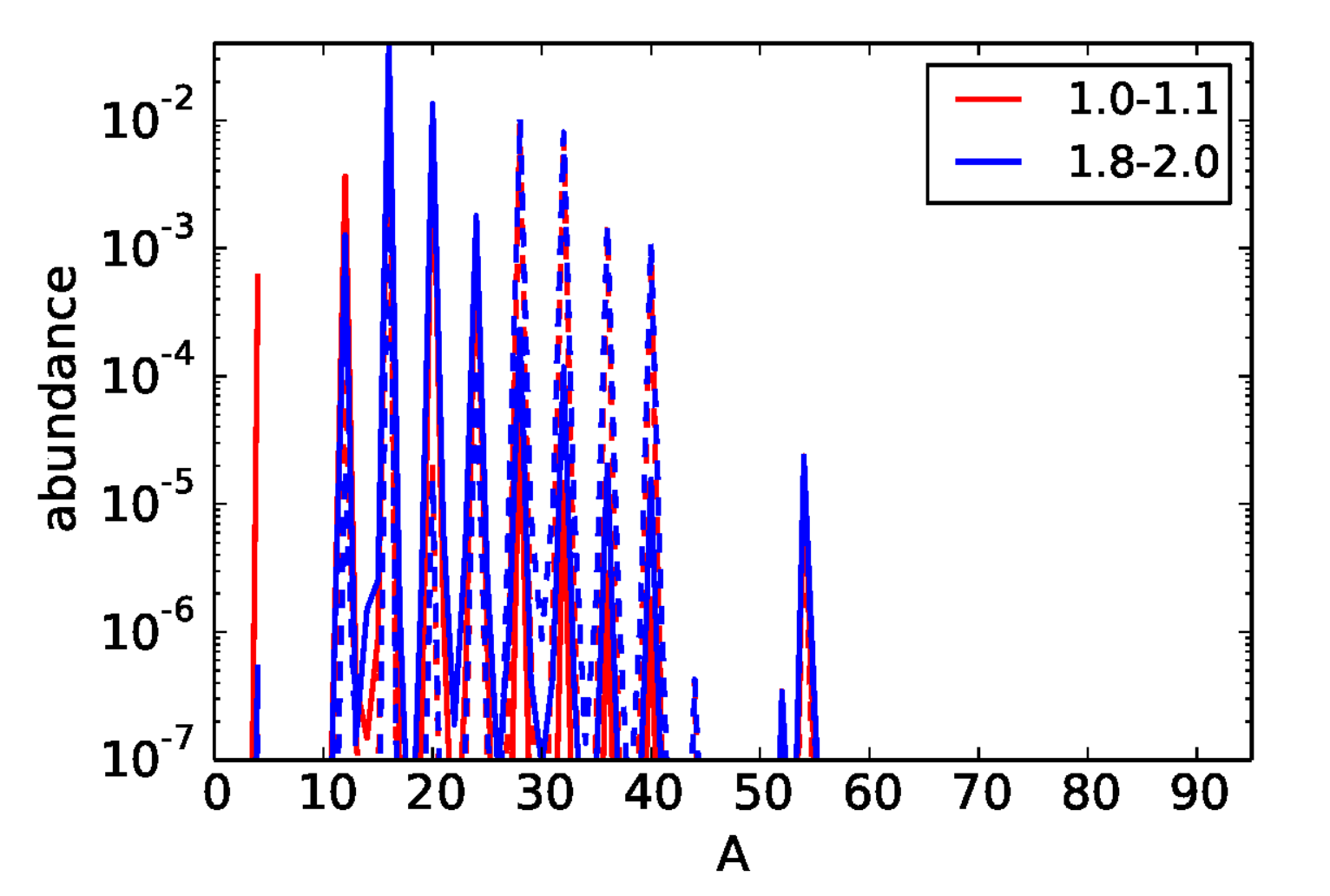}
\includegraphics[width=0.34\textwidth]{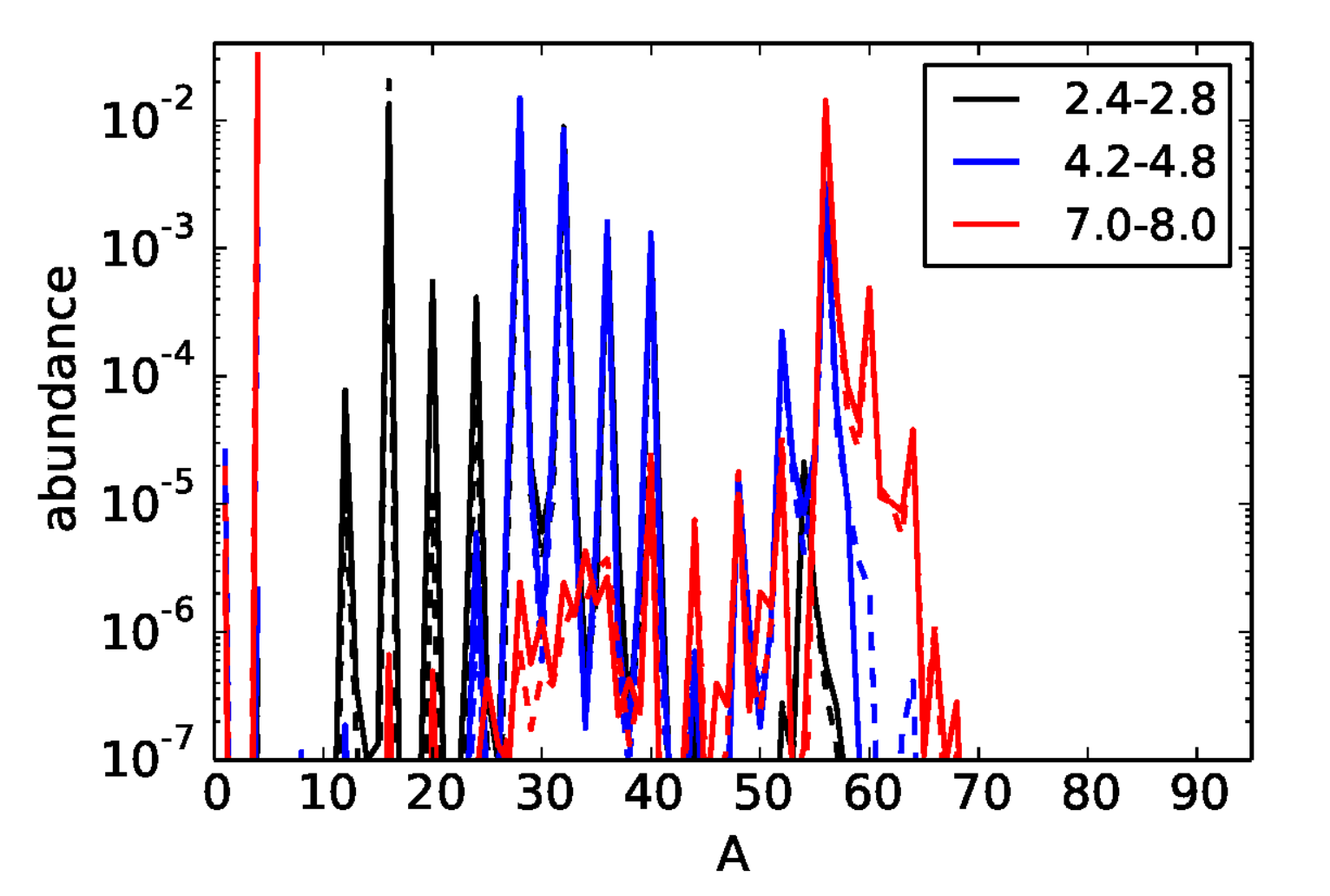}
\includegraphics[width=0.345\textwidth]{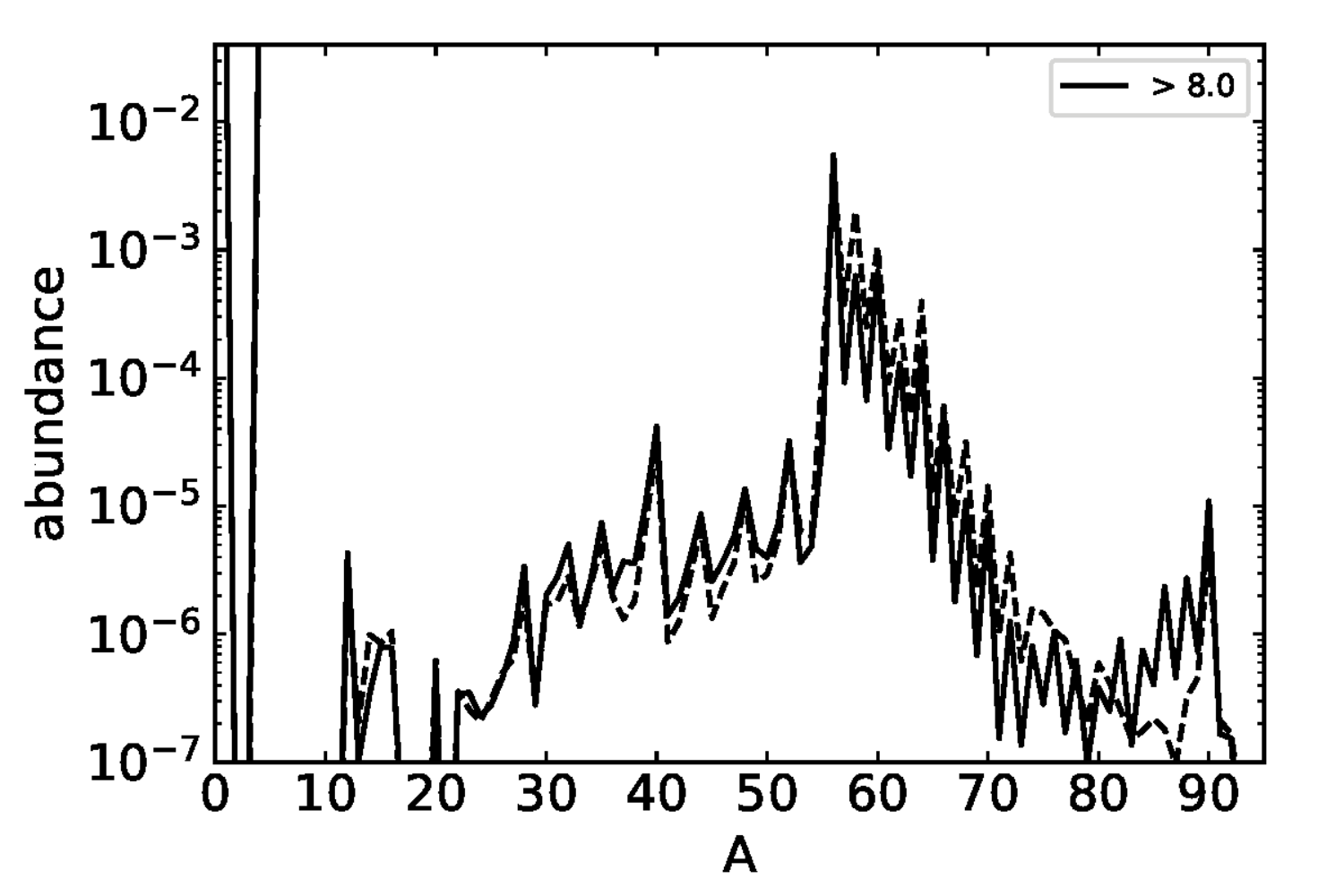}
\caption{Abundance patterns for individual peak temperature bins for the 11.2~M$_{\odot}$ (dashed lines) and the 17.0~M$_{\odot}$ (solid lines) models. Left: low-temperature bins. The different progenitor compositions for the low-temperature bins are evident. While for the 17.0~M$_{\odot}$ progenitor the low-temperature tracers originate mainly in the C-O shells, the same bins in the 17.0~M$_{\odot}$ model also eject unprocessed Si shell material. Middle: compositions from intermediate-temperature bins. Explosive nucleosynthesis at electron fractions close to $Y_e = 0.5$ leads to a good agreement across CCSN simulations. Right: highest-temperature bin. The inhomogeneous $Y_e$ distributions of tracers results in contributions from different nucleosynthesis processes and different nuclear compositions.}
\label{fig:comp_diffbins} 
\end{figure}

In a next step, we compare the individual bin abundance patterns across our two supernova models (see figure~\ref{fig:comp_diffbins}). The differences in $\alpha$-nuclei in the low-temperature bins (left panel) can be explained by the different initial compositions which are inherited from the progenitor for these tracer particles (see Sec.~\ref{sec:nucleomethod}). The abundance patterns for all bins of intermediate temperatures agrees remarkably well (middle panel), but there is a discrepancy for the highest-temperature bin which is due to the heterogenic nature of these tracer particles and the varying contribution of the $\nu$p-process (see also figure~\ref{fig:xoverfe}). Although the predictive power of our analysis is restricted to two models, we can identify a trend: for tracer particles with a peak temperature between 2.4~GK and 8.0~GK the nuclear composition can be predicted for different progenitors with good precision. At lower peak temperatures the composition is determined by the initial conditions (i.e., the progenitor abundances), while the tracer particles with peak temperatures above 8~GK usually pass close to the proto-neutron star at some point of their trajectory and therefore encompass a wide range of $Y_e$ values which has a direct impact on the nuclear reaction flow. In the following sections we further discuss and examine the nucleosynthesis in the highest-temperature bin.


\subsection{Highest temperature bin: subdivision in $Y_e$ bins}
\label{sec:yebins}

\begin{figure*}
\includegraphics[width=0.54\textwidth]{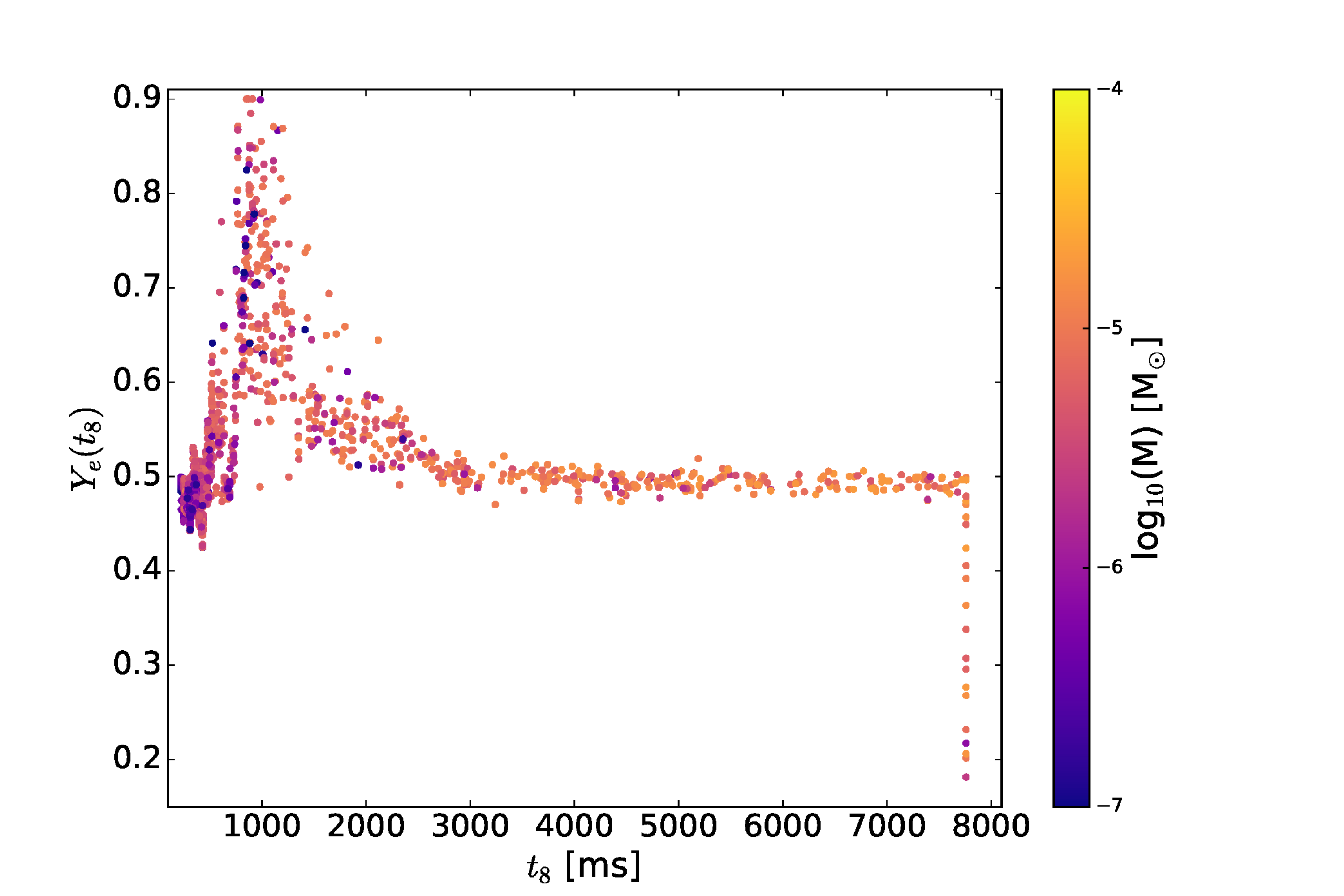}
\includegraphics[width=0.54\textwidth]{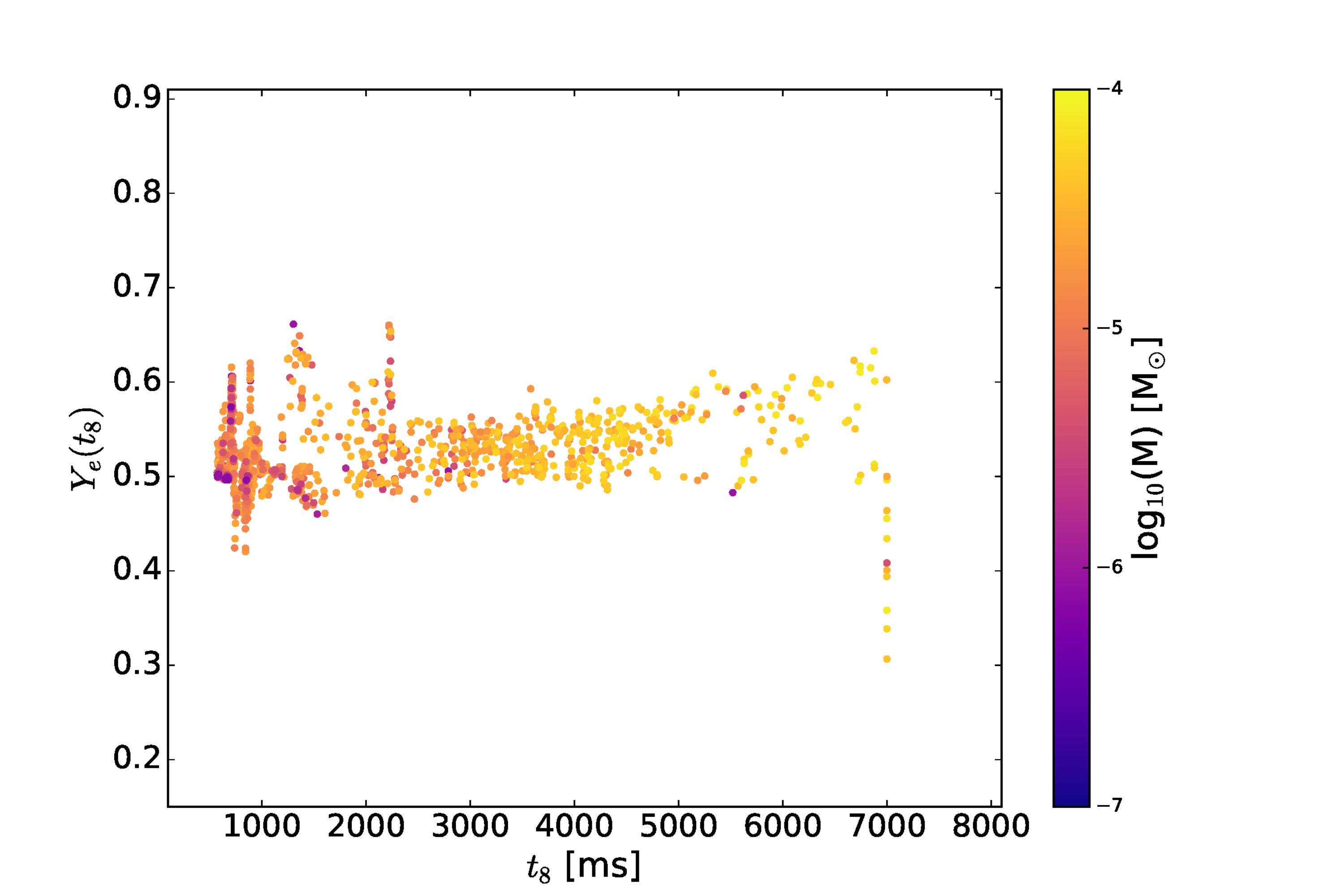}      \\
\includegraphics[width=0.54\textwidth]{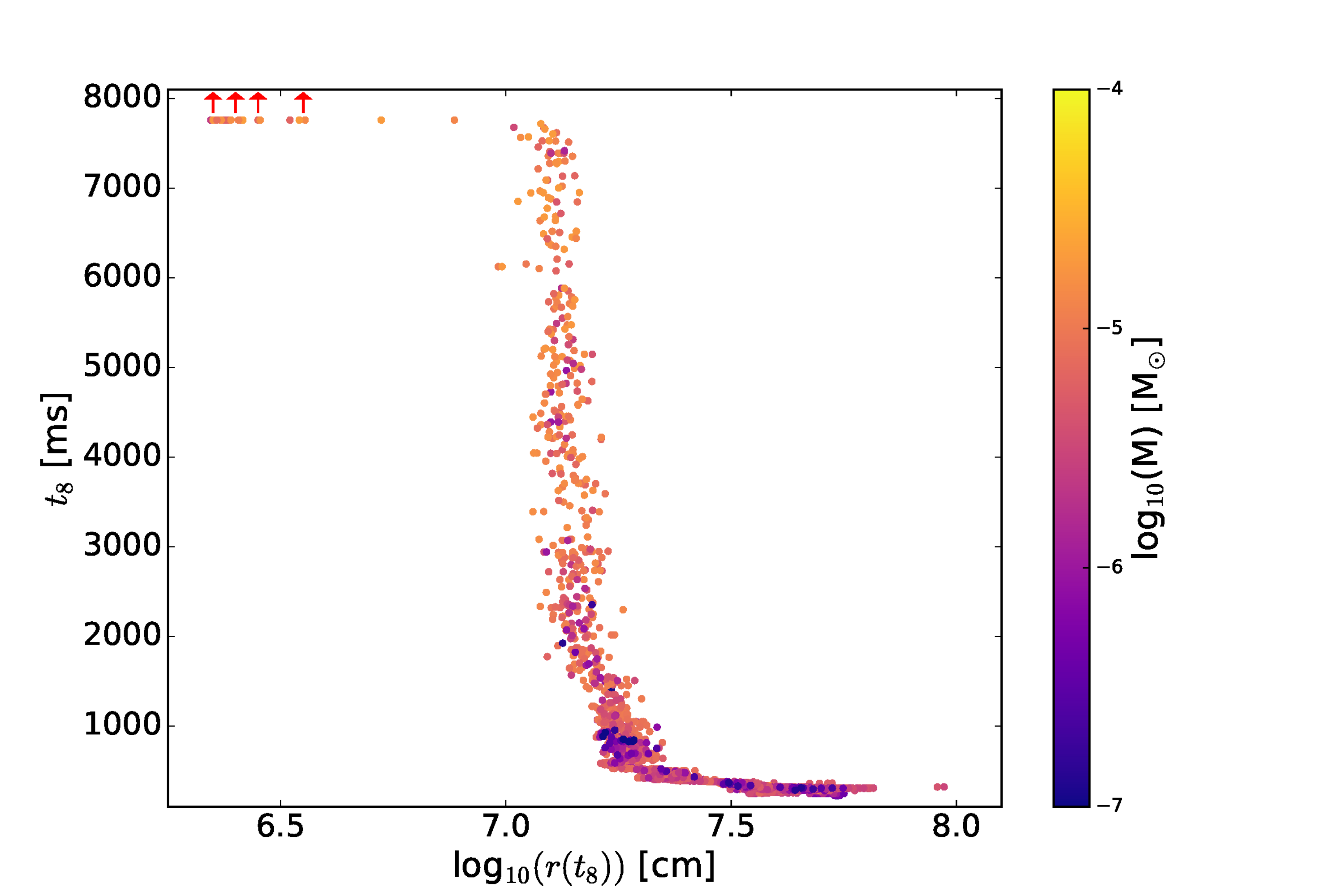} 
\includegraphics[width=0.54\textwidth]{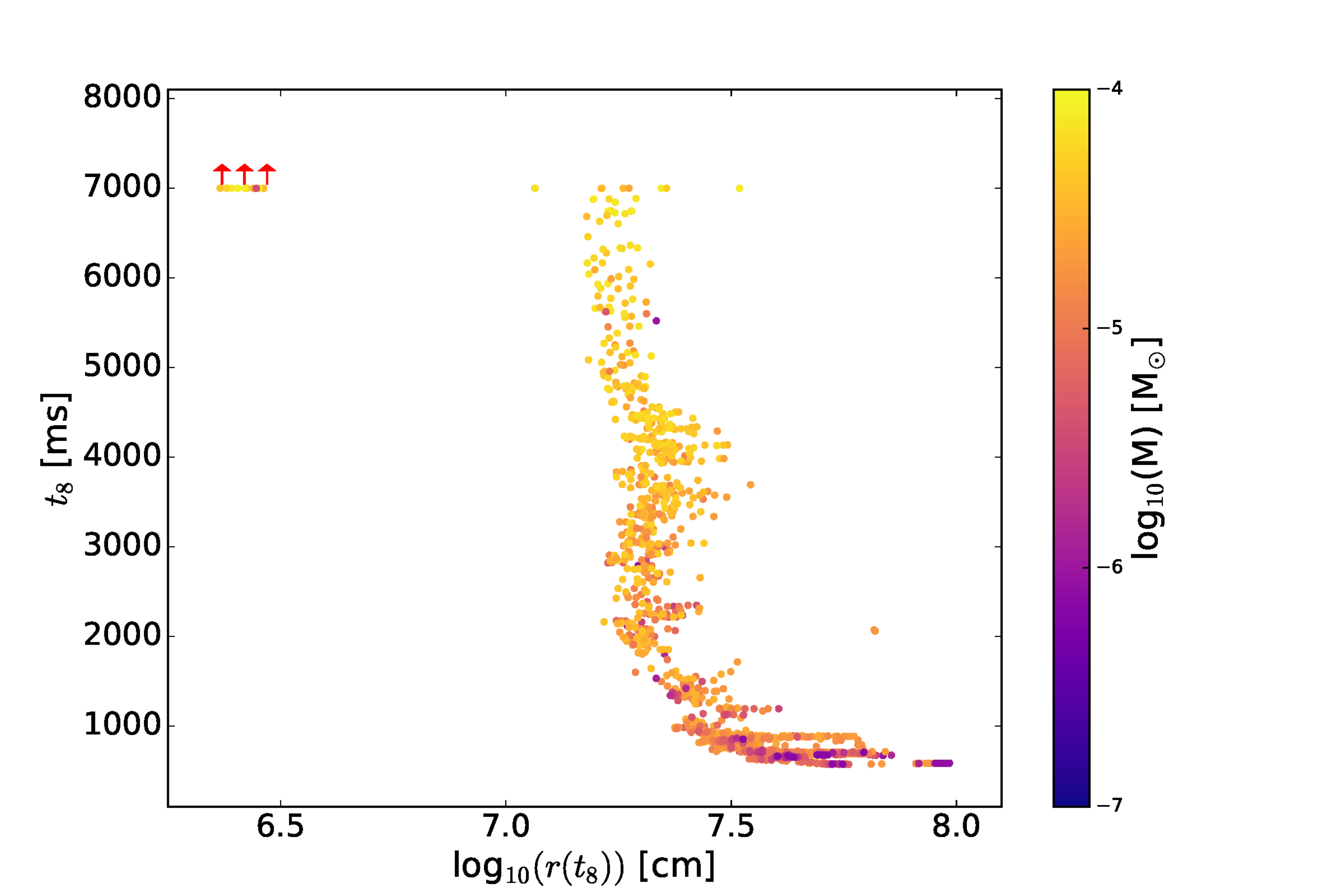}
\caption{Top: $Y_e$ values of the tracers in the hottest temperature bin at the time $t=t_8$ vs. $t_8$ for the 11.2~M$_{\odot}$ model (left) and the 17.0~M$_{\odot}$ model (right). Bottom: $t_8$ vs. position at $t=t_8$. The red arrows indicate that some tracers have a temperature higher than $T_{NSE} = 8$~GK at the end of the simulation.}
\label{fig:yefor8gk} 
\end{figure*}

The properties of the tracer particles in the highest temperature bin are extremely heterogenic and highly dependent on the shock evolution in the supernova simulation. Therefore, we cannot expect a robust abundance pattern for this bin across different explosion models. Instead, we make a secondary distinction by dividing (only) these particles into different bins according to their initial $Y_e$ value (i.e., the $Y_e$ value at the time where the temperature drops below $T_9=8$, in the following labelled as $t_8$).
Figure~\ref{fig:yefor8gk} correlates $t_8$, $Y_e\left(t=t_8\right)$, and the position at $t_8$, $r\left(t=t_8\right)$ in the simulation for both the 11.2~M$_{\odot}$ (left) and the 17.0~M$_{\odot}$ (right) progenitors. For model 11.2, it can be clearly seen that the particles with proton-rich conditions at $t_8$ reach $T_9 = 8$ in a limited time interval $500~\textrm{ms}~<~t~<~2500~\textrm{ms}$, while all other tracers have a $Y_e\left(t=t_8\right)$ close to 0.5. In some tracers $Y_e$ can attain very high values up to 0.9 (see section~\ref{sec:highYe} for further discussion). For 17.0, the distribution is more homogenic, with only few particles having a $Y_e\left(T_9=8\right)$ value below 0.45 or above 0.6. A handful of particles that we count as ejected are hotter than 8~GK at the end of the simulations. They are visible in the top panels as the few particles with a $Y_e~<<0.5$ at $t~=~7760~\left(7000\right)$~ms, the time where the simulations stop. The special case of these tracers will be discussed below. The rough correlation between the masses of the individual particles and $t_8$ also provides a hint about the origin of the tracer particles: The particles that come from further outward (representing smaller masses) drop below 8~GK early, at $Y_e$ values generally around 0.5 (or slightly below in the 11.2~M$_{\odot}$ model).


One can also study the relationship of $t_8$ and $r\left(t=t_8\right)$ (bottom panels in figure~\ref{fig:yefor8gk}). This graph shows that apart from the ones with a small $t_8$, all particles drop below 8~GK roughly at the same distance to the PNS. This argues for a small change with time in the temperature profiles of the 2D simulations after 0.5~s. The particles that do not fall below 8~GK stand out again at the upper edge with a small radius. As they have a temperature well above 8~GK at the end of the simulation, their $t_8$ (and possibly $r\left(t=t_8\right)$) would be larger than the one shown here, assuming that they become gravitationally unbound at all. This is indicated by red arrows.


\begin{figure}[b]
\includegraphics[width=0.54\textwidth]{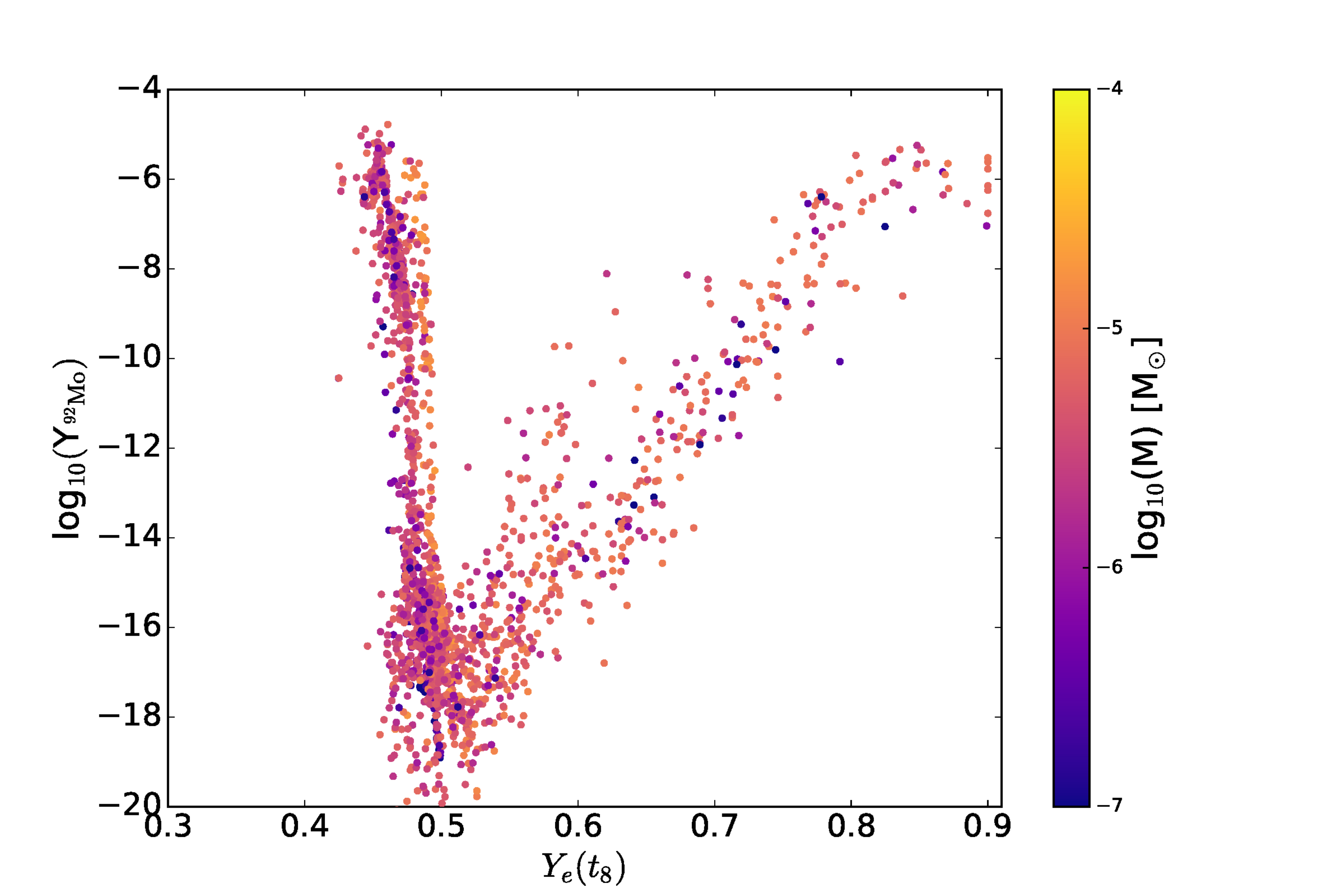}
\includegraphics[width=0.54\textwidth]{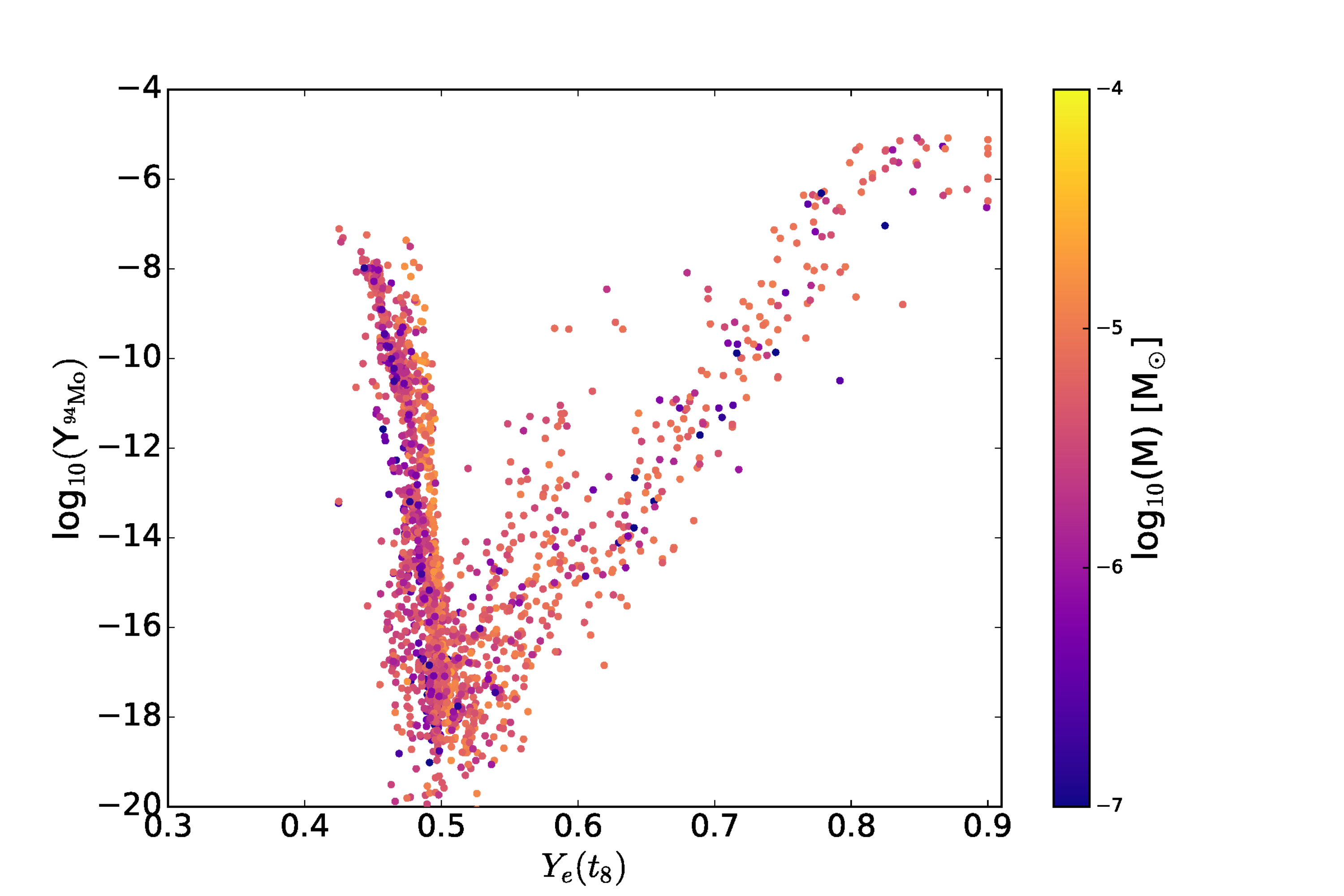} \\
\includegraphics[width=0.54\textwidth]{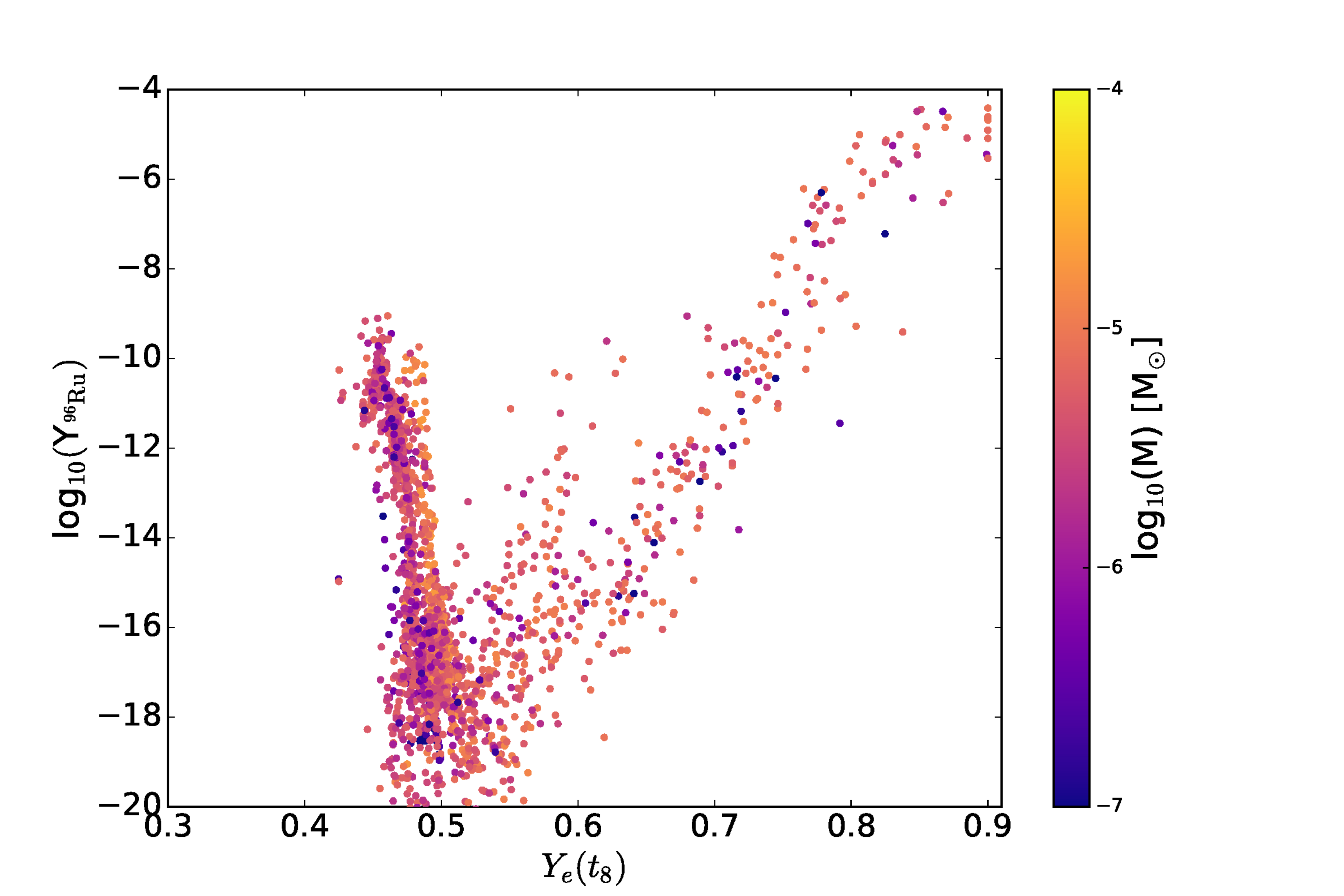}
\includegraphics[width=0.54\textwidth]{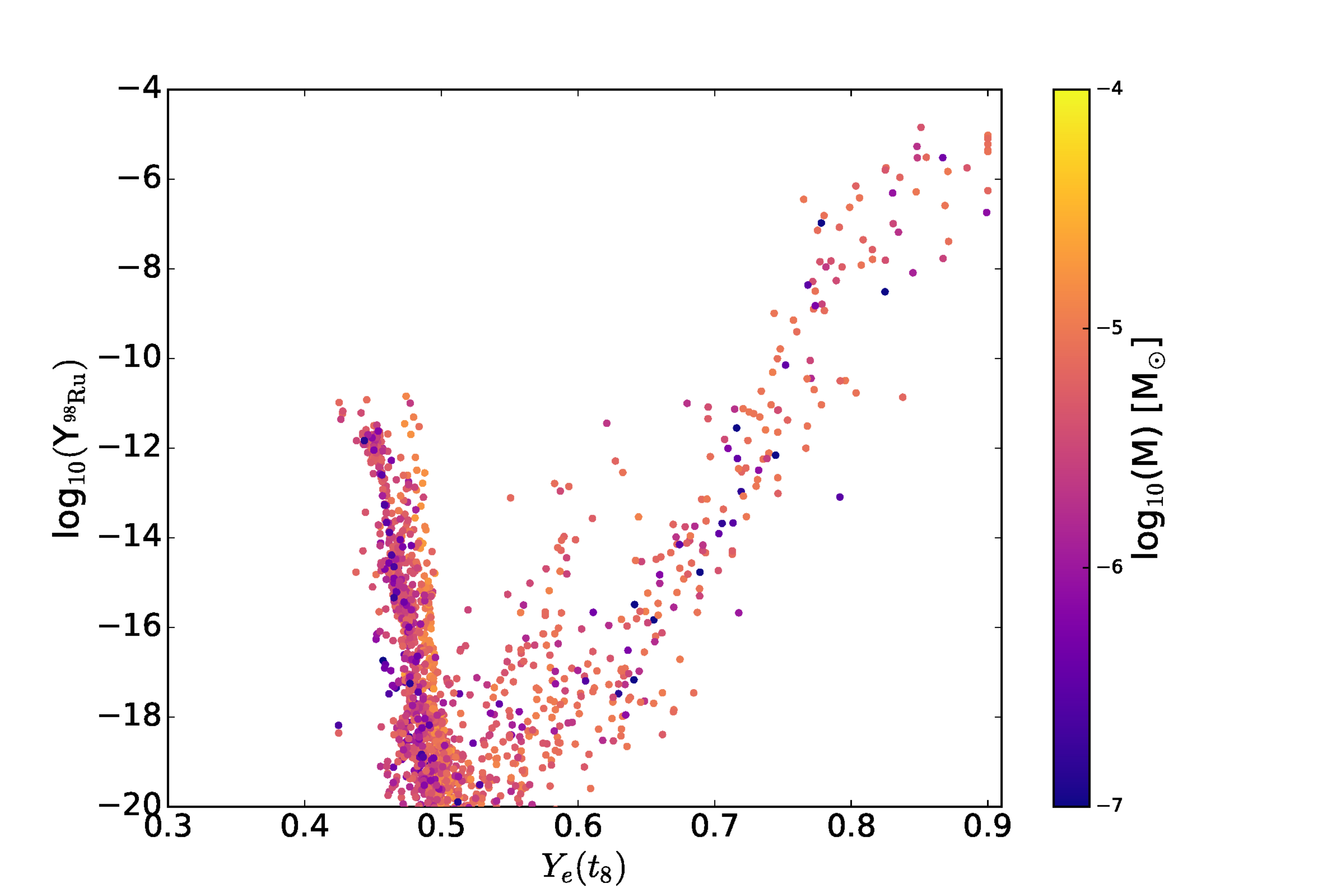}
\caption{Abundances of $^{92}$Mo (top left), $^{94}$Mo (top right), $^{96}$Ru (bottom left), and $^{98}$Ru (bottom right) plotted versus initial $Y_e$ for all ejected tracer particles with $T_9^{\rm peak}>8$ for model 11.2.}
\label{fig:11.2_morupeaks} 
\end{figure}


\begin{figure*}
\includegraphics[width=0.54\textwidth]{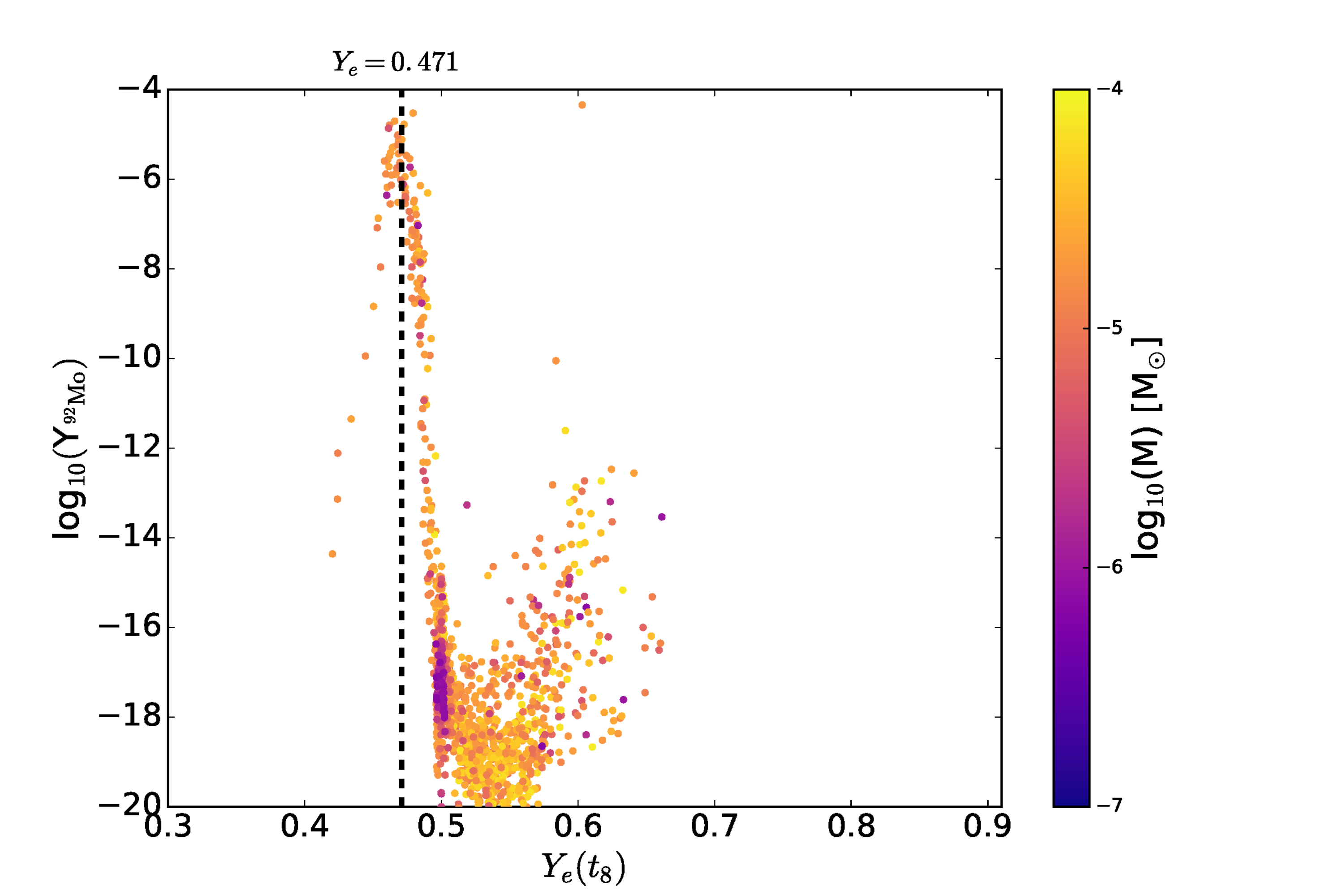}
\includegraphics[width=0.54\textwidth]{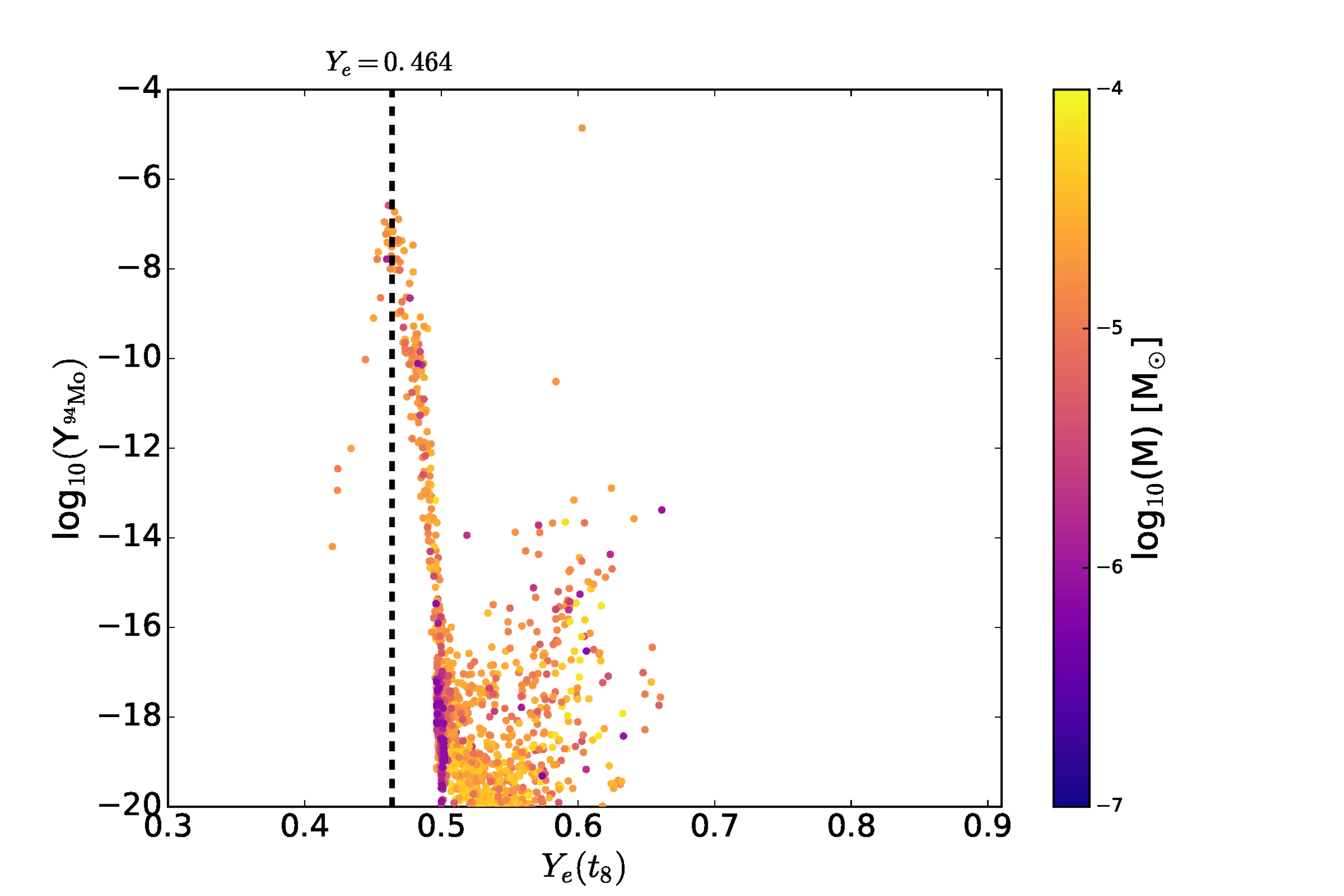} \\
\includegraphics[width=0.54\textwidth]{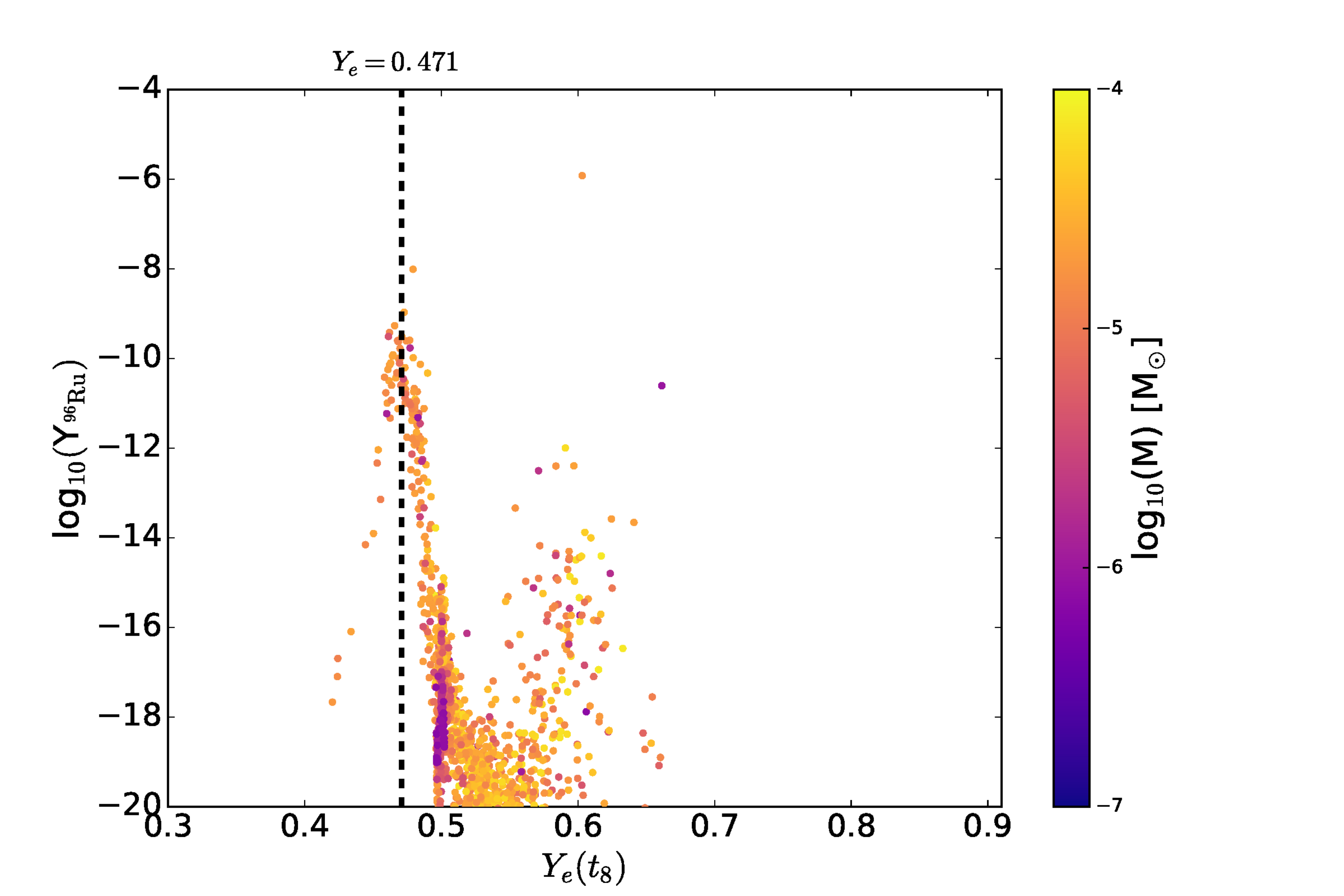}
\includegraphics[width=0.54\textwidth]{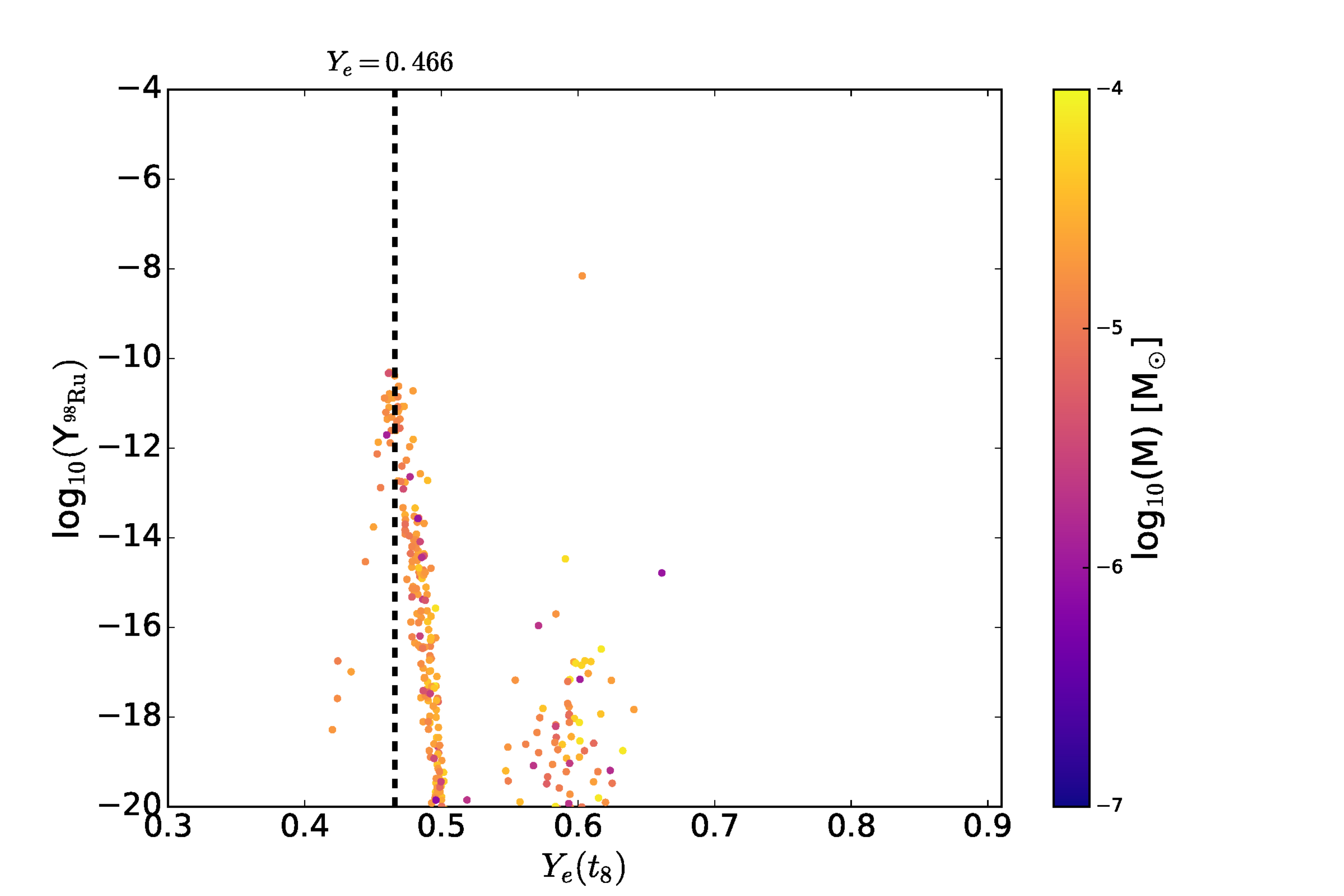}
\caption{Same as figure~\ref{fig:11.2_morupeaks}, but for model 17.0. The vertical lines indicate the positions of the abundance peaks.}
\label{fig:17.0_morupeaks} 
\end{figure*}

Figure~\ref{fig:yefor8gk} provides a concise picture of the ejection behaviour of the hottest tracers in the 11.2~M$_{\odot}$ (17.0~M$_{\odot}$) simulation as follows: The particles that represent a relatively small mass (M~$\leq 10^{-6}$~M$_{\odot}$) drop below 8~GK first, at a large distance to the centre. Since they are ejected straight away, they are not strongly affected by neutrino reactions, and their $Y_e$ value is slightly lower than 0.5 (around 0.5). In the first hundreds of miliseconds $r\left(t=t_8\right)$ quickly moves inwards and then stays more or less constant at about 150~(200)~km. In the 11.2~M$_{\odot}$ model, $t_8$ for all the proton-rich particles is between 500~ms and 2500~ms (and at a radius $1.3~\times~10^7~\textrm{cm}~<~r~<~2.8~\times~10^7~\textrm{cm}$), while for the 17.0~M$_{\odot}$ model no clear trend can be observed. The path of the twelve (eight) tracers with $T_9~>~8$ at the end of the simulation can also be traced. In the first few miliseconds after core bounce, they are drawn towards the centre and heated up well above 8~GK, where they remain until the end of the simulation. In addition, their density and electron fraction evolutions suggest that they could very well be accreted onto the PNS and not ejected at all. We will therefore not consider them ejected, since they also have a very high density at the end of the simulation (e.g., $\rho_{\rm fin}~\approx~5~\times~10^{10}$~g~cm$^{-3}$).

In the following, we try to find reasonable boundary values for the $Y_e$ bins, with the aim to group tracers with similar compositions together. One obvious choice is to group tracers with non-negligible abundances of proton-rich isotopes together. In order to do this, we plot the final abundances of $^{92}$Mo, $^{94}$Mo, $^{96}$Ru, and $^{98}$Ru, respectively, as a function of the initial $Y_e$ in figures~\ref{fig:11.2_morupeaks}~\&~\ref{fig:17.0_morupeaks}, where it becomes clear that there are two $Y_e$ regimes to produce these proton-rich isotopes: While the high-$Y_e$ particles ($Y_e~\geq~0.65$) contribute to the production of all four isotopes via the $\nu$p-process \cite{froehlich2006,pruet2006,wanajo2006}, slightly neutron-rich tracers with $Y_e~\leq~0.49$ produce the four isotopes to a varying degree, with a strong contribution to $^{92}$Mo and $^{94}$Mo, and much less $^{96}$Ru and $^{98}$Ru. In the low-$Y_e$ regime the four isotopes of interest here are produced only for a narrow range of $Y_e$ values, with the abundance peak at slightly different positions for every isotope. The abundances in the 11.2~M$_{\odot}$ model display a double-peak pattern in the low-$Y_e$ regime, indicating that $Y_e$ is not the only factor determining the $^{92,94}$Mo and $^{96,98}$Ru abundances. The significance of these results will be discussed in section~\ref{sec:mo92}. The 17.0~M$_{\odot}$ model does not expel tracer particles with $Y_e~>~0.67$, which naturally explains the high $^{92,94}$Mo abundances and the relatively low $^{96,98}$Ru abundances for this model as seen in figure~\ref{fig:xoverfe}.

Figures~\ref{fig:11.2_morupeaks}~\&~\ref{fig:17.0_morupeaks} provide natural boundaries for the $Y_e$ bins:
\begin{itemize}
\item bin \textsc{I}: $0.42<Y_e<0.49$; chosen such that tracers on the neutron-rich side which produce a lot of Mo are included.
\item bin \textsc{II}: $0.49<Y_e<0.55$; low Mo \& Ru abundances in Figures~\ref{fig:11.2_morupeaks}~\&~\ref{fig:17.0_morupeaks}, composition expected to be dominated by $\alpha$- and iron group nuclei.
\item bin \textsc{III}: $0.55<Y_e<0.67$; upper boundary chosen such that this bin includes all proton-rich tracers in the 17.0~M$_{\odot}$ model.
\item bin \textsc{IV}: $0.67<Y_e$; tracers with extremely high $Y_e$; lower boundary chosen such that no tracers in the 17.0~M$_{\odot}$ model appear in this bin.
\end{itemize}

Tables~\ref{tab:11.2yebins}~\&~\ref{tab:17.0yebins} list the number of tracer particles, the summed-up mass of the tracers, as well as the ejected $^{92}$Mo mass per bin for the two progenitors.

\begin{figure}[t]
\includegraphics[width=0.45\textwidth]{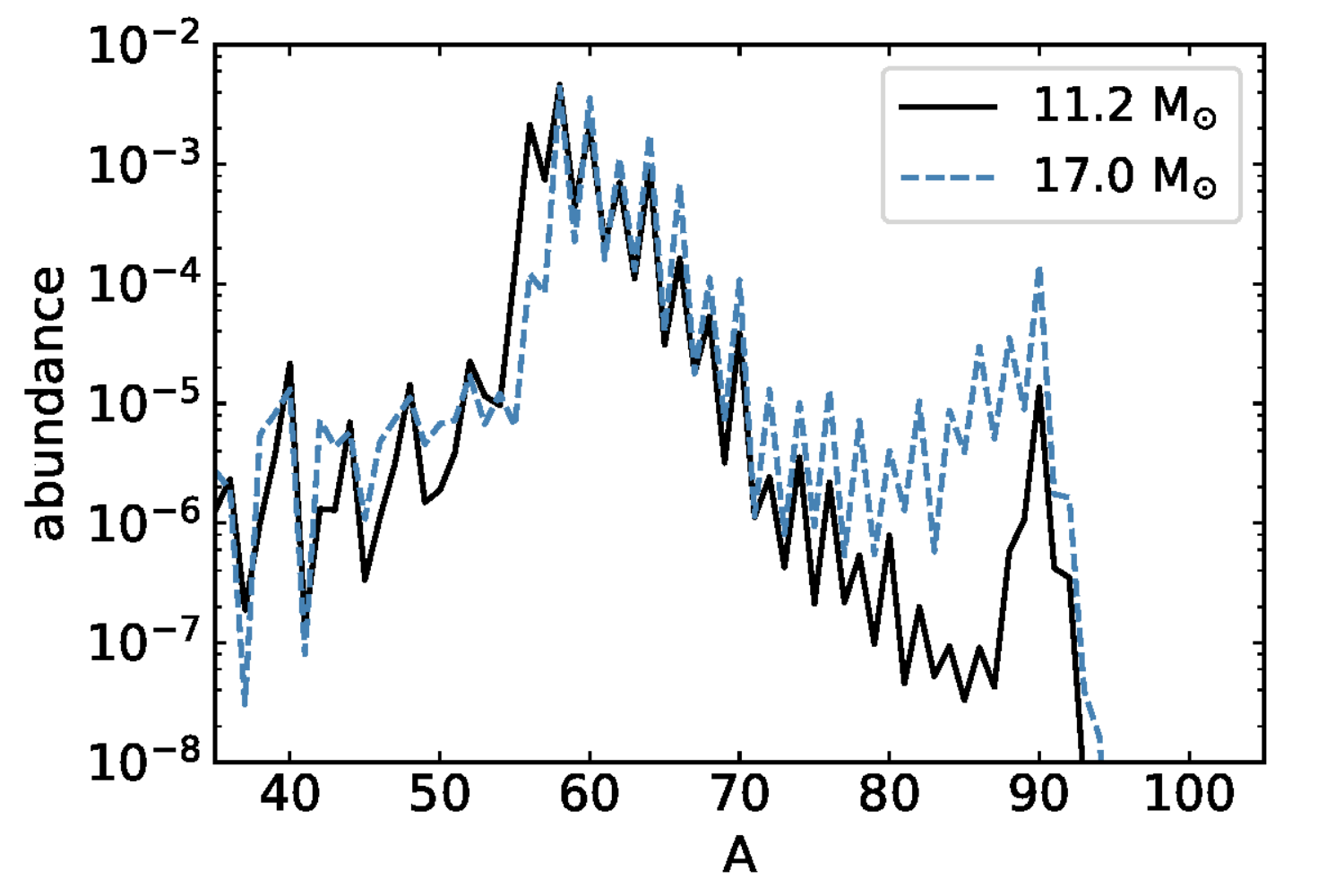}
\includegraphics[width=0.45\textwidth]{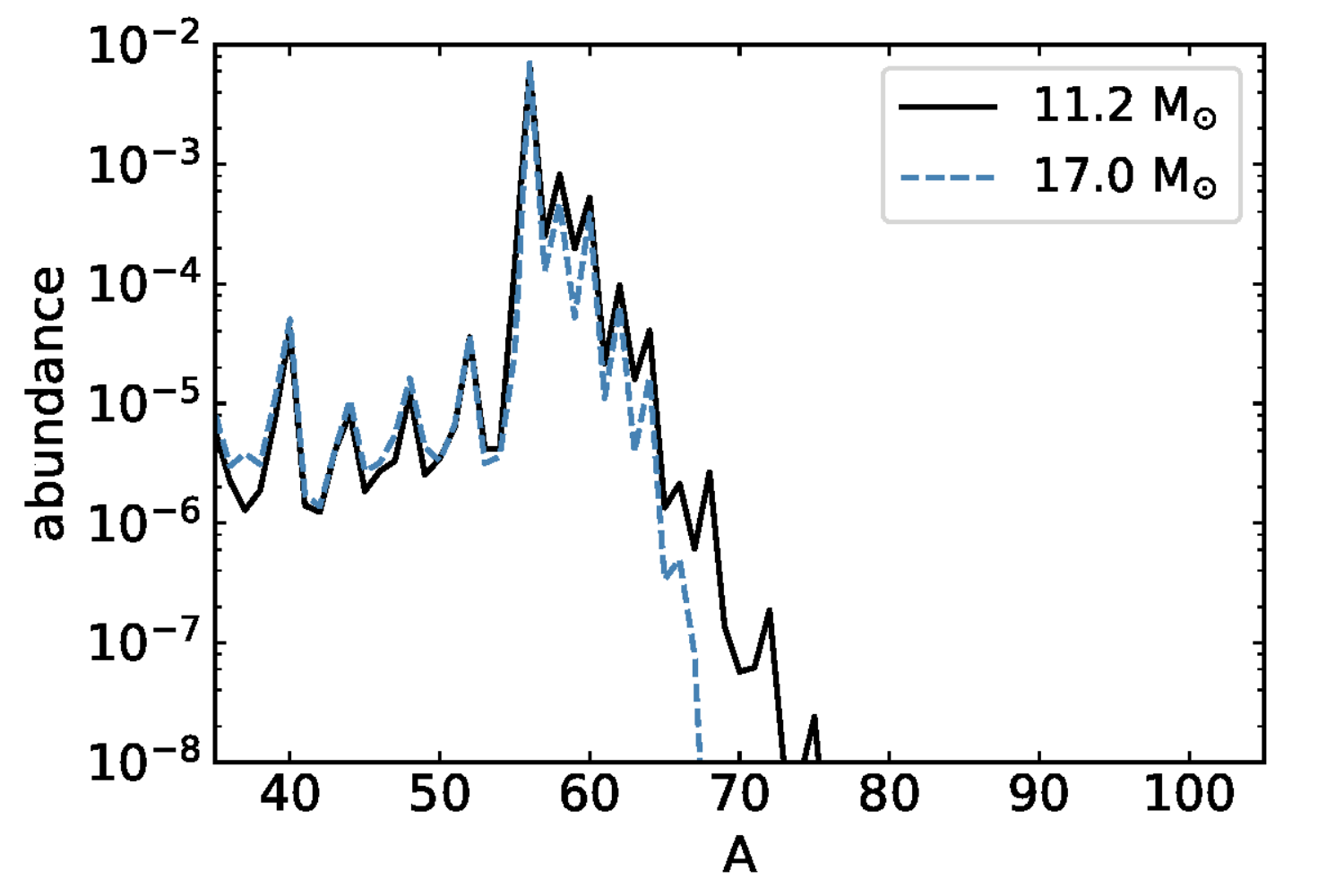} \\
\includegraphics[width=0.45\textwidth]{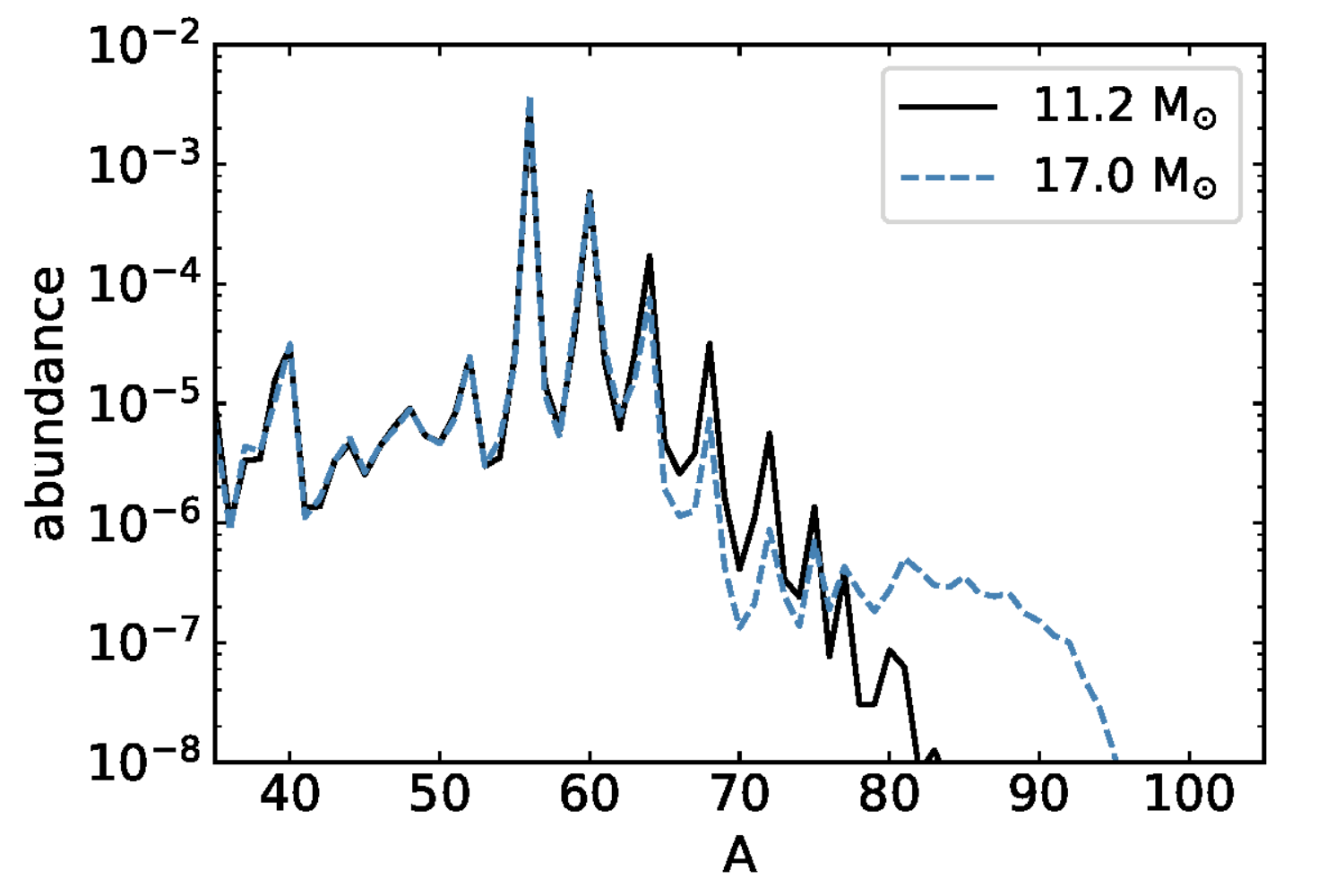}
\includegraphics[width=0.45\textwidth]{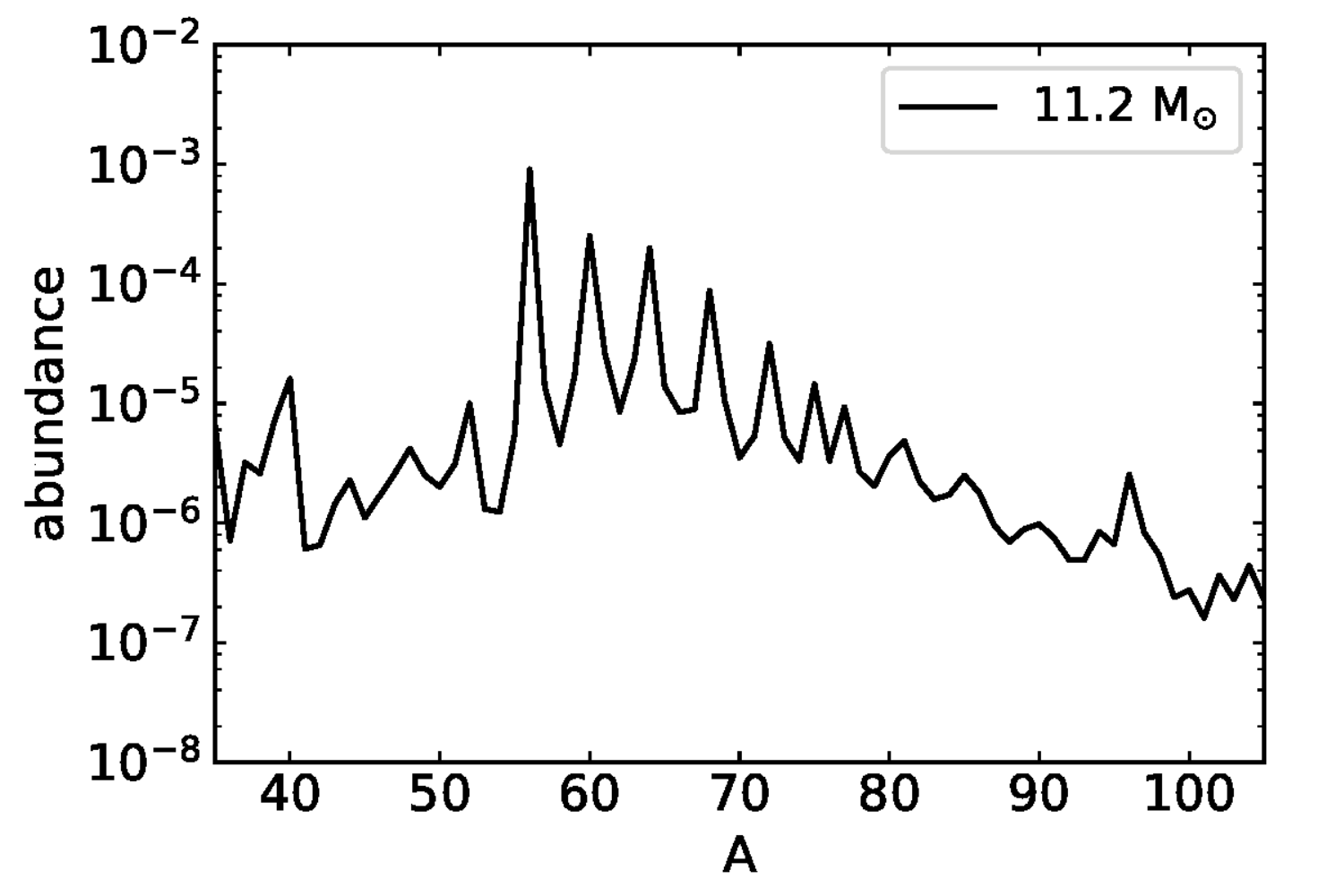}
\caption{Integrated abundances as a function of mass number for the individual $Y_e$ bins, and for both progenitors. Top left: $0.42~\leq~Y_e~<~0.49$, top right: $0.49~\leq~Y_e~<~0.55$, bottom left: $0.55~\leq~Y_e~<~0.67$, bottom right: $0.67~\leq~Y_e$. Note that the 17.0~M$_{\odot}$ progenitor does not have a contribution from the last bin.}
\label{fig:yebins_abunds} 
\end{figure}

More than 80\% (in both models) of the ejected $^{92}$Mo is produced by tracers from the low-$Y_e$ bin $0.42<Y_e<0.49$, although this bin makes up only about 36\% (8\%) of the total mass of the highest temperature bin. The production mechanism of $^{92}$Mo in slightly neutron-rich environments is similar to the one found in neutrino-driven winds \cite{hoffman1996} and is further discussed in section~\ref{sec:mo92}. In conditions with $Y_e \approx 0.5$ almost no Mo is produced, as the reaction flux stalls in the iron group nuclei. Only with $Y_e~>~0.6$ can the heavier nuclei on the proton-rich side of stability be produced again by means of the $\nu$p-process. Figure~\ref{fig:yebins_abunds} shows a quantitative comparison of the abundances of heavy nuclei, where the heterogenic nature of the highest-temperature bin becomes particularly clear. Nuclei beyond the iron group cannot be produced in conditions with $Y_e$ close to~0.5. An interesting contrast between the two bins producing the heavy nuclei is the sharp drop in abundances after $A=90$ in the first ($0.42~\leq~Y_e~<~0.49$) bin, while there is no clear limit in mass number for the $0.67~\leq~Y_e$ bin.

\begin{table}[h]
\caption{\label{tab:11.2yebins} Properties of the individual $Y_e$ bins, such as number of tracer particles N, integrated mass M$_{ej}$, and $^{92}$Mo yield M($^{92}$Mo), in the case of the 11.2~M$_{\odot}$ progenitor.}
\begin{indented}
\item[] \begin{tabular}{p{2.8cm} p{1cm} p{1.9cm} p{2cm}}
\br
Bin & N & M$_{\rm ej}$ [M$_{\odot}$] & M$_{^{92}\rm Mo}$ [M$_{\odot}$] \vspace{1px} \\
\mr
$\hspace{1.1cm} Y_e<0.42$ & 0 & 0 & 0 \\
$0.42<Y_e<0.49$ & 858 & $3.87\times10^{-3}$ & $1.23\times10^{-7}$ \\
$0.49<Y_e<0.55$ & 708 & $4.59\times10^{-3}$ & $8.75\times10^{-12}$ \\
$0.55<Y_e<0.67$ & 207 & $1.32\times10^{-3}$ & $2.61\times10^{-12}$ \\
$0.67<Y_e$ & 142 & $8.29\times10^{-4}$ & $3.04\times10^{-8}$ \\
\br
\end{tabular}
\end{indented}
\end{table}

\begin{table}[h]
\caption{\label{tab:17.0yebins} Same as table~\ref{tab:11.2yebins}, this time for the 17.0~M$_{\odot}$ progenitor.}
\begin{indented}
\item[] \begin{tabular}{p{2.8cm} p{1cm} p{1.9cm} p{2cm}}
\br
Bin & N & M$_{\rm ej}$ [M$_{\odot}$] & M$_{^{92}\rm Mo}$ [M$_{\odot}$] \vspace{1px}\\
\mr
$\hspace{1.1cm} Y_e<0.42$ & 0 & 0 & 0 \\
$0.42<Y_e<0.49$ & 132 & $2.44\times10^{-3}$ & $3.62\times10^{-7}$ \\
$0.49<Y_e<0.55$ & 926 & $1.91\times10^{-2}$ & $6.87\times10^{-12}$ \\
$0.55<Y_e<0.67$ & 347 & $9.27\times10^{-3}$ & $7.73\times10^{-8}$ \\
$0.67<Y_e$ & 0 & 0 & 0 \\
\br
\end{tabular}
\end{indented}
\vspace{0.5cm}
\end{table}


\subsection{Production of $^{92,94}$Mo in neutron-rich conditions}
\label{sec:mo92}
In the previous section we have shown that while the proton-rich isotopes of Ru are almost exclusively produced in the $\nu$p-process, the final abundances of $^{92,94}$Mo are the result of an overlay of very proton-rich as well as slightly neutron-rich tracers in the 11.2~M$_{\odot}$ model. For the 17.0~M$_{\odot}$ model, on the other hand, the $\nu$p-process does not occur, and therefore the only contribution to the production of $^{92,94}$Mo comes from the neutron-rich tracers. The production mechanism in $0.4~<Y_e~<0.49$ environments will be analyzed in this section.
The build-up of $^{92,94}$Mo proceeds in several consecutive phases:

\noindent 1) It is a well-known feature of multi-D simulations that the expansion of tracer particles occurs on slow timescales due to the gradual onset of the explosion. The hottest tracers in particular spend a long time in the dense central region, followed by a slow expansion after the normal (charged-particle) freeze-out. This means that the $\alpha$-particles are converted into heavier $\alpha$-isotopes (and especially iron group nuclei) quite efficiently.

\noindent 2) Starting just below the iron group nuclei, the valley of stability experiences a kink in the $N-Z$ plane, turning away from the $N=Z$ symmetry towards the $N>Z$ side. Since nuclei along the valley of stability are by default the tightest bound nuclei in each isotopic chain, captures of free neutrons (in addition to $\alpha$-captures) are required to synthesize heavier nuclei. Through a series of neutron, $\alpha$-, and proton captures the nucleosynthesis proceeds to higher proton numbers up to $^{90}$Zr along the line of Z/A~$\approx Y_e$. In this phase the free neutrons are depleted, and the proton-to-neutron ratio $Y_p / Y_n$ evolves from the initial value of $1/2$ to $10^3 / 1$. Although free neutrons are depleted, the electron fraction $Y_e$ does not change greatly and is still below 0.5. The temperature in this phase drops from $T_9=4$ to $T_9=2.8$.

\noindent 3) Yttrium ($Z=39$) has only one stable isotope, $^{89}$Y. Therefore, $^{89}$Y and $^{90}$Zr represent a bottleneck in the reaction flow and virtually all the nuclei reaching a mass number of 90 pass through $^{90}$Zr. At the time when the reaction flow has reached $^{90}$Zr, the neutron abundance is so low that $^{90}$Zr($n,\gamma$)$^{91}$Zr and $^{90}$Zr($p,\gamma$)$^{91}$Nb compete at comparable rates. Note that $\alpha$-captures are greatly reduced for $N=50$ nuclei \cite{moeller1997}. At $^{91}$Nb another branching occurs between ($n,\gamma$) and ($p,\gamma$), with the branching ratio again around 1:1. Therefore, around 25\% of the flow going out from $^{90}$Zr ends up at $^{92}$Mo via $^{90}$Zr($p,\gamma$)$^{91}$Nb($p,\gamma$)$^{92}$Mo.

\noindent 4) Once synthesized, $^{92}$Mo is not destroyed easily, since the temperatures at this point are around 1~GK, such that proton captures and photodissociations are not effective anymore and the other major destruction channel, ($n,\gamma$), is suppressed by the low neutron abundance. This is also the reason why the reaction flux stops at Mo, i.e. for the production of Ru another mechanism is required (such as the $\nu$p-process).

In the search of the origin of the solar $^{92,94}$Mo \& $^{96,98}$Ru abundances, slightly neutron-rich conditions in neutrino-driven winds of CCSNe have been considered before \cite{hoffman1996,arconesbliss2014,arconesmontes2011,bliss2014, blissinprep}. It is of particular interest to understand the conditions that lead to large abundances for these isotopes. Moreover, comparison of the isotopic ratios $Y_{^{94} \rm Mo} / Y_{^{92} \rm Mo}$ and $Y_{^{98} \rm Ru} / Y_{^{96} \rm Ru}$ with observational values can provide valuable insights into the mechanism that leads to the production of these isotopes. The abundance peaks at low $Y_e$ in figures~\ref{fig:11.2_morupeaks}~\&~\ref{fig:17.0_morupeaks} are different for any given isotope: $^{92}$Mo and $^{96}$Ru have their peak around $Y_e~=~0.471$, $^{94}$Mo around $Y_e~=~0.464$, and $^{98}$Ru around $Y_e~=~0.466$. Together with the fact that the peaks are very narrow, this may prove an interesting aspect in the search for the origin of the solar isotopic ratio of Mo and Ru, as different distributions of $Y_e$ values in the ejected tracer particles can easily lead to varying isotopic $Y_{^{92} \rm Mo} / Y_{^{94} \rm Mo}$ ratios and, to a lesser degree, $Y_{^{96} \rm Ru} / Y_{^{98} \rm Ru}$ ratios. The isotopic ratios can also be influenced by the presence or absence and the $Y_e$ distribution of proton-rich tracer particles hosting the $\nu$p-process. This can be verified by comparing figures~\ref{fig:11.2_morupeaks}~\&~\ref{fig:17.0_morupeaks}, where the high-$Y_e$ material directly determines the [Ru/Fe] value (see figure~\ref{fig:xoverfe}).


\begin{figure}[t]
\centering
\includegraphics[width=0.8\textwidth]{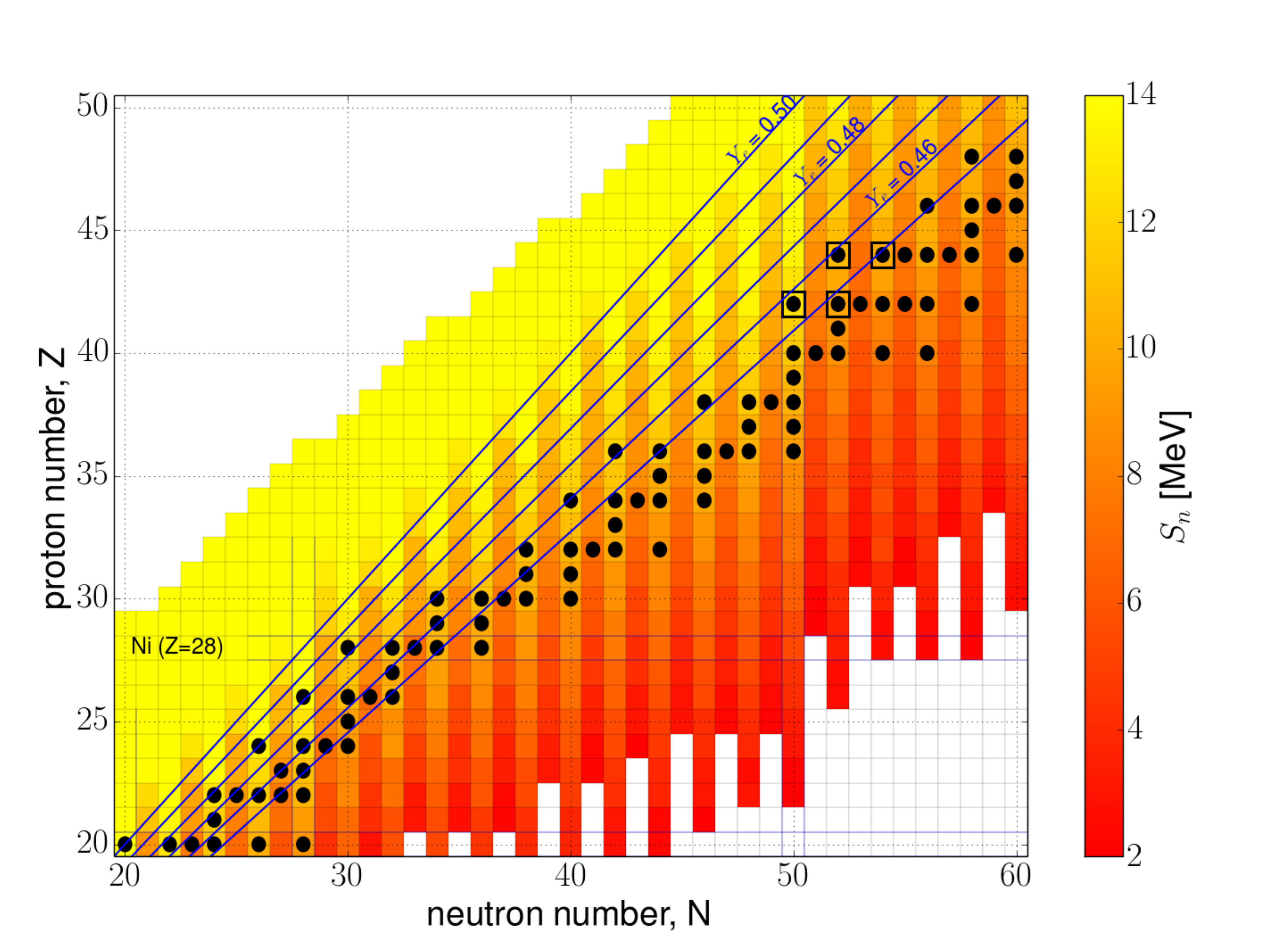}
\caption{Neutron separation energies from Ca to Sn. The diagonal lines represent lines of equal electron fraction, with a difference of~0.01 between two lines. The black dots mark stable nuclei and the four isotopes $^{92,94}$Mo and $^{96,98}$Ru can be identified by the black frames.}
\label{fig:sn}
\end{figure}

Comparing the absolute abundances for the four isotopes discussed here (figures~\ref{fig:11.2_morupeaks}~\&~\ref{fig:17.0_morupeaks}), one can see that the low-$Y_e$ abundances of $^{92}$Mo are larger by several orders of magnitude than the abundances of the other three species. The reason for this lies in the fact that in neutron-rich conditions all heavy nuclei are produced from lighter nuclei, i.e., almost all the reaction flow towards $^{94}$Mo and $^{96,98}$Ru passes through $^{92}$Mo. In figure~\ref{fig:sn}, the neutron separation energies $S_n$ are plotted in this region of the nuclear chart. As an overlay, lines of equal $Y_e$ are plotted, demonstrating the conditions under which $^{92}$Mo can be produced easily. The neutron separation energy experiences a sharp drop at $N=51$, thus making neutron captures on $N=50$ nuclei less likely than on lighter isotopes and, at the same time, considerably reducing the lifetime of $^{93}$Mo (and $^{91}$Zr) against ($\gamma,n$). The production of $N~\geq~51$ nuclei is therefore countered effectively by ($\gamma,n$) reactions at high temperatures and by the depletion of free neutrons by the time the temperature has decreased to lower values.


\subsection{Isotopic Mo and Ru Ratios in Presolar Grains}
In our calculations we find efficient production of $^{92}$Mo and $^{94}$Mo, in the case of the 11.2~M$_{\odot}$ model combined with large yields of $^{96}$Ru and $^{98}$Ru (see figure~\ref{fig:xoverfe}). The origin in the universe of these relatively neutron-deficient stable isotopes is still uncertain \cite{fisker2009}. Recent comparisons of measured elemental abundances in metal-poor stars and isotopic abundances in presolar grains \cite{hansen2014} suggest that Mo has several astrophysical sources, while the elemental abundance of Ru correlates with the Ag abundance in metal-poor stars, which points to the (weak) r-process as the major production mechanism, albeit only for the neutron-rich stable isotopes $^{99,101,102,104}$Ru. For $^{96}$Ru and $^{98}$Ru, the $\nu$p-process in core-collapse supernovae and the $\gamma$-process in supernovae type Ia are expected to be the main contributors. Recent models of SNe Ia show the production of these isotopes via a p-process \cite{travaglio2015}, leading to the conclusion that the solar abundance ratios between these isotopes are the result of an interplay of CCSNe and SNe Ia, while other sources cannot be excluded. The best observational evidence of CCSN ejecta compositions are presolar grains in meteorites, as they formed before the solar system and probably experienced less pollution from SNe Ia or r-process events. The subclass X of SiC grains, in particular, are thought to have formed from dust ejected in CCSNe, since they show a large $^{48}$Ca enrichment, a decay product of $^{48}$Ti which is produced exclusively in CCSNe. In the following we compare our yields to SiC X grains from the \textit{presolar database}\footnote{http://presolar.wustl.edu/$\sim$pgd/welcome.html} \cite{hynesgyngard2009} with respect to abundance ratios between $^{92}$Mo, $^{94}$Mo, $^{96}$Ru, and $^{98}$Ru. All grains for which these isotopes have been measured come from the \textit{Murchison} meteorite and we summarize their relative abundances (as reported in \cite{pellin2006}) in table~\ref{tab:grains}. The error ranges given here take into account the errors for both isotopes of the ratio, and should therefore represent an absolute upper and lower limit, respectively, since systematic errors enter twice. 

\begin{table}
\caption{\label{tab:grains} Measured isotopic Mo and Ru ratios in presolar SiC grains of class X as reported in \cite{pellin2006} with the terrestrial (standard) ratios taken from \cite{nicolussi1998} in the case of Mo and \cite{andersgrevesse1989} for Ru. For completeness the solar ratios are also given.}
\begin{indented}
\item[] \begin{tabular}{ccc}
\br
label & $\frac{\rm Y \left(^{94}\rm Mo \right)}{\rm Y \left(^{92}\rm Mo \right)}$ & $\frac{\rm Y \left(^{98}\rm Ru \right)}{\rm Y \left(^{96}\rm Ru \right)}$ \vspace{0.1cm} \\
\mr
100-2 & 0.715$^{+0.137}_{-0.116}$ &   \vspace{0.1cm}                        \\
113-2 & 0.560$^{+0.079}_{-0.069}$ &   \vspace{0.1cm}                        \\
113-3 & 0.512$^{+0.079}_{-0.069}$ &   \vspace{0.1cm}                         \\
133-1 & 0.714$^{+0.072}_{-0.066}$ &   \vspace{0.1cm}                         \\
153-8 & 0.550$^{+0.136}_{-0.111}$ &   \vspace{0.1cm}                         \\
209-1 & 0.622$^{+0.087}_{-0.078}$ &   \vspace{0.1cm}                         \\
B2-05 & 0.462$^{+0.067}_{-0.059}$ &   \vspace{0.1cm}                         \\
E2-10 & 0.739$^{+0.152}_{-0.127}$ &   \vspace{0.1cm}                         \\
322-1 &                           & 0.190$^{+0.071}_{-0.055}$ \vspace{0.1cm} \\
H - 2 &                           & 0.304$^{+0.048}_{-0.043}$ \vspace{0.1cm} \\
\hline
solar & 0.630 & 0.337 \\
\br
\end{tabular}
\end{indented}
\end{table}

We can also derive values for $Y \left(^{96,98}\rm Ru \right)$/$Y \left(^{92,94}\rm Mo \right)$ using
\begin{equation}
\frac{Y \left(^{96,98}\rm Ru \right)}{Y \left(^{92,94}\rm Mo \right)} = \frac{ \left(\frac{Y \left(^{98}\rm Ru \right)}{Y \left(^{96}\rm Ru \right)} + 1 \right)~Y \left(^{96}\rm Ru \right)}{ \left(\frac{Y \left(^{94}\rm Mo \right)}{Y \left(^{92}\rm Mo \right)} + 1 \right)~Y \left(^{92}\rm Mo \right)},
\end{equation}
with the $Y \left(^{96}\rm Ru \right)$ and $Y \left(^{92}\rm Mo \right)$ values from Anders~\&~Grevesse \cite{andersgrevesse1989} and the Ru and Mo isotopic ratios from table~\ref{tab:grains}. All the calculated values are close to 0.225, which is also the solar ratio.


\begin{figure}
\centering
\includegraphics[width=0.8\textwidth]{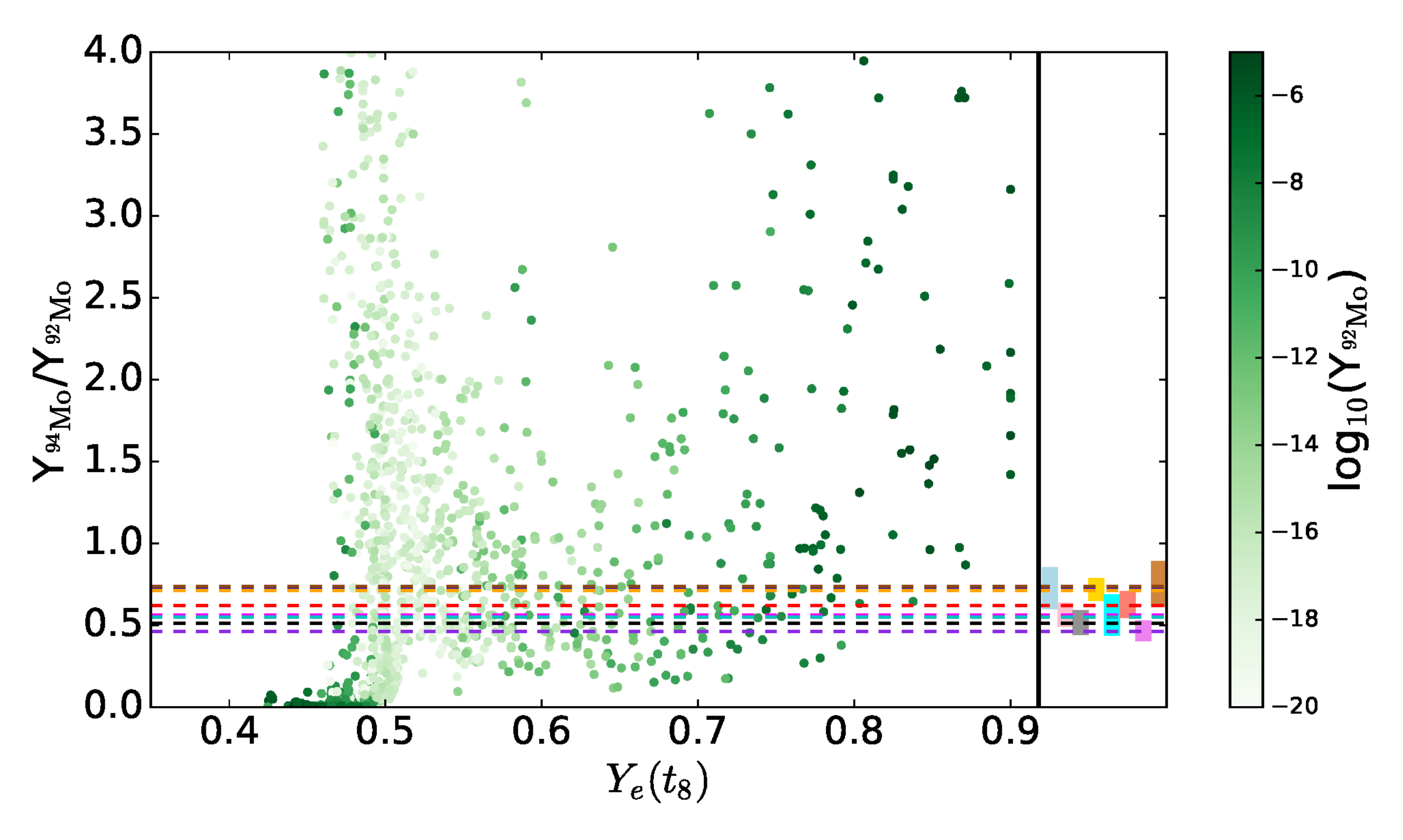} \\
\includegraphics[width=0.8\textwidth]{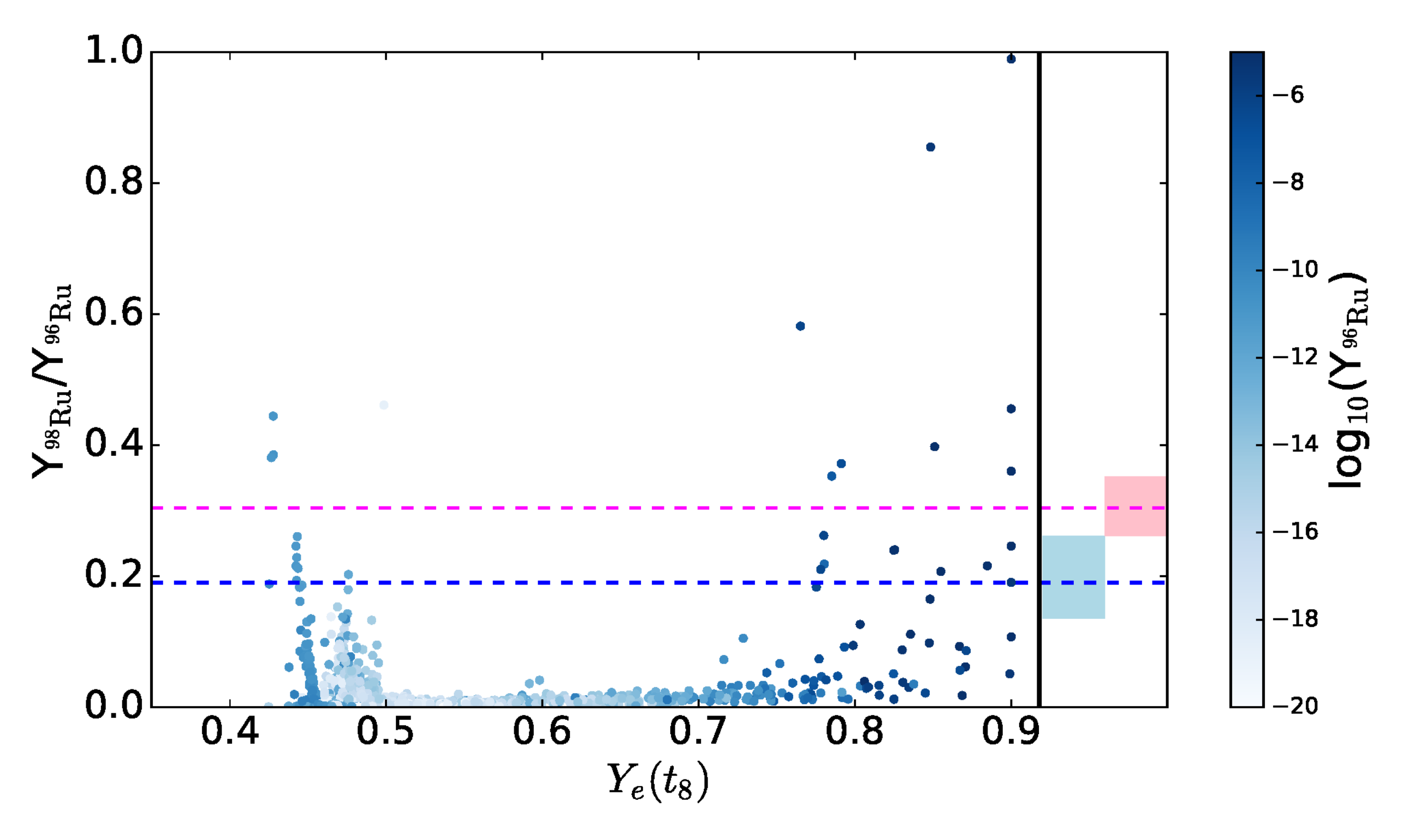}
\caption{$Y_{^{94}\rm Mo} / Y_{^{92}\rm Mo}$ (top) and $Y_{^{98}\rm Ru} / Y_{^{96}\rm Ru}$ (bottom) ratios plotted against Y$_e$ at $t=t_8$ (i.e., where T$_9 = 8$) for tracer particles in the 11.2~M$_{\odot}$ model. The dashed horizontal lines indicate the isotopic ratios measured in presolar grains (see table~\ref{tab:grains}), with the individual error ranges given as coloured boxes at the right-hand side of the Figure.}
\label{fig:moru_ratio} 
\end{figure}

For our analysis we first want to identify the corresponding isotopic ratios for our tracers depending on their initial electron fraction. Figure~\ref{fig:moru_ratio} shows the isotopic ratios in our tracer particles for the 11.2~M$_{\odot}$ model, along with the measured ratios from the presolar SiC X grains which are indicated by horizontal dashed lines. The $Y_{^{94}\rm Mo} / Y_{^{92}\rm Mo}$ ratio in the presolar grains can be reached for $Y_e > 0.5$, with a large scatter between the different tracer particles. The two regimes where $^{92}$Mo can be produced efficiently are characterized by very low $Y_{^{94}\rm Mo} / Y_{^{92}\rm Mo}$ ratios ($Y_e < 0.49$), and values larger than the ratios in the presolar grains ($Y_e > 0.67$), respectively. The $Y_{^{98}\rm Ru}/Y_{^{96}\rm Ru}$ ratio (bottom panel) shows a different behaviour, with some low-Y$_e$ tracer particles showing a ratio comparable to the measured ratios. However, since the absolute Ru abundances in this regime are extremely low, only the high-Y$_e$ regime is left as a viable production site in CCSNe. Under the conditions encountered in our 11.2~M$_{\odot}$ model, the $Y_{^{98}\rm Ru}/Y_{^{96}\rm Ru}$ ratio produced in the $\nu$p-process matches the measured ratio in presolar grains well.

In the following we describe a method to recombine tracer particles from our CCSN models in order to explore the possibilities to reproduce the isotopic $Y_{^{94}\rm Mo}/Y_{^{92}\rm Mo}$ and $Y_{^{98}\rm Ru}/Y_{^{96}\rm Ru}$ ratios found in presolar grains. To this end, we further divide our tracer particles into smaller Y$_e$ bins with a resolution $\Delta Y_e$ smaller than the differences in abundance peak positions in figures~\ref{fig:11.2_morupeaks}~\&~\ref{fig:17.0_morupeaks}, i.e. $\Delta Y_e \leq 0.002$. Similar to the procedure above, we introduce smaller $Y_e$ bins in the range $0.445~<~Y_e~<~0.477$, focusing on the $Y_{^{94}\rm Mo}/Y_{^{92}\rm Mo}$, $Y_{^{98}\rm Ru}/Y_{^{96}\rm Ru}$, and the $Y_{^{96,98}\rm Ru}/Y_{^{92,94}\rm Mo}$ ratios in the individual bins and testing if the ratios are robust across SN models. By combining tracer particles from different bins~$i$ with (arbitrarily) chosen contribution factors~$\alpha_i$, we can construct a large variety of isotopic ratios:

\begin{equation}
\label{eq:ratios}
\begin{aligned}
\frac{Y_{^{94}\rm Mo}}{Y_{^{92}\rm Mo}} &= \frac{ \sum_{i} \alpha_i \left( \frac{Y_{^{94}\rm Mo}}{Y_{^{92}\rm Mo}} \right)_i Y_{^{92}\rm Mo, i}}{ \sum_{i} \alpha_i Y_{^{92}\rm Mo, i}}  \\
\\
\frac{Y_{^{96,98}\rm Ru}}{Y_{^{92,94}\rm Mo}} &= \frac{ \sum_{i} \alpha_i \left( \frac{Y_{^{96,98}\rm Ru}}{Y_{^{92,94}\rm Mo}} \right)_i \left( Y_{^{92}\rm Mo, i} + Y_{^{94}\rm Mo, i} \right) }{ \sum_{i} \alpha_i \left( Y_{^{92}\rm Mo, i} + Y_{^{94}\rm Mo, i} \right)}
\end{aligned}
\end{equation}
where $i$ is the bin index (henceforth referred to as \textit{particle type}) and the $\alpha_i$ are weighting factors which can be interpreted as the summed-up mass share of each bin. The Ru ratio can be calculated the same way as the Mo ratio. For both the 11.2 and the 17.0~M$_{\odot}$ model the abundances and their ratios for each bin are given in tables~\ref{tab:ratios11.2}~\&~\ref{tab:ratios17.0}.

\begin{table}[b]
\caption{\label{tab:ratios11.2} Integrated $^{92}$Mo and $^{96}$Ru abundances as well as $Y_{^{94}\rm Mo} / Y_{^{92}\rm Mo}$ (labelled $^{94}\rm Mo / ^{92}\rm Mo$), $Y_{^{98}\rm Ru} / Y_{^{96}\rm Ru}$ (labelled $^{98}\rm Ru / ^{96}\rm Ru$), and $Y_{^{96,98}\rm Ru}$/$Y_{^{92,94}\rm Mo}$ (labelled $^{96,98}\rm Ru / ^{92,94}\rm Mo$) ratios for more refined $Y_e$ bins on the neutron-rich side and the highest-$Y_e$ bin $0.67~<~Y_e~<~0.90$ for the 11.2~M$_{\odot}$ model. For the second expression in Eq.~\eqref{eq:ratios} $Y_{^{94}\rm Mo, i}$ is needed. It can be calculated via $Y_{^{94}\rm Mo, i} = \left( Y_{^{94}\rm Mo} / Y_{^{92}\rm Mo} \right)_i Y_{^{92}\rm Mo, i}$.}
\begin{indented}
\item[] \begin{tabular}{ccccccc}
\br
$i$ & $Y_e$ bin & $\frac{^{94}\rm Mo}{^{92}\rm Mo}$ & $Y_{^{92}\rm Mo}$ & $\frac{^{98}\rm Ru}{^{96}\rm Ru}$ & $Y_{^{96}\rm Ru}$ & $\frac{^{96,98}\rm Ru}{^{92,94}\rm Mo}$\\
\mr
1 & $0.445-0.447$ & 0.011 & $2.1 \times 10^{-6}$  & 0.094 & $5.9 \times 10^{-11}$ & $3.0 \times 10^{-5}$\\
2 & $0.447-0.449$ & 0.006 & $8.1 \times 10^{-7}$  & 0.046 & $2.1 \times 10^{-11}$ & $2.7 \times 10^{-5}$\\
3 & $0.449-0.451$ & 0.005 & $1.2 \times 10^{-6}$  & 0.035 & $3.8 \times 10^{-11}$ & $3.2 \times 10^{-5}$\\
4 & $0.451-0.453$ & 0.005 & $9.9 \times 10^{-7}$  & 0.038 & $2.7 \times 10^{-11}$ & $2.9 \times 10^{-5}$\\
5 & $0.453-0.455$ & 0.002 & $2.3 \times 10^{-6}$  & 0.010 & $9.6 \times 10^{-11}$ & $4.3 \times 10^{-5}$\\
6 & $0.455-0.457$ & 0.001 & $1.4 \times 10^{-6}$  & 0.006 & $6.1 \times 10^{-11}$ & $4.4 \times 10^{-5}$\\
7 & $0.457-0.459$ & <0.001 & $8.2 \times 10^{-7}$ & 0.003 & $3.9 \times 10^{-11}$ & $4.7 \times 10^{-5}$\\
8 & $0.459-0.461$ & <0.001 & $2.0 \times 10^{-6}$ & 0.003 & $1.1 \times 10^{-10}$ & $5.3 \times 10^{-5}$\\
9 & $0.461-0.463$ & <0.001 & $6.5 \times 10^{-7}$ & 0.001 & $3.9 \times 10^{-11}$ & $6.0 \times 10^{-5}$\\
10 & $0.463-0.465$ & <0.001 & $5.8 \times 10^{-8}$ & 0.002 & $4.0 \times 10^{-12}$ & $6.8 \times 10^{-5}$\\
11 & $0.465-0.467$ & 0.001 & $2.3 \times 10^{-8}$  & 0.001 & $1.8 \times 10^{-12}$ & $8.1 \times 10^{-5}$\\
12 & $0.467-0.469$ & <0.001 & $5.7 \times 10^{-8}$ & 0.001 & $5.1 \times 10^{-12}$ & $9.0 \times 10^{-5}$\\
13 & $0.469-0.471$ & 0.002 & $1.6 \times 10^{-8}$  & 0.013 & $1.2 \times 10^{-12}$ & $7.9 \times 10^{-5}$\\
14 & $0.471-0.473$ & 0.001 & $1.2 \times 10^{-8}$  & 0.006 & $1.2 \times 10^{-12}$ & $1.1 \times 10^{-4}$\\
15 & $0.473-0.475$ & 0.013 & $2.0 \times 10^{-7}$  & 0.101 & $9.1 \times 10^{-12}$ & $4.9 \times 10^{-5}$\\
16 & $0.475-0.477$ & 0.001 & $7.7 \times 10^{-9}$  & 0.011 & $6.3 \times 10^{-13}$ & $8.3 \times 10^{-5}$ \vspace{0.05cm}\\
\hline \vspace{-0.25cm}\\
17 & $0.67 - 0.90$   & 2.109 & $3.9 \times 10^{-7}$ & 0.213 & $2.1 \times 10^{-6}$ & 2.464\\
\br
\end{tabular}
\end{indented}
\end{table}

\begin{table}[t]
\caption{\label{tab:ratios17.0} Same as table~\ref{tab:ratios11.2}, but for the 17.0~M$_{\odot}$ model. Note that for bins 1-4~and~17 there exist no tracer particles for this model.}
\begin{indented}
\item[] \begin{tabular}{ccccccc}
\br
$i$ & $Y_e$ bin & $\frac{^{94}\rm Mo}{^{92}\rm Mo}$ & $Y_{^{92}\rm Mo}$ & $\frac{^{98}\rm Ru}{^{96}\rm Ru}$ & $Y_{^{96}\rm Ru}$ & $\frac{^{96,98}\rm Ru}{^{92,94}\rm Mo}$\\
\mr
5 & $0.453-0.455$ & 0.181 & $1.2 \times 10^{-7}$  & 1.504 & $7.8 \times 10^{-13}$ & $1.4 \times 10^{-5}$ \\
6 & $0.455-0.457$ & 0.206 & $1.1 \times 10^{-8}$  & 1.832 & $7.2 \times 10^{-14}$ & $1.5 \times 10^{-5}$ \\
7 & $0.457-0.459$ & 0.044 & $2.5 \times 10^{-6}$  & 0.341 & $3.8 \times 10^{-11}$ & $1.9 \times 10^{-5}$ \\
8 & $0.459-0.461$ & 0.035 & $1.6 \times 10^{-6}$  & 0.264 & $2.9 \times 10^{-11}$ & $2.2 \times 10^{-5}$ \\
9 & $0.461-0.463$ & 0.020 & $4.8 \times 10^{-6}$  & 0.163 & $1.1 \times 10^{-10}$ & $2.5 \times 10^{-5}$ \\
10 & $0.463-0.465$ & 0.015 & $3.5 \times 10^{-6}$  & 0.122 & $7.9 \times 10^{-11}$ & $2.5 \times 10^{-5}$ \\
11 & $0.465-0.467$ & 0.009 & $9.7 \times 10^{-6}$  & 0.075 & $2.7 \times 10^{-10}$ & $2.9 \times 10^{-5}$ \\
12 & $0.467-0.469$ & 0.009 & $4.8 \times 10^{-6}$  & 0.059 & $1.4 \times 10^{-10}$ & $3.1 \times 10^{-5}$ \\
13 & $0.469-0.471$ & 0.004 & $2.5 \times 10^{-6}$  & 0.027 & $1.0 \times 10^{-10}$ & $4.2 \times 10^{-5}$ \\
14 & $0.471-0.473$ & 0.003 & $7.5 \times 10^{-6}$  & 0.012 & $4.2 \times 10^{-10}$ & $5.6 \times 10^{-5}$ \\
15 & $0.473-0.475$ & 0.001 & $8.5 \times 10^{-7}$  & 0.002 & $5.0 \times 10^{-11}$ & $5.9 \times 10^{-5}$ \\
16 & $0.475-0.477$ & 0.002 & $2.0 \times 10^{-6}$  & 0.004 & $1.7 \times 10^{-10}$ & $8.8 \times 10^{-5}$ \\
\br
\end{tabular}
\end{indented}
\end{table}

The trend in both the $Y_{^{94}\rm Mo}$/$Y_{^{92}\rm Mo}$ and the $Y_{^{98}\rm Ru}/Y_{^{96}\rm Ru}$ ratios is the same in the two CCSN models, as both ratios have a minimum around $Y_e \approx 0.47$. We have already established that Ru is almost exclusively produced by the $\nu$p-process, and the isotopic ratios for this bin differ a lot from the other bins. Therefore, the ratios required by presolar grains can only be achieved with a contribution from type~17 tracers. For instance, a combination of $\alpha_{17}/\alpha_1 = 2$ (with other $\alpha_i=0$) leads to ratios comparable to the values in SiC~X grains. While the $Y_{^{98}\rm Ru}~/~Y_{^{96}\rm Ru}$ ratio is determined exclusively by the value of bin~17, the overall Mo ratio for this combination is $Y_{^{94}\rm Mo}~/~Y_{^{92}\rm Mo}~=~0.670$.

No simultaneous $^{92,94}$Mo and $^{96,98}$Ru abundance measurements exist in any presolar grains, leaving only artificially derived ratios from the presolar grains and the solar $Y_{^{96,98}\rm Ru}/Y_{^{92,94}\rm Mo}$ ratio for the comparison across chemical elements. As tables~\ref{tab:ratios11.2}~\&~\ref{tab:ratios17.0} show, a very specific combination of our CCSN $Y_e$ bins would be required to simultaneously explain $Y_{^{94}\rm Mo}/Y_{^{92}\rm Mo}$ and the solar $Y_{^{96,98}\rm Ru}/Y_{^{92,94}\rm Mo}$ ratio, thus making it unlikely that the solar composition of these four isotopes originates exclusively from CCSNe of the type presented here.


\section{Discussion}
\label{sec:discussion}

\subsection{The high-$Y_e$ tracers in the 11.2~M$_{\odot}$ model}
\label{sec:highYe}
In our 11.2~M$_{\odot}$ model, we find tracer particles with very high $Y_e$ values at $t=t_8$ (see figure~\ref{fig:yefor8gk}). These $Y_e$ values are a consequence of the neutrino and anti-neutrino luminosities in the hydrodynamic simulation. The $\nu_e$ and $\overline{\nu}_e$ luminosities vary with time and with radial direction. In particular, we find a strong peak in $L_{\nu_e}$ around 1~s in the direction of the north pole, where most of the material is ejected, while the northward $L_{\overline{\nu}_e}$ experiences a minimum at the same time. In order to estimate the electron fraction, we can use a formula that is given in \cite{qianwoosley1996} for the neutrino-driven wind:
\begin{equation}
\label{eq:ye}
Y_e = \left[1 + \frac{L_{\overline{\nu}_e} \left( \epsilon_{\overline{\nu}_e} - 2\Delta + 1.2\Delta^2 / \epsilon_{\overline{\nu}_e} \right) }{L_{\nu_e} \left( \epsilon_{\nu_e} + 2\Delta + 1.2\Delta^2 / \epsilon_{\nu_e} \right) } \right]^{-1},
\end{equation}
with the electron neutrino and electron anti-neutrino luminosities, mean energies $L_{\nu_e}$, $L_{\overline{\nu}_e}$, $\epsilon_{\nu_e}$, $\epsilon_{\overline{\nu}_e}$, and the neutron-proton mass difference $\Delta = m_n-m_p = 1.293$~MeV. Note that this approach gives a good approximation of the electron fraction only if the expansion is slow enough such that a weak equilibrium is established and as long as the temperatures are high enough for the composition to be dominated by free nucleons. Using luminosities and mean energies from our 11.2~M$_{\odot}$ model, we find $Y_e=0.9$ for a short time in the direction of the north pole. Overall the equilibrium $Y_e$ evolves in a very similar fashion to the evolution of $Y_e \left(t=t_8 \right)$ in figure~\ref{fig:yefor8gk}. This gives us confidence about the $Y_e$ evolution in the simulations.

Nevertheless, we want to explore the possible uncertainties in $Y_e$ values arising from the employed neutrino transport and the EOS.
The IDSA, which was used in the simulations, has been well tested in the accretion phase up to several 100 ms post-bounce by Liebend\"orfer \etal \cite{liebendoerfer2009}. In this phase, it can reproduce the neutrino spectra and luminosities of detailed Boltztran simulations quite accurately. However, such a detailed comparison has not been done yet for the phase after the onset of the explosion and for the neutrino-driven wind. To estimate uncertainties due to the neutrino transport in the post-explosion phase of the supernova, we compared results from the codes AGILE-IDSA \cite{liebendoerfer2009} and AGILE-BOLTZTRAN \cite{liebendoerfer2004}, which use the same description for the hydrodynamics part AGILE and only differ in the method for the neutrino transport. In both codes we triggered explosions in spherical symmetry by artificially increasing the neutrino absorption rates at low densities and compared the results for up to three seconds after the onset of the explosion. To achieve a meaningful comparison of the two neutrino transport prescriptions, we chose a minimal setup, where in Boltztran only the same kind of neutrino-matter interactions are included as in IDSA, Newtonian relativity is used, and only electron-flavor neutrinos are considered. The luminosities agree very well (with a deviation below 20\% over the entire simulation) and for the mean energies a rather small maximum deviation on the order of 2~MeV was observed, where IDSA has a tendency to give slightly higher mean energies. This means IDSA works also well for the post-explosion phase, at least for this simplified setup. Note, however, that the electron-flavor neutrino quantities are also affected by the components which were not considered in the comparison, i.e.\ general relativity, treatment of $\mu / \tau$-neutrino, and of course the chosen set of neutrino-matter interactions. Especially the missing electron-neutrino scattering in IDSA could become relevant for the post-explosion phase, see \cite{fischer2012}. Its absence means that the neutrinos are less thermalized, the spectra are formed at slightly higher densities, and have somewhat higher mean energies. It is beyond the scope of the present study to estimate these uncertainties quantitatively, as the neutrino spectra evolution depends on many aspects of the simulation setup as mentioned already above. Neither would it be feasible for the very expensive multi-dimensional long-term simulations presented here.


\begin{figure}
\centering
\includegraphics[width=0.7\textwidth]{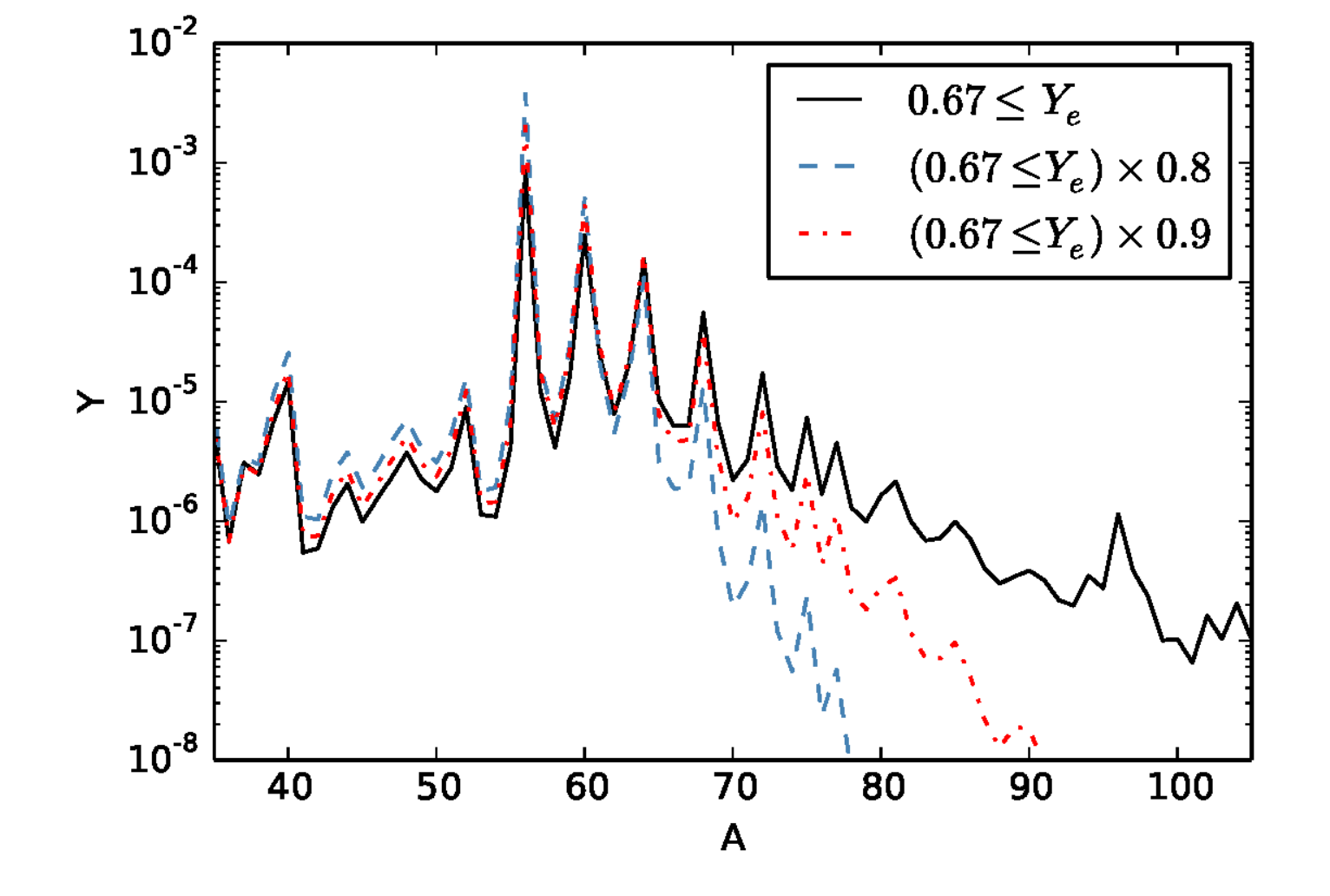}
\caption{Integrated composition of the tracers in the $0.67 \leq Y_e$ bin from the 11.2~M$_{\odot}$ model (see figure~\ref{fig:yebins_abunds}), compared to the compositions obtained when reducing the initial $Y_e$ of the same tracers by a factor of 0.8 and 0.9, respectively.}
\label{fig:ye_sensitivity} 
\end{figure}

In our simulation setup, the EOS allows for $Y_e$ values up to 0.9, but the proton fraction does not go beyond 0.6, the maximum proton fraction of the original EOS (LS220 \cite{lattimerswesty1991}). This leads to an underestimation of the $\overline{\nu}_e + p \rightarrow e^+ + n$ rate, which would decrease the electron fraction, and a runaway effect could occur, driving $Y_e$ closer to the maximum value of 0.9. We have tested the effect of a reduced initial electron fraction on the nucleosynthesis results for the tracer particles with the highest $Y_e$ in the 11.2~M$_{\odot}$ model (i.e., the $0.67 \leq Y_e$ bin). Multiplying $Y_e \left(t_8 \right)$ of the individual tracer particles with a factor of 0.8 and 0.9, respectively, the aim is to explore the sensitivity of the nuclear yields on the strength of the $\nu$p-process which is operating in these proton-rich conditions. Note that in our post-processing approach, the electron fraction is evolved in the nuclear network independently from the numerical simulation. Figure~\ref{fig:ye_sensitivity} presents the results of this test, demonstrating that the heaviest nuclei are only produced in the most proton-rich tracers. The new $Y_e$ range of the bin with a factor of~0.8 is $0.536 \leq Y_e \leq 0.72$, such that the $ \left(0.67 \leq Y_e \right) \times 0.8$ bin is comparable to the $0.55 \leq Y_e < 0.67$ bin in figure~\ref{fig:yebins_abunds}. This is confirmed by a comparison of the nuclear compositions of these two bins, which shows very good agreement. Figure~\ref{fig:ye_sensitivity} also demonstrates that the amount of produced Ru in our models directly depends on the contribution of very proton-rich ejecta. In order to produce Ru (and Mo) in the $\nu$p-process, electron fractions around $Y_e = 0.8$ or higher are needed. This is further confirmed by figure~\ref{fig:xoverfe}, where it can be seen that the Ru production in the 17.0~M$_{\odot}$ model is negligible, since there are no tracer particles with $Y_e > 0.67$.

\subsection{Low Ni Yields}
Both models discussed eject very small amounts of material. This is a direct consequence of the use of 2D simulations, which here are found to develop strong inflows around the equator where accretion onto the proto-neutron star occurs. As this accretion persists over the entire simulation time, its effect on the nucleosynthesis yields is enhanced by the long simulation time. Figure~\ref{fig:criteria_yields} demonstrates that almost all of the ejection criteria discussed would yield larger ejecta if the simulation stopped after 5s instead of 7s. This is especially true for the (heavier) products of the explosive nucleosynthesis, for which the yields are converged after 7s. In 3D models, the effect of late-time accretion is typically diminished, since three-dimensional effects (e.g., turbulence, SASI) tend to make the explosion more spherical (see, e.g., \cite{takiwaki2012,hanke2012,janka2016}). Thus, from the use of 2D models the yields of all the heavy nuclear species is rather underestimated. However, we show that all the heavy nuclei are affected to the same degree (\ref{app:ejeccrit}), which means that comparisons between the yields or abundances of heavy species are still viable.

On the other hand, the low mass of ejected $^{56}$Ni can be correlated to a special class of type~II~SNe, so-called \textit{faint SNe}, which are observed in nature. The low luminosities of these objects set upper limits between 10$^{-3} - 10^{-2}$~M$_{\odot}$ for the ejected $^{56}$Ni mass \cite{turatto1998,zampieri2003,pastorello2004,nomoto2006}. The first observation of such an event was SN~1997D, which, as reported in \cite{turatto1998}, was characterized by a low luminosity, low ejecta velocities, and an unusually red spectrum. They compared the observed light curve to theoretical models and obtained the best fit with a CCSN of a 26~M$_{\odot}$ with massive fallback, such that the ejected $^{56}$Ni mass was $2 \times 10^{-3}$. Smartt \etal \cite{smartt2009} conducted a survey of known SN type~II-P progenitors and concluded that faint CCSNe could originate from stars close to the low-mass end of the CCSN progenitor mass range (8.5~M$_{\odot}$ in their study). Moreover, they defined the \textit{red supergiant problem}, which states that no SN~II-P progenitor with $M~>~16.5$~M$_{\odot}$ has been identified in their sample, although they should not be harder to detect. If this value proves to be a true upper limit for type~II-P SNe, there would be a transition region where progenitors around that mass also produce a faint SN, as the explosion is successful, but a lot of fallback occurs. In this context, our results show faint CCSN at the low end of the mass spectrum (11.2~M$_{\odot}$) and an example of a faint CCSN close to the cutoff mass for successful type~II-P SNe (17.0~M$_{\odot}$).

\subsection{The Road to Solar Isotopic Mo and Ru Ratios}
The origin of the solar $^{92,94}$Mo and $^{96,98}$Ru abundances is still an unsolved problem. It has been shown that different astrophysical sites provide favourable conditions for the production of these isotopes: in (single-degenerate) SNe Ia a $\gamma$-process can operate, photodissociating seed nuclei previously produced by the s-process in the companion star \cite{kusakabe2011,travaglio2011,travaglio2015}. Neutrino-driven winds in CCSNe have also been studied extensively, with Hoffman \etal \cite{hoffman1996}  showing that neutron-rich conditions can lead to the production of the $^{92,94}$Mo isotopes and Bliss \& Arcones \cite{bliss2014} showing that this is not possible for $^{96,98}$Ru and that the solar ratio between the two Mo isotopes could not be achieved under any conditions. However, the observed solar abundance ratios cannot be achieved in either of the proposed scenarios. This is true especially for the $^{92}\textrm{Mo}/^{94}\textrm{Mo}$ ratio. What makes the problem even more pressing is that most nuclei involved are stable and therefore most nuclear reactions in question are experimentally well constrained \cite{dillmann2006}. Since $^{93}$Mo is unstable, $^{93}\textrm{Mo} \left(n,\gamma \right)^{94}\textrm{Mo}$ and its (reverse reaction) bear the largest uncertainty, as is also discussed in Ref.~\cite{travaglio2015}, who give an estimate of a factor of two as an upper limit to the uncertainty of this rate.

\subsection{Elemental compositions in extremely metal-poor stars as signatures of the first CCSNe}
\begin{figure}[p]
\centering
\includegraphics[width=0.44\textwidth]{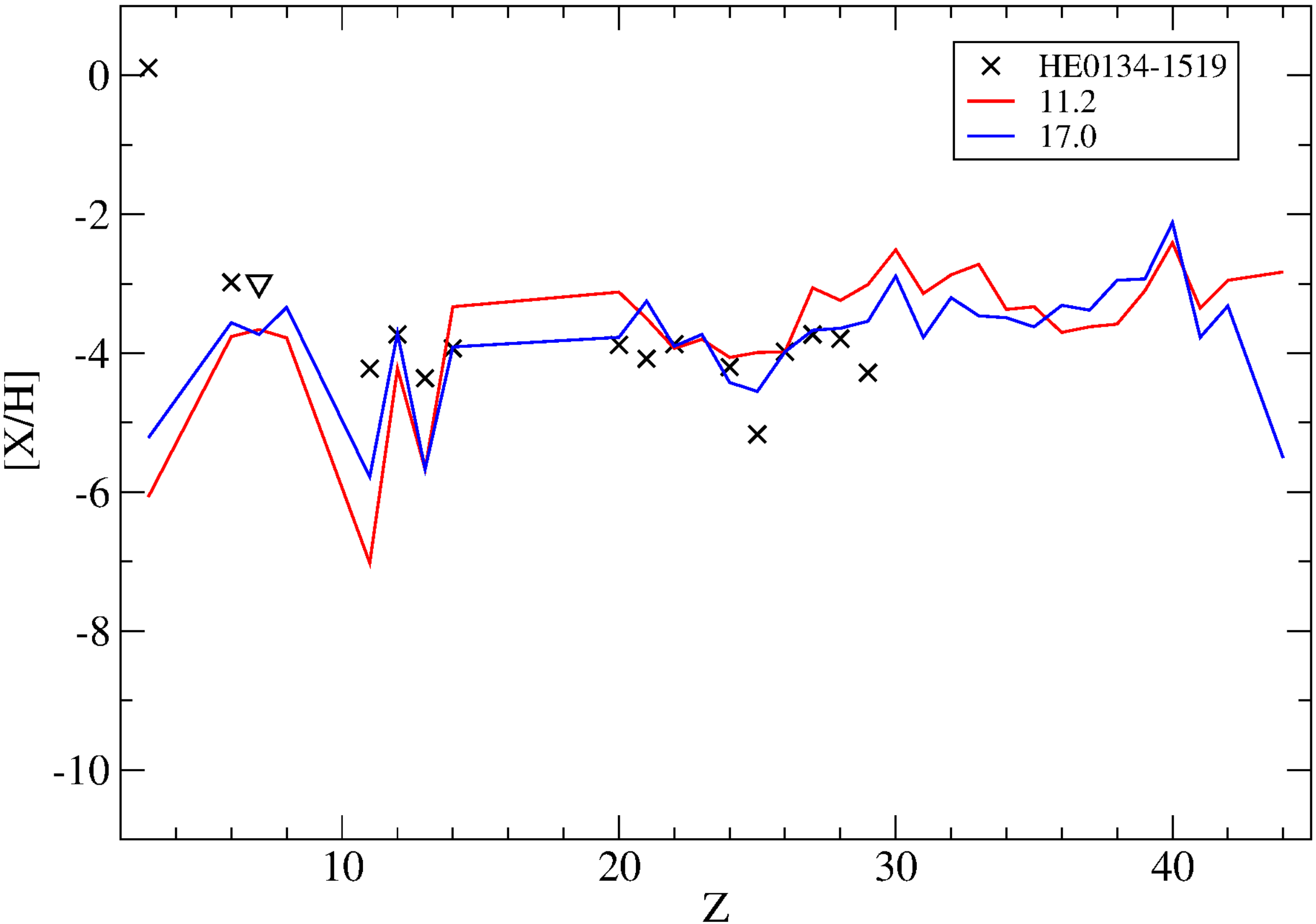}
\hspace{0.4cm} \includegraphics[width=0.44\textwidth]{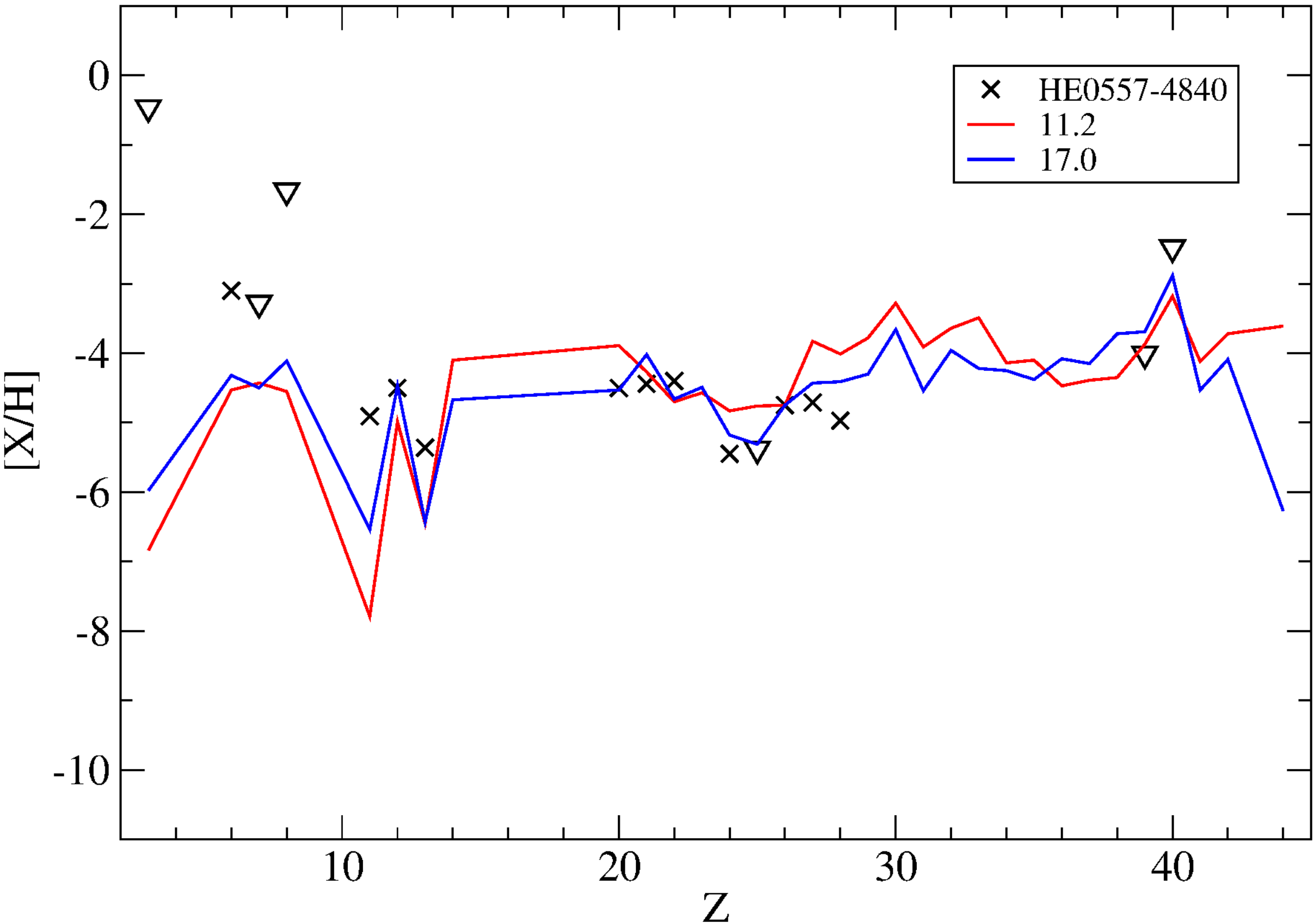}\\
\vspace{0.4cm} \includegraphics[width=0.44\textwidth]{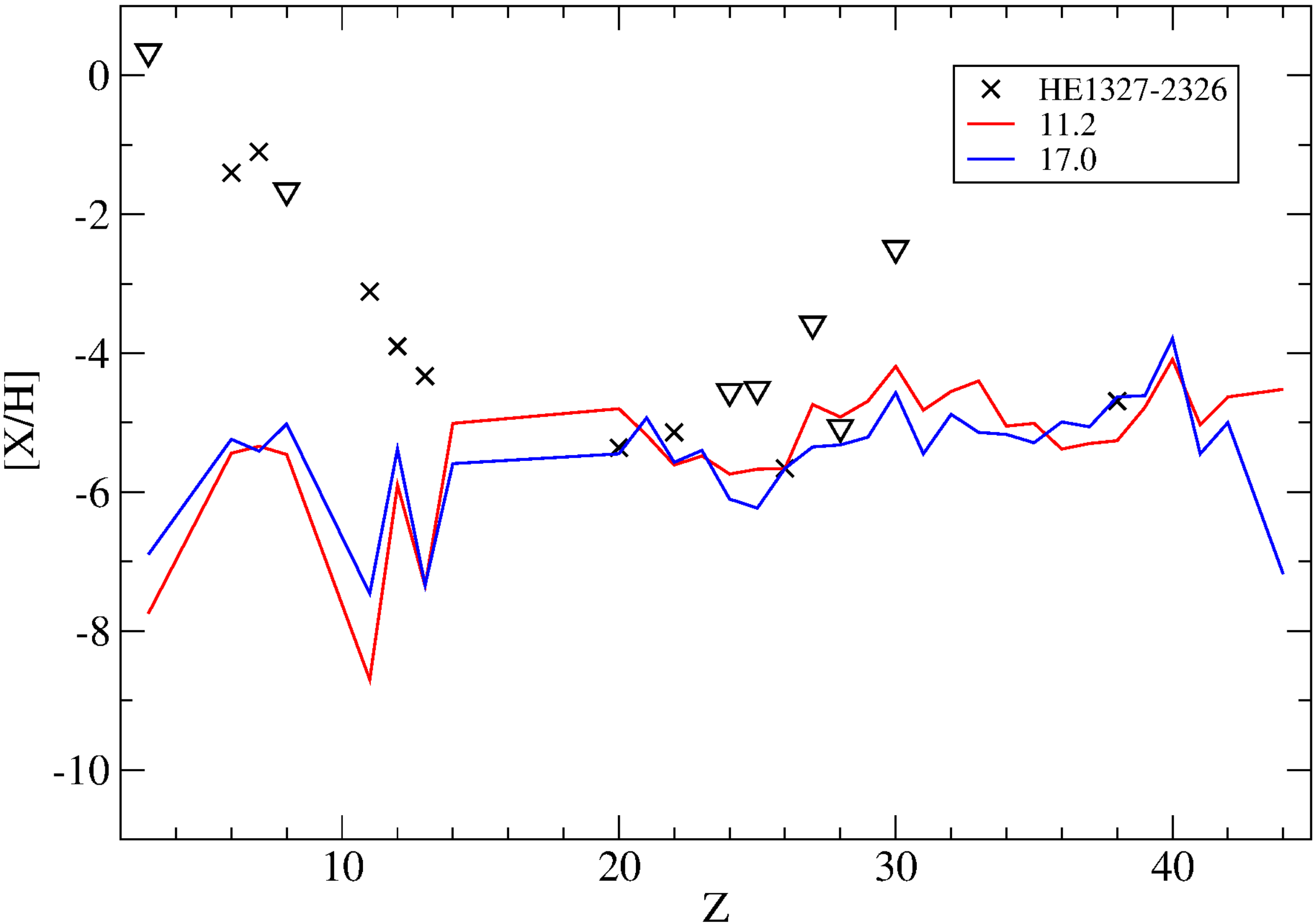}
\hspace{0.4cm} \includegraphics[width=0.44\textwidth]{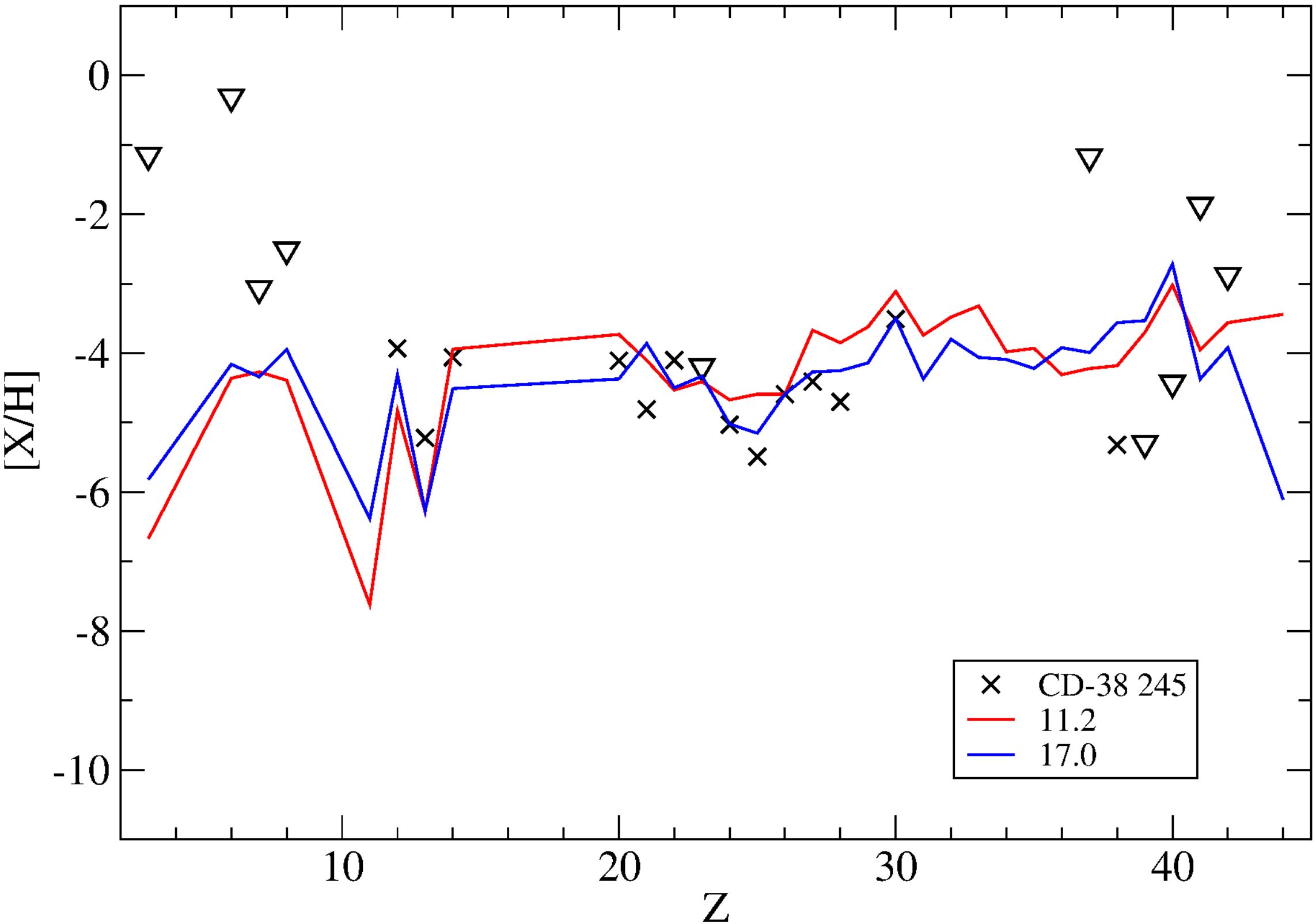}\\
\vspace{0.4cm} \includegraphics[width=0.44\textwidth]{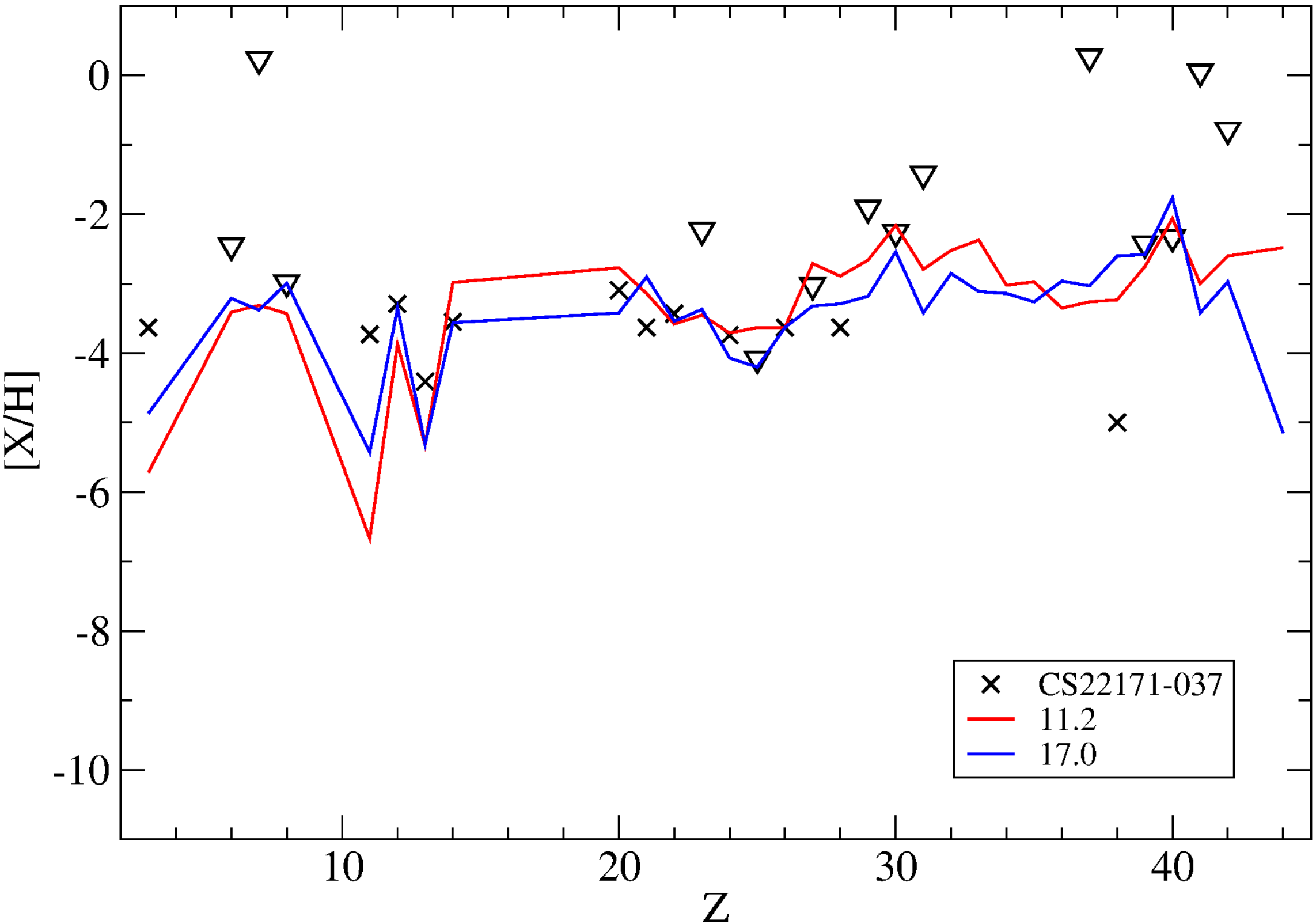}
\hspace{0.4cm} \includegraphics[width=0.44\textwidth]{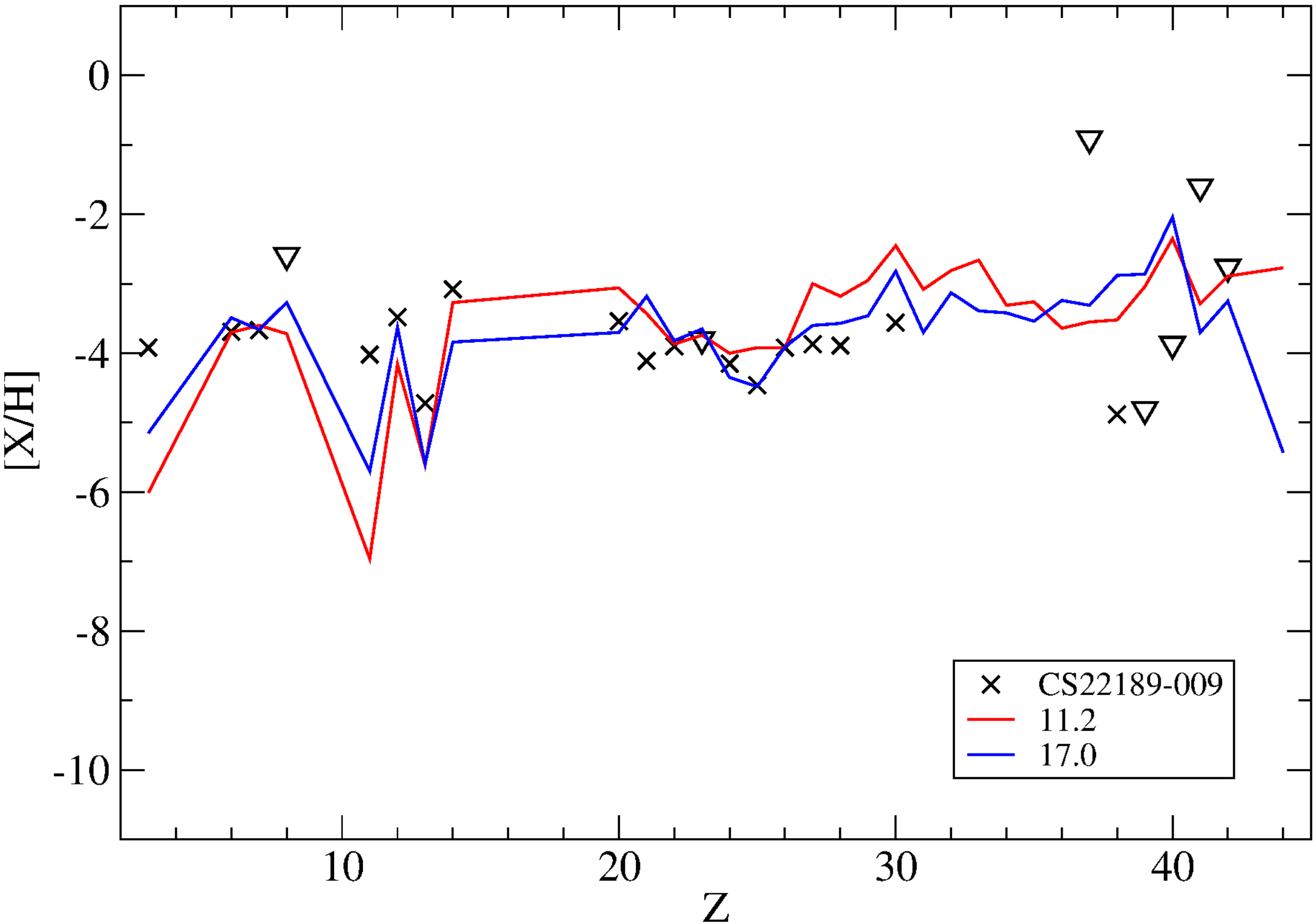}\\
\vspace{0.4cm} \includegraphics[width=0.44\textwidth]{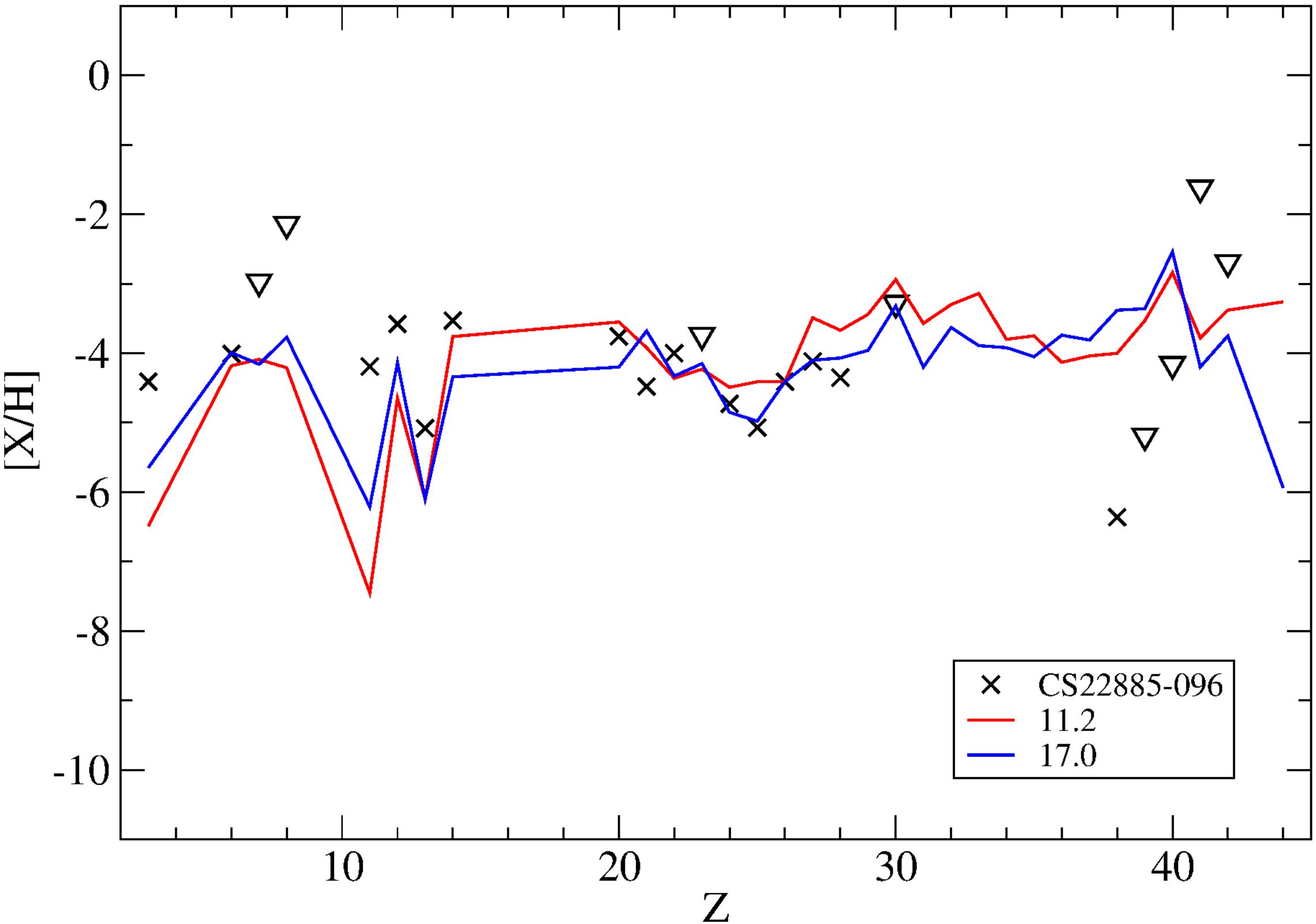}
\hspace{0.4cm} \includegraphics[width=0.44\textwidth]{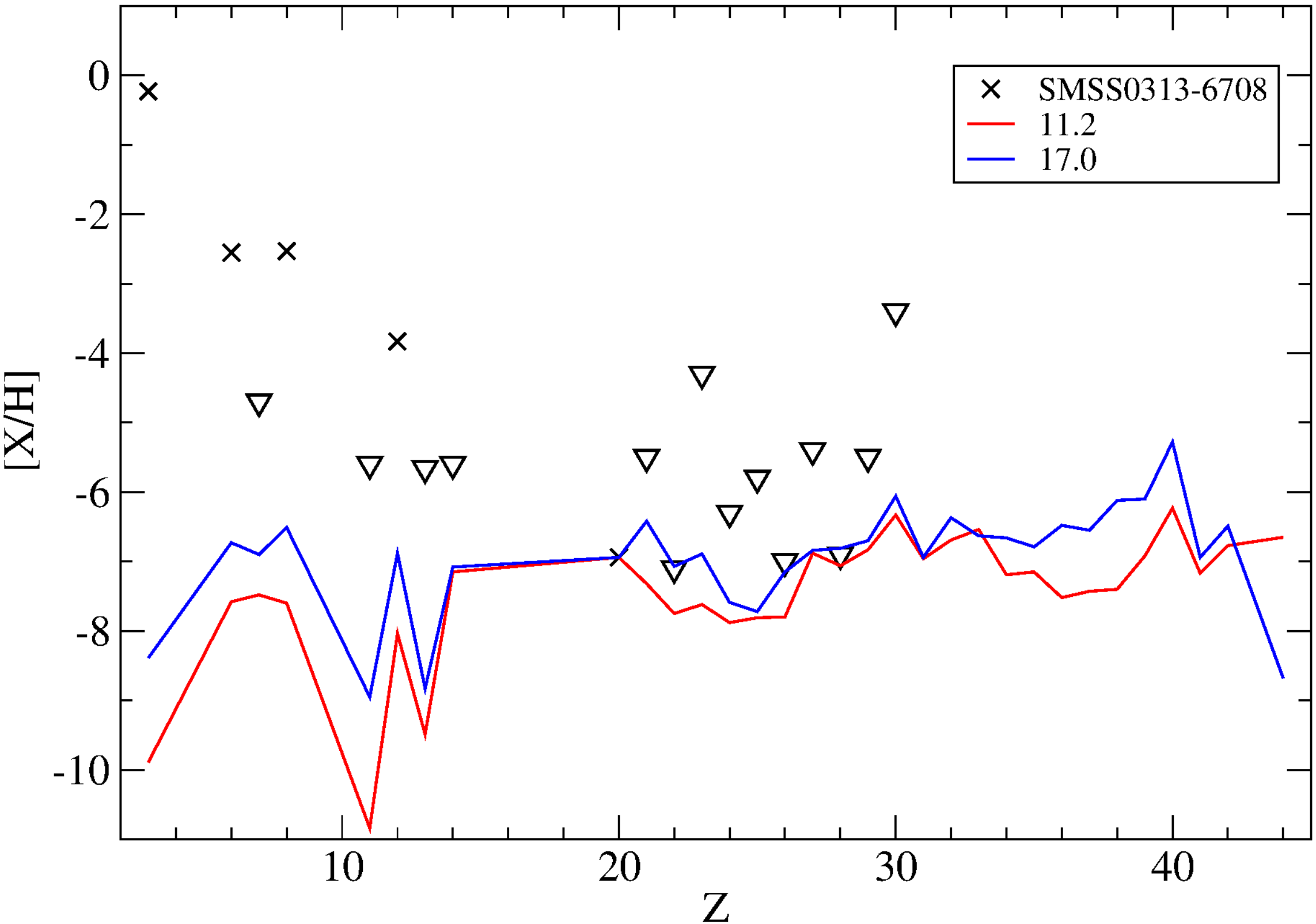}
\caption{Comparisons of elemental compositions in the ejecta of our two models with observed abundances in EMP and UMP stars (crosses: measured abundances; triangles: upper limits). The stellar abundances have been taken from the SAGA database, but originate from \cite{hansenT2014} (HE0134-1519), \cite{norris2007} (HE0557-4840), \cite{frebel2005} (HE1327-2326), \cite{roederer2014} (CD-38~245, CS22171-037, CS22189-009, CS22885-096), and \cite{nordlander2017} (SMSS0313-6708).}
\label{fig:oldstars} 
\end{figure}

The most metal-poor stars currently known are SMSS0313-6708 \cite{keller2014} and HE1327-2326 \cite{frebel2005} with reported metallicities of [Fe/H]$~<~-7$ and [Fe/H]$~\approx~-5.4$, respectively. In the range $-5~<~\mathrm{[Fe/H]}~<~-3$, many stars are known and their atmospheric nuclear compositions are well studied. It is possible that these extremely metal-poor (EMP) and ultra-metal-poor stars (UMP) stars contain the signatures of the ejecta of a single CCSN, which makes them valuable reference points for comparisons with CCSN nucleosynthesis models (e.g., \cite{nordlander2017}). To that end we have selected some of the most metal-poor stars from the Stellar Abundances for Galactic Archeology (SAGA) database \cite{suda2008,suda2011,yamada2013} for which the abundances of sufficient elements have been measured, and compared our yields from section~\ref{sec:yields} to their compositions. Note that the validity of this comparison may be weakened by the fact that our CCSNe have been simulated using progenitors with solar metallicity. Figure~\ref{fig:oldstars} shows the comparisons between the ejecta compositions of both our models (red: 11.2~M$_{\odot}$; blue: 17.0~M$_{\odot}$) to several EMP stars, where crosses denote measured values and triangles are upper limits given in the SAGA database. We have scaled our yields to the observed [Fe/H] value in each star, except for SMSS01313-6708, where only an upper limit for [Fe/H] is known. For this star, we used [Ca/H] instead.
Compared to most stars in this sample, the compositions in our CCSNe are deficient in the lightest $\alpha$-elements (especially C~and~O), but contain too much Sr ($\mathrm{Z}=38$). The discrepancy in the light nuclear species could at least partially be attributed to the uncertain amount of ejected progenitor material from outside the computational domain in our models due to the unipolar (bipolar) explosions (see also \ref{app:ejeccrit}). If we take into account this uncertainty, we find very good agreement between the ejecta of our 17.0~M$_{\odot}$ model with HE1327-2326, which seems to have an unusually large [Sr/H] value compared to the other stars in this sample. Both our models also agree reasonably well with the measured composition of HE0557-4840. Furthermore, they both very well reproduce the measured elemental abundances between Ca and Zn ($20 \leq \mathrm{A} \leq 30$) for all the stars in this sample.


\section{Conclusions}
\label{sec:conclusions}

We have presented nucleosynthesis results of two 2D CCSN models with masses of 11.2 and 17.0~M$_{\odot}$. We find that both models are characterized by low $^{56}$Ni yields, independent of the ejection criterion employed. Although $^{56}$Ni is produced in larger amounts, it is accreted onto the PNS within the first few seconds after bounce. With the final amount of ejected $^{56}$Ni, both CCSNe would classify as faint SNe.

A detailed analysis of the hottest (innermost) ejected tracer particles reveals that they experience a very broad range of $Y_e$ values, depending on the time the particles stay in the neutrino heating region. This results in a large variety of nuclear compositions and calls for a distinction of these tracers according to $Y_e$. For tracers with $0.49~\leq~Y_e\leq0.55$ the reaction flux stops at the iron group nuclei, while for smaller and larger electron fractions also trans-iron nuclei can be produced.

In both models studied here, trans-iron nuclei up to $^{92}$Mo can be efficiently produced in slightly neutron-rich conditions (see also \cite{wanajo2017}), but the reaction flux towards $^{94}$Mo is suppressed by the low neutron separation energies beyond the $N=50$ shell closure. The final composition of the ejecta in the case of the 11.2~M$_{\odot}$ model also contains high abundances of $^{96,98}$Ru, which is produced almost exclusively in conditions with $Y_e>0.7$ in the $\nu$p-process. We have shown that the amount of produced Ru directly depends on the contribution of very proton-rich ejecta, and that electron fractions around $Y_e = 0.8$ or higher are needed for efficient Ru (and Mo) production in the $\nu$p-process. The low-$Y_e$ regimes with enhanced Mo production are very narrow and the abundance peaks are at different $Y_e$ values for the different isotopes, which could prove an interesting aspect in the search for the origin of the solar isotopic ratio of Mo and Ru, as different distributions of $Y_e$ values in the ejected tracer particles can easily lead to varying isotopic ratios.

Furthermore, we performed an analysis of the isotopic ratios between the four isotopes $^{92,94}$Mo and $^{96,98}$Ru. As a reference, we used the solar ratios and SiC X presolar grains. The $Y_{^{98}\rm Ru}$/$Y_{^{96}\rm Ru}$ ratio in the final composition of the 11.2~M$_{\odot}$ model matches the ratio measured in SiC X grains reasonably well, and also the $Y_{^{94}\rm Mo}$/$Y_{^{92}\rm Mo}$ ratio can be reproduced by a combination of ejecta with different $Y_e$. The required $Y_e$ distribution however does not correspond to the $Y_e$ distribution of the integrated ejecta in our models.

Finally, we have explored the possibility that CCSN ejecta compositions with a strong contribution of trans-iron nuclei and low [Fe/H] may have occurred in the early universe by comparing the yields from our two models to the observed atmospheric compositions of the most metal-poor stars currently known. Most EMP and UMP stars from the employed sample show lower [Sr/H] as well as higher [C/H] and [O/H] compared to our CCSN ejecta. However, taking into account the uncertainty in ejected C and O contained in the progenitor material outside the computational domain of the CCSN simulations, it is possible that CCSNe of the type presented here have occurred in the early universe.

\section*{Acknowledgements}
The authors would like to thank J.~Bliss, A.~Arcones, C.~J.~Hansen, A. Perego, and K. Ebinger for enlightening and interesting discussions. M.~E. is supported by the Swiss National Foundation and the Helmholtz-University Investigator grant No.~VH-NG-825. M.~H., R.~C., M.~L., and F.-K.~T. are supported by the Swiss National Foundation. K.~N. acknowledges support by the JSPS KAKENHI Grant Number 16K17668. This study was also supported by the Ministry of Education, Science and Culture of Japan (Nos. 24103006, 24244036,  26707013, 26870823, 15KK0173, 15H01039, 15H00789), by the HPCI Strategic Program of Japanese MEXT, and the EU-FP7-ERC Advanced Grant 321263 FISH.

\section*{References}
\bibliographystyle{iopart-num}
\bibliography{Masterbib_thesis}

\newpage
\newpage
\appendix

\section{Ejection Criteria}
\label{app:ejeccrit}
This section presents a thorough investigation of the ejection criterion used in section~\ref{sec:ejeccrits}. In particular, we test the influence of different prescriptions of ejection criteria on the nuclear yields of isotopes of different masses. One commonly used criterion considers particles with a positive energy and a positive radial velocity component ejected, i.e., $e_{\rm tot}^{\rm fin}~>~0$ \& $v_{\rm rad}^{\rm fin}~>~0$.
Particles that have not encountered the shock at the end of the simulation do not count as ejected according to this criterion. As soon as the shock front reaches them, however, they will be turned around and their velocity vector will point outward. The total mass of ejecta according to the $e~\&~ v_{\rm rad}$ criterion is therefore sensitive to the simulation time at which the criterion is applied. Ideally, it would be applied at a time when the shock front has reached the outer layers of the star. This is not possible in this case, because our simulation stops before that point.

In order to obtain a better idea of the amount (and the composition) of the ejecta, we have tested several different criteria for the 17.0~M$_{\odot}$ simulation. One of them is the condition $\rho_{\rm fin} < 10^{11}$~g~cm$^{-3}$, henceforth labelled ``$\rho$''. Clearly, following this prescription all the particles outside the PNS count towards the final ejecta and are considered ejected even if they are infalling and ultimately accreted onto the PNS. Another option is to use a purely geometrical approach: The shock propagates mainly in a bipolar fashion along the $z$-axis, while matter around the equator is mostly infalling. Therefore, we can choose an angle $\theta$ above and below the equator that distinguishes between ejected and accreted material. Obviously, in order to avoid counting PNS material, we also ask for the $\rho$ criterion to be fulfilled.

We have already established that the ejecta mass is sensitive to the moment when the criteria are applied. The simulation stops at 7s and it can be expected that the most precise results are obtained when the criteria are applied at this moment. However, particles might not fulfill an ejection criterion when the simulation stops, but only later on, for instance when they encounter the shock front and are accelerated away from the center. Thus, we also apply all our criteria at a simulation time of $t=5$s. By doing this, we are able to identify trends for the behaviour of the ejecta mass and the individual isotopic yields according to all our criteria. The total mass of the ejecta (M$_{\rm ej}$) as well as the ejected masses of some isotopes are summarized in table~\ref{tab:criteria5vs7} for a criterion that requires the specific energy to be positive ($e > 0$), the energy with radial velocity criterion ($e+v_{\rm rad}$), the density ($\rho$), and the two angle criteria ($\theta^{30}~\&~\theta^{45}$) taken at 5s and 7s (indicated by the supplements ``5'' and ``7'').

\begin{figure*}
   \includegraphics[width=0.5\textwidth]{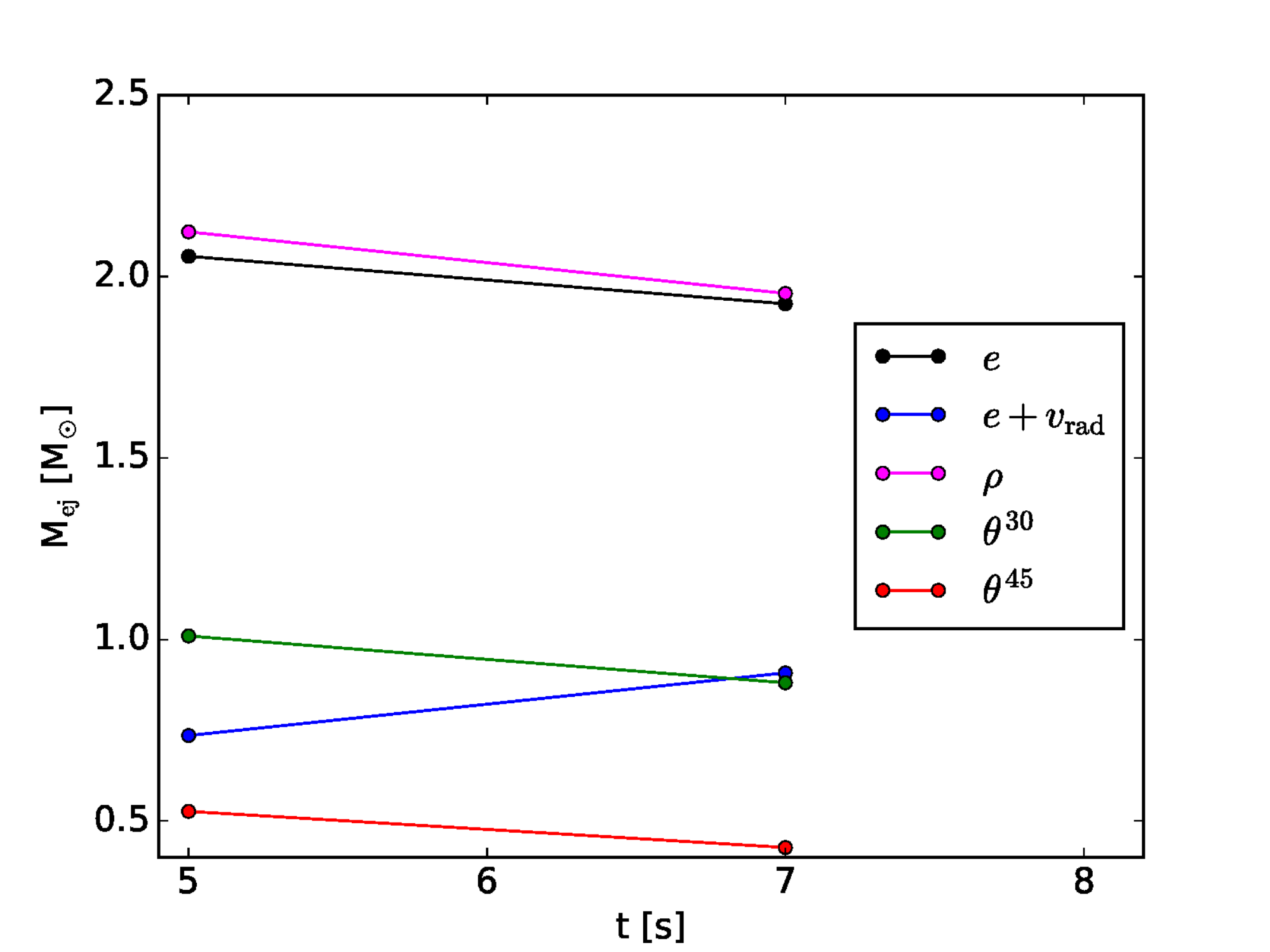}
   \includegraphics[width=0.5\textwidth]{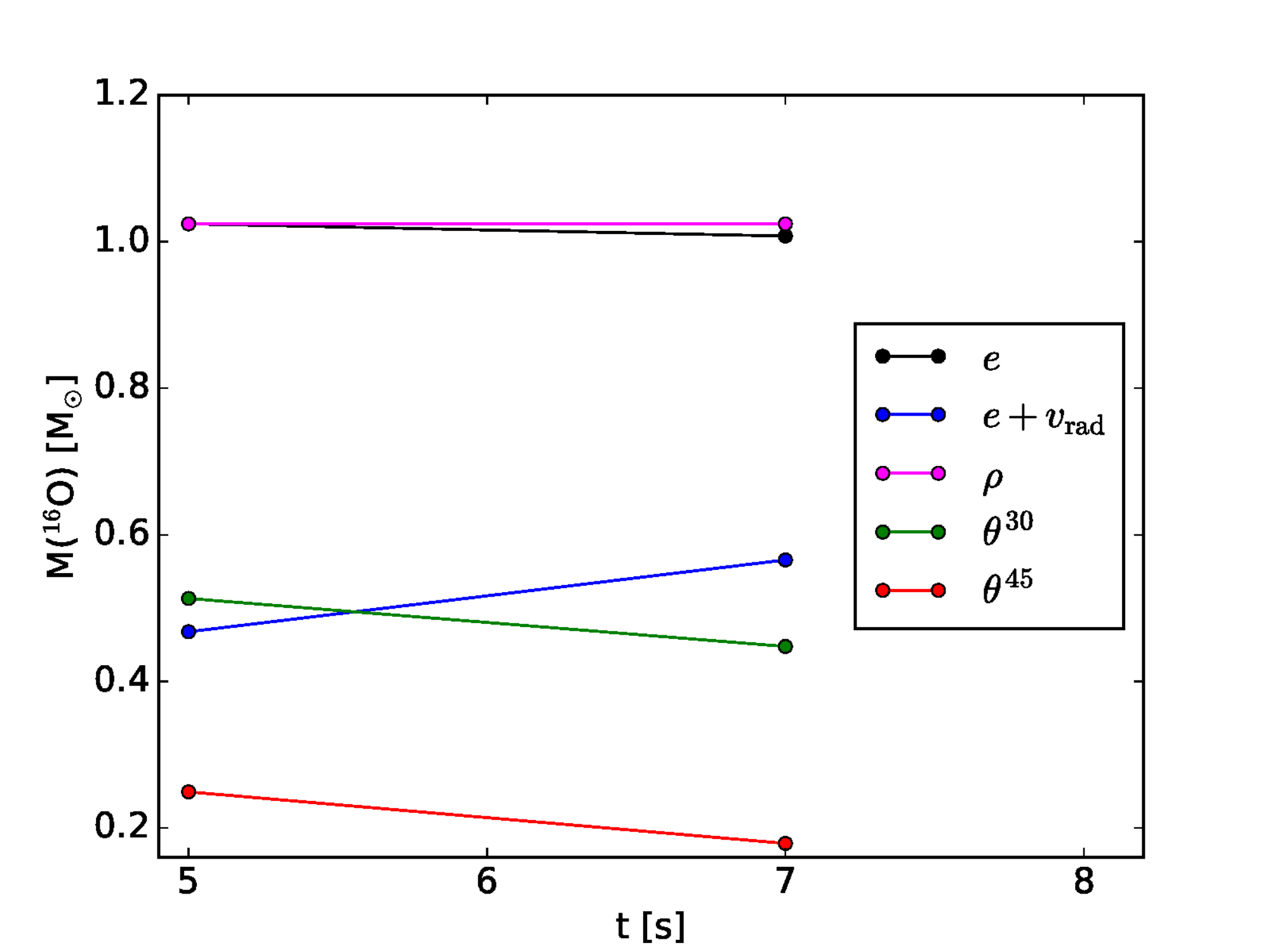}   \\
   \includegraphics[width=0.5\textwidth]{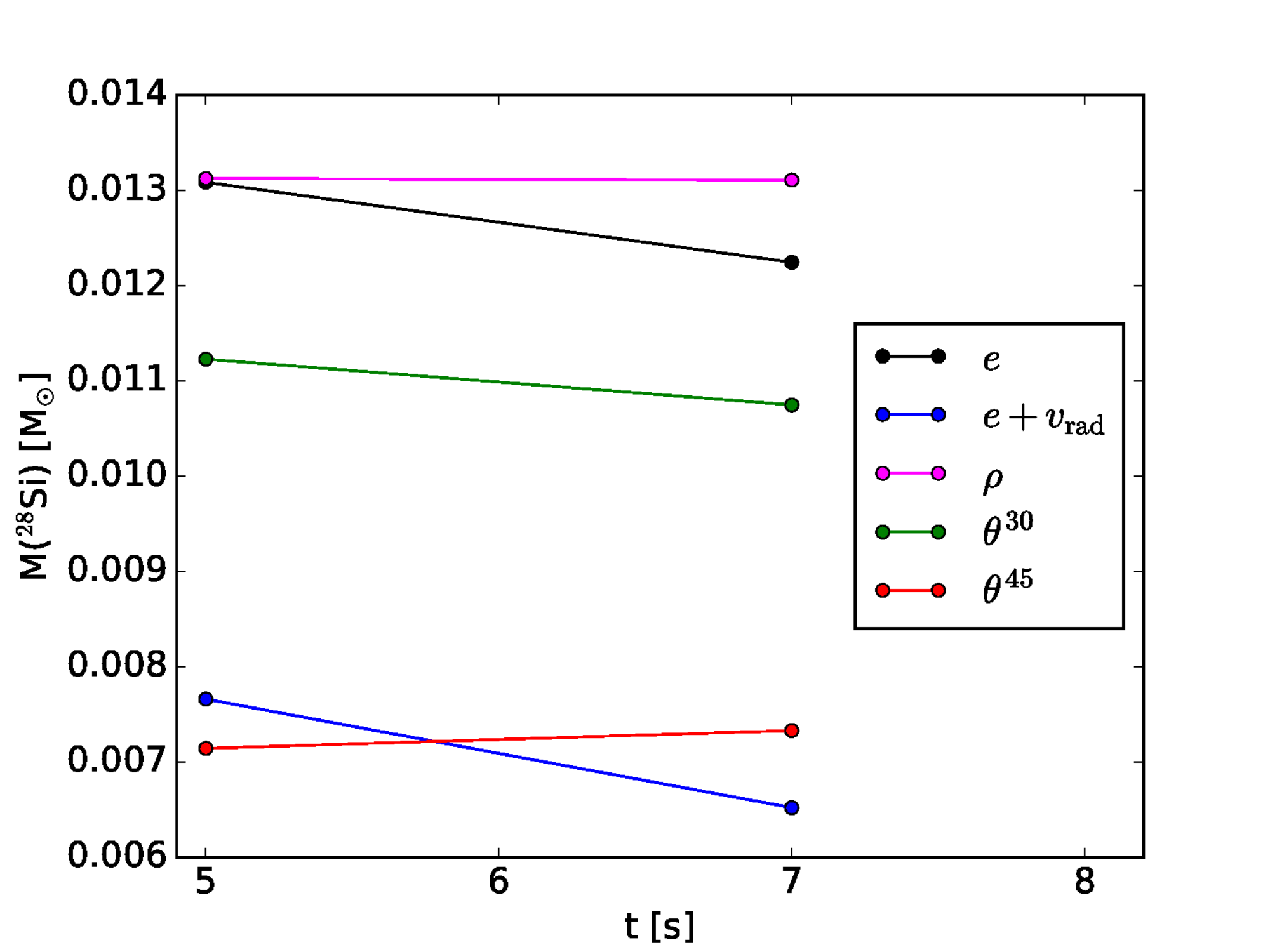}
   \includegraphics[width=0.5\textwidth]{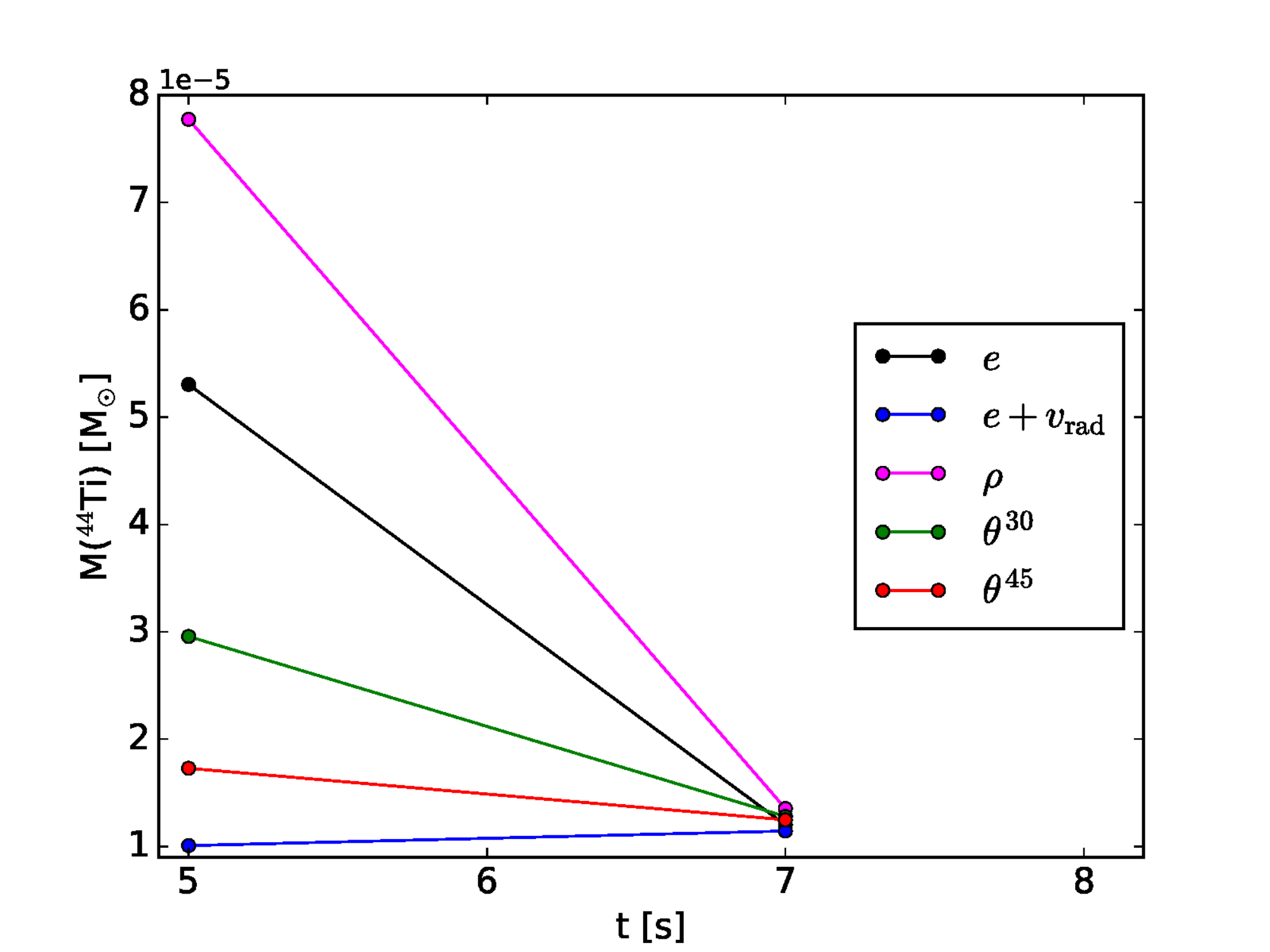}   \\
   \includegraphics[width=0.5\textwidth]{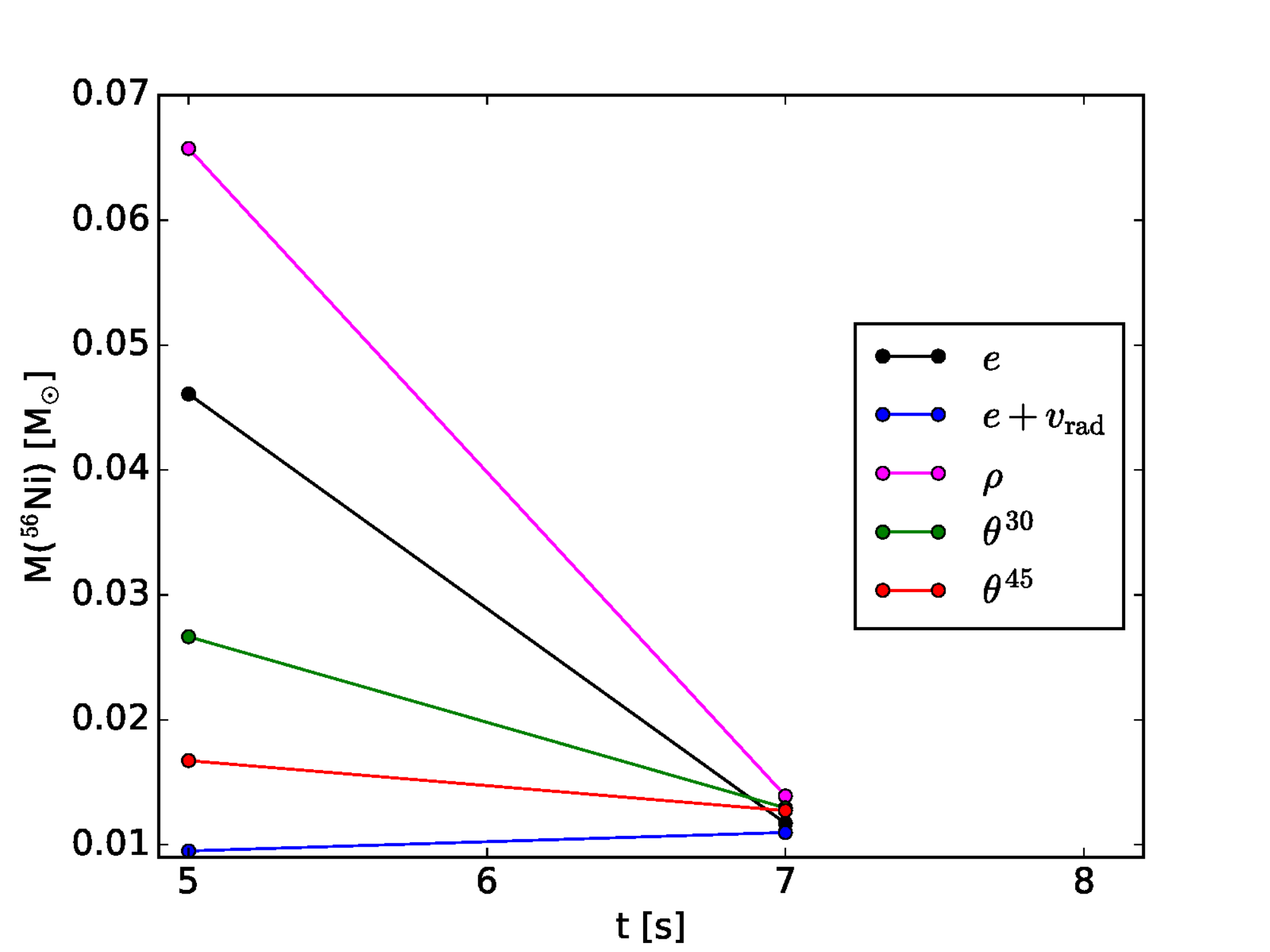}
   \includegraphics[width=0.5\textwidth]{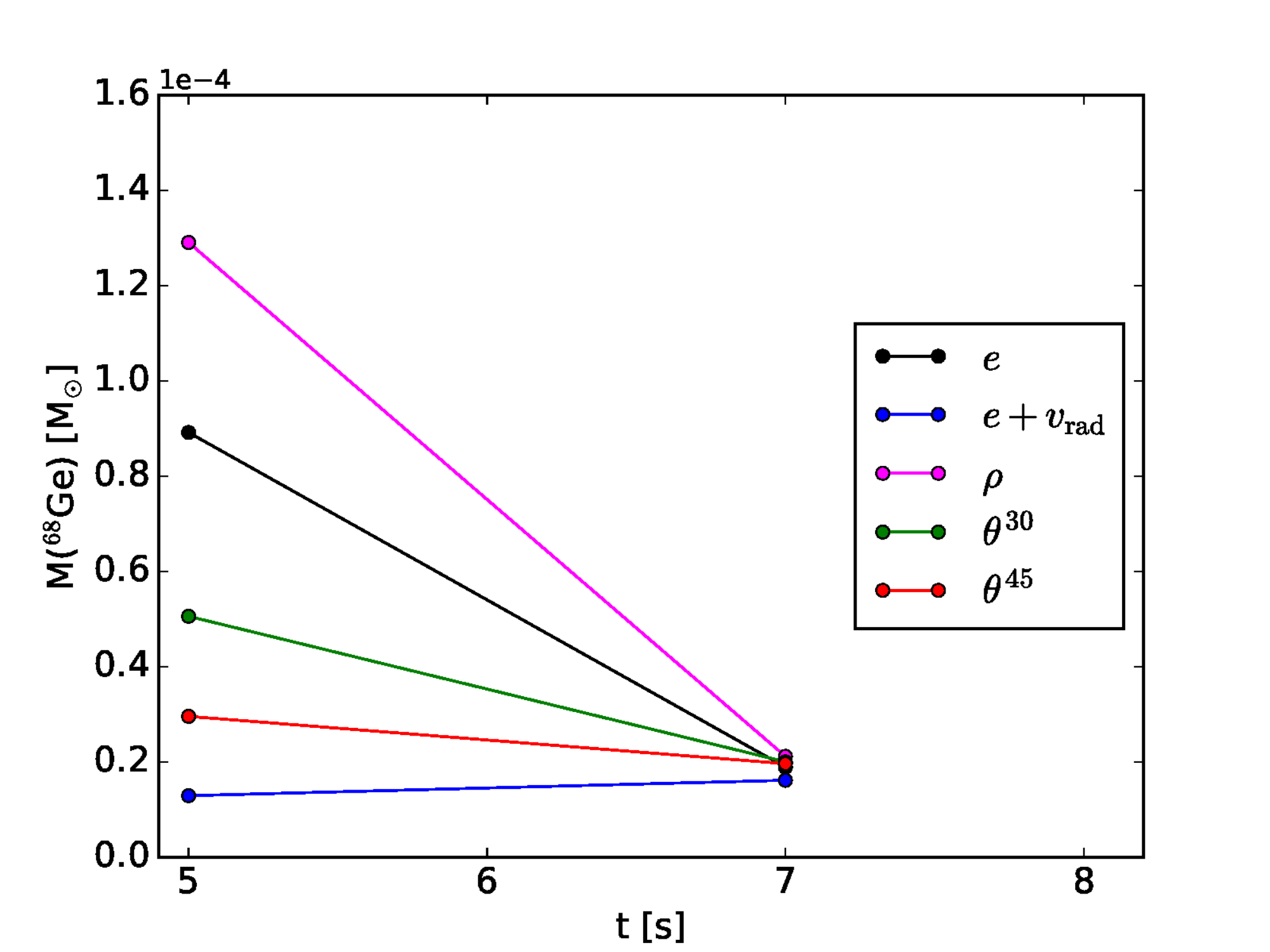}   \\
   \caption{Ejected total mass (top left) and yields of several selected isotopes (top right: $^{16}$O, middle left: $^{28}$Si, middle right: $^{44}$Ti, bottom left: $^{56}$Ni, bottom right: $^{68}$Ge) according to different ejection criteria for the 17.0~M$_{\odot}$ model. See text for explanation.}
   \label{fig:criteria_yields}
\end{figure*}

\begin{figure*}
   \includegraphics[width=0.5\textwidth]{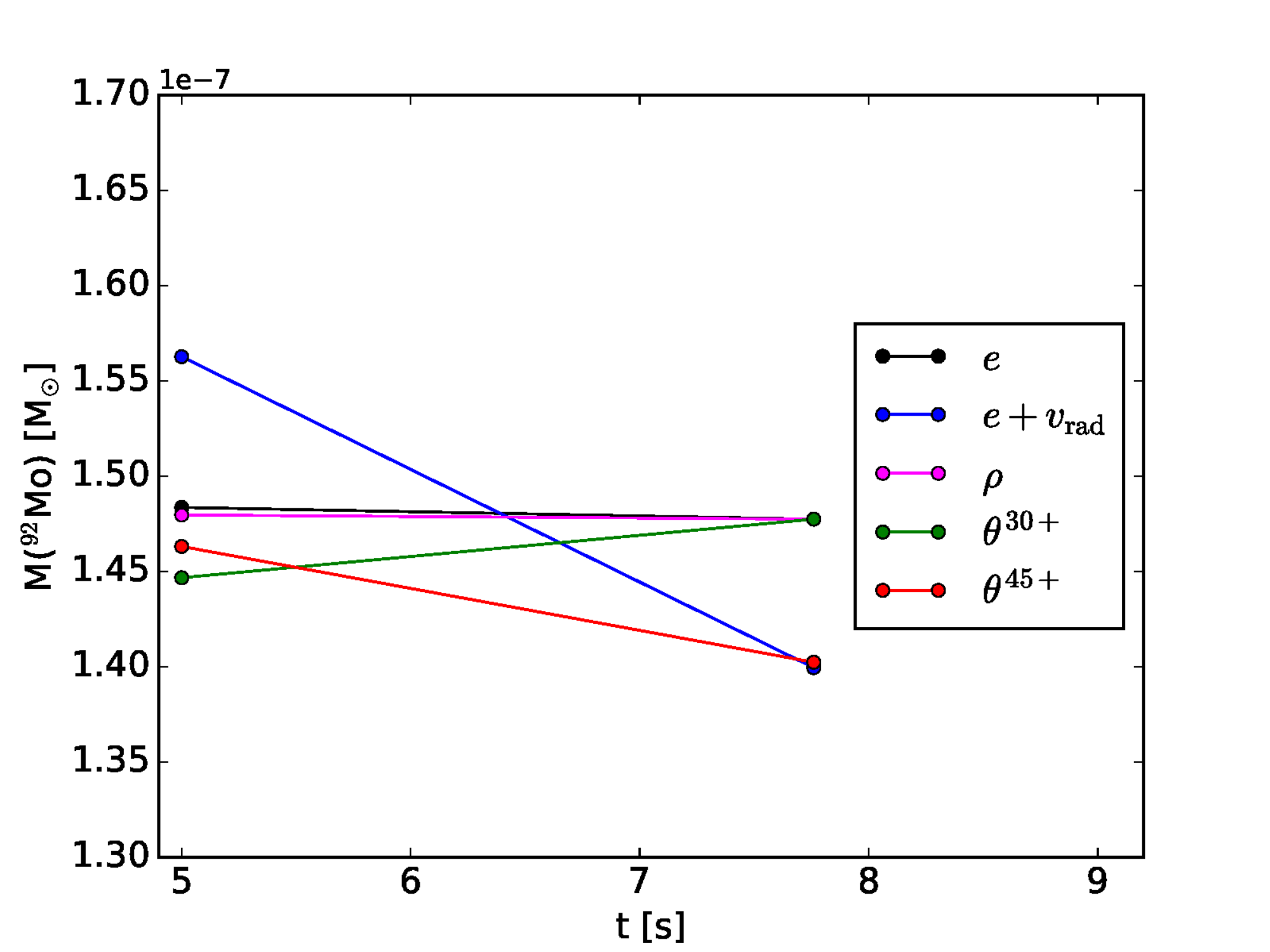}
   \includegraphics[width=0.5\textwidth]{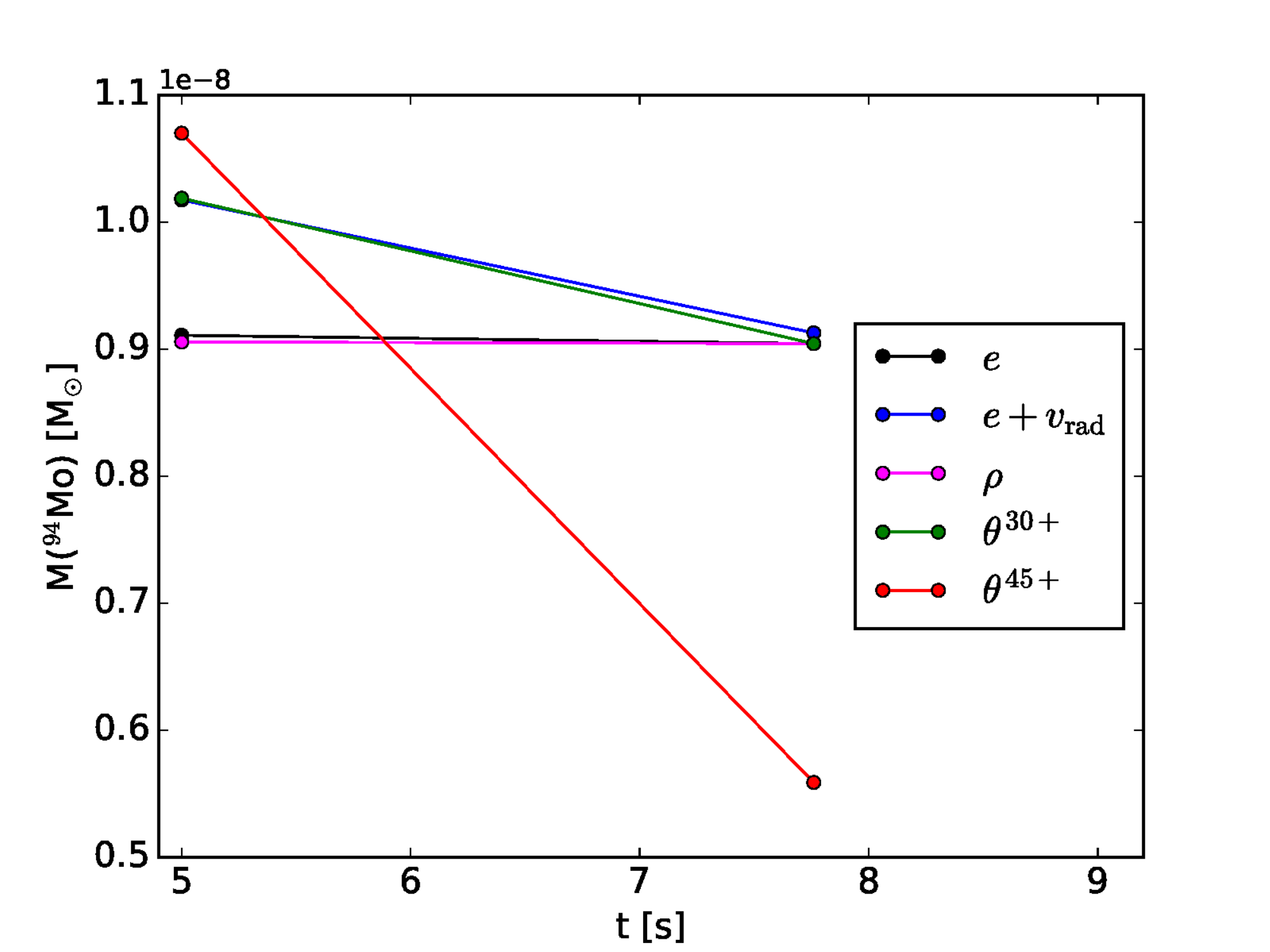}   \\
   \includegraphics[width=0.5\textwidth]{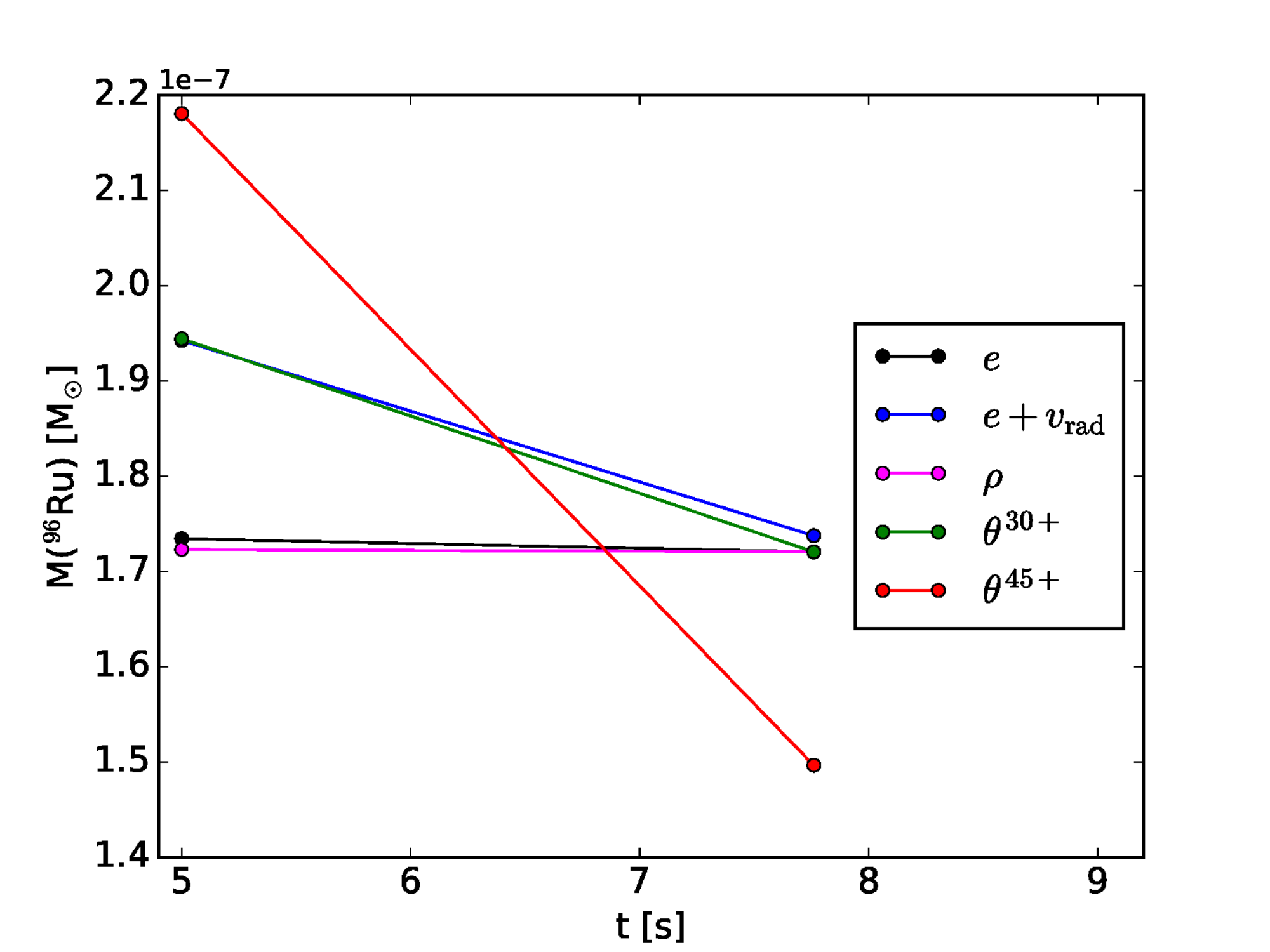}
   \includegraphics[width=0.5\textwidth]{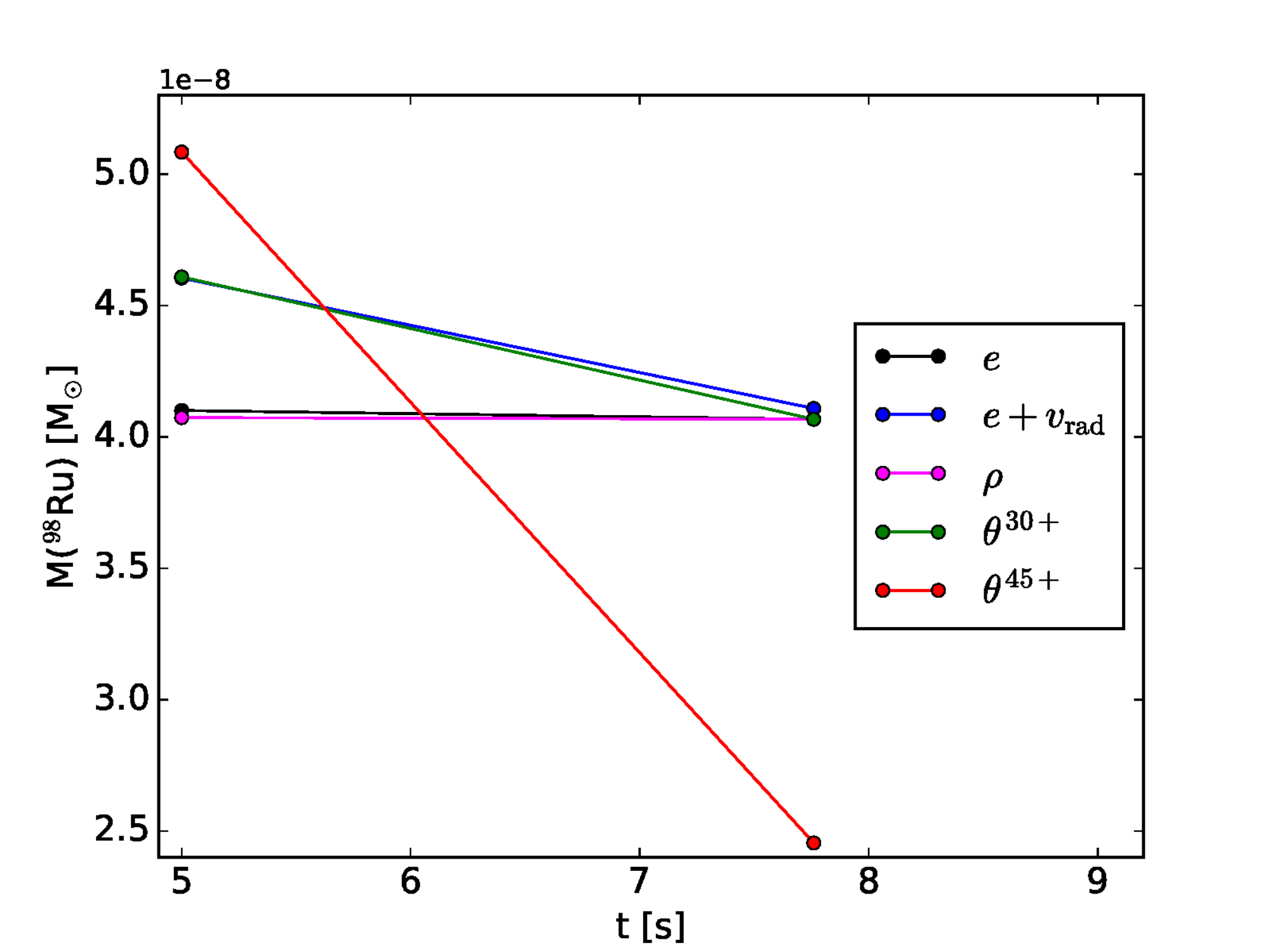}   \\
   \caption{Same as figure~\ref{fig:criteria_yields}, but for the 11.2~M$_{\odot}$ model and the isotopes $^{92}$Mo, $^{94}$Mo, $^{96}$Ru, and $^{98}$Ru.}
   \label{fig:criteria_yields2}
\end{figure*}

Table~\ref{tab:criteria5vs7} reveals that the different ejection criteria give very different predictions for the total mass of the ejecta and the yields for $^{16}$O and $^{28}$Si. However, the values for M($^{44}$Ti), M($^{56}$Ni), and M($^{68}$Ge) are very similar for all the criteria when applied at 7s. The predictions for the ejected $^{44}$Ti mass lie within a range of $1.14 \times 10^{-5}$~M$_{\odot}$ and $1.35 \times 10^{-5}$~M$_{\odot}$, the predicted $^{56}$Ni mass is between $1.10 \times 10^{-2}$~M$_{\odot}$ and $1.39 \times 10^{-2}$~M$_{\odot}$, and the $^{68}$Ge yield is between $1.62 \times 10^{-5}$~M$_{\odot}$ and $2.12 \times 10^{-5}$~M$_{\odot}$. The convergence of the yields for nuclei with higher mass numbers becomes even clearer in figure~\ref{fig:criteria_yields}, where the data from table~\ref{tab:criteria5vs7} are plotted as a function of the moment of application. The ejecta masses of $^{92}$Mo, $^{94}$Mo, $^{96}$Ru, and $^{98}$Ru for the 11.2~M$_{\odot}$ model are shown in table~\ref{tab:criteria5vs7_moru} and in figure~\ref{fig:criteria_yields2}. The yields of these isotopes converge as well at the end of the simulation, with the exception of the $\theta^{45+}$ criterion and the $e+v_{\rm rad}$ criterion for $^{92}$Mo which diverges from the other criteria by about 5 \%.

\begin{table}
\caption{Total ejecta masses (M$_{\rm ej}$) and individual isotopic yields according to different ejection criteria, which are applied at two different times in the simulation (5s and 7s) of the 17.0~M$_{\odot}$ progenitor. \label{tab:criteria5vs7}}
\begin{indented}
\item[] \begin{tabular}{ccccccc} 
\br
  Crit & M$_{\rm ej}$ & M($^{16}$O) & M($^{28}$Si) & M($^{44}$Ti) & M($^{56}$Ni) & M($^{68}$Ge) \\
       & [M$_{\odot}$]  & [M$_{\odot}$] & [M$_{\odot}$] & [M$_{\odot}$] & [M$_{\odot}$] & [M$_{\odot}$] \vspace{1px} \\
\mr
    $e \left(5 \right)$              & ~~~~2.06~~~   & ~~~~1.02~~~   & $1.31 \times 10^{-2}$ & $5.30 \times 10^{-5}$ & $4.61 \times 10^{-2}$ & $8.92 \times 10^{-5}$ \\
    $e \left(7 \right)$              & ~~~~1.93~~~   & ~~~~1.01~~~   & $1.22 \times 10^{-2}$ & $1.20 \times 10^{-5}$ & $1.17 \times 10^{-2}$ & $1.89 \times 10^{-5}$ \\
    $e+v_{\rm rad} \left(5 \right)$  & ~~~~0.74~~~   & ~~~~0.47~~~   & $7.67 \times 10^{-3}$ & $1.01 \times 10^{-5}$ & $9.51 \times 10^{-3}$ & $1.29 \times 10^{-5}$ \\
    $e+v_{\rm rad} \left(7 \right)$  & ~~~~0.91~~~   & ~~~~0.57~~~   & $6.52 \times 10^{-3}$ & $1.14 \times 10^{-5}$ & $1.10 \times 10^{-2}$ & $1.62 \times 10^{-5}$ \\
    $\rho \left(5 \right)$           & ~~~~2.12~~~   & ~~~~1.02~~~   & $1.31 \times 10^{-2}$ & $7.77 \times 10^{-5}$ & $6.57 \times 10^{-2}$ & $1.29 \times 10^{-4}$ \\
    $\rho \left(7 \right)$           & ~~~~1.95~~~   & ~~~~1.02~~~   & $1.31 \times 10^{-2}$ & $1.35 \times 10^{-5}$ & $1.39 \times 10^{-2}$ & $2.12 \times 10^{-5}$ \\
    $\theta^{30} \left(5 \right)$    & ~~~~1.01~~~   & ~~~~0.51~~~   & $1.12 \times 10^{-2}$ & $2.96 \times 10^{-5}$ & $2.67 \times 10^{-2}$ & $5.06 \times 10^{-5}$ \\
    $\theta^{30} \left(7 \right)$    & ~~~~0.88~~~   & ~~~~0.45~~~   & $1.07 \times 10^{-2}$ & $1.28 \times 10^{-5}$ & $1.30 \times 10^{-2}$ & $2.00 \times 10^{-5}$ \\
    $\theta^{45} \left(5 \right)$    & ~~~~0.53~~~   & ~~~~0.25~~~   & $7.14 \times 10^{-3}$ & $1.73 \times 10^{-5}$ & $1.67 \times 10^{-2}$ & $2.96 \times 10^{-5}$ \\
    $\theta^{45} \left(7 \right)$    & ~~~~0.43~~~   & ~~~~0.18~~~   & $7.33 \times 10^{-3}$ & $1.25 \times 10^{-5}$ & $1.27 \times 10^{-2}$ & $1.97 \times 10^{-5}$ \\
\br
\end{tabular}
\end{indented}
\end{table}

\begin{table}
\caption{Same as table~\ref{tab:criteria5vs7}, but for ejected $^{92}$Mo, $^{94}$Mo, $^{96}$Ru, and $^{98}$Ru masses in the case of the 11.2~M$_{\odot}$ progenitor. \label{tab:criteria5vs7_moru}}
\begin{indented}
\item[] \begin{tabular}{ccccc} 
\br
  Crit & M($^{92}$Mo) & M($^{94}$Mo) & M($^{96}$Ru) & M($^{98}$Ru) \\
       & [M$_{\odot}$] & [M$_{\odot}$] & [M$_{\odot}$] & [M$_{\odot}$] \vspace{1px} \\
\mr
    $e \left(5 \right)$                & $1.48 \times 10^{-7}$ & $9.11 \times 10^{-9}$ & $1.73 \times 10^{-7}$ & $4.10 \times 10^{-8}$ \\
    $e \left(7.76 \right)$             & $1.48 \times 10^{-7}$ & $9.04 \times 10^{-9}$ & $1.72 \times 10^{-7}$ & $4.07 \times 10^{-8}$ \\
    $e+v_{\rm rad} \left(5 \right)$    & $1.56 \times 10^{-7}$ & $1.02 \times 10^{-8}$ & $1.94 \times 10^{-7}$ & $4.60 \times 10^{-8}$ \\
    $e+v_{\rm rad} \left(7.76 \right)$ & $1.40 \times 10^{-7}$ & $9.13 \times 10^{-9}$ & $1.74 \times 10^{-7}$ & $4.11 \times 10^{-8}$ \\
    $\rho \left(5 \right)$             & $1.48 \times 10^{-7}$ & $9.06 \times 10^{-9}$ & $1.72 \times 10^{-7}$ & $4.07 \times 10^{-8}$ \\
    $\rho \left(7.76 \right)$          & $1.48 \times 10^{-7}$ & $9.04 \times 10^{-9}$ & $1.72 \times 10^{-7}$ & $4.07 \times 10^{-8}$ \\
    $\theta^{30+} \left(5 \right)$     & $1.45 \times 10^{-7}$ & $1.02 \times 10^{-8}$ & $1.94 \times 10^{-7}$ & $4.61 \times 10^{-8}$ \\
    $\theta^{30+} \left(7.76 \right)$  & $1.48 \times 10^{-7}$ & $9.04 \times 10^{-9}$ & $1.72 \times 10^{-7}$ & $4.07 \times 10^{-8}$ \\
    $\theta^{45+} \left(5 \right)$     & $1.46 \times 10^{-7}$ & $1.07 \times 10^{-8}$ & $2.18 \times 10^{-7}$ & $5.08 \times 10^{-8}$ \\
    $\theta^{45+} \left(7.76 \right)$  & $1.40 \times 10^{-7}$ & $5.59 \times 10^{-9}$ & $1.50 \times 10^{-7}$ & $2.45 \times 10^{-8}$ \\
\br
\end{tabular}
\end{indented}
\end{table}


Several trends are visible in figure~\ref{fig:criteria_yields}. First, the predicted total mass of the ejecta slowly decreases with increasing simulation time for three of the four criteria (top left panel). This can be explained by accretion onto the PNS, i.e., the density of accreting particles increases above $10^{11}$~g~cm$^{-3}$ (failing the $\rho$ criterion). The decreasing trend for the $\theta$ criteria is due to particles that are first accelerated by the shock close to the boundary and, after a while, fall into the equatorial plane, thus not counting towards the ejecta anymore. For the $e+v_{\rm rad}$ criterion the ejected mass is growing, since the shock is still moving outward and accelerating infalling particles. Furthermore, there is a striking difference between the temporal behaviour of the $^{16}$O and $^{28}$Si yields on one hand and the $^{44}$Ti, $^{56}$Ni, and $^{68}$Ge yields on the other hand: The former group includes nuclei that are already present in the pre-SN progenitor (mainly in the outer layers), which means that their yields generally follow the trends of the total mass. $^{28}$Si shows a slightly deviating behaviour from the total mass, because it is also the fuel for explosive Si burning in high enough temperatures (according to Thielemann \etal \cite{thielemann1996}, complete Si exhaustion is reached in tracers that reach $T = 5$~GK). Therefore, it is most abundant in tracer particles with an intermediate peak temperature within the range of $2~<~T_9~<~5$. The latter nuclei belong to the group of freshly synthesized products of the SN with relatively high mass number. They are produced mainly in the regions with the highest peak temperatures, and the convergence of their yields at $t=7$s means that most of the corresponding tracer particles are considered ejected in all criteria.

For the final criteria we choose a geometric prescription that takes into account the shape of the shock a the end of the simulation. For the 17.0~M$_{\odot}$ model, we use $\theta^{30}$. This can be justified by the notion that all particles with an angle $\theta \geq 30^{\circ}$ or $\theta \leq -30^{\circ}$ from the equator will encounter the shock at one point after the end of the simulation and obtain a positive radial velocity.
For the 11.2~M$_{\odot}$ model, we observe a strong shock progression in only one polar direction (in the positive $z$-direction). We therefore constrain the geometric criterion further in this case, requiring successfully ejected particles to be at least $30^{\circ}$ above the equator, and we call the corresponding criterion $\theta^{30+}$. Since the ejected masses of the intermediate- and high-mass nuclei are not sensitive to the applied ejection criterion (see figures~\ref{fig:criteria_yields}~\&~\ref{fig:criteria_yields2}), we expect that our choice of ejection criterion results only in a minor error in the yields of the nuclei that are produced in explosive nucleosynthesis.

\section{Tabulated Ejecta Compositions}
\label{app:tables}
Here we present the full ejecta compositions for both models in units of solar masses. The mass fractions $X_i$ and abundances $Y_i$ can be obtained by

\begin{equation}
\begin{aligned}
X_i &= \frac{M_i}{M_{tot}} \\
Y_i &= \frac{M_i}{A_i M_{tot}} ,
\end{aligned}
\end{equation}
where $M_i$ and $A_i$ are the ejected mass (i.e., the value given in the table) and the mass number of isotope $i$, and $M_{tot}$ is the total ejecta mass, which is equal to 2.337~M$_{\odot}$ in the case of the 11.2~M$_{\odot}$ progenitor and 5.744~M$_{\odot}$ for the 17.0~M$_{\odot}$ model.

\newpage
\begingroup
\fontsize{10}{10.7}\selectfont
\topcaption{Isotopic yields for the 11.2~M$_{\odot}$ progenitor. Only isotopes with  ejected masses greater than $10^{-15}$~M$_{\odot}$ are included.} \label{tab:11.2_yields}
\tablefirsthead{\toprule Iso & M/M$_{\odot}$ & Iso & M/M$_{\odot}$ & Iso & M/M$_{\odot}$ & Iso & M/M$_{\odot}$ & Iso & M/M$_{\odot}$ \\ \midrule}
\tablehead{\multicolumn{10}{c}{{\bfseries \tablename\ \thetable{} -- continued from previous page}} \\
\toprule Iso & M/M$_{\odot}$ & Iso & M/M$_{\odot}$ & Iso & M/M$_{\odot}$ & Iso & M/M$_{\odot}$ & Iso & M/M$_{\odot}$ \\ \midrule}
\begin{xtabular}{cl|cl|cl|cl|cl}
\shrinkheight{-4.2cm}
$^{1}$H  &   $1.34$ &$^{29}$Al  &   $1.83 \times 10^{-10}$ &$^{41}$Ca  &   $1.65 \times 10^{-6}$ &$^{52}$Fe  &   $6.86 \times 10^{-5}$ &$^{65}$Zn  &   $6.83 \times 10^{-6}$  \\
$^{2}$H  &   $9.58 \times 10^{-1}$ &$^{30}$Al  &   $4.55 \times 10^{-15}$ &$^{42}$Ca  &   $5.48 \times 10^{-7}$ &$^{53}$Fe  &   $8.41 \times 10^{-7}$ &$^{66}$Zn  &   $3.48 \times 10^{-5}$  \\
$^{3}$H  &   $3.28 \times 10^{-10}$ &$^{27}$Si  &   $7.17 \times 10^{-14}$ &$^{43}$Ca  &   $1.08 \times 10^{-7}$ &$^{54}$Fe  &   $2.99 \times 10^{-3}$ &$^{67}$Zn  &   $6.82 \times 10^{-8}$  \\
$^{3}$He  &   $4.38 \times 10^{-5}$ &$^{28}$Si  &   $1.70 \times 10^{-2}$ &$^{44}$Ca  &   $5.30 \times 10^{-9}$ &$^{55}$Fe  &   $1.90 \times 10^{-6}$ &$^{68}$Zn  &   $7.34 \times 10^{-8}$  \\
$^{4}$He  &   $8.56 \times 10^{-1}$ &$^{29}$Si  &   $6.08 \times 10^{-6}$ &$^{45}$Ca  &   $9.20 \times 10^{-11}$ &$^{56}$Fe  &   $1.66 \times 10^{-6}$ &$^{69}$Zn  &   $2.59 \times 10^{-10}$  \\
$^{6}$Li  &   $3.70 \times 10^{-12}$ &$^{30}$Si  &   $1.30 \times 10^{-6}$ &$^{46}$Ca  &   $6.01 \times 10^{-12}$ &$^{57}$Fe  &   $4.90 \times 10^{-8}$ &$^{70}$Zn  &   $1.06 \times 10^{-10}$ \\
$^{7}$Li  &   $4.24 \times 10^{-10}$ &$^{31}$Si  &   $4.21 \times 10^{-9}$ &$^{47}$Ca  &   $5.36 \times 10^{-13}$ &$^{58}$Fe  &   $1.18 \times 10^{-7}$ &$^{71}$Zn  &   $1.08 \times 10^{-14}$ \\
$^{7}$Be  &   $5.14 \times 10^{-12}$ &$^{32}$Si  &   $1.21 \times 10^{-7}$ &$^{48}$Ca  &   $2.60 \times 10^{-13}$ &$^{59}$Fe  &   $6.72 \times 10^{-9}$ &$^{72}$Zn  &   $2.40 \times 10^{-12}$ \\
$^{9}$Be  &   $5.00 \times 10^{-12}$ &$^{28}$P  &   $5.40 \times 10^{-12}$ &$^{40}$Sc  &   $4.02 \times 10^{-13}$ &$^{60}$Fe  &   $1.87 \times 10^{-8}$ &$^{62}$Ga  &   $2.21 \times 10^{-13}$ \\
$^{10}$Be  &   $2.03 \times 10^{-14}$ &$^{29}$P  &   $9.02 \times 10^{-14}$ &$^{41}$Sc  &   $1.85 \times 10^{-15}$ &$^{61}$Fe  &   $9.83 \times 10^{-12}$ &$^{63}$Ga  &   $1.80 \times 10^{-9}$  \\
$^{10}$B  &   $1.59 \times 10^{-13}$ &$^{30}$P  &   $1.88 \times 10^{-8}$ &$^{42}$Sc  &   $2.74 \times 10^{-15}$ &$^{62}$Fe  &   $5.80 \times 10^{-14}$ &$^{64}$Ga  &   $1.84 \times 10^{-6}$  \\
$^{11}$B  &   $3.96 \times 10^{-8}$ &$^{31}$P  &   $8.15 \times 10^{-6}$ &$^{43}$Sc  &   $1.13 \times 10^{-6}$ &$^{52}$Co  &   $1.43 \times 10^{-15}$ &$^{65}$Ga  &   $2.20 \times 10^{-6}$  \\
$^{12}$B  &   $8.86 \times 10^{-14}$ &$^{32}$P  &   $2.02 \times 10^{-5}$ &$^{44}$Sc  &   $1.23 \times 10^{-8}$ &$^{54}$Co  &   $1.43 \times 10^{-13}$ &$^{66}$Ga  &   $3.41 \times 10^{-6}$  \\
$^{11}$C  &   $2.80 \times 10^{-8}$ &$^{33}$P  &   $1.23 \times 10^{-8}$ &$^{45}$Sc  &   $8.37 \times 10^{-8}$ &$^{55}$Co  &   $6.62 \times 10^{-5}$ &$^{67}$Ga  &   $4.53 \times 10^{-6}$  \\
$^{12}$C  &   $1.96 \times 10^{-2}$ &$^{34}$P  &   $1.41 \times 10^{-14}$ &$^{46}$Sc  &   $1.60 \times 10^{-9}$ &$^{56}$Co  &   $2.56 \times 10^{-5}$ &$^{68}$Ga  &   $8.47 \times 10^{-8}$  \\
$^{13}$C  &   $1.99 \times 10^{-7}$ &$^{27}$S  &   $1.35 \times 10^{-14}$ &$^{47}$Sc  &   $1.84 \times 10^{-10}$ &$^{57}$Co  &   $4.80 \times 10^{-6}$ &$^{69}$Ga  &   $1.26 \times 10^{-7}$  \\
$^{14}$C  &   $2.04 \times 10^{-7}$ &$^{31}$S  &   $6.76 \times 10^{-14}$ &$^{48}$Sc  &   $1.57 \times 10^{-11}$ &$^{58}$Co  &   $1.06 \times 10^{-5}$ &$^{70}$Ga  &   $1.41 \times 10^{-8}$  \\
$^{12}$N  &   $5.94 \times 10^{-14}$ &$^{32}$S  &   $1.48 \times 10^{-2}$ &$^{49}$Sc  &   $5.47 \times 10^{-13}$ &$^{59}$Co  &   $7.28 \times 10^{-7}$ &$^{71}$Ga  &   $1.90 \times 10^{-9}$  \\
$^{13}$N  &   $1.32 \times 10^{-8}$ &$^{33}$S  &   $7.95 \times 10^{-6}$ &$^{44}$Ti  &   $4.32 \times 10^{-6}$ &$^{60}$Co  &   $3.02 \times 10^{-6}$ &$^{72}$Ga  &   $7.97 \times 10^{-10}$  \\
$^{14}$N  &   $8.46 \times 10^{-3}$ &$^{34}$S  &   $1.03 \times 10^{-6}$ &$^{45}$Ti  &   $5.58 \times 10^{-7}$ &$^{61}$Co  &   $2.96 \times 10^{-8}$ &$^{73}$Ga  &   $2.21 \times 10^{-11}$  \\
$^{15}$N  &   $1.53 \times 10^{-7}$ &$^{35}$S  &   $3.56 \times 10^{-9}$ &$^{46}$Ti  &   $1.10 \times 10^{-6}$ &$^{62}$Co  &   $5.58 \times 10^{-11}$ &$^{74}$Ga  &   $9.53 \times 10^{-14}$  \\
$^{16}$N  &   $5.09 \times 10^{-11}$ &$^{36}$S  &   $3.54 \times 10^{-8}$ &$^{47}$Ti  &   $7.20 \times 10^{-7}$ &$^{63}$Co  &   $1.75 \times 10^{-14}$ &$^{64}$Ge  &   $1.61 \times 10^{-7}$  \\
$^{14}$O  &   $1.21 \times 10^{-11}$ &$^{37}$S  &   $2.67 \times 10^{-13}$ &$^{48}$Ti  &   $1.92 \times 10^{-8}$ &$^{55}$Ni  &   $3.51 \times 10^{-15}$ &$^{65}$Ge  &   $5.75 \times 10^{-12}$ \\
$^{15}$O  &   $3.94 \times 10^{-13}$ &$^{38}$S  &   $3.15 \times 10^{-13}$ &$^{49}$Ti  &   $1.44 \times 10^{-9}$ &$^{56}$Ni  &   $3.03 \times 10^{-3}$ &$^{66}$Ge  &   $3.56 \times 10^{-6}$  \\
$^{16}$O  &   $5.38 \times 10^{-2}$ &$^{32}$Cl  &   $5.93 \times 10^{-12}$ &$^{50}$Ti  &   $3.42 \times 10^{-11}$ &$^{57}$Ni  &   $2.34 \times 10^{-4}$ &$^{67}$Ge  &   $9.80 \times 10^{-7}$  \\
$^{17}$O  &   $3.75 \times 10^{-9}$ &$^{33}$Cl  &   $5.79 \times 10^{-14}$ &$^{51}$Ti  &   $9.54 \times 10^{-14}$ &$^{58}$Ni  &   $1.22 \times 10^{-3}$ &$^{68}$Ge  &   $2.21 \times 10^{-5}$  \\
$^{18}$O  &   $9.79 \times 10^{-8}$ &$^{34}$Cl  &   $5.97 \times 10^{-15}$ &$^{52}$Ti  &   $3.92 \times 10^{-15}$ &$^{59}$Ni  &   $1.54 \times 10^{-4}$ &$^{69}$Ge  &   $1.22 \times 10^{-6}$  \\
$^{17}$F  &   $1.58 \times 10^{-13}$ &$^{35}$Cl  &   $3.34 \times 10^{-6}$ &$^{46}$V  &   $3.13 \times 10^{-15}$ &$^{60}$Ni  &   $5.07 \times 10^{-4}$ &$^{70}$Ge  &   $9.75 \times 10^{-6}$  \\
$^{18}$F  &   $4.68 \times 10^{-7}$ &$^{36}$Cl  &   $4.62 \times 10^{-6}$ &$^{47}$V  &   $1.05 \times 10^{-6}$ &$^{61}$Ni  &   $6.50 \times 10^{-6}$ &$^{71}$Ge  &   $4.98 \times 10^{-8}$  \\
$^{19}$F  &   $7.57 \times 10^{-9}$ &$^{37}$Cl  &   $4.93 \times 10^{-8}$ &$^{48}$V  &   $1.01 \times 10^{-7}$ &$^{62}$Ni  &   $1.64 \times 10^{-5}$ &$^{72}$Ge  &   $2.27 \times 10^{-7}$  \\
$^{20}$F  &   $1.34 \times 10^{-11}$ &$^{38}$Cl  &   $1.35 \times 10^{-10}$ &$^{49}$V  &   $5.00 \times 10^{-7}$ &$^{63}$Ni  &   $3.98 \times 10^{-8}$ &$^{73}$Ge  &   $1.09 \times 10^{-9}$  \\
$^{20}$Ne  &   $7.89 \times 10^{-3}$ &$^{39}$Cl  &   $1.41 \times 10^{-12}$ &$^{50}$V  &   $5.25 \times 10^{-9}$ &$^{64}$Ni  &   $2.31 \times 10^{-8}$ &$^{74}$Ge  &   $1.46 \times 10^{-9}$  \\
$^{21}$Ne  &   $5.29 \times 10^{-9}$ &$^{40}$Cl  &   $5.22 \times 10^{-15}$ &$^{51}$V  &   $1.94 \times 10^{-9}$ &$^{65}$Ni  &   $2.55 \times 10^{-10}$ &$^{75}$Ge  &   $3.60 \times 10^{-11}$  \\
$^{22}$Ne  &   $1.29 \times 10^{-9}$ &$^{35}$Ar  &   $8.17 \times 10^{-15}$ &$^{52}$V  &   $5.71 \times 10^{-12}$ &$^{66}$Ni  &   $2.79 \times 10^{-10}$ &$^{76}$Ge  &   $1.89 \times 10^{-11}$  \\
$^{23}$Ne  &   $4.53 \times 10^{-15}$ &$^{36}$Ar  &   $3.27 \times 10^{-3}$ &$^{53}$V  &   $5.27 \times 10^{-14}$ &$^{56}$Cu  &   $2.96 \times 10^{-14}$ &$^{77}$Ge  &   $7.56 \times 10^{-13}$  \\
$^{24}$Ne  &   $2.14 \times 10^{-13}$ &$^{37}$Ar  &   $2.52 \times 10^{-6}$ &$^{48}$Cr  &   $1.37 \times 10^{-5}$ &$^{57}$Cu  &   $1.65 \times 10^{-14}$ &$^{78}$Ge  &   $1.70 \times 10^{-13}$  \\
$^{20}$Na  &   $6.04 \times 10^{-13}$ &$^{38}$Ar  &   $6.53 \times 10^{-7}$ &$^{49}$Cr  &   $8.52 \times 10^{-7}$ &$^{58}$Cu  &   $3.62 \times 10^{-11}$ &$^{66}$As  &   $2.99 \times 10^{-15}$  \\
$^{21}$Na  &   $1.99 \times 10^{-15}$ &$^{39}$Ar  &   $4.75 \times 10^{-9}$ &$^{50}$Cr  &   $1.56 \times 10^{-6}$ &$^{59}$Cu  &   $1.29 \times 10^{-6}$ &$^{67}$As  &   $1.84 \times 10^{-10}$  \\
$^{22}$Na  &   $8.16 \times 10^{-8}$ &$^{40}$Ar  &   $1.90 \times 10^{-8}$ &$^{51}$Cr  &   $9.60 \times 10^{-7}$ &$^{60}$Cu  &   $1.49 \times 10^{-4}$ &$^{68}$As  &   $2.86 \times 10^{-7}$  \\
$^{23}$Na  &   $1.65 \times 10^{-7}$ &$^{41}$Ar  &   $1.91 \times 10^{-12}$ &$^{52}$Cr  &   $7.71 \times 10^{-8}$ &$^{61}$Cu  &   $5.23 \times 10^{-5}$ &$^{69}$As  &   $3.08 \times 10^{-7}$  \\
$^{24}$Na  &   $2.06 \times 10^{-7}$ &$^{42}$Ar  &   $2.67 \times 10^{-12}$ &$^{53}$Cr  &   $2.86 \times 10^{-9}$ &$^{62}$Cu  &   $1.24 \times 10^{-6}$ &$^{70}$As  &   $4.85 \times 10^{-7}$  \\
$^{25}$Na  &   $1.85 \times 10^{-14}$ &$^{36}$K  &   $1.55 \times 10^{-12}$ &$^{54}$Cr  &   $3.07 \times 10^{-9}$ &$^{63}$Cu  &   $1.57 \times 10^{-5}$ &$^{71}$As  &   $6.59 \times 10^{-7}$  \\
$^{23}$Mg  &   $2.17 \times 10^{-15}$ &$^{37}$K  &   $7.12 \times 10^{-15}$ &$^{55}$Cr  &   $3.00 \times 10^{-13}$ &$^{64}$Cu  &   $1.33 \times 10^{-6}$ &$^{72}$As  &   $5.31 \times 10^{-8}$ \\
$^{24}$Mg  &   $1.92 \times 10^{-3}$ &$^{38}$K  &   $1.38 \times 10^{-7}$ &$^{56}$Cr  &   $4.69 \times 10^{-13}$ &$^{65}$Cu  &   $9.72 \times 10^{-8}$ &$^{73}$As  &   $9.74 \times 10^{-8}$  \\
$^{25}$Mg  &   $2.53 \times 10^{-7}$ &$^{39}$K  &   $4.28 \times 10^{-6}$ &$^{50}$Mn  &   $3.20 \times 10^{-15}$ &$^{66}$Cu  &   $6.39 \times 10^{-9}$ &$^{74}$As  &   $1.42 \times 10^{-8}$  \\
$^{26}$Mg  &   $3.57 \times 10^{-10}$ &$^{40}$K  &   $2.76 \times 10^{-6}$ &$^{51}$Mn  &   $2.04 \times 10^{-6}$ &$^{67}$Cu  &   $2.77 \times 10^{-10}$ &$^{75}$As  &   $6.27 \times 10^{-9}$  \\
$^{27}$Mg  &   $2.99 \times 10^{-10}$ &$^{41}$K  &   $4.31 \times 10^{-9}$ &$^{52}$Mn  &   $1.89 \times 10^{-6}$ &$^{68}$Cu  &   $4.51 \times 10^{-15}$ &$^{76}$As  &   $4.10 \times 10^{-9}$  \\
$^{28}$Mg  &   $1.06 \times 10^{-8}$ &$^{42}$K  &   $6.51 \times 10^{-10}$ &$^{53}$Mn  &   $3.49 \times 10^{-6}$ &$^{69}$Cu  &   $1.29 \times 10^{-14}$ &$^{77}$As  &   $6.86 \times 10^{-10}$ \\
$^{24}$Al  &   $1.06 \times 10^{-12}$ &$^{43}$K  &   $1.99 \times 10^{-11}$ &$^{54}$Mn  &   $4.00 \times 10^{-7}$ &$^{60}$Zn  &   $2.93 \times 10^{-5}$ &$^{78}$As  &   $8.90 \times 10^{-11}$  \\
$^{25}$Al  &   $3.34 \times 10^{-15}$ &$^{44}$K  &   $1.95 \times 10^{-13}$ &$^{55}$Mn  &   $8.59 \times 10^{-9}$ &$^{61}$Zn  &   $1.97 \times 10^{-7}$ &$^{79}$As  &   $2.58 \times 10^{-13}$  \\
$^{26}$Al  &   $1.41 \times 10^{-7}$ &$^{45}$K  &   $4.98 \times 10^{-14}$ &$^{56}$Mn  &   $1.44 \times 10^{-9}$ &$^{62}$Zn  &   $1.75 \times 10^{-4}$ &$^{68}$Se  &   $1.85 \times 10^{-8}$  \\
$^{27}$Al  &   $6.28 \times 10^{-6}$ &$^{39}$Ca  &   $3.05 \times 10^{-15}$ &$^{57}$Mn  &   $4.65 \times 10^{-13}$ &$^{63}$Zn  &   $1.86 \times 10^{-5}$ &$^{69}$Se  &   $2.67 \times 10^{-12}$ \\
$^{28}$Al  &   $3.73 \times 10^{-8}$ &$^{40}$Ca  &   $2.37 \times 10^{-3}$ &$^{58}$Mn  &   $1.08 \times 10^{-15}$ &$^{64}$Zn  &   $2.63 \times 10^{-4}$ &$^{70}$Se  &   $1.82 \times 10^{-7}$  \\
$^{71}$Se  &   $4.10 \times 10^{-8}$ &$^{78}$Rb  &   $1.16 \times 10^{-8}$ &$^{86}$Zr  &   $1.02 \times 10^{-7}$ &$^{97}$Tc  &   $1.81 \times 10^{-9}$ &$^{106}$Pd  &   $1.07 \times 10^{-8}$ \\
$^{72}$Se  &   $2.93 \times 10^{-6}$ &$^{79}$Rb  &   $5.18 \times 10^{-8}$ &$^{87}$Zr  &   $4.52 \times 10^{-8}$ &$^{98}$Tc  &   $1.11 \times 10^{-9}$ &$^{107}$Pd  &   $6.62 \times 10^{-10}$ \\
$^{73}$Se  &   $3.67 \times 10^{-7}$ &$^{80}$Rb  &   $8.72 \times 10^{-10}$ &$^{88}$Zr  &   $7.01 \times 10^{-8}$ &$^{99}$Tc  &   $9.68 \times 10^{-10}$ &$^{108}$Pd  &   $3.47 \times 10^{-10}$ \\
$^{74}$Se  &   $1.07 \times 10^{-6}$ &$^{81}$Rb  &   $1.87 \times 10^{-7}$ &$^{89}$Zr  &   $1.76 \times 10^{-7}$ &$^{100}$Tc  &   $1.04 \times 10^{-13}$ &$^{109}$Pd  &   $5.73 \times 10^{-11}$ \\
$^{75}$Se  &   $2.49 \times 10^{-7}$ &$^{82}$Rb  &   $2.88 \times 10^{-8}$ &$^{90}$Zr  &   $3.60 \times 10^{-6}$ &$^{101}$Tc  &   $4.14 \times 10^{-12}$ &$^{110}$Pd  &   $2.87 \times 10^{-11}$ \\
$^{76}$Se  &   $5.95 \times 10^{-7}$ &$^{83}$Rb  &   $2.44 \times 10^{-8}$ &$^{91}$Zr  &   $7.88 \times 10^{-9}$ &$^{102}$Tc  &   $3.38 \times 10^{-14}$ &$^{111}$Pd  &   $4.03 \times 10^{-13}$ \\
$^{77}$Se  &   $1.33 \times 10^{-8}$ &$^{84}$Rb  &   $5.68 \times 10^{-9}$ &$^{92}$Zr  &   $2.95 \times 10^{-10}$ &$^{104}$Tc  &   $2.49 \times 10^{-13}$ &$^{112}$Pd  &   $1.08 \times 10^{-12}$ \\
$^{78}$Se  &   $3.42 \times 10^{-8}$ &$^{85}$Rb  &   $7.65 \times 10^{-9}$ &$^{93}$Zr  &   $1.74 \times 10^{-11}$ &$^{105}$Tc  &   $5.53 \times 10^{-15}$ &$^{98}$Ag  &   $5.86 \times 10^{-15}$ \\
$^{79}$Se  &   $1.44 \times 10^{-9}$ &$^{86}$Rb  &   $2.48 \times 10^{-9}$ &$^{94}$Zr  &   $4.45 \times 10^{-12}$ &$^{92}$Ru  &   $2.81 \times 10^{-10}$ &$^{99}$Ag  &   $2.03 \times 10^{-12}$ \\
$^{80}$Se  &   $8.82 \times 10^{-10}$ &$^{87}$Rb  &   $3.95 \times 10^{-9}$ &$^{95}$Zr  &   $7.24 \times 10^{-13}$ &$^{93}$Ru  &   $5.95 \times 10^{-13}$ &$^{100}$Ag  &   $1.03 \times 10^{-11}$ \\
$^{81}$Se  &   $2.52 \times 10^{-12}$ &$^{88}$Rb  &   $3.98 \times 10^{-10}$ &$^{96}$Zr  &   $2.90 \times 10^{-13}$ &$^{94}$Ru  &   $4.03 \times 10^{-8}$ &$^{101}$Ag  &   $2.04 \times 10^{-9}$ \\
$^{82}$Se  &   $4.93 \times 10^{-12}$ &$^{89}$Rb  &   $4.40 \times 10^{-12}$ &$^{97}$Zr  &   $1.27 \times 10^{-13}$ &$^{95}$Ru  &   $3.88 \times 10^{-8}$ &$^{102}$Ag  &   $1.12 \times 10^{-8}$ \\
$^{83}$Se  &   $3.64 \times 10^{-14}$ &$^{90}$Rb  &   $1.87 \times 10^{-14}$ &$^{85}$Nb  &   $4.59 \times 10^{-10}$ &$^{96}$Ru  &   $1.72 \times 10^{-7}$ &$^{103}$Ag  &   $2.38 \times 10^{-8}$ \\
$^{71}$Br  &   $1.64 \times 10^{-13}$ &$^{77}$Sr  &   $1.68 \times 10^{-15}$ &$^{86}$Nb  &   $1.64 \times 10^{-11}$ &$^{97}$Ru  &   $4.85 \times 10^{-8}$ &$^{104}$Ag  &   $2.57 \times 10^{-8}$ \\
$^{72}$Br  &   $2.07 \times 10^{-9}$ &$^{78}$Sr  &   $2.14 \times 10^{-12}$ &$^{87}$Nb  &   $1.77 \times 10^{-10}$ &$^{98}$Ru  &   $4.07 \times 10^{-8}$ &$^{105}$Ag  &   $2.84 \times 10^{-8}$ \\
$^{73}$Br  &   $8.84 \times 10^{-9}$ &$^{79}$Sr  &   $5.38 \times 10^{-10}$ &$^{88}$Nb  &   $9.38 \times 10^{-9}$ &$^{99}$Ru  &   $6.38 \times 10^{-9}$ &$^{106}$Ag  &   $3.77 \times 10^{-9}$ \\
$^{74}$Br  &   $1.02 \times 10^{-7}$ &$^{80}$Sr  &   $1.87 \times 10^{-7}$ &$^{89}$Nb  &   $8.24 \times 10^{-8}$ &$^{100}$Ru  &   $5.59 \times 10^{-9}$ &$^{107}$Ag  &   $1.80 \times 10^{-8}$ \\
$^{75}$Br  &   $8.18 \times 10^{-7}$ &$^{81}$Sr  &   $1.38 \times 10^{-7}$ &$^{90}$Nb  &   $8.14 \times 10^{-7}$ &$^{101}$Ru  &   $1.10 \times 10^{-9}$ &$^{108}$Ag  &   $1.80 \times 10^{-11}$ \\
$^{76}$Br  &   $8.97 \times 10^{-8}$ &$^{82}$Sr  &   $1.17 \times 10^{-7}$ &$^{91}$Nb  &   $1.72 \times 10^{-7}$ &$^{102}$Ru  &   $9.03 \times 10^{-10}$ &$^{109}$Ag  &   $6.17 \times 10^{-9}$ \\
$^{77}$Br  &   $2.13 \times 10^{-7}$ &$^{83}$Sr  &   $1.02 \times 10^{-7}$ &$^{92}$Nb  &   $9.83 \times 10^{-9}$ &$^{103}$Ru  &   $1.41 \times 10^{-10}$ &$^{110}$Ag  &   $8.13 \times 10^{-13}$ \\
$^{78}$Br  &   $1.56 \times 10^{-10}$ &$^{84}$Sr  &   $1.42 \times 10^{-7}$ &$^{93}$Nb  &   $3.51 \times 10^{-9}$ &$^{104}$Ru  &   $6.46 \times 10^{-11}$ &$^{111}$Ag  &   $1.07 \times 10^{-9}$ \\
$^{79}$Br  &   $1.59 \times 10^{-8}$ &$^{85}$Sr  &   $3.73 \times 10^{-8}$ &$^{94}$Nb  &   $8.39 \times 10^{-10}$ &$^{105}$Ru  &   $8.80 \times 10^{-12}$ &$^{112}$Ag  &   $4.34 \times 10^{-10}$ \\
$^{80}$Br  &   $6.66 \times 10^{-10}$ &$^{86}$Sr  &   $3.94 \times 10^{-8}$ &$^{95}$Nb  &   $5.27 \times 10^{-10}$ &$^{106}$Ru  &   $2.48 \times 10^{-12}$ &$^{113}$Ag  &   $2.93 \times 10^{-10}$ \\
$^{81}$Br  &   $3.09 \times 10^{-9}$ &$^{87}$Sr  &   $7.13 \times 10^{-9}$ &$^{96}$Nb  &   $2.46 \times 10^{-10}$ &$^{94}$Rh  &   $3.39 \times 10^{-14}$ &$^{114}$Ag  &   $3.46 \times 10^{-15}$ \\
$^{82}$Br  &   $1.39 \times 10^{-9}$ &$^{88}$Sr  &   $1.64 \times 10^{-7}$ &$^{97}$Nb  &   $6.11 \times 10^{-11}$ &$^{95}$Rh  &   $1.04 \times 10^{-9}$ &$^{96}$Cd  &   $1.20 \times 10^{-12}$ \\
$^{83}$Br  &   $2.93 \times 10^{-10}$ &$^{89}$Sr  &   $2.26 \times 10^{-9}$ &$^{98}$Nb  &   $1.19 \times 10^{-15}$ &$^{96}$Rh  &   $3.45 \times 10^{-8}$ &$^{97}$Cd  &   $1.95 \times 10^{-11}$ \\
$^{84}$Br  &   $1.21 \times 10^{-11}$ &$^{90}$Sr  &   $2.58 \times 10^{-10}$ &$^{88}$Mo  &   $2.33 \times 10^{-9}$ &$^{97}$Rh  &   $1.92 \times 10^{-8}$ &$^{100}$Cd  &   $1.27 \times 10^{-14}$ \\
$^{85}$Br  &   $3.65 \times 10^{-15}$ &$^{91}$Sr  &   $8.19 \times 10^{-13}$ &$^{89}$Mo  &   $1.23 \times 10^{-10}$ &$^{98}$Rh  &   $1.09 \times 10^{-9}$ &$^{101}$Cd  &   $1.20 \times 10^{-12}$ \\
$^{72}$Kr  &   $1.30 \times 10^{-14}$ &$^{92}$Sr  &   $1.48 \times 10^{-14}$ &$^{90}$Mo  &   $2.28 \times 10^{-7}$ &$^{99}$Rh  &   $1.38 \times 10^{-8}$ &$^{102}$Cd  &   $1.65 \times 10^{-9}$ \\
$^{73}$Kr  &   $7.47 \times 10^{-14}$ &$^{80}$Y  &   $1.31 \times 10^{-13}$ &$^{91}$Mo  &   $2.23 \times 10^{-8}$ &$^{100}$Rh  &   $1.03 \times 10^{-8}$ &$^{103}$Cd  &   $2.13 \times 10^{-9}$ \\
$^{74}$Kr  &   $3.81 \times 10^{-8}$ &$^{81}$Y  &   $4.12 \times 10^{-11}$ &$^{92}$Mo  &   $1.48 \times 10^{-7}$ &$^{101}$Rh  &   $3.42 \times 10^{-9}$ &$^{104}$Cd  &   $4.83 \times 10^{-8}$ \\
$^{75}$Kr  &   $6.53 \times 10^{-8}$ &$^{82}$Y  &   $1.44 \times 10^{-10}$ &$^{93}$Mo  &   $8.72 \times 10^{-9}$ &$^{102}$Rh  &   $2.34 \times 10^{-9}$ &$^{105}$Cd  &   $2.72 \times 10^{-8}$ \\
$^{76}$Kr  &   $1.58 \times 10^{-7}$ &$^{83}$Y  &   $1.27 \times 10^{-9}$ &$^{94}$Mo  &   $9.04 \times 10^{-9}$ &$^{103}$Rh  &   $2.35 \times 10^{-9}$ &$^{106}$Cd  &   $2.35 \times 10^{-7}$ \\
$^{77}$Kr  &   $4.72 \times 10^{-7}$ &$^{84}$Y  &   $7.39 \times 10^{-11}$ &$^{95}$Mo  &   $3.89 \times 10^{-10}$ &$^{104}$Rh  &   $6.64 \times 10^{-13}$ &$^{107}$Cd  &   $9.60 \times 10^{-8}$ \\
$^{78}$Kr  &   $2.93 \times 10^{-7}$ &$^{85}$Y  &   $1.22 \times 10^{-7}$ &$^{96}$Mo  &   $1.65 \times 10^{-10}$ &$^{105}$Rh  &   $6.52 \times 10^{-10}$ &$^{108}$Cd  &   $1.07 \times 10^{-7}$ \\
$^{79}$Kr  &   $1.00 \times 10^{-7}$ &$^{86}$Y  &   $1.85 \times 10^{-8}$ &$^{97}$Mo  &   $8.02 \times 10^{-11}$ &$^{106}$Rh  &   $1.01 \times 10^{-13}$ &$^{109}$Cd  &   $3.46 \times 10^{-8}$ \\
$^{80}$Kr  &   $3.06 \times 10^{-7}$ &$^{87}$Y  &   $3.00 \times 10^{-8}$ &$^{98}$Mo  &   $3.49 \times 10^{-11}$ &$^{107}$Rh  &   $1.56 \times 10^{-11}$ &$^{110}$Cd  &   $2.82 \times 10^{-8}$ \\
$^{81}$Kr  &   $2.85 \times 10^{-8}$ &$^{88}$Y  &   $4.64 \times 10^{-8}$ &$^{99}$Mo  &   $4.56 \times 10^{-12}$ &$^{108}$Rh  &   $1.48 \times 10^{-13}$ &$^{111}$Cd  &   $6.12 \times 10^{-9}$ \\
$^{82}$Kr  &   $6.36 \times 10^{-8}$ &$^{89}$Y  &   $1.66 \times 10^{-7}$ &$^{100}$Mo  &   $1.50 \times 10^{-12}$ &$^{96}$Pd  &   $2.37 \times 10^{-10}$ &$^{112}$Cd  &   $2.71 \times 10^{-8}$ \\
$^{83}$Kr  &   $2.35 \times 10^{-9}$ &$^{90}$Y  &   $3.67 \times 10^{-8}$ &$^{101}$Mo  &   $1.11 \times 10^{-14}$ &$^{97}$Pd  &   $1.77 \times 10^{-10}$ &$^{113}$Cd  &   $1.24 \times 10^{-9}$ \\
$^{84}$Kr  &   $2.44 \times 10^{-9}$ &$^{91}$Y  &   $3.27 \times 10^{-10}$ &$^{102}$Mo  &   $1.48 \times 10^{-15}$ &$^{98}$Pd  &   $4.17 \times 10^{-9}$ &$^{114}$Cd  &   $4.76 \times 10^{-10}$ \\
$^{85}$Kr  &   $1.56 \times 10^{-10}$ &$^{92}$Y  &   $1.19 \times 10^{-11}$ &$^{90}$Tc  &   $2.34 \times 10^{-14}$ &$^{99}$Pd  &   $2.52 \times 10^{-9}$ &$^{115}$Cd  &   $4.26 \times 10^{-11}$ \\
$^{86}$Kr  &   $4.35 \times 10^{-11}$ &$^{93}$Y  &   $3.87 \times 10^{-12}$ &$^{91}$Tc  &   $2.09 \times 10^{-10}$ &$^{100}$Pd  &   $1.46 \times 10^{-8}$ &$^{116}$Cd  &   $1.95 \times 10^{-11}$ \\
$^{87}$Kr  &   $1.46 \times 10^{-11}$ &$^{94}$Y  &   $1.81 \times 10^{-14}$ &$^{92}$Tc  &   $2.64 \times 10^{-9}$ &$^{101}$Pd  &   $1.16 \times 10^{-8}$ &$^{102}$In  &   $2.68 \times 10^{-15}$ \\
$^{88}$Kr  &   $3.61 \times 10^{-12}$ &$^{82}$Zr  &   $8.49 \times 10^{-9}$ &$^{93}$Tc  &   $2.99 \times 10^{-8}$ &$^{102}$Pd  &   $3.15 \times 10^{-8}$ &$^{103}$In  &   $2.72 \times 10^{-13}$ \\
$^{75}$Rb  &   $1.24 \times 10^{-14}$ &$^{83}$Zr  &   $1.44 \times 10^{-13}$ &$^{94}$Tc  &   $1.89 \times 10^{-8}$ &$^{103}$Pd  &   $1.16 \times 10^{-8}$ &$^{104}$In  &   $2.19 \times 10^{-11}$ \\
$^{76}$Rb  &   $1.03 \times 10^{-12}$ &$^{84}$Zr  &   $4.76 \times 10^{-9}$ &$^{95}$Tc  &   $1.35 \times 10^{-8}$ &$^{104}$Pd  &   $1.20 \times 10^{-8}$ &$^{105}$In  &   $6.14 \times 10^{-10}$ \\
$^{77}$Rb  &   $2.26 \times 10^{-8}$ &$^{85}$Zr  &   $2.57 \times 10^{-8}$ &$^{96}$Tc  &   $2.93 \times 10^{-9}$ &$^{105}$Pd  &   $1.67 \times 10^{-9}$ &$^{106}$In  &   $1.94 \times 10^{-8}$ \\
$^{107}$In  &   $9.40 \times 10^{-8}$ &$^{111}$Sn  &   $3.99 \times 10^{-8}$ &$^{115}$Sb  &   $1.17 \times 10^{-8}$ &$^{121}$Te  &   $1.63 \times 10^{-9}$ &$^{124}$Xe  &   $3.60 \times 10^{-8}$ \\
$^{108}$In  &   $2.02 \times 10^{-7}$ &$^{112}$Sn  &   $1.44 \times 10^{-7}$ &$^{116}$Sb  &   $4.28 \times 10^{-9}$ &$^{122}$Te  &   $5.90 \times 10^{-9}$ &$^{125}$Xe  &   $2.16 \times 10^{-8}$ \\
$^{109}$In  &   $1.80 \times 10^{-7}$ &$^{113}$Sn  &   $3.83 \times 10^{-8}$ &$^{117}$Sb  &   $1.25 \times 10^{-8}$ &$^{123}$Te  &   $5.12 \times 10^{-10}$ &$^{126}$Xe  &   $4.49 \times 10^{-8}$ \\
$^{110}$In  &   $1.03 \times 10^{-7}$ &$^{114}$Sn  &   $6.57 \times 10^{-8}$ &$^{118}$Sb  &   $7.12 \times 10^{-11}$ &$^{124}$Te  &   $1.88 \times 10^{-10}$ &$^{127}$Xe  &   $1.38 \times 10^{-8}$ \\
$^{111}$In  &   $1.39 \times 10^{-7}$ &$^{115}$Sn  &   $2.13 \times 10^{-8}$ &$^{119}$Sb  &   $5.32 \times 10^{-9}$ &$^{115}$I  &   $2.54 \times 10^{-15}$ &$^{128}$Xe  &   $1.04 \times 10^{-8}$ \\
$^{112}$In  &   $4.20 \times 10^{-9}$ &$^{116}$Sn  &   $2.28 \times 10^{-8}$ &$^{120}$Sb  &   $2.68 \times 10^{-10}$ &$^{116}$I  &   $1.46 \times 10^{-15}$ &$^{126}$Cs  &   $4.87 \times 10^{-10}$ \\
$^{113}$In  &   $6.05 \times 10^{-8}$ &$^{117}$Sn  &   $4.65 \times 10^{-9}$ &$^{121}$Sb  &   $2.50 \times 10^{-9}$ &$^{117}$I  &   $2.38 \times 10^{-13}$ &$^{127}$Cs  &   $6.47 \times 10^{-9}$ \\
$^{114}$In  &   $2.40 \times 10^{-11}$ &$^{118}$Sn  &   $8.75 \times 10^{-9}$ &$^{122}$Sb  &   $8.41 \times 10^{-10}$ &$^{118}$I  &   $8.98 \times 10^{-11}$ &$^{128}$Cs  &   $2.88 \times 10^{-9}$ \\
$^{115}$In  &   $1.09 \times 10^{-8}$ &$^{119}$Sn  &   $4.34 \times 10^{-10}$ &$^{112}$Te  &   $7.87 \times 10^{-13}$ &$^{119}$I  &   $3.14 \times 10^{-10}$ &$^{129}$Cs  &   $2.54 \times 10^{-8}$ \\
$^{116}$In  &   $7.84 \times 10^{-13}$ &$^{120}$Sn  &   $1.73 \times 10^{-9}$ &$^{113}$Te  &   $2.80 \times 10^{-13}$ &$^{120}$I  &   $2.97 \times 10^{-10}$ &$^{130}$Cs  &   $1.39 \times 10^{-8}$ \\
$^{117}$In  &   $6.92 \times 10^{-10}$ &$^{108}$Sb  &   $1.10 \times 10^{-14}$ &$^{114}$Te  &   $1.94 \times 10^{-9}$ &$^{121}$I  &   $2.89 \times 10^{-9}$ &$^{130}$Ba  &   $3.25 \times 10^{-8}$ \\
$^{118}$In  &   $1.50 \times 10^{-14}$ &$^{109}$Sb  &   $2.13 \times 10^{-14}$ &$^{115}$Te  &   $3.40 \times 10^{-10}$ &$^{122}$I  &   $8.43 \times 10^{-11}$ &$^{131}$Ba  &   $2.02 \times 10^{-8}$ \\
$^{106}$Sn  &   $1.33 \times 10^{-10}$ &$^{110}$Sb  &   $2.03 \times 10^{-13}$ &$^{116}$Te  &   $1.21 \times 10^{-8}$ &$^{123}$I  &   $8.35 \times 10^{-9}$ &$^{132}$Ba  &   $7.54 \times 10^{-8}$ \\
$^{107}$Sn  &   $3.56 \times 10^{-10}$ &$^{111}$Sb  &   $1.84 \times 10^{-13}$ &$^{117}$Te  &   $5.18 \times 10^{-9}$ &$^{124}$I  &   $6.94 \times 10^{-9}$ &$^{136}$Ce  &   $1.92 \times 10^{-9}$ \\
$^{108}$Sn  &   $3.90 \times 10^{-8}$ &$^{112}$Sb  &   $8.68 \times 10^{-13}$ &$^{118}$Te  &   $1.17 \times 10^{-8}$ &$^{125}$I  &   $9.33 \times 10^{-9}$ \\
$^{109}$Sn  &   $3.11 \times 10^{-8}$ &$^{113}$Sb  &   $9.56 \times 10^{-10}$ &$^{119}$Te  &   $3.89 \times 10^{-9}$ &$^{126}$I  &   $3.90 \times 10^{-9}$ \\
$^{110}$Sn  &   $2.15 \times 10^{-7}$ &$^{114}$Sb  &   $6.58 \times 10^{-10}$ &$^{120}$Te  &   $4.20 \times 10^{-9}$ &$^{123}$Xe  &   $1.24 \times 10^{-9}$ \\
\end{xtabular}
\endgroup
\newpage

\begingroup
\fontsize{10}{10.7}\selectfont
\topcaption{Isotopic yields for the 17.0~M$_{\odot}$ progenitor. Only isotopes with  ejected masses greater than $10^{-15}$~M$_{\odot}$ are included.} \label{tab:17.0_yields}
\tablefirsthead{\toprule Iso & M/M$_{\odot}$ & Iso & M/M$_{\odot}$ & Iso & M/M$_{\odot}$ & Iso & M/M$_{\odot}$ & Iso & M/M$_{\odot}$ \\ \midrule}
\tablehead{\multicolumn{10}{c}{{\bfseries \tablename\ \thetable{} -- continued from previous page}} \\
\toprule Iso & M/M$_{\odot}$ & Iso & M/M$_{\odot}$ & Iso & M/M$_{\odot}$ & Iso & M/M$_{\odot}$ & Iso & M/M$_{\odot}$ \\ \midrule}
\begin{xtabular}{cl|cl|cl|cl|cl}
\shrinkheight{-4.2cm}
$^{1}$H  & $ 2.81$ & $^{24}$Al  & $ 1.62 \times 10^{-11}$ & $^{43}$K  & $ 7.74 \times 10^{-12}$ & $^{51}$Mn  & $ 5.87 \times 10^{-06}$ & $^{68}$Cu  & $ 1.47 \times 10^{-14}$ \\
$^{2}$H  & $ 1.06 \times 10^{-10}$ & $^{25}$Al  & $ 9.77 \times 10^{-14}$ & $^{44}$K  & $ 6.19 \times 10^{-14}$ & $^{52}$Mn  & $ 1.31 \times 10^{-05}$ & $^{69}$Cu  & $ 4.74 \times 10^{-15}$ \\
$^{3}$H  & $ 2.83 \times 10^{-12}$ & $^{26}$Al  & $ 4.10 \times 10^{-07}$ & $^{45}$K  & $ 8.58 \times 10^{-15}$ & $^{53}$Mn  & $ 9.56 \times 10^{-06}$ & $^{70}$Cu  & $ 1.62 \times 10^{-15}$ \\
$^{3}$He  & $ 7.43 \times 10^{-05}$ & $^{27}$Al  & $ 2.08 \times 10^{-05}$ & $^{39}$Ca  & $ 4.87 \times 10^{-15}$ & $^{54}$Mn  & $ 3.73 \times 10^{-06}$ & $^{60}$Zn  & $ 2.85 \times 10^{-05}$ \\
$^{4}$He  & $ 2.15$ & $^{28}$Al  & $ 5.30 \times 10^{-08}$ & $^{40}$Ca  & $ 1.71 \times 10^{-03}$ & $^{55}$Mn  & $ 2.82 \times 10^{-07}$ & $^{61}$Zn  & $ 7.46 \times 10^{-08}$ \\
$^{6}$Li  & $ 4.12 \times 10^{-11}$ & $^{29}$Al  & $ 5.34 \times 10^{-10}$ & $^{41}$Ca  & $ 4.38 \times 10^{-06}$ & $^{56}$Mn  & $ 1.45 \times 10^{-07}$ & $^{62}$Zn  & $ 1.68 \times 10^{-04}$ \\
$^{7}$Li  & $ 2.10 \times 10^{-10}$ & $^{30}$Al  & $ 4.48 \times 10^{-15}$ & $^{42}$Ca  & $ 2.52 \times 10^{-06}$ & $^{57}$Mn  & $ 2.93 \times 10^{-13}$ & $^{63}$Zn  & $ 8.91 \times 10^{-06}$ \\
$^{7}$Be  & $ 9.68 \times 10^{-09}$ & $^{27}$Si  & $ 8.09 \times 10^{-14}$ & $^{43}$Ca  & $ 6.41 \times 10^{-07}$ & $^{52}$Fe  & $ 2.13 \times 10^{-04}$ & $^{64}$Zn  & $ 3.14 \times 10^{-04}$ \\
$^{9}$Be  & $ 3.87 \times 10^{-11}$ & $^{28}$Si  & $ 1.45 \times 10^{-02}$ & $^{44}$Ca  & $ 1.26 \times 10^{-08}$ & $^{53}$Fe  & $ 9.95 \times 10^{-07}$ & $^{65}$Zn  & $ 6.34 \times 10^{-06}$ \\
$^{10}$Be  & $ 2.47 \times 10^{-10}$ & $^{29}$Si  & $ 1.61 \times 10^{-05}$ & $^{45}$Ca  & $ 1.25 \times 10^{-10}$ & $^{54}$Fe  & $ 7.36 \times 10^{-03}$ & $^{66}$Zn  & $ 8.25 \times 10^{-05}$ \\
$^{11}$Be  & $ 4.48 \times 10^{-15}$ & $^{30}$Si  & $ 5.50 \times 10^{-06}$ & $^{46}$Ca  & $ 2.67 \times 10^{-12}$ & $^{55}$Fe  & $ 1.01 \times 10^{-05}$ & $^{67}$Zn  & $ 4.66 \times 10^{-07}$ \\
$^{10}$B  & $ 2.61 \times 10^{-11}$ & $^{31}$Si  & $ 1.02 \times 10^{-08}$ & $^{47}$Ca  & $ 2.75 \times 10^{-11}$ & $^{56}$Fe  & $ 6.31 \times 10^{-06}$ & $^{68}$Zn  & $ 3.37 \times 10^{-06}$ \\
$^{11}$B  & $ 4.48 \times 10^{-06}$ & $^{32}$Si  & $ 2.33 \times 10^{-07}$ & $^{48}$Ca  & $ 1.12 \times 10^{-07}$ & $^{57}$Fe  & $ 2.12 \times 10^{-07}$ & $^{69}$Zn  & $ 9.36 \times 10^{-08}$ \\
$^{12}$B  & $ 3.80 \times 10^{-13}$ & $^{28}$P  & $ 1.44 \times 10^{-12}$ & $^{49}$Ca  & $ 2.84 \times 10^{-14}$ & $^{58}$Fe  & $ 8.28 \times 10^{-07}$ & $^{70}$Zn  & $ 3.46 \times 10^{-07}$ \\
$^{11}$C  & $ 3.01 \times 10^{-07}$ & $^{29}$P  & $ 7.91 \times 10^{-14}$ & $^{40}$Sc  & $ 8.95 \times 10^{-14}$ & $^{59}$Fe  & $ 2.64 \times 10^{-08}$ & $^{71}$Zn  & $ 3.69 \times 10^{-15}$ \\
$^{12}$C  & $ 1.01 \times 10^{-01}$ & $^{30}$P  & $ 1.16 \times 10^{-07}$ & $^{41}$Sc  & $ 1.41 \times 10^{-15}$ & $^{60}$Fe  & $ 1.90 \times 10^{-06}$ & $^{72}$Zn  & $ 2.99 \times 10^{-07}$ \\
$^{13}$C  & $ 5.11 \times 10^{-06}$ & $^{31}$P  & $ 2.05 \times 10^{-05}$ & $^{42}$Sc  & $ 1.01 \times 10^{-15}$ & $^{61}$Fe  & $ 3.68 \times 10^{-14}$ & $^{62}$Ga  & $ 2.26 \times 10^{-14}$ \\
$^{14}$C  & $ 7.08 \times 10^{-07}$ & $^{32}$P  & $ 2.68 \times 10^{-05}$ & $^{43}$Sc  & $ 4.58 \times 10^{-06}$ & $^{52}$Co  & $ 1.80 \times 10^{-15}$ & $^{63}$Ga  & $ 6.35 \times 10^{-09}$ \\
$^{12}$N  & $ 2.55 \times 10^{-13}$ & $^{33}$P  & $ 9.30 \times 10^{-08}$ & $^{44}$Sc  & $ 1.14 \times 10^{-08}$ & $^{54}$Co  & $ 3.37 \times 10^{-13}$ & $^{64}$Ga  & $ 1.27 \times 10^{-06}$ \\
$^{13}$N  & $ 1.91 \times 10^{-08}$ & $^{34}$P  & $ 4.94 \times 10^{-15}$ & $^{45}$Sc  & $ 6.11 \times 10^{-07}$ & $^{55}$Co  & $ 5.02 \times 10^{-05}$ & $^{65}$Ga  & $ 3.14 \times 10^{-07}$ \\
$^{14}$N  & $ 3.16 \times 10^{-04}$ & $^{35}$P  & $ 1.12 \times 10^{-15}$ & $^{46}$Sc  & $ 1.11 \times 10^{-09}$ & $^{56}$Co  & $ 6.22 \times 10^{-05}$ & $^{66}$Ga  & $ 2.34 \times 10^{-06}$ \\
$^{15}$N  & $ 5.76 \times 10^{-06}$ & $^{27}$S  & $ 7.26 \times 10^{-12}$ & $^{47}$Sc  & $ 1.02 \times 10^{-10}$ & $^{57}$Co  & $ 1.03 \times 10^{-05}$ & $^{67}$Ga  & $ 2.43 \times 10^{-06}$ \\
$^{16}$N  & $ 2.49 \times 10^{-10}$ & $^{31}$S  & $ 5.23 \times 10^{-14}$ & $^{48}$Sc  & $ 3.93 \times 10^{-09}$ & $^{58}$Co  & $ 4.92 \times 10^{-06}$ & $^{68}$Ga  & $ 3.62 \times 10^{-08}$ \\
$^{14}$O  & $ 5.23 \times 10^{-10}$ & $^{32}$S  & $ 9.50 \times 10^{-03}$ & $^{49}$Sc  & $ 5.79 \times 10^{-09}$ & $^{59}$Co  & $ 5.73 \times 10^{-07}$ & $^{69}$Ga  & $ 6.83 \times 10^{-07}$ \\
$^{15}$O  & $ 2.17 \times 10^{-08}$ & $^{33}$S  & $ 1.79 \times 10^{-05}$ & $^{50}$Sc  & $ 2.24 \times 10^{-15}$ & $^{60}$Co  & $ 2.09 \times 10^{-06}$ & $^{70}$Ga  & $ 1.67 \times 10^{-09}$ \\
$^{16}$O  & $ 4.82 \times 10^{-01}$ & $^{34}$S  & $ 3.47 \times 10^{-06}$ & $^{44}$Ti  & $ 1.28 \times 10^{-05}$ & $^{61}$Co  & $ 1.10 \times 10^{-07}$ & $^{71}$Ga  & $ 9.25 \times 10^{-08}$ \\
$^{17}$O  & $ 1.77 \times 10^{-06}$ & $^{35}$S  & $ 2.08 \times 10^{-08}$ & $^{45}$Ti  & $ 3.03 \times 10^{-06}$ & $^{62}$Co  & $ 1.73 \times 10^{-12}$ & $^{72}$Ga  & $ 2.52 \times 10^{-08}$ \\
$^{18}$O  & $ 5.16 \times 10^{-06}$ & $^{36}$S  & $ 8.43 \times 10^{-08}$ & $^{46}$Ti  & $ 5.32 \times 10^{-06}$ & $^{63}$Co  & $ 3.83 \times 10^{-15}$ & $^{73}$Ga  & $ 1.15 \times 10^{-08}$ \\
$^{19}$O  & $ 5.00 \times 10^{-14}$ & $^{37}$S  & $ 7.06 \times 10^{-13}$ & $^{47}$Ti  & $ 4.93 \times 10^{-06}$ & $^{55}$Ni  & $ 6.41 \times 10^{-15}$ & $^{74}$Ga  & $ 2.05 \times 10^{-11}$ \\
$^{20}$O  & $ 8.70 \times 10^{-15}$ & $^{38}$S  & $ 1.06 \times 10^{-13}$ & $^{48}$Ti  & $ 3.79 \times 10^{-08}$ & $^{56}$Ni  & $ 1.30 \times 10^{-02}$ & $^{76}$Ga  & $ 1.68 \times 10^{-15}$ \\
$^{17}$F  & $ 9.38 \times 10^{-11}$ & $^{32}$Cl  & $ 1.21 \times 10^{-12}$ & $^{49}$Ti  & $ 1.36 \times 10^{-08}$ & $^{57}$Ni  & $ 2.22 \times 10^{-04}$ & $^{64}$Ge  & $ 3.40 \times 10^{-07}$ \\
$^{18}$F  & $ 1.20 \times 10^{-05}$ & $^{33}$Cl  & $ 3.96 \times 10^{-14}$ & $^{50}$Ti  & $ 8.90 \times 10^{-08}$ & $^{58}$Ni  & $ 1.15 \times 10^{-03}$ & $^{65}$Ge  & $ 7.70 \times 10^{-13}$ \\
$^{19}$F  & $ 8.05 \times 10^{-06}$ & $^{34}$Cl  & $ 2.47 \times 10^{-15}$ & $^{51}$Ti  & $ 1.89 \times 10^{-13}$ & $^{59}$Ni  & $ 1.23 \times 10^{-04}$ & $^{66}$Ge  & $ 1.86 \times 10^{-06}$ \\
$^{20}$F  & $ 2.41 \times 10^{-10}$ & $^{35}$Cl  & $ 1.16 \times 10^{-05}$ & $^{52}$Ti  & $ 2.48 \times 10^{-15}$ & $^{60}$Ni  & $ 1.04 \times 10^{-03}$ & $^{67}$Ge  & $ 2.27 \times 10^{-07}$ \\
$^{21}$F  & $ 9.19 \times 10^{-15}$ & $^{36}$Cl  & $ 5.64 \times 10^{-06}$ & $^{46}$V  & $ 1.09 \times 10^{-15}$ & $^{61}$Ni  & $ 1.64 \times 10^{-05}$ & $^{68}$Ge  & $ 2.00 \times 10^{-05}$ \\
$^{19}$Ne  & $ 3.28 \times 10^{-11}$ & $^{37}$Cl  & $ 1.57 \times 10^{-07}$ & $^{47}$V  & $ 3.59 \times 10^{-06}$ & $^{62}$Ni  & $ 8.53 \times 10^{-05}$ & $^{69}$Ge  & $ 6.77 \times 10^{-07}$ \\
$^{20}$Ne  & $ 1.26 \times 10^{-01}$ & $^{38}$Cl  & $ 9.99 \times 10^{-11}$ & $^{48}$V  & $ 3.52 \times 10^{-07}$ & $^{63}$Ni  & $ 4.05 \times 10^{-07}$ & $^{70}$Ge  & $ 1.76 \times 10^{-05}$ \\
$^{21}$Ne  & $ 1.85 \times 10^{-05}$ & $^{39}$Cl  & $ 6.04 \times 10^{-13}$ & $^{49}$V  & $ 3.90 \times 10^{-06}$ & $^{64}$Ni  & $ 2.23 \times 10^{-05}$ & $^{71}$Ge  & $ 1.05 \times 10^{-07}$ \\
$^{22}$Ne  & $ 1.72 \times 10^{-07}$ & $^{35}$Ar  & $ 8.88 \times 10^{-15}$ & $^{50}$V  & $ 6.08 \times 10^{-09}$ & $^{65}$Ni  & $ 3.45 \times 10^{-08}$ & $^{72}$Ge  & $ 1.73 \times 10^{-06}$ \\
$^{23}$Ne  & $ 1.95 \times 10^{-12}$ & $^{36}$Ar  & $ 1.95 \times 10^{-03}$ & $^{51}$V  & $ 8.51 \times 10^{-09}$ & $^{66}$Ni  & $ 2.27 \times 10^{-05}$ & $^{73}$Ge  & $ 3.38 \times 10^{-08}$ \\
$^{24}$Ne  & $ 9.05 \times 10^{-11}$ & $^{37}$Ar  & $ 7.89 \times 10^{-06}$ & $^{52}$V  & $ 1.13 \times 10^{-11}$ & $^{67}$Ni  & $ 1.86 \times 10^{-15}$ & $^{74}$Ge  & $ 2.20 \times 10^{-07}$ \\
$^{20}$Na  & $ 1.07 \times 10^{-11}$ & $^{38}$Ar  & $ 4.01 \times 10^{-06}$ & $^{53}$V  & $ 5.87 \times 10^{-14}$ & $^{56}$Cu  & $ 6.40 \times 10^{-14}$ & $^{75}$Ge  & $ 2.14 \times 10^{-08}$ \\
$^{21}$Na  & $ 2.74 \times 10^{-12}$ & $^{39}$Ar  & $ 3.36 \times 10^{-08}$ & $^{54}$V  & $ 7.98 \times 10^{-15}$ & $^{57}$Cu  & $ 3.89 \times 10^{-14}$ & $^{76}$Ge  & $ 4.00 \times 10^{-07}$ \\
$^{22}$Na  & $ 4.11 \times 10^{-07}$ & $^{40}$Ar  & $ 3.39 \times 10^{-08}$ & $^{48}$Cr  & $ 4.14 \times 10^{-05}$ & $^{58}$Cu  & $ 3.64 \times 10^{-12}$ & $^{77}$Ge  & $ 5.33 \times 10^{-09}$ \\
$^{23}$Na  & $ 9.33 \times 10^{-06}$ & $^{41}$Ar  & $ 4.85 \times 10^{-12}$ & $^{49}$Cr  & $ 3.44 \times 10^{-06}$ & $^{59}$Cu  & $ 1.54 \times 10^{-06}$ & $^{78}$Ge  & $ 4.11 \times 10^{-07}$ \\
$^{24}$Na  & $ 1.88 \times 10^{-05}$ & $^{42}$Ar  & $ 7.18 \times 10^{-12}$ & $^{50}$Cr  & $ 6.28 \times 10^{-06}$ & $^{60}$Cu  & $ 2.46 \times 10^{-04}$ & $^{67}$As  & $ 3.64 \times 10^{-09}$ \\
$^{25}$Na  & $ 2.09 \times 10^{-11}$ & $^{36}$K  & $ 2.83 \times 10^{-13}$ & $^{51}$Cr  & $ 5.76 \times 10^{-06}$ & $^{61}$Cu  & $ 3.80 \times 10^{-05}$ & $^{68}$As  & $ 8.10 \times 10^{-08}$ \\
$^{23}$Mg  & $ 9.69 \times 10^{-13}$ & $^{37}$K  & $ 5.21 \times 10^{-15}$ & $^{52}$Cr  & $ 4.52 \times 10^{-07}$ & $^{62}$Cu  & $ 3.61 \times 10^{-07}$ & $^{69}$As  & $ 1.89 \times 10^{-08}$ \\
$^{24}$Mg  & $ 2.03 \times 10^{-02}$ & $^{38}$K  & $ 5.28 \times 10^{-07}$ & $^{53}$Cr  & $ 2.46 \times 10^{-08}$ & $^{63}$Cu  & $ 2.52 \times 10^{-05}$ & $^{70}$As  & $ 6.73 \times 10^{-08}$ \\
$^{25}$Mg  & $ 1.17 \times 10^{-05}$ & $^{39}$K  & $ 1.64 \times 10^{-05}$ & $^{54}$Cr  & $ 1.89 \times 10^{-07}$ & $^{64}$Cu  & $ 1.04 \times 10^{-06}$ & $^{71}$As  & $ 1.33 \times 10^{-07}$ \\
$^{26}$Mg  & $ 1.83 \times 10^{-07}$ & $^{40}$K  & $ 3.92 \times 10^{-06}$ & $^{55}$Cr  & $ 1.70 \times 10^{-12}$ & $^{65}$Cu  & $ 1.06 \times 10^{-06}$ & $^{72}$As  & $ 5.99 \times 10^{-08}$ \\
$^{27}$Mg  & $ 1.04 \times 10^{-09}$ & $^{41}$K  & $ 2.07 \times 10^{-08}$ & $^{56}$Cr  & $ 1.20 \times 10^{-11}$ & $^{66}$Cu  & $ 3.56 \times 10^{-08}$ & $^{73}$As  & $ 2.15 \times 10^{-07}$ \\
$^{28}$Mg  & $ 1.97 \times 10^{-08}$ & $^{42}$K  & $ 6.01 \times 10^{-10}$ & $^{50}$Mn  & $ 1.05 \times 10^{-15}$ & $^{67}$Cu  & $ 6.42 \times 10^{-07}$ & $^{74}$As  & $ 9.92 \times 10^{-09}$ \\
$^{75}$As  & $ 1.12 \times 10^{-07}$ & $^{86}$Br  & $ 3.84 \times 10^{-14}$ & $^{83}$Sr  & $ 5.37 \times 10^{-09}$ & $^{92}$Zr  & $ 1.23 \times 10^{-09}$  & $^{98}$Tc  & $ 9.96 \times 10^{-13}$  \\
$^{76}$As  & $ 2.05 \times 10^{-08}$ & $^{72}$Kr  & $ 3.20 \times 10^{-12}$ & $^{84}$Sr  & $ 2.72 \times 10^{-07}$ & $^{93}$Zr  & $ 1.07 \times 10^{-10}$  & $^{99}$Tc  & $ 5.18 \times 10^{-15}$  \\
$^{77}$As  & $ 1.57 \times 10^{-08}$ & $^{73}$Kr  & $ 3.42 \times 10^{-12}$ & $^{85}$Sr  & $ 2.37 \times 10^{-07}$ & $^{94}$Zr  & $ 5.04 \times 10^{-11}$  & $^{92}$Ru  & $ 3.34 \times 10^{-14}$  \\
$^{78}$As  & $ 3.40 \times 10^{-07}$ & $^{74}$Kr  & $ 1.59 \times 10^{-09}$ & $^{86}$Sr  & $ 2.95 \times 10^{-07}$ & $^{95}$Zr  & $ 2.24 \times 10^{-12}$  & $^{93}$Ru  & $ 3.49 \times 10^{-15}$  \\
$^{79}$As  & $ 2.19 \times 10^{-11}$ & $^{75}$Kr  & $ 1.55 \times 10^{-09}$ & $^{87}$Sr  & $ 2.39 \times 10^{-07}$ & $^{96}$Zr  & $ 8.02 \times 10^{-14}$  & $^{94}$Ru  & $ 3.16 \times 10^{-10}$  \\
$^{82}$As  & $ 2.85 \times 10^{-15}$ & $^{76}$Kr  & $ 1.73 \times 10^{-08}$ & $^{88}$Sr  & $ 7.07 \times 10^{-06}$ & $^{77}$Nb  & $ 2.71 \times 10^{-14}$  & $^{95}$Ru  & $ 4.54 \times 10^{-11}$  \\
$^{68}$Se  & $ 2.87 \times 10^{-08}$ & $^{77}$Kr  & $ 5.95 \times 10^{-09}$ & $^{89}$Sr  & $ 1.05 \times 10^{-08}$ & $^{78}$Nb  & $ 1.44 \times 10^{-14}$  & $^{96}$Ru  & $ 2.22 \times 10^{-09}$  \\
$^{69}$Se  & $ 5.04 \times 10^{-11}$ & $^{78}$Kr  & $ 4.06 \times 10^{-07}$ & $^{90}$Sr  & $ 2.01 \times 10^{-09}$ & $^{85}$Nb  & $ 2.00 \times 10^{-12}$  & $^{97}$Ru  & $ 3.48 \times 10^{-12}$  \\
$^{70}$Se  & $ 2.18 \times 10^{-08}$ & $^{79}$Kr  & $ 3.50 \times 10^{-08}$ & $^{91}$Sr  & $ 2.67 \times 10^{-12}$ & $^{86}$Nb  & $ 3.67 \times 10^{-13}$  & $^{98}$Ru  & $ 1.36 \times 10^{-11}$  \\
$^{71}$Se  & $ 1.08 \times 10^{-08}$ & $^{80}$Kr  & $ 7.85 \times 10^{-07}$ & $^{92}$Sr  & $ 7.58 \times 10^{-13}$ & $^{87}$Nb  & $ 1.69 \times 10^{-13}$  & $^{99}$Ru  & $ 3.53 \times 10^{-13}$  \\
$^{72}$Se  & $ 7.40 \times 10^{-07}$ & $^{81}$Kr  & $ 4.03 \times 10^{-07}$ & $^{80}$Y  & $ 1.50 \times 10^{-12}$ & $^{88}$Nb  & $ 8.62 \times 10^{-12}$  & $^{100}$Ru  & $ 1.14 \times 10^{-14}$  \\
$^{73}$Se  & $ 5.12 \times 10^{-08}$ & $^{82}$Kr  & $ 3.80 \times 10^{-07}$ & $^{81}$Y  & $ 7.60 \times 10^{-12}$ & $^{89}$Nb  & $ 1.85 \times 10^{-09}$  & $^{95}$Rh  & $ 4.00 \times 10^{-14}$  \\
$^{74}$Se  & $ 1.64 \times 10^{-06}$ & $^{83}$Kr  & $ 6.42 \times 10^{-08}$ & $^{82}$Y  & $ 1.42 \times 10^{-13}$ & $^{90}$Nb  & $ 8.92 \times 10^{-07}$  & $^{96}$Rh  & $ 5.54 \times 10^{-14}$  \\
$^{75}$Se  & $ 5.13 \times 10^{-07}$ & $^{84}$Kr  & $ 1.45 \times 10^{-06}$ & $^{83}$Y  & $ 8.19 \times 10^{-12}$ & $^{91}$Nb  & $ 4.36 \times 10^{-07}$  & $^{97}$Rh  & $ 2.46 \times 10^{-14}$ \\
$^{76}$Se  & $ 2.01 \times 10^{-06}$ & $^{85}$Kr  & $ 7.47 \times 10^{-07}$ & $^{84}$Y  & $ 5.52 \times 10^{-14}$  & $^{92}$Nb  & $ 1.14 \times 10^{-08}$  & $^{98}$Rh  & $ 1.07 \times 10^{-15}$ \\
$^{77}$Se  & $ 3.18 \times 10^{-07}$ & $^{86}$Kr  & $ 5.80 \times 10^{-06}$ & $^{85}$Y  & $ 3.38 \times 10^{-10}$  & $^{93}$Nb  & $ 2.52 \times 10^{-09}$  & $^{99}$Rh  & $ 2.20 \times 10^{-13}$ \\
$^{78}$Se  & $ 3.68 \times 10^{-07}$ & $^{87}$Kr  & $ 7.04 \times 10^{-10}$ & $^{86}$Y  & $ 2.66 \times 10^{-09}$  & $^{94}$Nb  & $ 1.23 \times 10^{-09}$  & $^{96}$Pd  & $ 8.18 \times 10^{-15}$ \\
$^{79}$Se  & $ 4.00 \times 10^{-08}$ & $^{88}$Kr  & $ 1.14 \times 10^{-10}$ & $^{87}$Y  & $ 9.72 \times 10^{-09}$  & $^{95}$Nb  & $ 1.15 \times 10^{-10}$  & $^{98}$Pd  & $ 4.36 \times 10^{-15}$ \\
$^{80}$Se  & $ 1.83 \times 10^{-07}$ & $^{75}$Rb  & $ 5.03 \times 10^{-13}$ & $^{88}$Y  & $ 2.70 \times 10^{-07}$  & $^{96}$Nb  & $ 2.04 \times 10^{-14}$  & $^{100}$Pd  & $ 2.59 \times 10^{-15}$ \\
$^{81}$Se  & $ 6.59 \times 10^{-09}$ & $^{76}$Rb  & $ 2.29 \times 10^{-12}$ & $^{89}$Y  & $ 1.67 \times 10^{-06}$  & $^{97}$Nb  & $ 4.39 \times 10^{-15}$  & $^{98}$Ag  & $ 2.23 \times 10^{-12}$ \\
$^{82}$Se  & $ 1.68 \times 10^{-06}$ & $^{77}$Rb  & $ 3.18 \times 10^{-10}$ & $^{90}$Y  & $ 6.09 \times 10^{-08}$  & $^{88}$Mo  & $ 1.60 \times 10^{-13}$  & $^{99}$Ag  & $ 2.64 \times 10^{-13}$ \\
$^{83}$Se  & $ 6.64 \times 10^{-10}$ & $^{78}$Rb  & $ 4.74 \times 10^{-11}$ & $^{91}$Y  & $ 7.74 \times 10^{-10}$  & $^{89}$Mo  & $ 1.73 \times 10^{-13}$  & $^{100}$Ag  & $ 2.03 \times 10^{-14}$ \\
$^{84}$Se  & $ 2.18 \times 10^{-14}$ & $^{79}$Rb  & $ 1.62 \times 10^{-10}$ & $^{92}$Y  & $ 4.21 \times 10^{-12}$  & $^{90}$Mo  & $ 1.06 \times 10^{-08}$  & $^{96}$Cd  & $ 1.04 \times 10^{-12}$ \\
$^{71}$Br  & $ 1.95 \times 10^{-11}$ & $^{80}$Rb  & $ 4.79 \times 10^{-12}$ & $^{93}$Y  & $ 2.59 \times 10^{-12}$  & $^{91}$Mo  & $ 4.24 \times 10^{-10}$  & $^{97}$Cd  & $ 8.50 \times 10^{-14}$ \\
$^{72}$Br  & $ 2.33 \times 10^{-09}$ & $^{81}$Rb  & $ 5.08 \times 10^{-09}$ & $^{94}$Y  & $ 1.66 \times 10^{-15}$  & $^{92}$Mo  & $ 4.32 \times 10^{-07}$  & $^{98}$Cd  & $ 4.58 \times 10^{-12}$ \\
$^{73}$Br  & $ 5.97 \times 10^{-10}$ & $^{82}$Rb  & $ 8.33 \times 10^{-09}$ & $^{76}$Zr  & $ 2.94 \times 10^{-13}$  & $^{93}$Mo  & $ 4.89 \times 10^{-08}$  & $^{99}$Cd  & $ 4.25 \times 10^{-13}$ \\
$^{74}$Br  & $ 2.54 \times 10^{-09}$ & $^{83}$Rb  & $ 2.36 \times 10^{-07}$ & $^{77}$Zr  & $ 4.11 \times 10^{-13}$  & $^{94}$Mo  & $ 2.80 \times 10^{-08}$  & $^{100}$Cd  & $ 4.39 \times 10^{-14}$ \\
$^{75}$Br  & $ 2.25 \times 10^{-08}$ & $^{84}$Rb  & $ 1.14 \times 10^{-08}$ & $^{78}$Zr  & $ 4.87 \times 10^{-13}$  & $^{95}$Mo  & $ 1.06 \times 10^{-08}$  & $^{101}$Cd  & $ 1.57 \times 10^{-15}$ \\
$^{76}$Br  & $ 4.96 \times 10^{-08}$ & $^{85}$Rb  & $ 9.11 \times 10^{-08}$ & $^{82}$Zr  & $ 8.00 \times 10^{-12}$  & $^{96}$Mo  & $ 2.77 \times 10^{-10}$  & $^{97}$In  & $ 1.37 \times 10^{-13}$ \\
$^{77}$Br  & $ 6.84 \times 10^{-08}$ & $^{86}$Rb  & $ 2.27 \times 10^{-07}$ & $^{83}$Zr  & $ 4.38 \times 10^{-13}$  & $^{97}$Mo  & $ 3.30 \times 10^{-12}$  & $^{98}$In  & $ 4.08 \times 10^{-12}$ \\
$^{78}$Br  & $ 7.36 \times 10^{-13}$ & $^{87}$Rb  & $ 1.00 \times 10^{-06}$ & $^{84}$Zr  & $ 3.15 \times 10^{-12}$  & $^{98}$Mo  & $ 6.75 \times 10^{-14}$  & $^{99}$In  & $ 2.69 \times 10^{-13}$ \\
$^{79}$Br  & $ 1.64 \times 10^{-07}$ & $^{88}$Rb  & $ 2.75 \times 10^{-10}$ & $^{85}$Zr  & $ 5.31 \times 10^{-12}$  & $^{91}$Tc  & $ 7.09 \times 10^{-14}$  & $^{100}$In  & $ 4.26 \times 10^{-14}$ \\
$^{80}$Br  & $ 1.57 \times 10^{-10}$ & $^{89}$Rb  & $ 4.06 \times 10^{-12}$ & $^{86}$Zr  & $ 7.21 \times 10^{-10}$  & $^{92}$Tc  & $ 9.00 \times 10^{-12}$  & $^{101}$In  & $ 2.34 \times 10^{-15}$ \\
$^{81}$Br  & $ 2.23 \times 10^{-07}$ & $^{78}$Sr  & $ 2.30 \times 10^{-12}$ & $^{87}$Zr  & $ 3.53 \times 10^{-10}$  & $^{93}$Tc  & $ 6.62 \times 10^{-10}$  & $^{99}$Sn  & $ 7.10 \times 10^{-13}$ \\
$^{82}$Br  & $ 1.04 \times 10^{-07}$ & $^{79}$Sr  & $ 1.61 \times 10^{-11}$ & $^{88}$Zr  & $ 1.92 \times 10^{-07}$  & $^{94}$Tc  & $ 1.59 \times 10^{-10}$  & $^{100}$Sn  & $ 3.23 \times 10^{-14}$ \\
$^{83}$Br  & $ 4.75 \times 10^{-08}$ & $^{80}$Sr  & $ 1.02 \times 10^{-09}$ & $^{89}$Zr  & $ 3.52 \times 10^{-07}$  & $^{95}$Tc  & $ 4.08 \times 10^{-11}$  & $^{101}$Sn  & $ 2.33 \times 10^{-15}$ \\
$^{84}$Br  & $ 2.39 \times 10^{-07}$ & $^{81}$Sr  & $ 2.34 \times 10^{-10}$ & $^{90}$Zr  & $ 3.00 \times 10^{-05}$  & $^{96}$Tc  & $ 1.27 \times 10^{-11}$  & & \\
$^{85}$Br  & $ 3.00 \times 10^{-14}$ & $^{82}$Sr  & $ 1.67 \times 10^{-07}$ & $^{91}$Zr  & $ 3.80 \times 10^{-08}$  & $^{97}$Tc  & $ 2.18 \times 10^{-10}$  & & \\
\end{xtabular}
\endgroup

\newpage
\section{Uncertainties in neutrino luminosities}
\label{app:lumintop}
\begin{figure}[h]
\centering
  \includegraphics[width=0.6\textwidth]{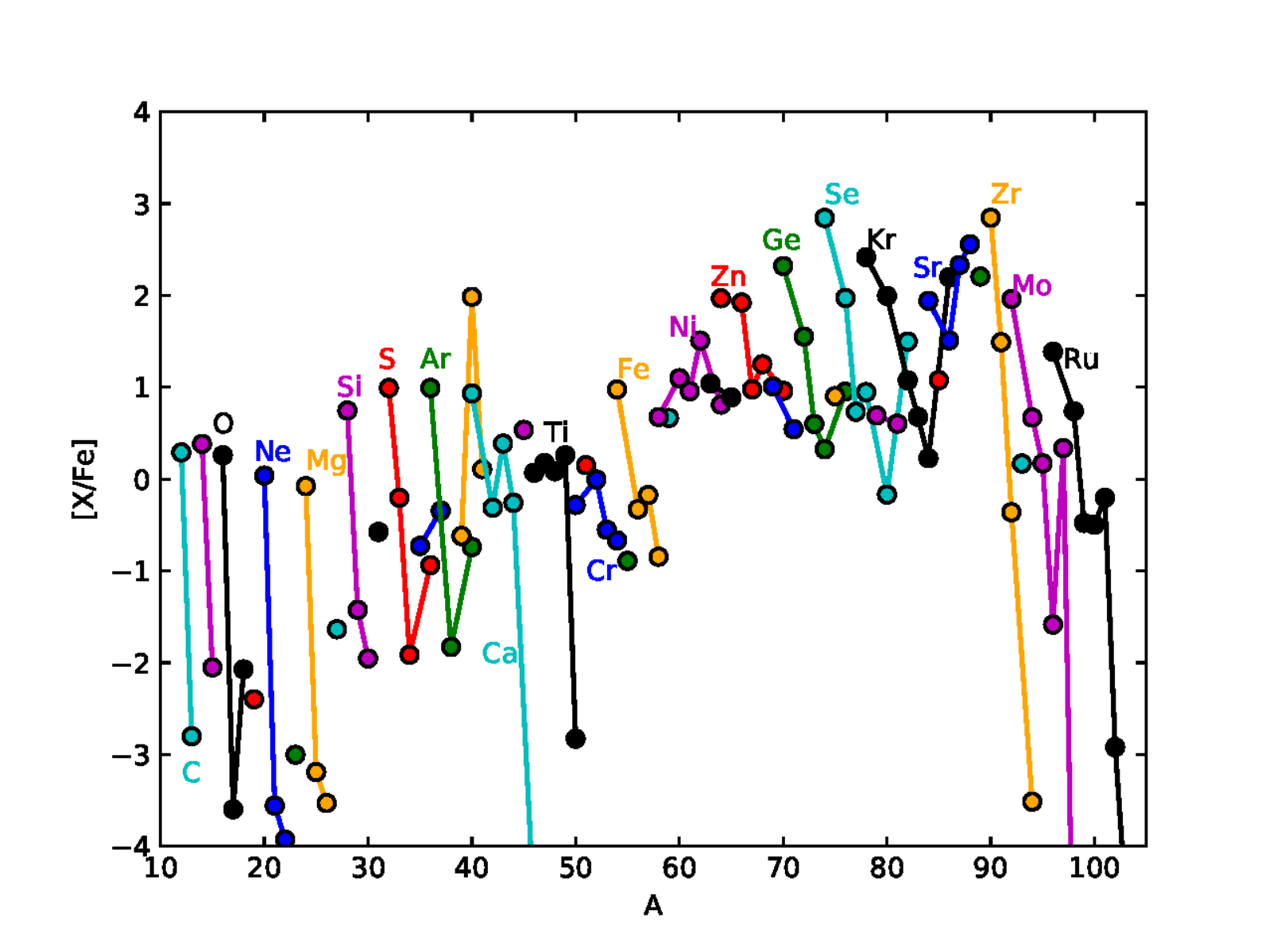}
  \caption{Same as figure~3, for the 11.2~M$_{\odot}$ model with neutrino luminosities and temperatures as seen from the north polar direction.}
  \label{fig:C1}
\end{figure}

For our calculations we have used angle-averaged neutrino luminosities and temperatures (see section~\ref{sec:nucleomethod}). While the differences for different angles in the 17.0~M$_{\odot}$ model are small, the neutrino properties for the 11.2~M$_{\odot}$ model show some angular dependence. Since this model also exhibits a unipolar outflow of material, the local neutrino luminosities and temperatures in the ejecta can differ from the angle-averaged values. Therefore, we have repeated the nucleosynthesis calculations for this CCSN model using neutrino properties as seen from the north pole of the simulation, i.e., the dominant direction of the ejecta. Figures~\ref{fig:C1} to~\ref{fig:C4} show that the different neutrino luminosities can affect the production of heavy nuclei and their abundance ratios in the high-, but also in the low-$Y_e$ tracer particles of the highest-temperature bin.


\begin{figure}[h]
\includegraphics[width=0.54\textwidth]{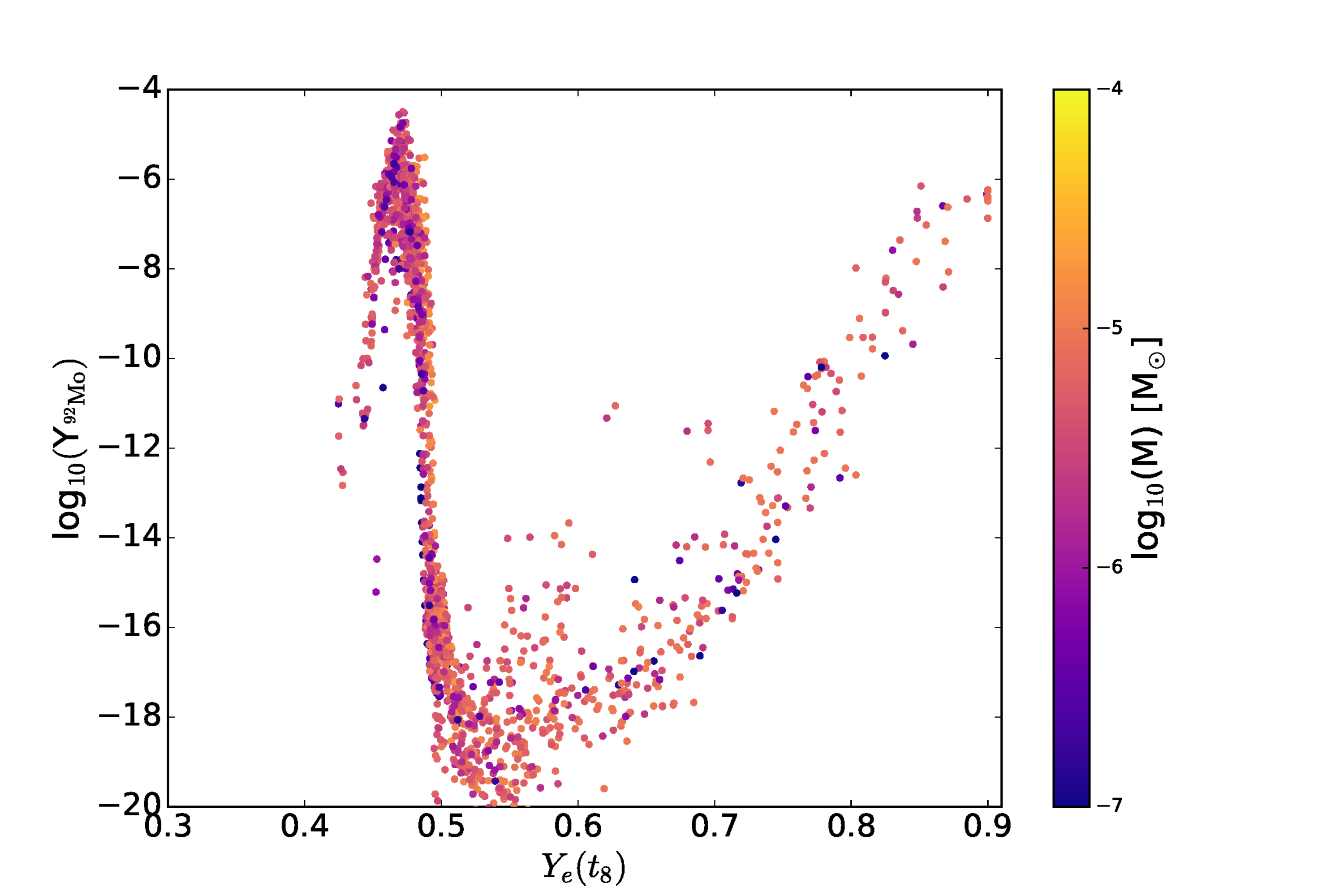}
\includegraphics[width=0.54\textwidth]{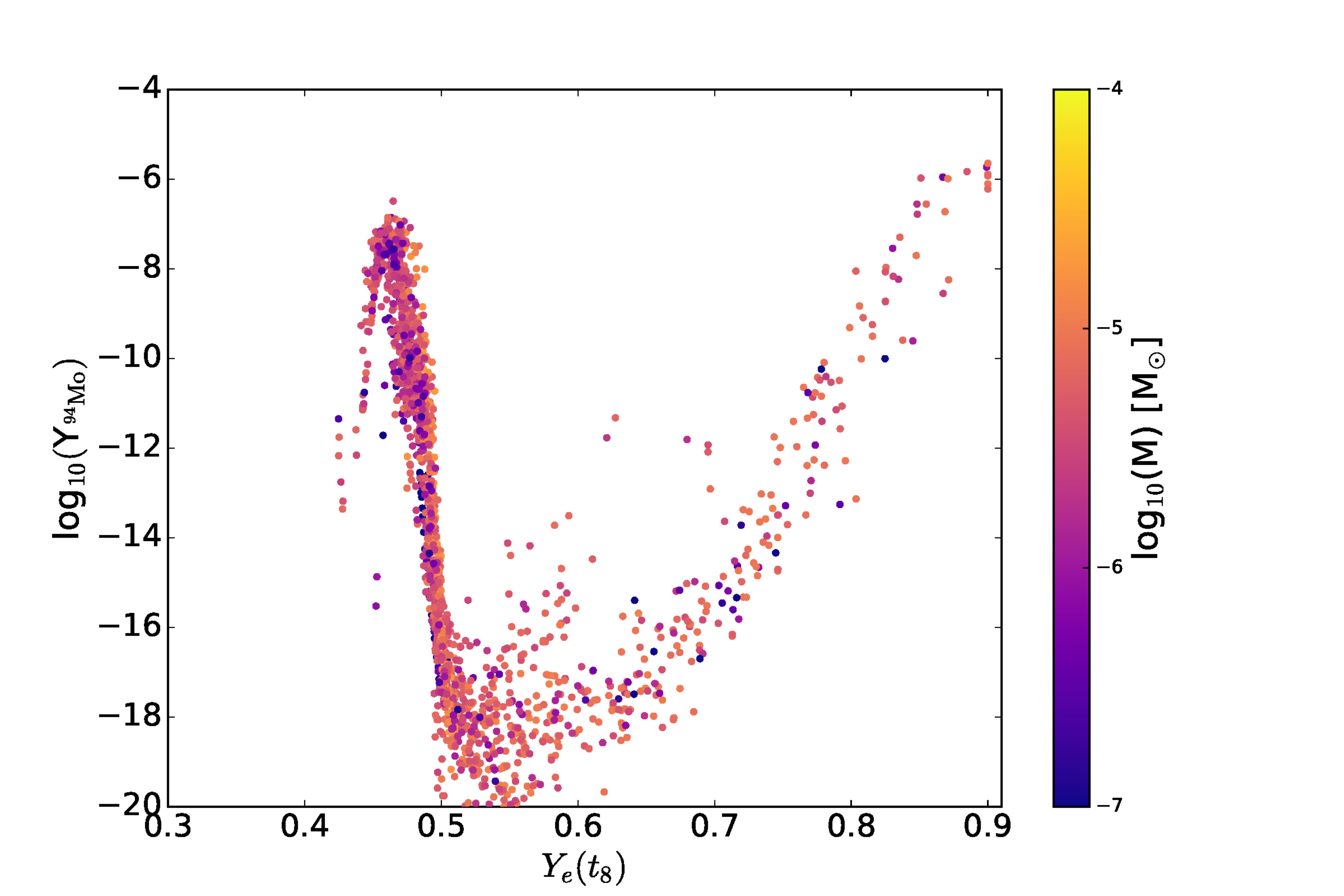} \\
\includegraphics[width=0.54\textwidth]{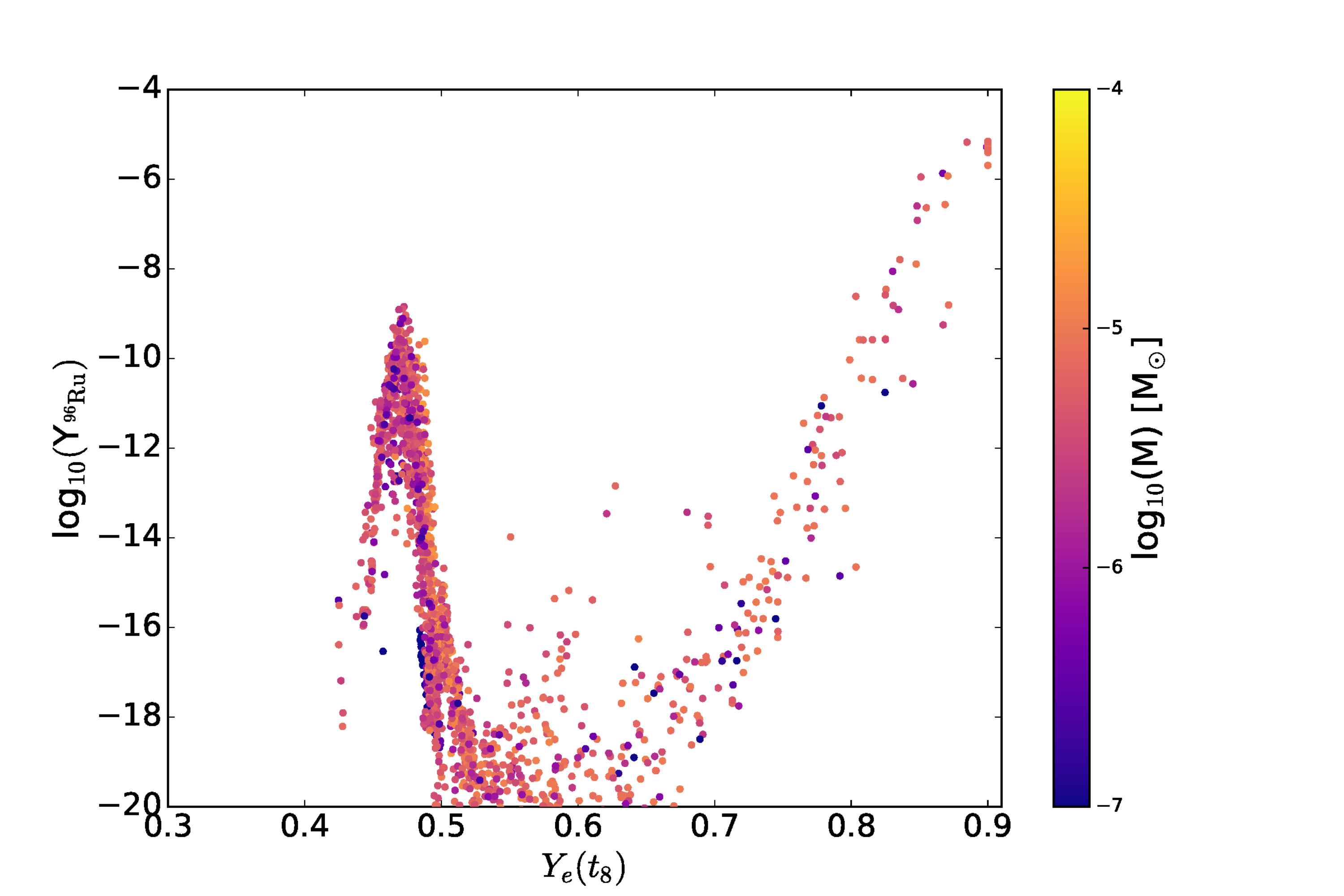}
\includegraphics[width=0.54\textwidth]{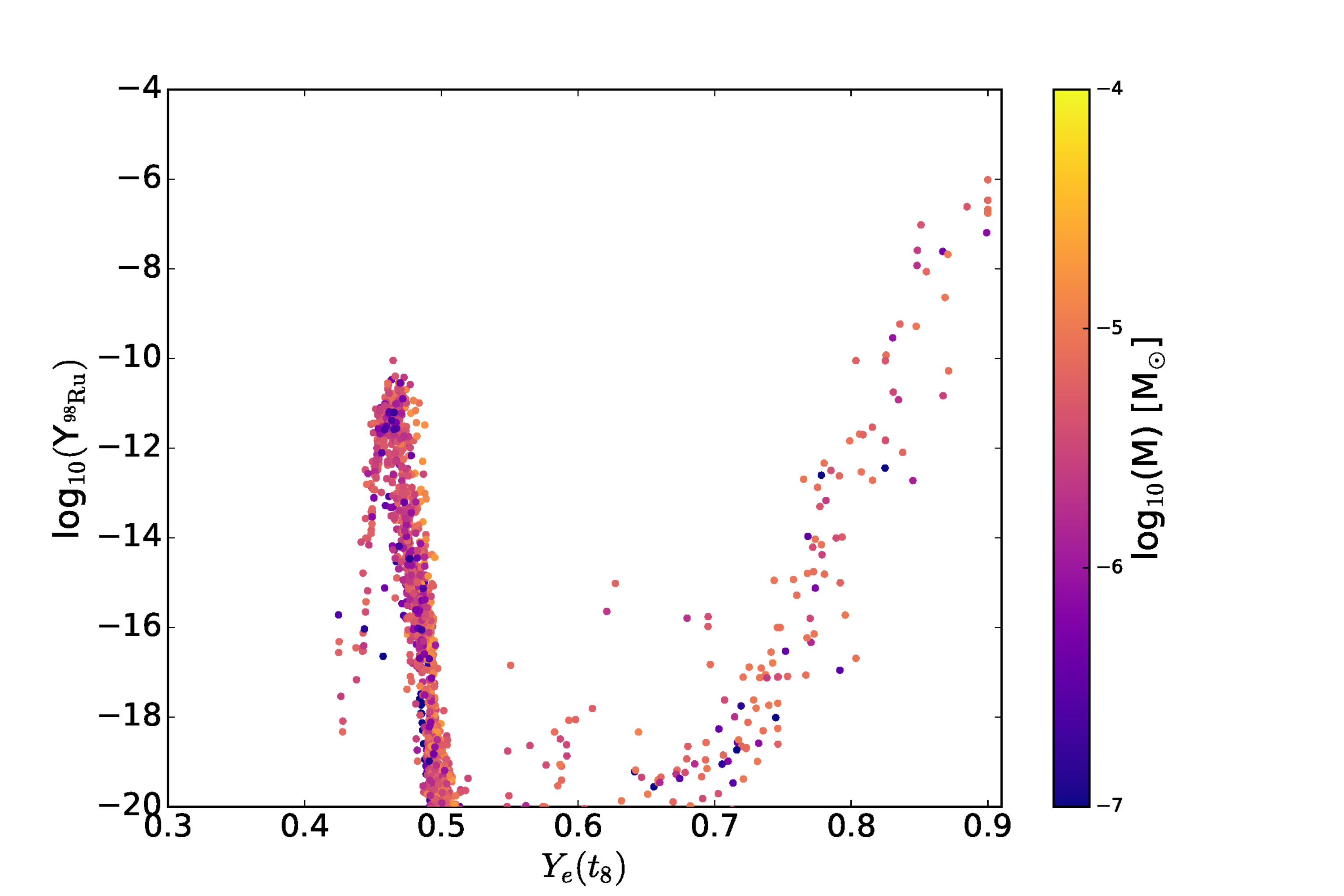}
\caption{Same as figure~\ref{fig:11.2_morupeaks}, but with neutrino luminosities and temperatures as seen from the north polar direction.}
\end{figure}


\begin{figure}[h]
\includegraphics[width=0.45\textwidth]{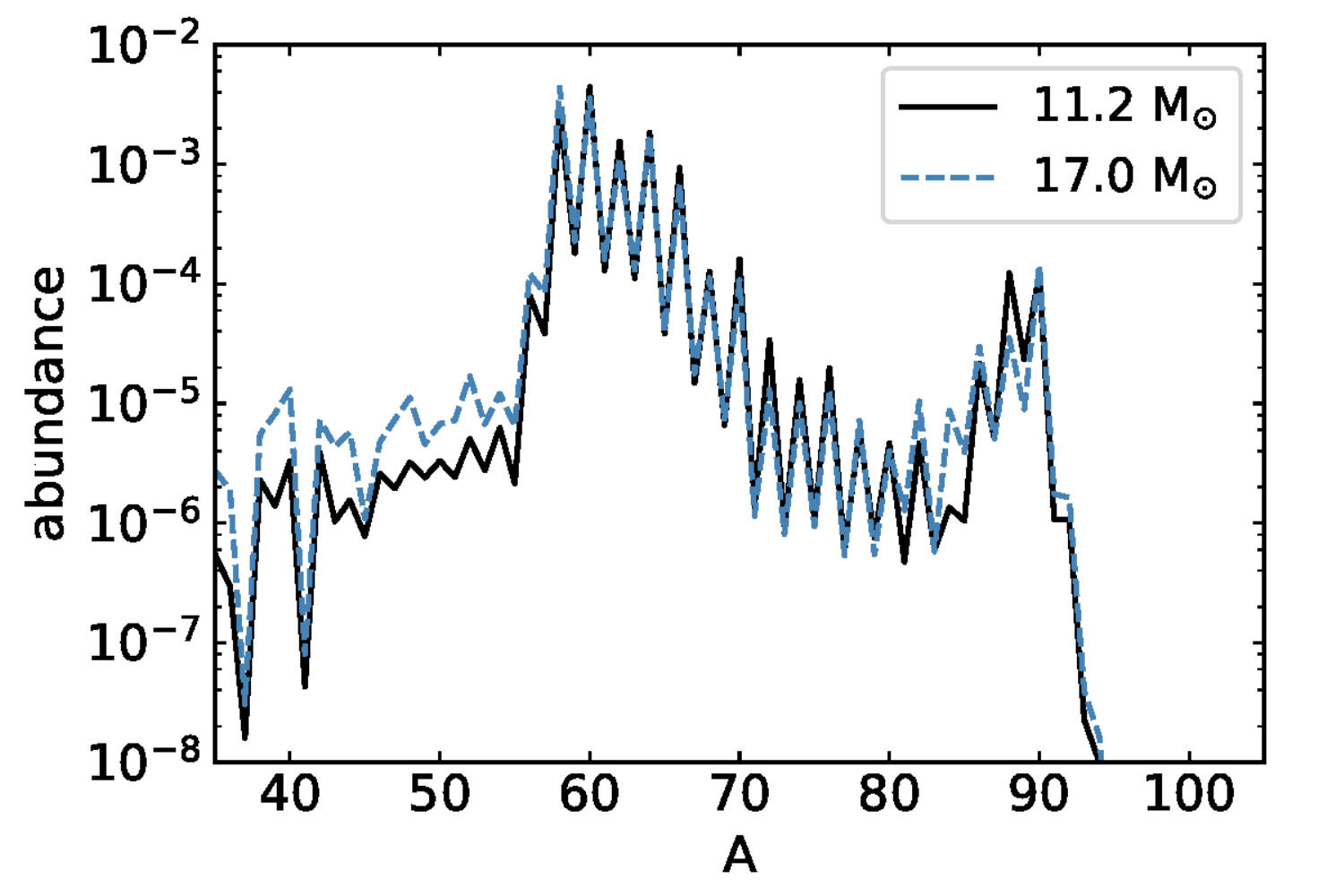}
\includegraphics[width=0.45\textwidth]{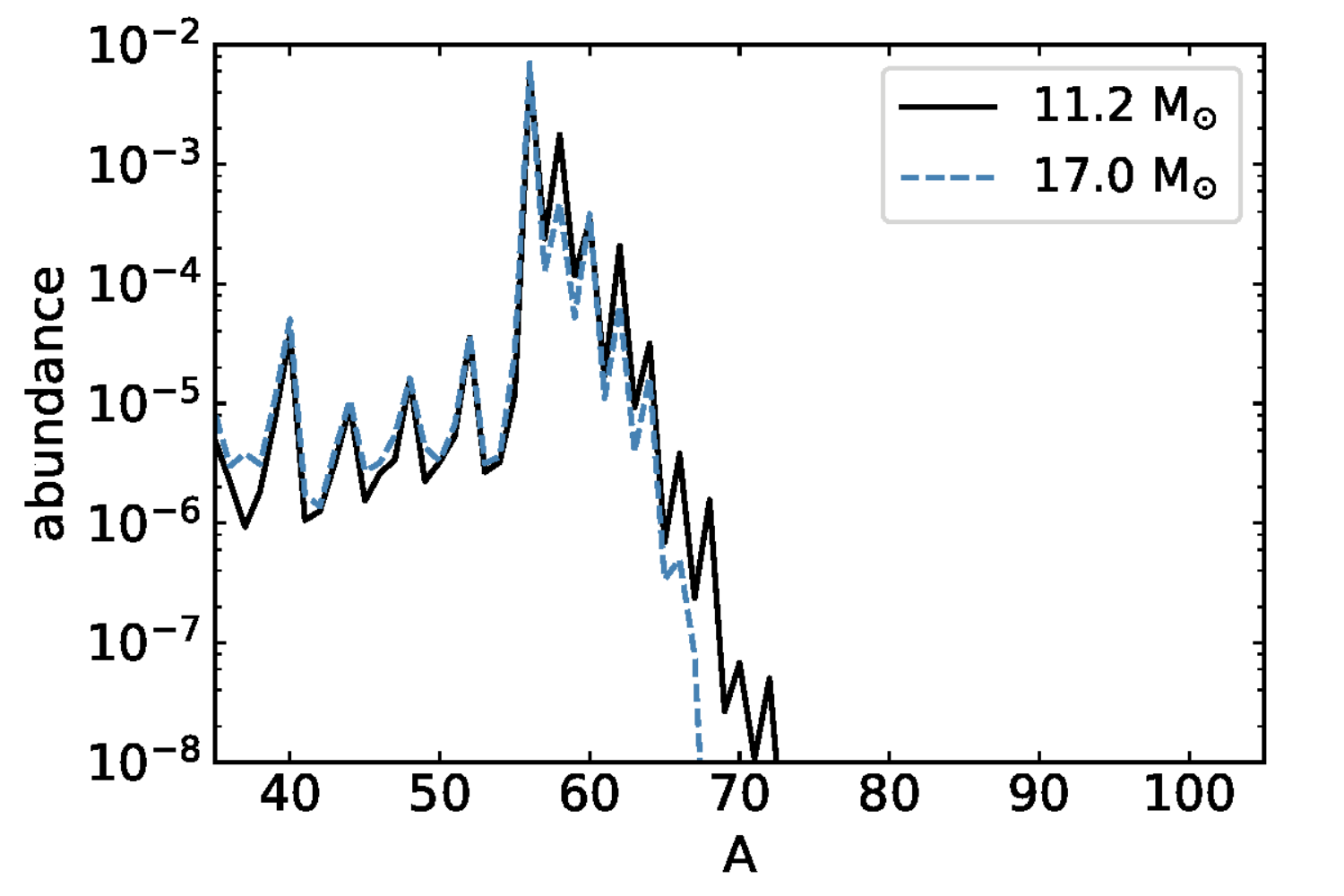} \\
\includegraphics[width=0.45\textwidth]{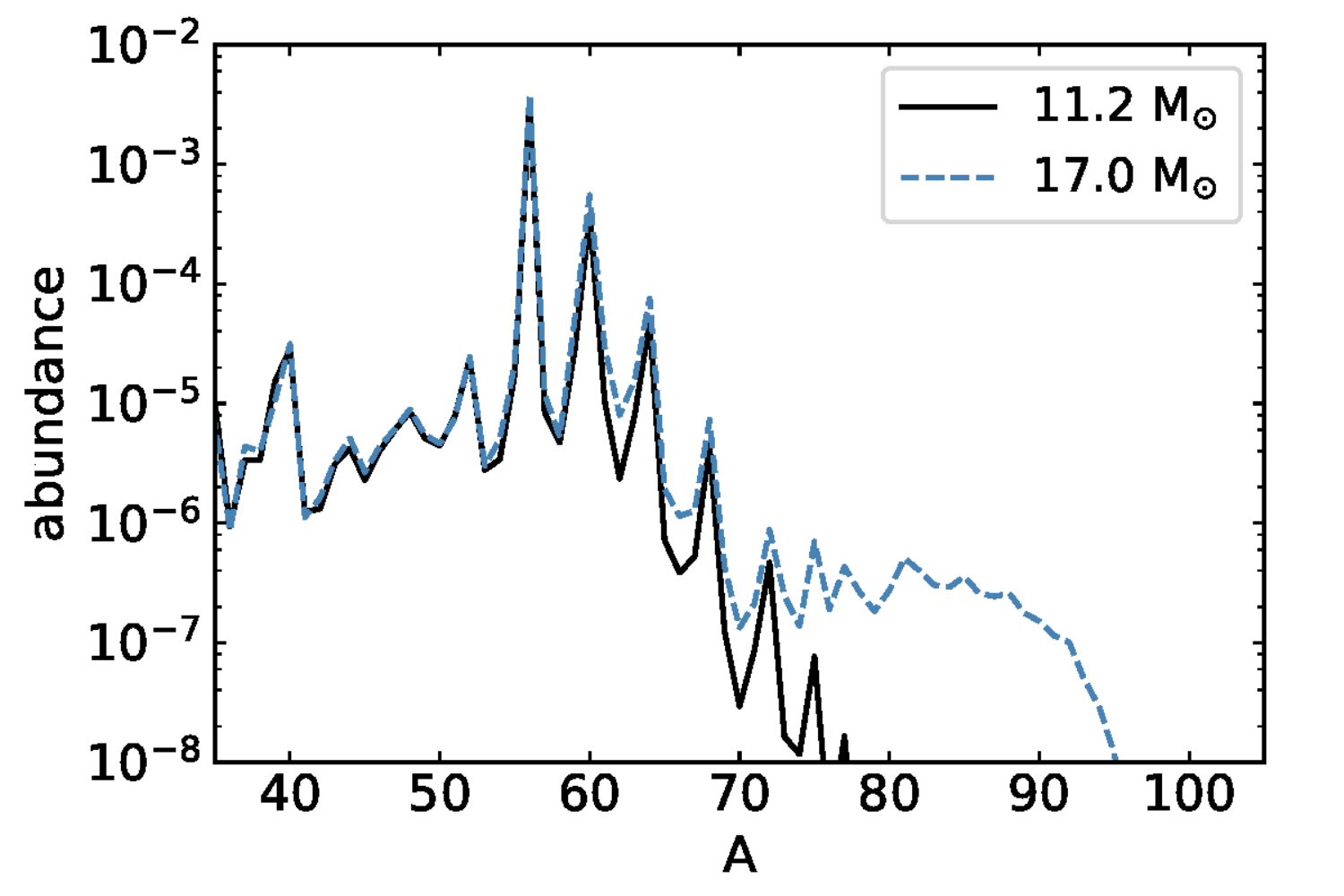}
\includegraphics[width=0.45\textwidth]{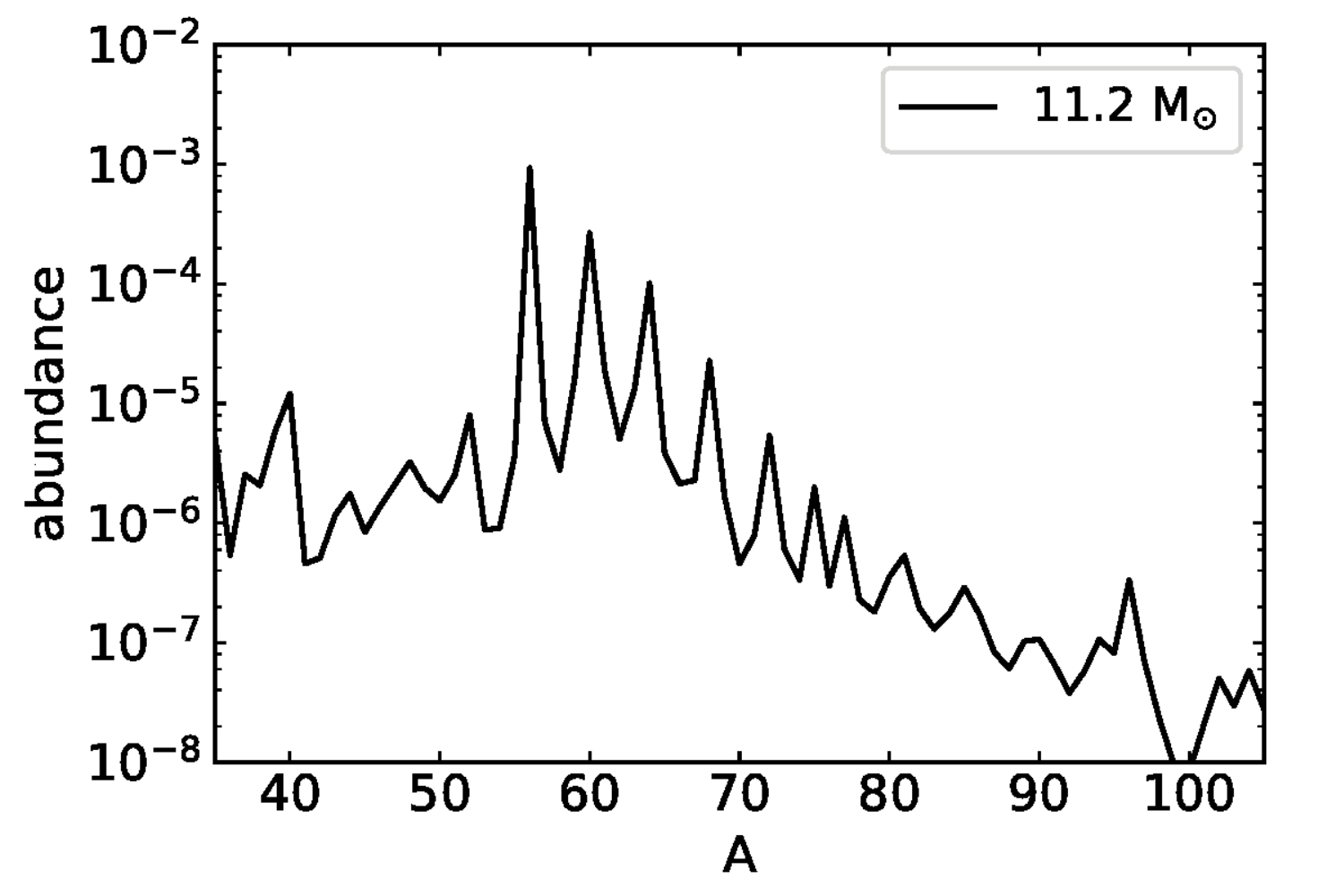}
\caption{Same as figure~\ref{fig:yebins_abunds}, but with neutrino luminosities and temperatures as seen from the north pole for the 11.2~M$_{\odot}$ model.}
\end{figure}


\begin{figure}
\centering
\includegraphics[width=0.8\textwidth]{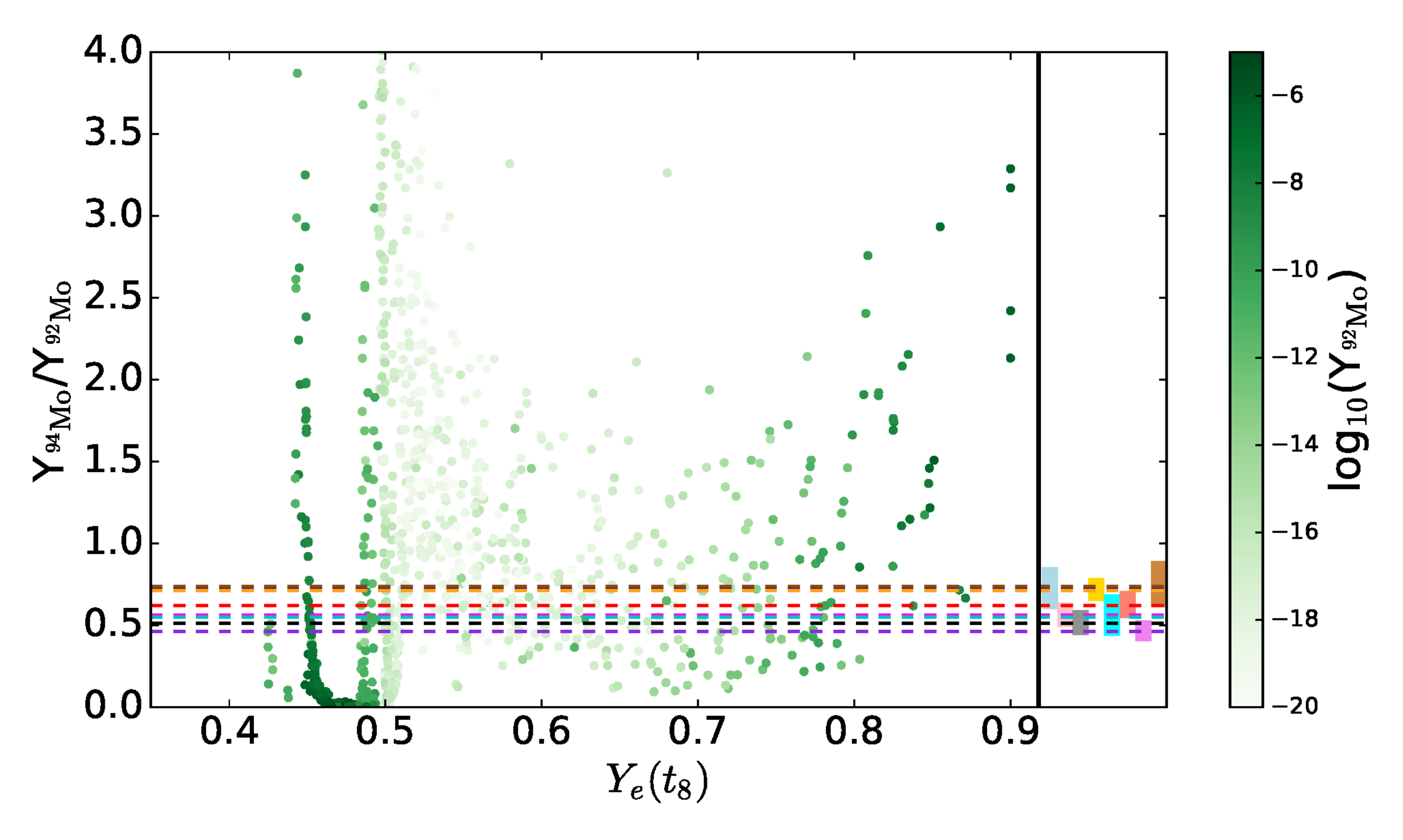} \\
\includegraphics[width=0.8\textwidth]{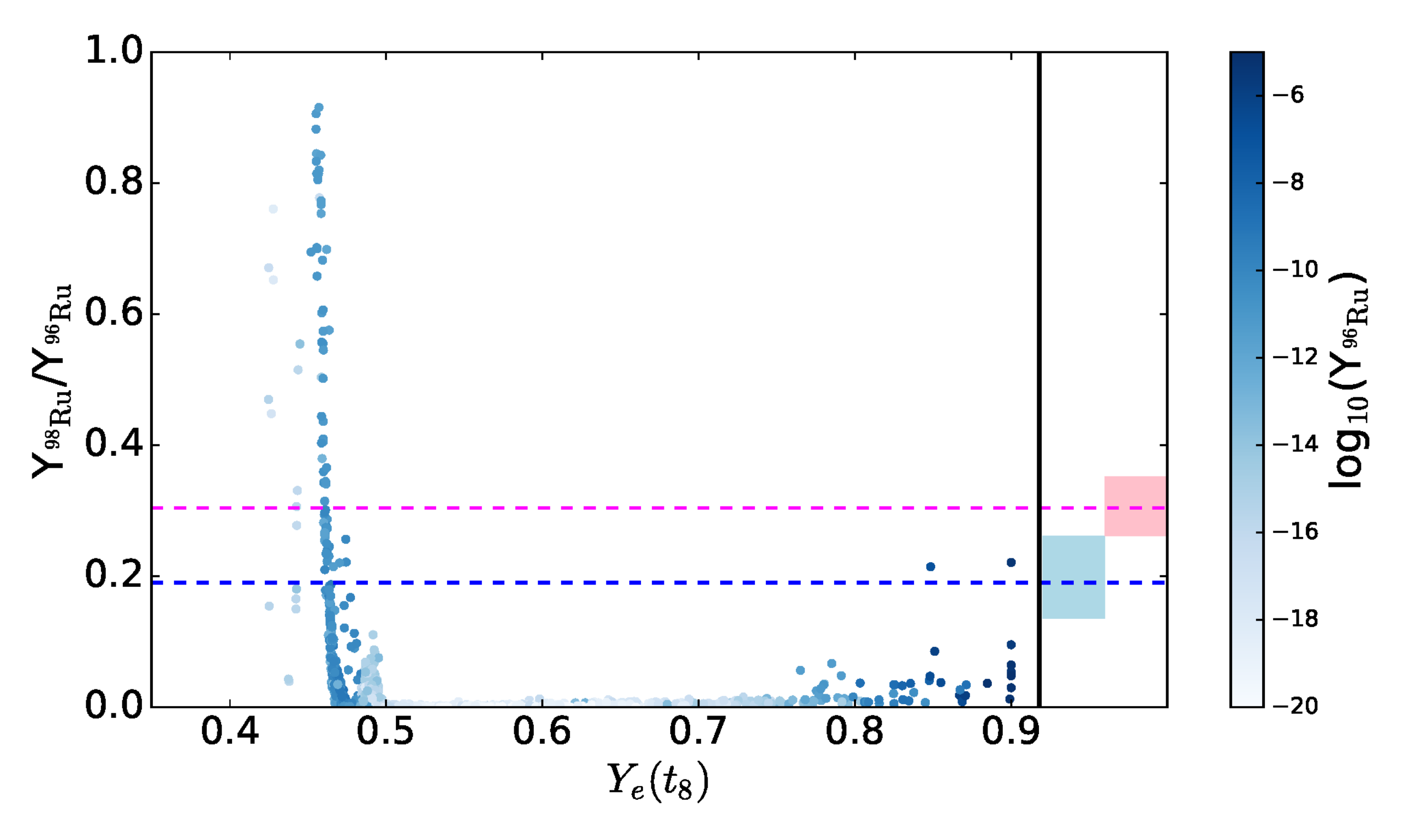}
\caption{Same as figure~\ref{fig:moru_ratio}, but with neutrino luminosities and temperatures as seen from the north pole.}
\label{fig:C4}
\end{figure}

\end{document}